\documentclass[12pt,twoside]{article}

\usepackage{amsmath}
\usepackage{graphicx}
\usepackage{epsfig}
\usepackage{rotating}
\usepackage{pstcol}
\newcommand{\mysection}[1]{\section[#1]{\boldmath #1}}
\newcommand{\mysubsection}[1]{\subsection[#1]{\boldmath #1}}
\newcommand{\mysubsubsection}[1]{\subsubsection[#1]{\boldmath #1}}
\newcommand{\mysubsubsubsection}[1]{\subsubsubsection{\boldmath #1}}

\newcommand{\lesssim}{\ensuremath{\raise-.5ex\hbox{$\buildrel<\over\sim$}\,}} 

\def\dof{{\rm dof}}

\newcommand\VCKM{{V}}
\newcommand\etacpf{{\eta_f}}
\newcommand\etacp{{\eta}}

\renewcommand\Im{{\rm Im}}

\newcommand\Abar{\kern 0.18em\overline{\kern -0.18em A}{}}
\newcommand\Af{A_f}
\newcommand\Abarf{\Abar_f}
\newcommand\Afbar{A_{\bar f}}
\newcommand\Abarfbar{\Abar_{\bar f}}
\newcommand\Acp{{\cal A}}
\newcommand\Adirnoncp{\ensuremath{\langle{\cal A}_{f\bar f}\rangle}\xspace}
\newcommand\mc{\multicolumn}
\newcommand\ph{\phantom}
\newcommand {\cbf}{\ensuremath{{\cal B}}}
\newcommand {\qq}{\ensuremath{q^2}}
\newcommand {\mx}{\ensuremath{m_X}}
\newcommand {\mxqq}{\ensuremath{(m_X, \qq)}}

\newcommand {\vcb}{\ensuremath{|V_{cb}|}}
\newcommand {\vub}{\ensuremath{|V_{ub}|}}

\newcommand {\Bxclnu}{\ensuremath{\Bb\to X_c\ell\nub}}

\newcommand {\breco}{\ensuremath{B_{reco}}}

\def\Bp      {\ensuremath{B^{+}}}
\def\Bm      {\ensuremath{B^{-}}}
\def\Bz      {\ensuremath{B^{0}}}
\def\Bs      {\ensuremath{B_{s}}}

\newcommand{\BzbDplnu}    {\ensuremath{\bar{B}^{0}\to D^{+}\ell^{-}\nub}}
\newcommand{\BzbDstarlnu} {\ensuremath{\bar{B}^{0}\to D^{*+}\ell^{-}\nub}}

\newcommand {\Bxulnu}{\ensuremath{B \to X_u \ell \bar{\nu}}\hbox{ }}

\usepackage{relsize}
\def\beq{\begin{equation}}
\def\eeq#1{\label{#1}\end{equation}}
\def\eeqn{\end{equation}}

\def\beqa{\begin{eqnarray}}
\def\eeqa#1{\label{#1}\end{eqnarray}}
\def\eeqan{\end{eqnarray}}

\let\bar=\overbar

\def\etal{{\it et al.}}
\def\ie{{\it i.e.}}
\def\eg{{\it e.g.}}
\def\etc{{\it etc.}}
\def\cf{{\it cf.}}

\def\O{{\cal O}}

\def\Dslash{\ensuremath{\not{\hbox{\kern-4pt $D$}}}\xspace}
\def\dslash{\not{\hbox{\kern-2pt $\del$}}}

\def\BR{\mbox{\rm BR}}
\def\ee{e^+e^-}

\def\msb{{\bar{\ssstyle M \kern -1pt S}}}

\RequirePackage{xspace}

\usepackage{relsize}
\def\babar{\mbox{\slshape B\kern-0.1em{\smaller A}\kern-0.1em
    B\kern-0.1em{\smaller A\kern-0.2em R}}\xspace}
\def\belle{\mbox{\normalfont Belle}\xspace}
\newcommand{\dzero}{D\O\xspace}

\def\ee         {\ensuremath{e^-e^-}\xspace}

\def\nub        {\ensuremath{\overline{\nu}}\xspace}

\def\nub        {\ensuremath{\overline{\nu}}\xspace}

\def\ubar  {\ensuremath{\overline u}\xspace}

\def\sbar  {\ensuremath{\overline s}\xspace}
\def\ssbar {\ensuremath{s\overline s}\xspace}

\def\b  {\ensuremath{b}\xspace}

\def\piz   {\ensuremath{\pi^0}\xspace}

\def\pip   {\ensuremath{\pi^+}\xspace}
\def\pim   {\ensuremath{\pi^-}\xspace}

\def\etapr {\ensuremath{\eta^{\prime}}\xspace}

\def\Kbar  {\kern 0.2em\overline{\kern -0.2em K}{}\xspace}

\def\Kmp   {\ensuremath{K^\mp}\xspace}
\def\Kp    {\ensuremath{K^+}\xspace}
\def\Km    {\ensuremath{K^-}\xspace}
\def\KS    {\ensuremath{K^0_{\scriptscriptstyle S}}\xspace} 
\def\KL    {\ensuremath{K^0_{\scriptscriptstyle L}}\xspace}

\def\Kstar   {\ensuremath{K^*}\xspace}

\def\Kstarmp   {\ensuremath{K^{*\mp}}\xspace}
\def\Kz   {\ensuremath{K^0}\xspace}
\def\Kzb   {\ensuremath{\Kbar^0}\xspace}
\def\KzKzb {\ensuremath{K^0 \kern -0.16em \Kzb}\xspace}

\def\Dz    {\ensuremath{D^0}\xspace}
\def\Dbar  {\kern 0.2em\overline{\kern -0.2em D}{}\xspace}

\def\Dzb   {\ensuremath{\Dbar^0}\xspace}
\def\DzDzb {\ensuremath{D^0 {\kern -0.16em \Dzb}}\xspace}
\def\Dp    {\ensuremath{D^+}\xspace}

\def\Dstar   {\ensuremath{D^*}\xspace}

\def\Dstarz  {\ensuremath{D^{*0}}\xspace}

\def\Dstarp  {\ensuremath{D^{*+}}}

\def\DorDstar   {\ensuremath{D^{(*)}}\xspace}
\def\DorDstarz  {\ensuremath{D^{(*)0}}\xspace}
\def\DorDstarzb {\ensuremath{\Dbar^{(*)0}}\xspace}

\def\Bz    {\ensuremath{B^0}\xspace}
\def\B     {\ensuremath{B}\xspace}
\def\Bbar  {\kern 0.18em\overline{\kern -0.18em B}{}\xspace}
\def\Bb    {\ensuremath{\Bbar}\xspace}
\def\Bzb   {\ensuremath{\Bbar^0}\xspace}
\def\Bu    {\ensuremath{B^+}\xspace}

\def\Bmp   {\ensuremath{B^\mp}\xspace}
\def\Bs    {\ensuremath{B_s}\xspace}
\def\Bsb   {\ensuremath{\Bbar_s}\xspace}
\def\BB    {\ensuremath{B\Bbar}\xspace} 
\def\BzBzb {\ensuremath{B^0 {\kern -0.16em \Bzb}}\xspace}

\def\jpsi  {\ensuremath{{J\mskip -3mu/\mskip -2mu\psi\mskip 2mu}}\xspace}

\mathchardef\Upsilon="7107
\def\Y#1S{\ensuremath{\Upsilon{(#1S)}}\xspace}% no space before {...}!

\def\FourS {\Y4S}

\mathchardef\Deltares="7101
\mathchardef\Xi="7104
\mathchardef\Lambda="7103
\mathchardef\Sigma="7106
\mathchardef\Omega="710A
\def\Deltabar   {\kern 0.25em\overline{\kern -0.25em \Deltares}{}\xspace}
\def\Lbar {\kern 0.2em\overline{\kern -0.2em\Lambda\kern 0.05em}\kern-0.05em{}\xspace}
\def\Sigbar{\kern 0.2em\overline{\kern -0.2em \Sigma}{}\xspace}
\def\Xibar{\kern 0.2em\overline{\kern -0.2em \Xi}{}\xspace}
\def\Obar{\kern 0.2em\overline{\kern -0.2em \Omega}{}\xspace}
\def\Nbar{\kern 0.2em\overline{\kern -0.2em N}{}\xspace}
\def\Xb{\kern 0.2em\overline{\kern -0.2em X}{}}

\def\BR{{\ensuremath{\cal B}}}

\newcommand{\tev}{\ensuremath{\mathrm{Te\kern -0.1em V}}\xspace}
\newcommand{\gev}{\ensuremath{\mathrm{Ge\kern -0.1em V}}\xspace}
\newcommand{\mev}{\ensuremath{\mathrm{Me\kern -0.1em V}}\xspace}
\newcommand{\kev}{\ensuremath{\mathrm{ke\kern -0.1em V}}\xspace}
\newcommand{\ev}{\ensuremath{\mathrm{e\kern -0.1em V}}\xspace}
\newcommand{\gevc}{\ensuremath{{\mathrm{Ge\kern -0.1em V\!/}c}}\xspace}
\newcommand{\mevc}{\ensuremath{{\mathrm{Me\kern -0.1em V\!/}c}}\xspace}
\newcommand{\gevcc}{\ensuremath{{\mathrm{Ge\kern -0.1em V\!/}c^2}}\xspace}
\newcommand{\mevcc}{\ensuremath{{\mathrm{Me\kern -0.1em V\!/}c^2}}\xspace}
\def\mus  {\ensuremath{\rm \,\mus}\xspace}

\def\ps   {\ensuremath{\rm \,ps}\xspace}

\def\mus        {\ensuremath{\,\mu{\rm s}}\xspace}    %% microsecond
\def\ps         {\ensuremath{{\rm \,ps}}\xspace}  %% picosecond
\def\gsim{{~\raise.15em\hbox{$>$}\kern-.85em
          \lower.35em\hbox{$\sim$}~}\xspace}
\def\lsim{{~\raise.15em\hbox{$<$}\kern-.85em
          \lower.35em\hbox{$\sim$}~}\xspace}

\def\CP                 {\ensuremath{C\!P}\xspace}
\def\CPT                {\ensuremath{C\!PT}\xspace}
\def\pep2{PEP-II}

\def\rhobar {\ensuremath{\overline{\rho}}\xspace}
\def\etabar {\ensuremath{\overline{\eta}}\xspace}
\def\stwob{\ensuremath{\sin\! 2 \beta   }\xspace}

\def\deltamd{\ensuremath{{\rm \Delta}m_d}\xspace}
\xspace

\def\jetset74   {\mbox{\tt Jetset \hspace{-0.5em}7.\hspace{-0.2em}4}}
\newcommand{\aerr}[4]   {\mbox{${{#1}^{+ #2}_{- #3}\pm #4}$}}
\newcommand{\berr}[4]   {\mbox{${{#1}\pm #2^{+ #3}_{- #4}}$}}
\newcommand{\cerr}[3]   {\mbox{${{#1}^{+ #2}_{- #3}}$}}
\newcommand{\aerrsy}[5] {\mbox{${{#1}^{+ #2 + #4}_{- #3 - #5}}$}}

\newcommand{\berrsyt}[6] {\mbox{${{#1}\pm #2^{+ #3 + #5}_{- #4 - #6}}$}}
\newcommand{\err}[3]   {\mbox{${{#1}\pm{#2}\pm{#3}}$}}
\newcommand{\nodata}{$$}
\def\etapr{{\eta^{\prime}}}
\newcommand{\ppbar}             {\mbox{${p\bar p}$}}
\newcommand{\pL}                {\mbox{${p \bar\Lambda}$}}
\newcommand{\LL}                {\mbox{${\Lambda\bar\Lambda}$}}

\def\sgline{\noalign{\vskip 0.10truecm\hrule\vskip 0.10truecm}}
\def\sglinespt{\noalign{\vskip 0.05truecm\hrule}}
\def\sglinespb{\noalign{\hrule\vskip 0.05truecm}}

\newcommand{\kz}    {\mbox{$K^0$}}

\begin{document}

\setcounter{page}{1}

\title{\begin{flushleft}
% \mbox{\normalsize Version 08}
\end{flushleft}
\vskip 20pt
Averages of $b$-hadron Properties \\
as of Summer 2004 }
\author{Heavy Flavor Averaging Group (HFAG)\footnote{
 The members of the HFAG and those involved in the subgroup activities are:
  J.~Alexander, M.~Artuso, M.~Battaglia, E.~Barberio, T.E.~Browder,
  P.~Chang, L.~Di~Ciaccio, H.~Evans, T.~Gershon, L.~Gibbons, A.~H\"ocker, 
  T.~Iijima, D.~Kirkby,   U.~Langenegger, A.~Limosani, O.~Long, D.~Lucchesi,
  V.~Luth, T.~Nozaki,  %removed by his request  P.~Roudeau,
  Y.~Sakai, O.~Schneider, C.~Schwanda,
  M.~Shapiro, J.~Smith, A.~Stocchi, R.~Van~Kooten, C.~Weiser, 
  and W.~Yao
  }
 }
\maketitle
\thispagestyle{empty}
\begin{abstract}
This article  reports world averages for measurements on
$b$-hadron properties obtained by the
Heavy Flavor Averaging Group (HFAG) using the available results as of
summer 2004 conferences. 
In the averaging, the input parameters used in the various analyses are 
adjusted (rescaled) to common values, and all known correlations are 
taken into account.  The averages include  $b$-hadron lifetimes,
$B$-oscillation (mixing) parameters, semileptonic decay parameters, 
rare decay branching fractions, and \CP violation measurements.
\end{abstract}

\newpage
\tableofcontents
\newpage

%\section{Introduction }
%\documentclass[12pt]{article}
%\begin{document}

\mysection{Introduction}
\label{sec:intro}

 The flavor dynamics is one of the important elements in understanding
the nature of particle physics.  The accurate knowledge of properties of
heavy flavor hadrons, especially $b$ hadrons, play an essential role for
determination of the Cabbibo-Kobayashi-Maskawa (CKM) matrix~\cite{ref_ckm}.
Since asymmetric $B$ factories started their operation, available amounts
of $B$ meson samples has been dramatically increased and the accuracies
of measurements have been improved.
 
 The Heavy Flavor Averaging Group (HFAG) has been formed, continuing the
activities of LEP Heavy Flavor Steering group~\cite{LEPHFS}, 
to provide the
averages for measurements dedicated to the $b$-flavor related quantities.
The HFAG consists of representatives and contacts from the experimental
groups: \babar, \belle, CDF, CLEO, \dzero, and LEP. %, and SLD. 

 The HFAG is currently organized into four subgroups.
\begin{itemize}
\item the ``Lifetime and mixing'' group provides
averages for $b$-hadron lifetimes, $b$-hadron fractions in $\Upsilon(4S)$ decay
and high energy collisions, and various parameters in $B^0$ and $B_s^0$
oscillation (mixing).

\item the ``Semileptonic $B$ decays'' group provides averages
for inclusive and exclusive $B$-decay branching fractions, and best values
of the CKM matrix elements $|V_{cb}|$ and $|V_{ub}|$. 

\item the ``$\CP(t)$ and Unitarity Triangle angles'' group provides averages for
time-dependent $\CP$ asymmetry parameters and angles of the unitarity
triangles.

\item the ``Rare decays'' group provides averages of branching fractions and
their asymmetries between $B$ and $\bar B$ for charmless mesonic,
radiative, leptonic, and baryonic $B$ decays.
\end{itemize}

The first two subgroups continue the activities from LEP working
groups with some reorganization (merging four groups into two groups).
The latter two groups are newly formed to take care of new results 
which are available from asymmetric $B$ factory experiments.

In this article, we report the world averages using the available results
as of summer 2004 conferences (ICHEP04 and FPCP04).  
All results that are publicly available, including 
recent preliminary results, are used in averages.  
We do not use preliminary results which remain unpublished for a long time
or for which no publication is planned. 
Close contacts have been established between representatives from
the experiments and members of different subgroups in charge of the
averages, to ensure that the data are prepared in a form suitable
for combinations.  

We do not scale % apply the rescaling factor on 
the error of an average 
% that is presently taken by the Particle Data Group (PDG)
(as is presently done by the Particle Data Group~\cite{Eidelman:2004wy})
in case $\chi^2/\dof > 1$, where $\dof$ is the number of 
degrees of freedom in the average calculation.
In this case, we examine the systematics of each measurement and 
try to understand them.  
Unless we find possible systematic discrepancies between the measurements, 
we do not make any special treatment for the calculated error.  
We provide the confidence level of the fit so that
one can know the consistency of the measurements included in the average.
We attach a warning message in case that some special treatment is done
or the approximation used in the average calculation may not be good enough
(\eg, Gaussian error is used in averaging though the likelihood 
indicates non-Gaussian behavior).

Section~\ref{sec:method} describes the methodology for
averaging various quantities in the HFAG.  
In the averaging, the input parameters used in the various analyses are 
adjusted (rescaled) to common values, and, where possible, known 
correlations are taken into account. 
The general philosophy and tools for calculations of averages are presented.

%Section~\ref{sec:com_inputs}  presents the values  of the common 
%input parameters that 
%contribute to the systematic uncertainties given in this article.

Sections~\ref{sec:life_mix}--\ref{sec:rare} describe the averaging of 
the quantities from each subgroup mentioned above.

A summary of the averages described in this article is given in
Sec.~\ref{sec:summary}.   
% In addition, Appendices give details of particular techniques etc.

 The complete listing of averages and plots described in this article
are also available on the HFAG Web page:
 
 {\tt http://www.slac.stanford.edu/xorg/hfag } and 

 {\tt http://belle.kek.jp/mirror/hfag } (KEK mirror site).

\section{Methodology } \label{sec:method} 
The general averaging problem that HFAG faces is to combine the
information provided by different measurements of the same parameter,
to obtain our best estimate of the parameter's value and
uncertainty. The methodology described here focuses on the problems of
combining measurements performed with different systematic assumptions
and with potentially-correlated systematic uncertainties. Our methodology
relies on the close involvement of the people performing the
measurements in the averaging process.

Consider two hypothetical measurements of a parameter $x$, which might
be summarized as
\begin{align*}
x &= x_1 \pm \delta x_1 \pm \Delta x_{1,1} \pm \Delta x_{2,1} \ldots \\
x &= x_2 \pm \delta x_2 \pm \Delta x_{1,2} \pm \Delta x_{2,2} \ldots
\; ,
\end{align*}
where the $\delta x_k$ are statistical uncertainties, and
the $\Delta x_{i,k}$ are contributions to the systematic
uncertainty. One popular approach is to combine statistical and
systematic uncertainties in quadrature
\begin{align*}
x &= x_1 \pm \left(\delta x_1 \oplus \Delta x_{1,1} \oplus \Delta
x_{2,1} \oplus \ldots\right) \\
x &= x_2 \pm \left(\delta x_2 \oplus \Delta x_{1,2} \oplus \Delta
x_{2,2} \oplus \ldots\right)
\end{align*}
and then perform a weighted average of $x_1$ and $x_2$, using their
combined uncertainties, as if they were independent. This approach
suffers from two potential problems that we attempt to address. First,
the values of the $x_k$ may have been obtained using different
systematic assumptions. For example, different values of the \Bz
lifetime may have been assumed in separate measurements of the
oscillation frequency $\deltamd$. The second potential problem is that
some contributions of the systematic uncertainty may be correlated
between experiments. For example, separate measurements of $\deltamd$
may both depend on an assumed Monte-Carlo branching fraction used to
model a common background.

The problems mentioned above are related since, ideally, any quantity $y_i$
that $x_k$ depends on has a corresponding contribution $\Delta x_{i,k}$ to the
systematic error which reflects the uncertainty $\Delta y_i$ on $y_i$
itself. We assume that this is the case, and use the values of $y_i$ and
$\Delta y_i$ assumed by each measurement explicitly in our
averaging (we refer to these values as $y_{i,k}$ and $\Delta y_{i,k}$
below). Furthermore, since we do not lump all the systematics
together,
we require that each measurement used in an average have a consistent
definition of the various contributions to the systematic uncertainty.
Different analyses often use different decompositions of their systematic
uncertainties, so achieving consistent definitions for any potentially
correlated contributions requires close coordination between HFAG and
the experiments. In some cases, a group of
systematic uncertainties must be lumped to obtain a coarser
description that is consistent between measurements. Systematic uncertainties
that are uncorrelated with any other sources of uncertainty appearing
in an average are lumped with the statistical error, so that the only
systematic uncertainties treated explicitly are those that are
correlated with at least one other measurement via a consistently-defined
external parameter $y_i$. When asymmetric statistical or systematic
uncertainties are quoted, we symmetrize them since our combination
method implicitly assumes parabolic likelihoods for each measurement.

\begin{figure}
\begin{center}
\includegraphics[width=6.0in]{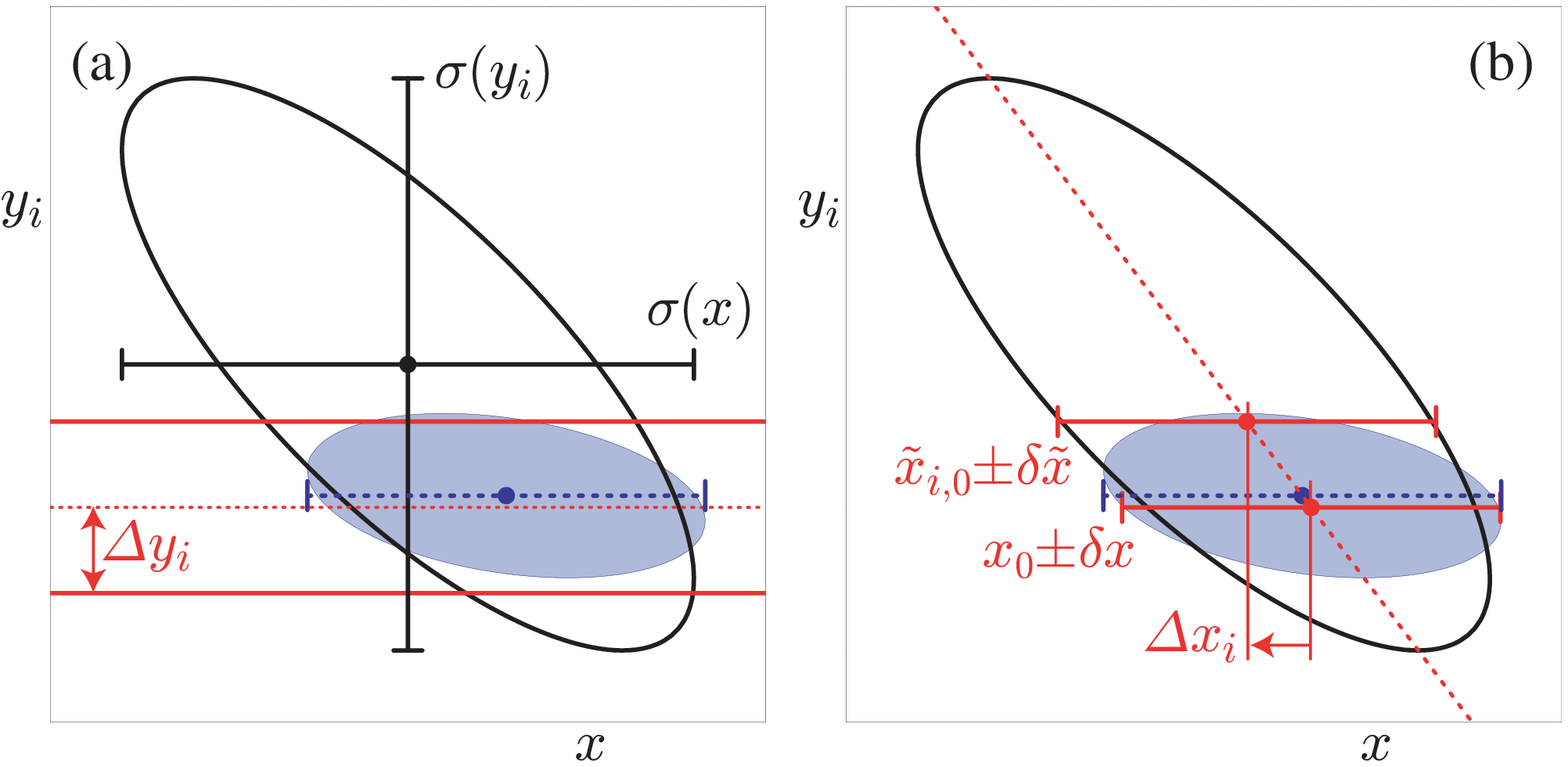}
\end{center}
\caption{The left-hand plot, (a), compares the 68\% confidence-level
  contours of a
  hypothetical measurement's unconstrained (large ellipse) and
  constrained (filled ellipse) likelihoods, using the Gaussian
  constraint on $y_i$ represented by the horizontal band. The solid
  error bars represent the statistical uncertainties, $\sigma(x)$ and
  $\sigma(y_i)$, of the unconstrained likelihood. The dashed
  error bar shows the statistical error on $x$ from a
  constrained simultaneous fit to $x$ and $y_i$. The right-hand plot,
  (b), illustrates the method described in the text of performing fits
  to $x$ only with $y_i$ fixed at different values. The dashed
  diagonal line between these fit results has the slope
  $\rho(x,y_i)\sigma(y_i)/\sigma(x)$ in the limit of a parabolic
  unconstrained likelihood. The result of the constrained simultaneous
  fit from (a) is shown as a dashed error bar on $x$.}
\label{fig:singlefit}
\end{figure}

The fact that a measurement of $x$ is sensitive to the value of $y_i$
indicates that, in principle, the data used to measure $x$ could
equally-well be used for a simultaneous measurement of $x$ and $y_i$, as
illustrated by the large contour in Fig.~\ref{fig:singlefit}(a) for a hypothetical
measurement. However, we often have an external constraint $\Delta
y_i$ on the value of $y_i$ (represented by the horizontal band in
Fig.~\ref{fig:singlefit}(a)) that is more precise than the constraint
$\sigma(y_i)$ from
our data alone. Ideally, in such cases we would perform a simultaneous
fit to $x$ and $y_i$, including the external constraint, obtaining the
filled $(x,y)$ contour and corresponding dashed one-dimensional estimate of
$x$ shown in Fig.~\ref{fig:singlefit}(a). Throughout, we assume that
the external constraint $\Delta y_i$ on $y_i$ is Gaussian.

In practice, the added technical complexity of a constrained fit with
extra free parameters is not justified by the small increase in
sensitivity, as long as the external constraints $\Delta y_i$ are
sufficiently precise when compared with the sensitivities $\sigma(y_i)$
to each $y_i$ of the data alone. Instead, the usual procedure adopted
by the experiments is to perform a baseline fit with all $y_i$ fixed
to nominal values $y_{i,0}$, obtaining $x = x_0 \pm \delta
x$. This baseline fit neglects the uncertainty due to $\Delta y_i$, but
this error can be mostly recovered by repeating the fit separately for
each external parameter $y_i$ with its value fixed at $y_i = y_{i,0} +
\Delta y_i$ to obtain $x = \tilde{x}_{i,0} \pm \delta\tilde{x}$, as
illustrated in Fig.~\ref{fig:singlefit}(b). The absolute shift,
$|\tilde{x}_{i,0} - x_0|$, in the central value of $x$ is what the
experiments usually quote as their systematic uncertainty $\Delta x_i$
on $x$ due to the unknown value of $y_i$. Our procedure requires that
we know not only the magnitude of this shift but also its sign. In the
limit that the unconstrained data is represented by a parabolic
likelihood, the signed shift is given by
\begin{equation}
\Delta x_i = \rho(x,y_i)\frac{\sigma(x)}{\sigma(y_i)}\,\Delta y_i \;,
\end{equation}
where $\sigma(x)$ and $\rho(x,y_i)$ are the statistical uncertainty on
$x$ and the correlation between $x$ and
$y_i$ in the unconstrained data.
While our procedure is not
equivalent to the constrained fit with extra parameters, it yields (in
the limit of a parabolic unconstrained likelihood) a central value
$x_0$ that agrees 
to ${\cal O}(\Delta y_i/\sigma(y_i))^2$ and an uncertainty $\delta x
\oplus \Delta x_i$ that agrees to ${\cal O}(\Delta y_i/\sigma(y_i))^4$.

In order to combine two or more measurements that share systematics
due to the same external parameters $y_i$, we would ideally perform a
constrained simultaneous fit of all data samples to obtain values of
$x$ and each $y_i$, being careful to only apply the constraint on each
$y_i$ once. This is not practical since we generally do not have
sufficient information to reconstruct the unconstrained likelihoods
corresponding to each measurement. Instead, we perform the two-step
approximate procedure described below.

\begin{figure}
\begin{center}
\includegraphics[width=6.0in]{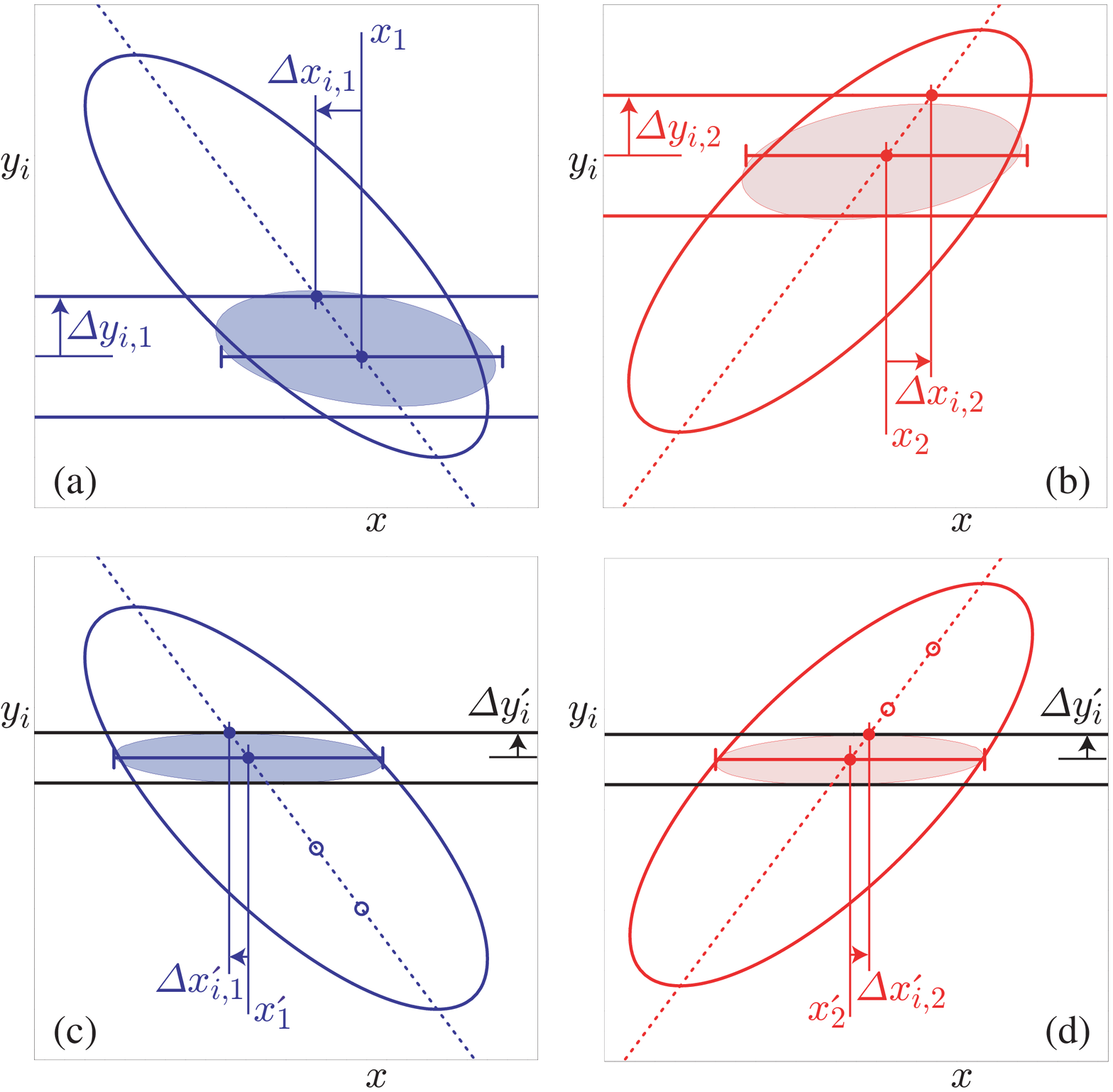}
\end{center}
\caption{The upper plots, (a) and (b), show examples of two individual
  measurements to be combined. The large ellipses represent their
  unconstrained likelihoods, and the filled ellipses represent their
  constrained likelihoods. Horizontal bands indicate the different
  assumptions about the value and uncertainty of $y_i$ used by each
  measurement. The error bars show the results of the approximate
  method described in the text for obtaining $x$ by performing fits
  with $y_i$ fixed to different values. The lower plots, (c) and (d),
  illustrate the adjustments to accommodate updated and consistent
  knowledge of $y_i$ described in the text. Hollow circles mark the
  central values of the unadjusted fits to $x$ with $y$ fixed, which
  determine the dashed line used to obtain the adjusted values. }
\label{fig:multifit}
\end{figure}

Figs.~\ref{fig:multifit}(a,b) illustrate two
statistically-independent measurements, $x_1 \pm (\delta x_1 \oplus
\Delta x_{i,1})$ and $x_2\pm(\delta x_i\oplus \Delta x_{i,2})$, of the same
hypothetical quantity $x$ (for simplicity, we only show the
contribution of a single correlated systematic due to an external
parameter $y_i$). As our knowledge of the external parameters $y_i$
evolves, it is natural that the different measurements of $x$ will
assume different nominal values and ranges for each $y_i$. The first
step of our procedure is to adjust the values of each measurement to
reflect the current best knowledge of the values $y_i'$ and ranges
$\Delta y_i'$ of the external parameters $y_i$, as illustrated in
Figs.~\ref{fig:multifit}(c,b). We adjust the
central values $x_k$ and correlated systematic uncertainties $\Delta
x_{i,k}$ linearly for each measurement (indexed by $k$) and each
external parameter (indexed by $i$):
\begin{align}
x_k' &= x_k + \sum_i\,\frac{\Delta x_{i,k}}{\Delta y_{i,k}}
\left(y_i'-y_{i,k}\right)\\
\Delta x_{i,k}'&= \Delta x_{i,k}\cdot \frac{\Delta y_i'}{\Delta
  y_{i,k}} \; .
\end{align}
This procedure is exact in the limit that the unconstrained
likelihoods of each measurement is parabolic.

\begin{figure}
\begin{center}
\includegraphics[width=3.5in]{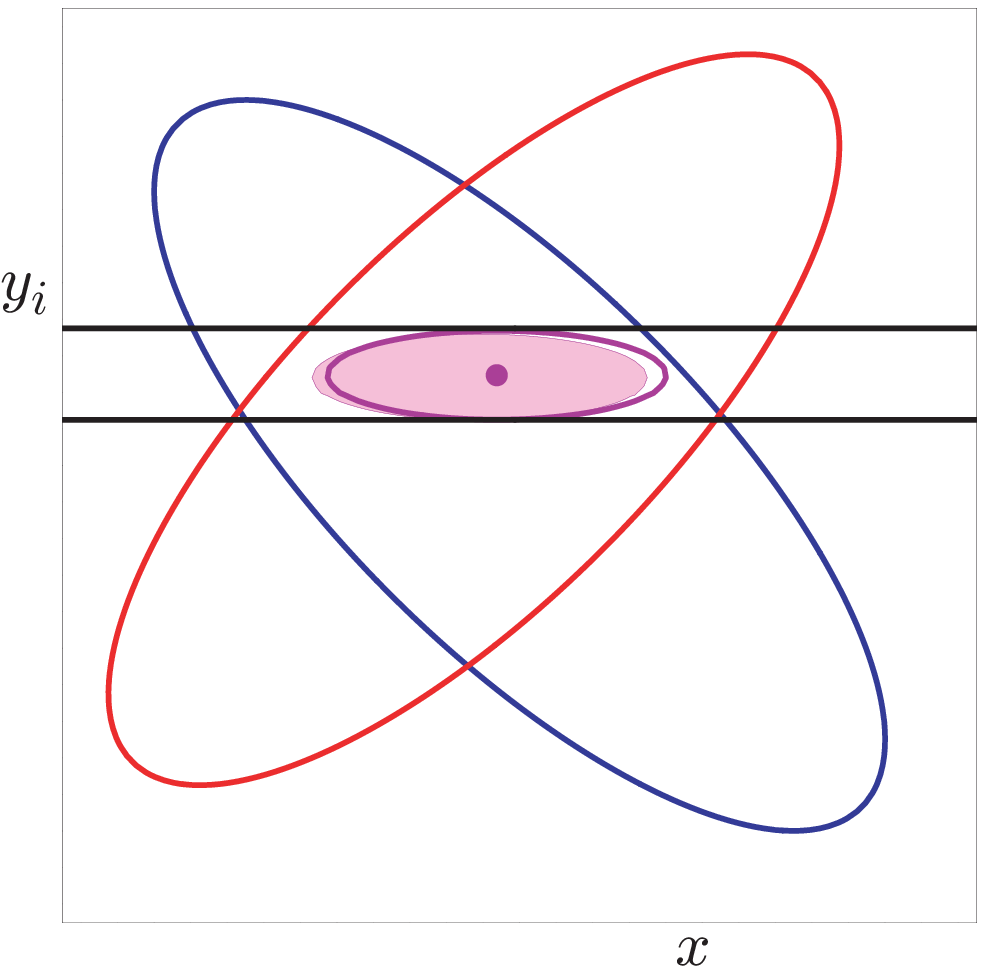}
\end{center}
\caption{An illustration of the combination of two hypothetical
  measurements of $x$ using the method described in the text. The
  ellipses represent the unconstrained likelihoods of each measurement
  and the horizontal band represents the latest knowledge about $y_i$ that
  is used to adjust the individual measurements. The filled small
  ellipse shows the result of the exact method using ${\cal
  L}_{\text{comb}}$ and the hollow small ellipse and dot show the
  result of the approximate method using $\chi^2_{\text{comb}}$.}
\label{fig:fit12}
\end{figure}

The second step of our procedure is to combine the adjusted
measurements, $x_k'\pm (\delta x_k\oplus \Delta x_{k,1}'\oplus \Delta
x_{k,2}'\oplus\ldots)$ using the chi-square 
\begin{equation}
\chi^2_{\text{comb}}(x,y_1,y_2,\ldots) \equiv \sum_k\,
\frac{1}{\delta x_k^2}\left[
x_k' - \left(x + \sum_i\,(y_i-y_i')\frac{\Delta x_{i,k}'}{\Delta y_i'}\right)
\right]^2 + \sum_i\,
\left(\frac{y_i - y_i'}{\Delta y_i'}\right)^2 \; ,
\end{equation}
and then minimize this $\chi^2$ to obtain the best values of $x$ and
$y_i$ and their uncertainties, as illustrated in
Fig.~\ref{fig:fit12}. Although this method determines new values for
the $y_i$, we do not report them since the $\Delta x_{i,k}$ reported
by each experiment are generally not intended for this purpose (for
example, they may represent a conservative upper limit rather than a
true reflection of a 68\% confidence level).

For comparison, the exact method we would
perform if we had the unconstrained likelihoods ${\cal L}_k(x,y_1,y_2,\ldots)$
available for each
measurement is to minimize the simultaneous constrained likelihood
\begin{equation}
{\cal L}_{\text{comb}}(x,y_1,y_2,\ldots) \equiv \prod_k\,{\cal
  L}_k(x,y_1,y_2,\ldots)\,\prod_{i}\,{\cal 
  L}_i(y_i) \; ,
\end{equation}
with an independent Gaussian external constraint on each $y_i$
\begin{equation}
{\cal L}_i(y_i) \equiv \exp\left[-\frac{1}{2}\,\left(\frac{y_i-y_i'}{\Delta
 y_i'}\right)^2\right] \; .
\end{equation}
The results of this exact method are illustrated by the filled ellipses
in Figs.~\ref{fig:fit12}(a,b), and agree with our method in the limit that
each ${\cal L}_k$ is parabolic and that each $\Delta
y_i' \ll \sigma(y_i)$. In the case of a non-parabolic unconstrained
likelihood, experiments would have to provide a description of ${\cal
  L}_k$ itself to allow an improved combination. In the case of some
$\sigma(y_i)\simeq \Delta y_i'$, experiments are advised to perform a
simultaneous measurement of both $x$ and $y$ so that their data will
improve the world knowledge about $y$. 

 The algorithm described above is used as a default in the averages
reported in the following sections.  For some cases, somewhat simplified
or more complex algorithms are used and noted in the corresponding 
sections. 

Following the prescription described above,
the central values and errors are rescaled
to a common set of input parameters in the averaging procedures, 
according to the dependency on
any of these input parameters.
We try to use the most up-to-date values for these common inputs and 
the same values among the HFAG subgroups.
For the parameters whose averages are produced by the HFAG, we use 
the updated values in the current update cycle.  For other external
parameters, we use the most recent PDG values. %\cite{Eidelman:2004wy}.

  The parameters and values used in this update cycle are listed in
each subgroup section.

%\section{Common inputs} \label{sec:com_inputs}
%% merged into Methodolgy section
% \input{hfag_com_inputs}

%\section{Averages of b-hadron lifetimes and B-oscillation parameters }
%%%%%%%%%%%%%%%%%%%%%%%%%%%%%%%%%%%%%%%%%%%%%%%%
%
% This is file life_mix.tex
% containing the chapter on b-hadron fractions,
% lifetimes and mixing parameters
%
% Olivier Schneider, EPFL, June 23, 2004
%
%%%%%%%%%%%%%%%%%%%%%%%%%%%%%%%%%%%%%%%%%%%%%%%
%

% -----------------------------
% LaTeX macros for this chapter
% -----------------------------
%
% There are some conflicts with Urs' definitions (to be resolved at  some point)

%%%%%%%%%%%%%%%%%%%%%%%%%%%%%%%%%%%%%%%%%%%%%%%
%
% This is file life_mix_defintions.tex
%
% Olivier Schneider, EPFL, Nov 6, 2004
%
%%%%%%%%%%%%%%%%%%%%%%%%%%%%%%%%%%%%%%%%%%%%%%%
%

% Experiments (-> defined in symbols.tex)
% \newcommand{\dzero}{D\O\xspace}
%
% \belle already defined
% \newcommand{\belle}{Belle\xspace}
% \babar already defined
% \def\babar{\mbox{\slshape B\kern-0.1em{\smaller A}\kern-0.1em B\kern-0.1em{\smaller A\kern-0.2em R}}\xspace}

\renewcommand{\topfraction}{0.9}

% References
\newcommand{\auth}[1]{#1,}
\newcommand{\coll}[1]{#1 Collaboration,}
\newcommand{\authcoll}[2]{#1 \etal\ (#2 Collaboration),}
\newcommand{\titl}[1]{``#1'',} 
\newcommand{\J}[4]{{#1} {\bf #2}, #3 (#4)}
\newcommand{\subJ}[1]{submitted to #1}
\newcommand{\PRL}[3]{\J{Phys.\ Rev.\ Lett.}{#1}{#2}{#3}}
\newcommand{\subPRL}{\subJ{Phys.\ Rev.\ Lett.}}
\newcommand{\PRD}[3]{\J{Phys.\ Rev.\ D}{#1}{#2}{#3}}
\newcommand{\subPRD}{\subJ{Phys.\ Rev.\ D}}
\newcommand{\ZPC}[3]{\J{Z.\ Phys.\ C}{#1}{#2}{#3}}
\newcommand{\PLB}[3]{\J{Phys.\ Lett.\ B}{#1}{#2}{#3}}
\newcommand{\subPLB}{\subJ{Phys.\ Lett.\ B}}
\newcommand{\EPJC}[3]{\J{Eur.\ Phys.\ J.\ C}{#1}{#2}{#3}}
\newcommand{\NPB}[3]{\J{Nucl.\ Phys.\ B}{#1}{#2}{#3}}
\newcommand{\subNPB}{\subJ{Nucl.\ Phys.\ B}}
\newcommand{\NIMA}[3]{\J{Nucl.\ Instrum.\ Methods A}{#1}{#2}{#3}}
\newcommand{\subNIMA}{\subJ{Nucl.\ Instrum.\ Methods A}}
\newcommand{\JHEP}[3]{\J{J.\ of High Energy Physics }{#1}{#2}{#3}}
\newcommand{\ARNS}[3]{\J{Ann.\ Rev.\ Nucl.\ Sci.}{#1}{#2}{#3}}
\newcommand{\newref}{\\}

%Particles
\newcommand{\particle}[1]{\ensuremath{#1}\xspace}
\renewcommand{\ee}{\particle{e^+e^-}}
\newcommand{\Ups}{\particle{\Upsilon(4S)}}
\renewcommand{\b}{\particle{b}}
\renewcommand{\B}{\particle{B}}
\newcommand{\Bd}{\particle{B^0}}
\renewcommand{\Bs}{\particle{B^0_s}}
\renewcommand{\Bu}{\particle{B^+}}
\newcommand{\Bc}{\particle{B^+_c}}
\newcommand{\Bdbar}{\particle{\bar{B}^0}}
\newcommand{\Bsbar}{\particle{\bar{B}^0_s}}
\newcommand{\Lb}{\particle{\Lambda_b^0}}
\newcommand{\Xib}{\particle{\Xi_b}}
\newcommand{\Lc}{\particle{\Lambda_c^+}}

% Fractions
\newcommand{\fBs}{\ensuremath{f_{\particle{s}}}\xspace}
\newcommand{\fBd}{\ensuremath{f_{\particle{d}}}\xspace}
\newcommand{\fBu}{\ensuremath{f_{\particle{u}}}\xspace}
\newcommand{\fbb}{\ensuremath{f_{\rm baryon}}\xspace}
%\newcommand{\fBs}{\ensuremath{f_{\Bs}}\xspace}
%\newcommand{\fBd}{\ensuremath{f_{\Bd}}\xspace}
%\newcommand{\fBu}{\ensuremath{f_{\Bu}}\xspace}
%\newcommand{\fbb}{\ensuremath{f_{\mbox{\scriptsize \b-baryon}}}\xspace}

% Mixing
\newcommand{\dmd}{\ensuremath{\Delta m_{\particle{d}}}\xspace}
\newcommand{\dms}{\ensuremath{\Delta m_{\particle{s}}}\xspace}
\newcommand{\xd}{\ensuremath{x_{\particle{d}}}\xspace}
\newcommand{\xs}{\ensuremath{x_{\particle{s}}}\xspace}
\newcommand{\yd}{\ensuremath{y_{\particle{d}}}\xspace}
\newcommand{\ys}{\ensuremath{y_{\particle{s}}}\xspace}
\newcommand{\chibar}{\ensuremath{\overline{\chi}}\xspace}
\newcommand{\chid}{\ensuremath{\chi_{\particle{d}}}\xspace}
\newcommand{\chis}{\ensuremath{\chi_{\particle{s}}}\xspace}
\newcommand{\Gd}{\ensuremath{\Gamma_{\particle{d}}}\xspace}
\newcommand{\DGd}{\ensuremath{\Delta\Gd}\xspace}
\newcommand{\DGGd}{\ensuremath{\DGd/\Gd}\xspace}
\newcommand{\Gs}{\ensuremath{\Gamma_{\particle{s}}}\xspace}
\newcommand{\DGs}{\ensuremath{\Delta\Gs}\xspace}
\newcommand{\DGGs}{\ensuremath{\Delta\Gs/\Gs}\xspace}

% Branching ratios, CL, ... and miscellaneous
%\renewcommand{\BR}[1]{\particle{\mbox{BR}(#1)}}
\renewcommand{\BR}[1]{\particle{{\cal B}(#1)}}
\newcommand{\CL}[1]{#1\%~\mbox{CL}}
\newcommand{\Qjet}{\ensuremath{Q_{\rm jet}}\xspace}

% Units
%\newcommand{\microns}{\mbox{$\mu$m}}
%\newcommand{\GeVc}{\mbox{GeV$/c$}}
%\newcommand{\GeVcc}{\mbox{GeV$/c^2$}}
%\newcommand{\MeVcc}{\mbox{MeV$/c^2$}}
%\newcommand{\GeV}{{\rm GeV}}
\newcommand{\unit}[1]{~\ensuremath{\rm #1}\xspace}
\renewcommand{\ps}{\unit{ps}}
\newcommand{\invps}{\unit{ps^{-1}}}
\newcommand{\TeV}{\unit{TeV}}
\newcommand{\MeVcc}{\unit{MeV/\mbox{$c$}^2}}
\newcommand{\MeV}{\unit{MeV}}

% Labels, and references (equations, figures, tables, sections, ...)
\newcommand{\labe}[1]{\label{equ:#1}}
\newcommand{\labs}[1]{\label{sec:#1}}
\newcommand{\labf}[1]{\label{fig:#1}}
\newcommand{\labt}[1]{\label{tab:#1}}
\newcommand{\refe}[1]{\ref{equ:#1}}
\newcommand{\refs}[1]{\ref{sec:#1}}
\newcommand{\reff}[1]{\ref{fig:#1}}
\newcommand{\reft}[1]{\ref{tab:#1}}
\newcommand{\Ref}[1]{Ref.~\cite{#1}}
\newcommand{\Refs}[1]{Refs.~\cite{#1}}
\newcommand{\Refss}[2]{Refs.~\cite{#1} and \cite{#2}}
\newcommand{\Refsss}[3]{Refs.~\cite{#1}, \cite{#2} and \cite{#3}}
\newcommand{\eq}[1]{(\refe{#1})}
\newcommand{\Eq}[1]{Eq.~(\refe{#1})}
\newcommand{\Eqs}[1]{Eqs.~(\refe{#1})}
\newcommand{\Eqss}[2]{Eqs.~(\refe{#1}) and (\refe{#2})}
\newcommand{\Eqssor}[2]{Eqs.~(\refe{#1}) or (\refe{#2})}
\newcommand{\Eqsss}[3]{Eqs.~(\refe{#1}), (\refe{#2}), and (\refe{#3})}
\newcommand{\Figure}[1]{Figure~\reff{#1}}
\newcommand{\Figuress}[2]{Figures~\reff{#1} and \reff{#2}}
\newcommand{\Fig}[1]{Fig.~\reff{#1}}
\newcommand{\Figs}[1]{Figs.~\reff{#1}}
\newcommand{\Figss}[2]{Figs.~\reff{#1} and \reff{#2}}
\newcommand{\Figsss}[3]{Figs.~\reff{#1}, \reff{#2}, and \reff{#3}}
\newcommand{\Section}[1]{Section~\refs{#1}}
\newcommand{\Sec}[1]{Sec.~\refs{#1}}
\newcommand{\Secs}[1]{Secs.~\refs{#1}}
\newcommand{\Secss}[2]{Secs.~\refs{#1} and \refs{#2}}
\newcommand{\Secsss}[3]{Secs.~\refs{#1}, \refs{#2}, and \refs{#3}}
\newcommand{\Table}[1]{Table~\reft{#1}}
\newcommand{\Tables}[1]{Tables~\reft{#1}}
\newcommand{\Tabless}[2]{Tables~\reft{#1} and \reft{#2}}
\newcommand{\Tablesss}[3]{Tables~\reft{#1}, \reft{#2}, and \reft{#3}}

% Smart titles with boldmath, already defined
%\newcommand{\mysection}[1]{\section[\boldmath #1]{\boldmath #1}}
%\newcommand{\mysubsection}[1]{\subsection[#1]{\boldmath #1}}
%\newcommand{\mysubsubsection}[1]{\subsubsection[#1]{\boldmath #1}}
\newcommand{\subsubsubsection}[1]{\vspace{2ex}\par\noindent {\bf\boldmath\em #1} \vspace{2ex}\par}

%%%%%%%%%%%%%%%%%%%%%%%%%%%%%%%%%%%%%%%%%%%%%%%
%
% This is file life_mix/life_mix_averages.tex
%
% Olivier Schneider, EPFL, June 23, 2004
%
%%%%%%%%%%%%%%%%%%%%%%%%%%%%%%%%%%%%%%%%%%%%%%%
%
%\input{life_mix/life_mix_averages}

% ==========================================================
% HFAG averages from lifetime/mixing sub-group (Summer 2004)
% ==========================================================
% checked on Dec 22, 2004
%

% b-hadron fractions at Upsilon(4S)
\newcommand{\HFAGfplusfzero}{\ensuremath{1.010 \pm 0.038}} %  R = f+-/f00
\newcommand{\HFAGfzero}{\ensuremath{0.486 \pm 0.010 \pm 0.009}} %  f00_direct measurement
\newcommand{\HFAGfplus}{\ensuremath{0.491 \pm 0.023}} %  f+- = R * f00_direct
\newcommand{\HFAGfsum}{\ensuremath{0.977 \pm 0.033}}  %  f+- + f00 = (1+R) * f00_direct

\newcommand{\HFAGfplusWorld}{\ensuremath{0.506 \pm 0.008}}      %  f+- (world avg)
\newcommand{\HFAGfzeroWorld}{\ensuremath{0.494 \pm 0.008}}      %  f00 (world avg)
\newcommand{\HFAGfplusfzeroWorld}{\ensuremath{1.026 \pm 0.032}} %  f+-/f00 (world avg)

% b-hadron fractions at high energy, using only direct BR measurements ("step 1" fractions)
\newcommand{\HFAGfBsBR}{\ensuremath{0.088 \pm 0.021}}   % f(Bs)
\newcommand{\HFAGfbbBR}{\ensuremath{0.107 \pm 0.019}}   % f(b-baryon)
\newcommand{\HFAGfBdBR}{\ensuremath{0.403 \pm 0.011}}   % f(Bd)
\newcommand{\HFAGfBuBR}{\HFAGfBuBR}                     % f(Bu)=f(Bd)
\newcommand{\HFAGcorrfbbfBsBR}{\ensuremath{-0.382}}     % corr(f(b-baryon),f(Bs))
%\newcommand{\HFAGcorrfBdfBsBR}{\ensuremath{-xxx}}      % corr(f(Bd),f(Bs))
%\newcommand{\HFAGcorrfBufBsBR}{\HFAGcorrfBdfBs}        % corr(f(Bu),f(Bs))
%\newcommand{\HFAGcorrfBdfbbBR}{\ensuremath{-xxx}}      % corr(f(Bd),f(b-baryon))
%\newcommand{\HFAGcorrfBufbbBR}{\HFAGcorrfBdfbbaryonBR} % corr(f(Bu),f(b-baryon))

% Bs fraction from mixing constraints (chibar, ...)
\newcommand{\HFAGfBsmix}{\ensuremath{0.120 \pm 0.021}}  % f(Bs) from mixing and f(b-baryon)

% b-hadron fractions at high energy, using only direct BR measurements and constraint from mixing
\newcommand{\HFAGfBs}{\ensuremath{0.104 \pm 0.015}}     % f(Bs)
\newcommand{\HFAGfbb}{\ensuremath{0.100 \pm 0.017}}     % f(b-baryon)
\newcommand{\HFAGfBd}{\ensuremath{0.398 \pm 0.010}}     % f(Bd)
\newcommand{\HFAGfBu}{\HFAGfBu}                         % f(Bu)=f(Bd)
\newcommand{\HFAGcorrfbbfBs}{\ensuremath{-0.176}}       % corr(f(b-baryon),f(Bs))
\newcommand{\HFAGcorrfBdfBs}{\ensuremath{-0.566}}       % corr(f(Bd),f(Bs))
\newcommand{\HFAGcorrfBufBs}{\HFAGcorrfBdfBs}           % corr(f(Bu),f(Bs))
\newcommand{\HFAGcorrfBdfbb}{\ensuremath{-0.712}}       % corr(f(Bd),f(b-baryon))
\newcommand{\HFAGcorrfBufbb}{\HFAGcorrfBdfbbaryon}      % corr(f(Bu),f(b-baryon))

% Lifetimes (summer 2003 + new Bs, Bc, Lambda_b and b-baryon lifetimes from Rick
%            + new b-hadron lifetime since new CDF measurement with Jpsi has been removed)
\newcommand{\HFAGtauBd}{\ensuremath{1.534 \pm 0.013\ps}}       % B0 lifetime
\newcommand{\HFAGtauBu}{\ensuremath{1.653 \pm 0.014\ps}}       % B+ lifetime
\newcommand{\HFAGtauBuBd}{\ensuremath{1.081 \pm 0.015}}        % B+/B0 lifetime ratio
\newcommand{\HFAGtauBsfs}{\ensuremath{1.442 \pm 0.066\ps}}     % Bs lifetime, flavor-specific decay
\newcommand{\HFAGtauBsjpsi}{\ensuremath{1.404 \pm 0.066\ps}}   % Bs lifetime, Jpsi phi decay %RvK
\newcommand{\HFAGtauBs}{\ensuremath{1.469 \pm 0.059\ps}}       % Bs lifetime
\newcommand{\HFAGtauBsfsBd}{\ensuremath{0.939 \pm 0.044}}        % Bs/B0 lifetime ratio (ratio to flavor-specific lifetime for definiteness) %RvK
\newcommand{\HFAGtauBc}{\ensuremath{0.45 \pm 0.12\ps}}         % Bc lifetime %RvK
\newcommand{\HFAGtauBcBd}{\ensuremath{0.293 \pm 0.078}}        % Bc/B0 lifetime %RvK
\newcommand{\HFAGtauLb}{\ensuremath{1.232 \pm 0.072\ps}}       % Lambda_b lifetime %RvK
\newcommand{\HFAGtauLbBd}{\ensuremath{0.803 \pm 0.047}}        % Lambda_b/B0 lifetime %RvK
\newcommand{\HFAGtauXib}{\ensuremath{1.39^{+0.34}_{-0.28}\ps}} % Xib0 and Xib- average lifetime
\newcommand{\HFAGtaubbaryon}{\ensuremath{1.210 \pm 0.048\ps}}  % b-baryon average lifetime %RvK
\newcommand{\HFAGtaubbaryonBd}{\ensuremath{0.789 \pm 0.032}}   % b-baryon/Bd lifetime %RvK

\newcommand{\HFAGtauball}{\ensuremath{1.574 \pm 0.008\ps}}     % b-hadron average lifetime (all)
\newcommand{\HFAGtaubvertex}{\ensuremath{1.576 \pm 0.008\ps}}  % b vertex average lifetime
\newcommand{\HFAGtaublepton}{\ensuremath{1.537 \pm 0.020\ps}}  % b->l average lifetime
\newcommand{\HFAGtaubJpsi}{\ensuremath{1.533 ^{+0.038}_{-0.034}\ps}} % b->J/psi average lifetime

% CP violation in B0-B0bar mixing
\newcommand{\HFAGASL}{\ensuremath{-0.0026 \pm 0.0067}}   % A_SL
\newcommand{\HFAGqp}{\ensuremath{1.0013 \pm 0.0034}}    % |q/p|
\newcommand{\HFAGreb}{\ensuremath{-0.0007 \pm 0.0017}}  % Re(epsB)/(1+|epsB|^2)

% Neutral B meson oscillations

\newcommand{\HFAGdmdtime}{\ensuremath{0.502 \pm 0.006\invps}}     % dmd (time-dependent)
\newcommand{\HFAGdmdtimeall}{\ensuremath{0.502 \pm 0.004 \pm 0.005\invps}} % dmd (time-dependent)
\newcommand{\HFAGchidint}{\ensuremath{0.182 \pm 0.015}}           % chid (time-integrated, ARGUS+CLEO)
\newcommand{\HFAGdmd}{\ensuremath{0.502 \pm 0.006\invps}}         % dmd (world average)
\newcommand{\HFAGxd}{\ensuremath{0.770 \pm 0.011}}                % xd (world average)
\newcommand{\HFAGchid}{\ensuremath{0.186 \pm 0.003}}              % chid (world average)
\newcommand{\HFAGdmslim}{\ensuremath{\rm 14.5\invps}}    % dms lower limit
\newcommand{\HFAGxslim}{\ensuremath{\rm 20.8}}           % xs lower limit
\newcommand{\HFAGchislim}{\ensuremath{\rm 0.49885}}      % chis lower limit
\newcommand{\HFAGdmssens}{\ensuremath{\rm 18.2\invps}} % dms sensitivity

\newcommand{\HFAGdmslimax}{\ensuremath{\rm 21.7\invps}}    % range \HFAGdmslim -- \HFAGdmslimax 
                                                           % cannot be excluded 
                                                           
\newcommand{\HFAGdmslimCL}{\ensuremath{\rm\HFAGdmslim~at~\CL{95}}}   % dms lower limit
\newcommand{\HFAGxslimCL}{\ensuremath{\rm\HFAGxslim~at~\CL{95}}}     % xs lower limit
\newcommand{\HFAGchislimCL}{\ensuremath{\rm\HFAGchislim~at~\CL{95}}} % chis lower limit
\newcommand{\HFAGdmssensCL}{\ensuremath{\rm\HFAGdmssens~at~\CL{95}}} % dms sensitivity

% chibar = average (semileptonic) time-integrated mixing probability at high energy
\newcommand{\HFAGchibarZ}{\ensuremath{0.1257 \pm 0.0042}} % chibar (from LEP EW WG)
    % Note: The above value was computed by the LEP EW WG for PDG 2004 (and is quoted by PDG 2004).
    %       It is different from the value found in Appendix B on page 176 of 
    %       the LEP EW WG preprint hep-ex/0312023, Dec 2003 (rev Feb 2004)
    %       because it does not include the asymmetries, nor some unpublished 
    %       Rb and Rc measurements from SLD.
\newcommand{\HFAGchibarTeV}{\ensuremath{0.152 \pm 0.013}} % chibar (from CDF)
                                                          % in fact $0.152 \pm 0.007 \pm \0.011$ 
\newcommand{\HFAGchibar}{\ensuremath{0.1282 \pm 0.0077}}  % chibar (world average)

% Bs decay width difference (PDG 2004)
\newcommand{\HFAGDGGs}{\ensuremath{0.16^{+0.15}_{-0.16}}}       % DGs/Gs 
\newcommand{\HFAGDGGslimCL}{\ensuremath{\rm 0.54~at~95\%~CL}}     % DGs/Gs upper limit
\newcommand{\HFAGDGGscons}{\ensuremath{0.07^{+0.09}_{-0.07}}}   % DGs/Gs with Bd lifetime constraint 
\newcommand{\HFAGDGGsconslimCL}{\ensuremath{\rm 0.29~at~95\%~CL}} % DGs/Gs upper limit with Bd constraint

% ---------------------
% Title of this chapter
% ---------------------

\mysection{Averages of \b-hadron fractions, lifetimes and mixing parameters}
\labs{life_mix}

% ----------------------------
% Introduction to this chapter
% ----------------------------

Quantities such as \b-hadron production fractions, \b-hadron lifetimes, 
and neutral \B-meson oscillation frequencies have been measured
for many years at high-energy colliders, namely LEP and SLC 
(\ee colliders at $\sqrt{s}=m_{\particle{Z}}$) as well as Tevatron Run I
(\particle{p\bar{p}} collider at $\sqrt{s}=1.8\TeV$). More recently, 
precise measurements of the \Bd and \Bu lifetimes, as well as of the 
\Bd oscillation frequency, have also been performed at the 
asymmetric \B factories, KEKB and PEPII
(\ee colliders at $\sqrt{s}=m_{\Ups}$).
In most cases, these basic quantities, although interesting by themselves,
can now be seen as necessary ingredients for the more complicated and 
refined analyses being currently performed at the asymmetric \B factories
and at the Tevatron Run II ($\sqrt{s}=2\TeV$), in particular the time-dependent
\CP measurements. It is therefore important that the best experimental
values of these quantities continue to be kept up-to-date and improved. 

In several cases, the averages presented in this chapter are indeed
needed and used as input for the results given in the subsequent chapters. 
However, within this chapter, some averages need the knowledge of other 
averages in a circular way. This ``coupling'', which appears through the 
\b-hadron fractions whenever inclusive or semi-exclusive measurements 
have to be considered, has been reduced significantly in the last years 
with increasingly precise exclusive measurements becoming available. 
To cope with this circularity,
a rather involved averaging procedure had been developed, in the framework 
of the former LEP Heavy Flavour Steering Group. This is still in use now
(details can be found in~\cite{LEPHFS}), 
although simplifications can be envisaged in the future when even more 
precise exclusive measurements become available. 

% ----------------------------------------
\mysubsection{\b-hadron production fractions}
% ----------------------------------------
\labs{fractions}
 
We consider here the relative fractions of the different \b-hadron 
species found in an unbiased sample of weakly-decaying \b hadrons 
produced under some specific conditions. The knowledge of these fractions
is useful to characterize the signal composition in inclusive \b-hadron 
analyses, or to predict the background composition in exclusive analyses.
Many analyses in \B physics need these fractions as input. We distinguish 
here the following two conditions: \Ups decays and 
high-energy collisions. 

% -------------------------------------------------------------
\mysubsubsection{\b-hadron fractions in \Ups decays}
% -------------------------------------------------------------
\labs{bfraction}

Only pairs of the two lightest (charged and neutral) \B mesons 
can be produced in \Ups decays, 
and it is enough to determine the following branching 
fractions:
\begin{eqnarray}
f^{+-} & = & \Gamma(\Ups \to \particle{B^+B^-})/
             \Gamma_{\rm tot}(\Ups)  \,, \\
f^{00} & = & \Gamma(\Ups \to \particle{B^0\bar{B}^0})/
             \Gamma_{\rm tot}(\Ups) \,.
\end{eqnarray}
In practice, most analyses measure their ratio
\begin{equation}
R^{+-/00} = f^{+-}/f^{00} = \Gamma(\Ups \to \particle{B^+B^-})/
             \Gamma(\Ups \to \particle{B^0\bar{B}^0}) \,,
\end{equation}
which is easier to access experimentally.
Since an inclusive (but separate) reconstruction of 
\Bu and \Bd is difficult, specific exclusive decay modes, 
${\Bu} \to x^+$ and ${\Bd} \to x^0$, are usually considered to perform 
a measurement of $R^{+-/00}$, whenever they can be related by 
isospin symmetry (for example \particle{\Bu \to J/\psi K^+} and 
\particle{\Bd \to J/\psi K^0}).
Under the assumption that $\Gamma(\Bu \to x^+) = \Gamma(\Bd \to x^0)$, 
\ie\ that isospin invariance holds in these \B decays,
the ratio of the number of reconstructed
$\Bu \to x^+$ and $\Bd \to x^0$ mesons is proportional to
\begin{equation}
\frac{f^{+-}\, B({\Bu}\to x^+)}{f^{00}\, B({\Bd}\to x^0)} 
= \frac{f^{+-}\, \Gamma({\Bu}\to x^+)\, \tau(\Bu)}%
{f^{00}\, \Gamma({\Bd}\to x^0)\,\tau(\Bd)}
= \frac{f^{+-}}{f^{00}} \, \frac{\tau(\Bu)}{\tau(\Bd)}  \,, 
\end{equation} 
where $\tau(\Bu)$ and $\tau(\Bd)$ are the \Bu and \Bd 
lifetimes respectively.
Hence the primary quantity measured in these analyses 
is $R^{+-/00} \, \tau(\Bu)/\tau(\Bd)$, 
and the extraction of $R^{+-/00}$ with this method therefore 
requires the knowledge of the $\tau(\Bu)/\tau(\Bd)$ lifetime ratio. 

\begin{table}
\caption{Published measurements of the $\Bu/\Bd$ production ratio
in \Ups decays, together with their average (see text).
Systematic uncertainties due to the imperfect knowledge of 
$\tau(\Bu)/\tau(\Bd)$ are included.}
\labt{R_data}
\begin{center}
\begin{tabular}{@{}l@{}c@{\,}cll@{}}
\hline
Experiment & Ref. & Decay modes & Published value of & Assumed value \\
and year & & or method & $R^{+-/00}=f^{+-}/f^{00}$ & of $\tau(\Bu)/\tau(\Bd)$ \\
\hline
%OS Note than an old and imprecise CLEO measurement from 1995, 
%OS B. Barish et al (CLEO), PRD 51 (1995) 407, is not listed here 
CLEO,   2001 & \cite{CLEO_R2001}  & \particle{J/\psi K^{(*)}} 
             & $1.04 \pm0.07 \pm0.04$ & $1.066 \pm0.024$ \\
\babar, 2002 & \cite{BABAR_R2002} & \particle{(c\bar{c})K^{(*)}}
             & $1.10 \pm0.06 \pm0.05$ & $1.062 \pm0.029$\\ 
CLEO,   2002 & \cite{CLEO_R2002}  & \particle{D^*\ell\nu}
             & $1.058 \pm0.084 \pm0.136$ & $1.074 \pm0.028$\\
\belle, 2003 & \cite{BELLE_dmd_dilepton} & dilepton events 
             & $1.01 \pm0.03 \pm0.09$ & $1.083 \pm0.017$\\
\babar, 2004 & \cite{BABAR_R2004} & \particle{J/\psi K}
             & $1.006 \pm0.036 \pm0.031$ & $1.083 \pm0.017$ \\
\hline
Average      & & & \HFAGfplusfzero~(tot) & \HFAGtauBuBd \\
\hline
\end{tabular}
\end{center}
\end{table}

The published measurements of $R^{+-/00}$ are listed 
in \Table{R_data} together with the corresponding assumed values of 
$\tau(\Bu)/\tau(\Bd)$.
All measurements are based on the above-mentioned method, 
except the one from \belle, which is a by-product of the 
\Bd mixing frequency analysis using dilepton events
(but note that it also assumes isospin invariance, 
namely $\Gamma(\Bu \to \ell^+{\rm X}) = \Gamma(\Bd \to \ell^+{\rm X})$).
The latter is therefore treated in a slightly different 
manner in the following procedure used to combine 
these measurements:
\begin{itemize} 
\item each published value of $R^{+-/00}$ from CLEO and \babar
      is first converted back to the original measurement of 
      $R^{+-/00} \, \tau(\Bu)/\tau(\Bd)$, using the value of the 
      lifetime ratio assumed in the corresponding analysis;
\item a simple weighted average of these original
      measurements of $R^{+-/00} \, \tau(\Bu)/\tau(\Bd)$ from 
      CLEO and \babar (which do not depend on the assumed value 
      of the lifetime ratio) is then computed, assuming no 
      statistical or systematic correlations between them;

% {\em ***** this may not be true in the case of \babar;
% waiting for more information 
% from David about averaging of the two \babar results ****}
% \marginpar{David}

\item the weighted average of $R^{+-/00} \, \tau(\Bu)/\tau(\Bd)$ 
      is converted into a value of $R^{+-/00}$, using the latest 
      average of the lifetime ratios, $\tau(\Bu)/\tau(\Bd)=\HFAGtauBuBd$ 
      (see \Sec{lifetime_ratio});
\item the \belle measurement of $R^{+-/00}$ is adjusted to the 
      current values of $\tau(\Bd)=\HFAGtauBd$ and 
      $\tau(\Bu)/\tau(\Bd)=\HFAGtauBuBd$ (see \Sec{lifetime_ratio}),
      using the quoted systematic uncertainties due to these parameters;
\item the combined value of $R^{+-/00}$ from CLEO and \babar is averaged 
      with the adjusted value of $R^{+-/00}$ from \belle, assuming a 100\% 
      correlation of the systematic uncertainty due to the limited 
      knowledge on $\tau(\Bu)/\tau(\Bd)$; no other correlation is considered. 
\end{itemize} 
The resulting global average, 
\begin{equation}
R^{+-/00} = \frac{f^{+-}}{f^{00}} =  \HFAGfplusfzero \,,
\labe{Rplusminus}
\end{equation}
is consistent with an equal production of charged and neutral \B mesons.

On the other hand, the \babar collaboration has 
recently performed a direct measurement of the $f^{00}$ fraction 
using a novel method, which does not rely on isospin symmetry nor requires 
the knowledge of $\tau(\Bu)/\tau(\Bd)$. Its preliminary analysis, 
based on a comparison between the number of events where a single 
$B^0 \to D^{*-} \ell^+ \nu$ decay could be reconstructed and the number 
of events where two such decays could be reconstructed, yields~\cite{BABAR_f00_preliminary}
\begin{equation}
f^{00}= 0.486 \pm 0.010\,\mbox{(stat)} \pm 0.009\,\mbox{(syst)} \,.
\labe{fzerozero}
\end{equation}

The two results of \Eqss{Rplusminus}{fzerozero} are of very different natures 
and completely independent of each other. 
Their product is equal to $f^{+-} = \HFAGfplus$, 
while another combination of them gives $f^{+-} + f^{00}= \HFAGfsum$, 
compatible with unity.
Assuming $f^{+-}+ f^{00}= 1$, also consistent with 
CLEO's observation that the fraction of \Ups decays 
to \BB pairs is larger than 0.96 at \CL{95}~\cite{CLEO_frac_limit},
the results of \Eqss{Rplusminus}{fzerozero}
can be averaged (first converting \Eq{Rplusminus} 
into a value of $f^{00}=1/(R^{+-/00}+1)$) 
to yield the following more precise estimates:
\begin{equation}
f^{00} = \HFAGfzeroWorld  \,,~~~ f^{+-} = 1 -f^{00} =  \HFAGfplusWorld \,,~~~
\frac{f^{+-}}{f^{00}} =  \HFAGfplusfzeroWorld \,.
\end{equation}

%------------------------------------------------
\mysubsubsection{\b-hadron fractions at high energy}
%------------------------------------------------
\labs{fractions_high_energy}

At high energy, all species of weakly-decaying \b hadrons 
can be produced.
We assume here that the fractions of these different species 
are the same in unbiased samples of high-$p_{\rm T}$ \b jets 
originating from \particle{Z^0} decays or from \particle{p\bar{p}} 
collisions at the Tevatron, either directly or in strong and electromagnetic 
decays of excited \b hadrons.
% ($\sqrt{s} =1.8-2\TeV$).
This hypothesis is plausible considering that, in both cases, 
the last step of the jet hadronization is a non-perturbative
QCD process occurring at a scale of order $\Lambda_{\rm QCD}$.
On the other hand, there is no strong argument to claim that these 
fractions should be strictly equal, so this assumption 
should be checked experimentally.
Although the available data is not quite sufficient at 
this time to perform a significant check, 
it is expected that the new data from 
Tevatron Run II will soon improve this situation and 
allow to confirm or infirm this assumption with reasonable 
confidence. Meanwhile, the attitude adopted here is that these 
fractions are assumed to be equal at all high-energy colliders
until demonstrated otherwise by experiment.\footnote{It is not unlikely
that the \b-hadron fractions in low-$p_{\rm T}$ jets 
at a hadronic machine be different; in particular, beam-remnant effects may
enhance the \b-baryon production.}

Contrary to what happens in the charm sector where the fractions of \particle{D^+} 
and \particle{D^0} are different, the relative amount of \Bu and \Bd is not affected by the 
electromagnetic decays of excited ${\Bu}^*$ and ${\Bd}^*$ states and strong decays of excited
${\Bu}^{**}$ and ${\Bd}^{**}$ states. Decays of the type \particle{{\Bs}^{**} \to B^{(*)}K}
also contribute to the \Bu and \Bd rates, but with the same magnitude if mass effects
can be neglected.
We therefore assume equal production of \Bu and \Bd. We also  
neglect the production of weakly-decaying states
made of several heavy quarks (like \Bc and other heavy baryons) 
which is known to be very small. Hence, for the purpose of determining 
the \b-hadron fractions, we use the constraints
\begin{equation}
\fBu = \fBd ~~~~\mbox{and}~~~ \fBu + \fBd + \fBs + \fbb = 1 \,,
\labe{constraints}
\end{equation}
where \fBu, \fBd, \fBs and \fbb
are the unbiased fractions of \Bu, \Bd, \Bs and \b-baryons, respectively.

The LEP experiments have measured
$\fBs \times \BR{\Bs\to\particle{D_s^-} \ell^+ \nu_\ell \mbox{$X$}}$~\cite{LEP_fs}, 
$\BR{\b\to\Lb} \times \BR{\Lb\to\Lc\ell^-\bar{\nu}_\ell \mbox{$X$}}$~\cite{DELPHI_fla,ALEPH_fla}
and $\BR{\b\to\Xib^-} \times \BR{\Xi_b^- \to \Xi^-\ell^-\overline\nu_\ell 
\mbox{$X$}}$~\cite{DELPHI_fxi,ALEPH_fxi}
from partially reconstructed final states 
including a lepton, \fbb
from protons identified in \b events~\cite{ALEPH-fbar}, and the 
production rate of charged \b hadrons~\cite{DELPHI-fch}. 
The various \b-hadron fractions 
have also been measured at CDF using electron-charm final states~\cite{CDF_f_ec}
and double semileptonic decays with \particle{\phi\ell} and 
\particle{K^*\ell} final states~\cite{CDF_f_phil_Kstl}.
All these published results have been combined 
following the procedure and assumptions described in~\cite{LEPHFS}
to yield $\fBu=\fBd=\HFAGfBdBR$, 
$\fBs=\HFAGfBsBR$ and $\fbb=\HFAGfbbBR$
under the constraints of \Eq{constraints}.
For this combination, other external inputs are used, \eg\ the branching 
ratios of \B mesons to final states with a \particle{D}, \particle{D^*} or 
\particle{D^{**}} in semileptonic decays, which are needed to evaluate the 
fraction of semileptonic \Bs decays with a \particle{D_s^-} in the final state.

%%% \marginpar{xxx}
%%% {\em Need to absorb new B factory results with $D_s\pi^0$ final states \ldots}

Time-integrated mixing analyses performed with lepton pairs 
from \particle{b\bar{b}} 
events produced at high-energy colliders measure the quantity 
\begin{equation}
\chibar = f'_{\particle{d}} \,\chid + f'_{\particle{s}} \,\chis \,,
\end{equation}
where $f'_{\particle{d}}$ and $f'_{\particle{s}}$ are 
the fractions of \Bd and \Bs hadrons 
in a sample of semileptonic \b-hadron decays, and where \chid and \chis 
are the \Bd and \Bs time-integrated mixing probabilities.
Assuming that all \b hadrons have the same semileptonic decay width implies 
$f'_i = f_i R_i$, where $R_i = \tau_i/\tau_{\particle{b}}$ is the ratio of the lifetime 
$\tau_i$ of species $i$ to the average \b-hadron lifetime 
$\tau_{\particle{b}} = \sum_i f_i \tau_i$.
Hence measurements of the mixing probabilities
\chibar, \chid and \chis can be used to improve our 
knowledge on the \fBu, \fBd, \fBs and \fbb fractions.
In practice, the above relations yield another determination of 
\fBs obtained from \fbb and mixing information, 
\begin{equation}
\fBs = \frac{1}{R_{\particle{s}}}
\frac{(1+r)\overline{\chi}-(1-\fbb R_{\rm baryon}) \chid}{(1+r)\chis - \chid} \,,
\labe{fBs-mixing}
\end{equation}
where $r=R_{\particle{u}}/R_{\particle{d}} = \tau(\Bu)/\tau(\Bd)$.

\labs{chibar}
The published measurements of \chibar performed by the LEP
experiments have been combined by the LEP Electroweak Working Group to yield 
$\chibar = \HFAGchibarZ$~\cite{LEPEWWG}. This can be compared with a 
recent measurement from CDF, $\chibar = \HFAGchibarTeV$~\cite{CDF-chibar}, 
obtained from an analysis of the Run I data. The two estimates deviate from each other 
by $1.9\,\sigma$, and could be an indication that the fractions of \b hadrons produced
at the \particle{Z} peak or at the Tevatron are not the same. Although this discrepancy 
is not very significant it should be carefully monitored in the future. 
We choose to combine these two results in a simple weighted average, assuming no correlations, 
and, following the PDG prescription, we multiply the combined uncertainty by 1.9 to account 
for the discrepancy. Our world average is then
\begin{equation}
\chibar = \HFAGchibar \,.
\end{equation}

\begin{table}
\caption{Fractions of the different \b-hadron species in an unbiased sample of 
weakly-decaying \b hadrons produced at high energy, obtained from both direct
and mixing measurements.}
\labt{fractions}
\begin{center}
\begin{tabular}{crcc}
\hline
\b-hadron & \multicolumn{1}{c}{Fraction} & \multicolumn{2}{l}{Correlation coefficients} \\
species   &          & with $\fBd=\fBu$ & and \fBs\\
\hline
\Bd, \Bu   & $\fBd=\fBu = \HFAGfBd$  & & \\
\Bs        & $\fBs = \HFAGfBs$       & \HFAGcorrfBdfBs & \\
\b baryons & $\fbb = \HFAGfbb$       & \HFAGcorrfBdfbb & \HFAGcorrfbbfBs \\
%OS all        & $\fBd+\fBu+\fBs+\fbb=1$ & & \\
\hline
\end{tabular}
\end{center}
\end{table}

Introducing the latter result in \Eq{fBs-mixing}, together with our world average 
$\chid = \HFAGchid$ (see \Eq{chid} of \Sec{dmd}), the assumption $\chis= 1/2$ 
(justified by the large value of \dms, see \Eq{chis} in \Sec{dms}), the 
best knowledge of the lifetimes (see \Sec{lifetimes}) and the estimate of \fbb given above, 
yields $\fBs = \HFAGfBsmix$, an estimate dominated by the mixing information. 
Taking into account all known correlations (including the one introduced by \fbb), 
this result is then combined with the set of fractions obtained from direct measurements 
(given above), to yield the % following improved estimates, 
improved estimates of \Table{fractions}, 
still under the constraints of \Eq{constraints}. %:
%OS \begin{eqnarray}
%OS \fBd=\fBu & = & \HFAGfBd \,, \labe{fBd} \\
%OS \fBs      & = & \HFAGfBs \,, \labe{fBs} \\
%OS \fbb      & = & \HFAGfbb \,. \labe{fbb}
%OS \end{eqnarray}
As can be seen, our knowledge on the mixing parameters 
substantially reduces the uncertainty on \fBs, despite the rather strong 
deweighting introduced in the computation of the world average of \chibar.
It should be noted that the results % of \Eqsss{fBd}{fBs}{fbb} 
are correlated, as indicated in \Table{fractions}.

%------------------------------------------------
%\mysubsection{\b-hadron lifetimes}
%------------------------------------------------

% Introduction, b-hadron lifetime, B0 lifetime, B+ lifetime, B+/B0 lifetime ratio
%%$$%%%%%%%%%%%%%%%%%%%%%%%%%%%%%%%%%%%%%%%%%%%%%%
%
% This is file life_mix_tau1.tex containing the
% first part of the chapter on the b-hadron life-
% times: introduction, average b-hadron, B0 and B+
% lifetimes as well as the B+/B0 lifetime ratio
%
%%%%%%%%%%%%%%%%%%%%%%%%%%%%%%%%%%%%%%%%%%%%%%%%%

%------------------------------------------------
\mysubsection{\b-hadron lifetimes}
%------------------------------------------------
\labs{lifetimes}

In the spectator model the decay of \b-flavored hadrons $H_b$ is
governed entirely by the flavor changing \particle{b\to Wq} transition
($\particle{q}=\particle{c,u}$).  For this very reason, lifetimes of all
\b-flavored hadrons are the same in the spectator approximation
regardless of the (spectator) quark content of the $H_b$.  In the early
1990's experiments became sophisticated enough to start seeing the
differences of the lifetimes among various $H_b$ species.  The first
theoretical calculations of the spectator quark effects on $H_b$
lifetime emerged only few years earlier.

Currently, most of such calculations are performed in the framework of
the Heavy Quark Expansion, HQE.  In the HQE, under certain assumptions
(most important of which is that of quark-hadron duality), the decay
rate of an $H_b$ to an inclusive final state $f$ is expressed as the sum
of a series of expectation values of operators of increasing dimension,
multiplied by the correspondingly higher powers of $\Lambda_{\rm
QCD}/m_b$:
\begin{equation}
\Gamma_{H_b\to f} = |CKM|^2\sum_n c_n^{(f)}
\Bigl(\frac{\Lambda_{\rm QCD}}{m_b}\Bigr)^n\langle H_b|O_n|H_b\rangle,
\labe{hqe}
\end{equation}
where $|CKM|^2$ is the relevant combination of the CKM matrix elements.
Coefficients $c_n^{(f)}$ of this expansion, known as Operator Product
Expansion~\cite{OPE}, can be calculated perturbatively.  Hence, the HQE
predicts $\Gamma_{H_b\to f}$ in the form of an expansion in both
$\Lambda_{\rm QCD}/m_{\b}$ and $\alpha_s(m_{\b})$.  The precision of
current experiments makes it mandatory to go to the next-to-leading
order in QCD, {\em i.e.}\ to include correction of the order of
$\alpha_s(m_{\b})$ to the $c_n^{(f)}$'s.  All non-perturbative physics
is shifted into the expectation values $\langle H_b|O_n|H_b\rangle$ of
operators $O_n$.  These can be calculated using lattice QCD or QCD sum
rules, or can be related to other observables via the
HQE~\cite{Bigi_1995}.  One may reasonably expect that powers of
$\Lambda_{\rm QCD}/m_{\b}$ provide enough suppression that only the
first few terms of the sum in \Eq{hqe} matter.

Theoretical predictions are usually made for the ratios of the lifetimes
(with $\tau(\Bd)$ chosen as the common denominator) rather than for the
individual lifetimes, for this allows several calculational
uncertainties to cancel.  Present HQE results including next-to-leading
order corrections
(see \Refs{Gabbiani_et_al,nlo_lifetimes,tarantino} for the
latest updates) are in some
instances already surpassed by the experimental measurements, 
\eg\ in the case of $\tau(\Bu)/\tau(\Bd)$.  Also, HQE calculations
are not assumption-free.  More accurate predictions are a matter of
progress in the evaluation of the non-perturbative hadronic matrix
elements and verifying the assumptions that the calculations are based
upon.  However, the HQE, even in its present shape, draws a number of
important conclusions, which are in agreement with experimental
observations:
\begin{itemize}
\item The heavier the mass of the heavy quark the smaller is the
  variation in the lifetimes among different hadrons containing this
  quark, which is to say that as $m_{\b}\to\infty$ we retrieve the
  spectator picture in which the lifetimes of all $H_b$'s are the same.
  This is well illustrated by the fact that lifetimes in the $b$ sector
  are all very similar, while in the $c$ sector
  ($m_{\particle{c}}<m_{\b}$) lifetimes differ by as much as a factor of
  2.
\item The non-perturbative corrections arise only at the order of
  $\Lambda_{\rm QCD}^2/m_{\b}^2$, which translates into 
  differences among $H_b$ lifetimes of only a few percent.
\item It is only the difference between meson and baryon lifetimes that
  appears at the $\Lambda_{\rm QCD}^2/m_{\b}^2$ level.  The splitting of the
  meson lifetimes occurs at the $\Lambda_{\rm QCD}^3/m_{\b}^3$ level, yet it is
  enhanced by a phase space factor $16\pi^2$ with respect to the leading
  free \b decay.
\end{itemize}

To ensure that certain sources of systematic uncertainty cancel, 
lifetime analyses are sometimes designed to measure a 
ratio of lifetimes.  However, because of the differences in decay
topologies, abundance (or lack thereof) of decays of a certain kind,
{\em etc.}, measurements of the individual lifetimes are more 
common.  In the following section we review the most common
types of the lifetime measurements.  This discussion is followed by the
presentation of the averaging of the various lifetime measurements, each
with a brief description of its particularities.

%% Experimental measurements too often benefit from a partial systematic
%% uncertainty cancellation if a measurement is that of the ratio of two
%% quantities of the same kind, which are affected similarly by one or
%% more systematic effect(s).  For this reason, rather often the lifetime
%% measurements are being designed to be those of the ratio of the
%% lifetimes.  However, because of the differences in decay topologies,
%% abundance (or lack thereof) of decays of a certain kind, {\em etc.}\
%% measurements of the individual lifetimes are not particularly rare.  In
%% the following section we review the most common types of the lifetime
%% measurements.  This discussion is followed by the presentation of the
%% averaging of the various lifetime measurements, each with a brief
%% description of its particularities.

%% Details of procedures used to combine the different measurements can be
%% found in \Ref{lifetime_details}. {\sc do we want this? HERE?}

\mysubsubsection{Lifetime measurements, uncertainties and correlations}

In most cases lifetime of an $H_b$ is estimated from a flight distance
and a $\beta\gamma$ factor which is used to convert the geometrical
distance into the proper decay time.  Methods of accessing lifetime
information can roughly be divided in the following five categories:
\begin{enumerate}
\item {\bf\em Inclusive (flavor blind) measurements}.  These
  measurements are aimed at extracting the lifetime from a mixture of
  \b-hadron decays, without distinguishing the decaying species.  Often
  the knowledge of the mixture composition is limited, which makes these
  measurements experiment or accelerator specific.  Also, these
  measurements have to rely on Monte Carlo for estimating the
  $\beta\gamma$ factor, because the decaying hadrons are not fully
  reconstructed.  On the bright side, these usually are the largest
  statistics \b-hadron lifetime measurements that are accessible to a
  given experiment, and can, therefore, serve as an important
  performance benchmark.
\item {\bf\em Measurements in semileptonic decays of a specific
  {\boldmath $H_b$\unboldmath}}.  \particle{W}from \particle{\b\to Wc}
  produces $\ell\nu_l$ pair (\particle{\ell=e,\mu}) in about 21\% of the
  cases.  Electron or muon from such decays is usually a well-detected
  signature, which provides for clean and efficient trigger.
  \particle{c} quark from \particle{b\to Wc} transition and the other
  quark(s) making up the decaying $H_b$ combine into a charm hadron,
  which is reconstructed in one or more exclusive decay channels.
  Knowing what this charmed hadron is allows one to separate, at least
  statistically, different $H_b$ species.  The advantage of these
  measurements is in statistics which is usually superior to that of the
  exclusively reconstructed $H_b$ decays.  Some of the main
  disadvantages are related to the difficulty of estimating lepton+charm
  sample composition and Monte Carlo reliance for the $\beta\gamma$
  factor estimate.
\item {\bf\em Measurements in exclusively reconstructed decays}.  These
  have the advantage of complete reconstruction of decaying $H_b$, which
  allows one to infer the decaying species as well as to perform precise
  measurement of the $\beta\gamma$ factor.  Both lead to generally
  smaller systematic uncertainties than in the above two categories.
  The downsides are smaller branching ratios, larger combinatoric
  backgrounds, especially in $H_b\rightarrow H_c\pi(\pi\pi)$ and
  multi-body $H_c$ decays, or in a hadron collider environment with
  non-trivial underlying event.  $H_b\to J/\psi H_s$ are relatively
  clean and easy to trigger on $J/\psi\to \ell^+\ell^-$, but their
  branching fraction is only about 1\%.
\item {\bf\em Measurements at asymmetric B factories}. In 
  the $\Ups\rightarrow B \bar{B}$ decay, the \B mesons (\Bu or \Bd) are
  essentially at rest in the \Ups rest frame.  This makes lifetime
  measurements impossible with experiments, such as CLEO, in which \Ups
  produced at rest.  At asymmetric \B factories \Ups is boosted
  resulting in \B and \particle{\bar{B}} moving nearly parallel to each
  other.  The lifetime is inferred from the distance $\Delta z$
  separating \B and \particle{\bar{B}} decay vertices and \Ups boost
  %% vertices is nt in many spell-checkers, but is correct according to
  %% most dictionaries
  known from colliding beam energies.  In order to maximize the
  precision of the measurement one \B meson is reconstructed in the
  \particle{D^{(*)}\ell\nu_{\ell}} decay.  The other \B is typically not
  fully reconstructed, only position of its decay vertex is determined.
  These measurements benefit from very large statistics, but suffer from
  poor $\Delta z$ resolution.
\item {\bf\em Direct measurement of lifetime ratios}.  This method has
  so far been only applied in the measurement of $\tau(\Bu)/\tau(\Bd)$.
  The ratio of the lifetimes is extracted from the dependence of the
  observed relative number of \Bu and \Bd candidates (both reconstructed
  in semileptonic decays) on the proper decay time.
\end{enumerate}

In some of the latest analyses, measurements of two (\eg\ $\tau(\Bu)$ and
$\tau(\Bu)/\tau(\Bd)$) or three (\eg\ $\tau(\Bu)$,
$\tau(\Bu)/\tau(\Bd)$, and \dmd) quantities are combined.  This
introduces correlations among measurements.  Another source of
correlations among the measurements are the systematic effects, which
could be common to an experiment or to an analysis technique across the
experiments.  When calculating the averages, such correlations are taken
into account per general procedure, described in
\Ref{lifetime_details}.

%% ====================================================================
\mysubsubsection{Inclusive \b-hadron lifetimes}
%% ====================================================================

The inclusive \b hadron lifetime is defined as $\tau_{\b} = \sum_i f_i
\tau_i$ where $\tau_i$ are the individual \b-hadron lifetimes and $f_i$
the fractions of the various species present in an unbiased sample of
weakly-decaying \b hadrons produced at a high-energy
collider.\footnote{In principle such a quantity could be slightly
different in \particle{Z} decays and a the Tevatron, in case the
fractions of \b-hadron species are not exactly the same; see the
discussion in \Sec{fractions_high_energy}.}  This quantity is certainly
less fundamental than the lifetimes of the individual \b hadron species,
the latter being much more useful in the comparison of measurements with
theoretical predictions.  Nonetheless, we present the measurements of
the inclusive lifetime for completeness.

\begin{table}[tp]
\caption{Measurements of average \b-hadron lifetimes.}
\labt{lifeincl}
\begin{center}
\begin{tabular}{lcccl} \hline
Experiment & Method            &Data set & $\tau_{\b}$ (ps)            & Ref. \\   
\hline
ALEPH      & Dipole            &   91    & $1.511 \pm 0.022 \pm 0.078$ & \cite{ALEIN2}\\
DELPHI     & All track i.p.\ (2D)&91--92 & $1.542 \pm 0.021 \pm 0.045$ & \cite{DELIN0}$^a$\\
DELPHI     & Sec.\ vtx         & 91--93  & $1.582 \pm 0.011 \pm 0.027$ & \cite{DELIN}$^a$\\
DELPHI     & Sec.\ vtx         & 94--95  & $1.570 \pm 0.005 \pm 0.008$ & \cite{DELB04}\\
L3         & Sec.\ vtx + i.p.  & 91--94  & $1.556 \pm 0.010 \pm 0.017$ & \cite{L3IN1}$^b$ \\
OPAL       & Sec.\ vtx         & 91--94  & $1.611 \pm 0.010 \pm 0.027$ & \cite{OPAIN2}\\
SLD        & Sec.\ vtx         &  93     & $1.564 \pm 0.030 \pm 0.036$ & \cite{SLDIN} \\ 
\hline
\multicolumn{2}{l}{Average set 1 (\b vertex)} && $1.576 \pm 0.008$ & \\   
\hline\hline
ALEPH      & Lepton i.p.\ (3D) & 91--93  & $1.533 \pm 0.013 \pm 0.022$ & \cite{ALEIN1}\\
L3         & Lepton i.p.\ (2D) & 91--94  & $1.544 \pm 0.016 \pm 0.021$ & \cite{L3IN1}$^b$\\
OPAL       & Lepton i.p.\ (2D) & 90--91  & $1.523 \pm 0.034 \pm 0.038$ & \cite{OPAIN1}\\ 
\hline
\multicolumn{2}{l}{Average set 2 ($\b\to\ell$)} && $1.537 \pm 0.020$ & \\   
\hline\hline
CDF        & \particle{J/\psi} vtx
                               &  92--95 & $1.533 \pm 0.015 ^{+0.035}_{-0.031}$ & \cite{CDFIN_BS1} \\ 
%% CDF        & \particle{J/\psi} vtx
%%                                &  02--03 & $1.526 \pm 0.034 \pm 0.035$ & \cite{CDFB03} \\ 
%% \hline
%% \multicolumn{2}{l}{Average set 3 (\particle{\b\to J/\psi})} && $1.533 \pm 0.036$ & \\ 
\hline\hline
%% \multicolumn{2}{l}{Average of all above} && $1.573 \pm 0.008$ & \\
\multicolumn{2}{l}{Average of all above} && $1.574 \pm 0.008$ & \\
\hline
\multicolumn{5}{l}{$^a$ \footnotesize The combined DELPHI result quoted in
\cite{DELIN} is 1.575 $\pm$ 0.010 $\pm$ 0.026 ps.} \\[-0.5ex]
\multicolumn{5}{l}{$^b$ \footnotesize The combined L3 result quoted in \cite{L3IN1} 
is 1.549 $\pm$ 0.009 $\pm$ 0.015 ps.}
\end{tabular}
\end{center}
\end{table}

In practice, an unbiased measurement of the inclusive lifetime is
difficult to achieve, because it would imply an efficiency which is
guaranteed to be the same across species.  So most of the measurements
are biased.
In an attempt to group analyses which are expected to select the same
mixture of \b hadrons, the available results (given in \Table{lifeincl})
are divided into the following three sets:
\begin{enumerate}
\item measurements at LEP and SLD that accept any \b-hadron decay, based 
      on topological reconstruction (secondary vertex or impact
      parameters) using charged tracks;
\item measurements at LEP based on the identification
      of a lepton from a \b decay; and
\item measurements at the Tevatron based on inclusive 
      \particle{\b\to J/\psi} reconstruction, where the
      \particle{J/\psi} is fully reconstructed.
\end{enumerate}

The measurements of the first set are generally considered as estimates
of $\tau_{\b}$, although the efficiency to reconstruct a secondary vertex 
most probably depends, in an analysis-dependent way, on the number charged tracks 
from a \b hadron decay, which in turn should depend on the type of \b hadron.
Even though these efficiency variations can in principle be accounted for 
using Monte Carlo simulations (which inevitably contain assumptions on 
branching fractions), the \b hadron mixture in that case can remain
somewhat ill-defined and could be slightly different between analyses 
in that set. 

On the contrary, the mixtures corresponding to the other two sets of
measurements are better defined, in the limit where the reconstruction
and selection efficiency of a lepton or a \particle{J/\psi} from a \b
hadron does not depend on the type of that \b hadron. These mixtures are
given by the production fractions and the inclusive branching fractions
for each \b hadron species to give a lepton or a \particle{J/\psi}. In
particular, under the assumption that all \b hadrons have the same
semileptonic decay width, the analyses of the second set should measure
$\tau(\b\to\ell) = (\sum_i f_i \tau_i^2) /(\sum_i f_i \tau_i)$ which is
necessarily larger than $\tau_{\b}$ if lifetime differences exist.
Given the present knowledge on $\tau_i$ and $f_i$,
$\tau(\b\to\ell)-\tau_{\b}$ is expected to be of the order of 0.01\ps.

For the averaging, correlated systematic are taken into account, which are due to 
\b and \particle{c} fragmentation, \b and \particle{c} decay models,
\BR{B\to\ell}, \BR{B\to c\to\ell}, $\tau_{\particle{c}}$,
\BR{c\to\ell}, and \B charged track decay multiplicity.
The averages for the sets defined above (also given in \Table{lifeincl}) are
\begin{eqnarray}
\tau(\b~\mbox{vertex}) &=& \HFAGtaubvertex \,,\\
\tau(\b\to\ell) &=& \HFAGtaublepton  \,, \\
\tau(\b\to\particle{J/\psi}) &=& \HFAGtaubJpsi\,,
\end{eqnarray}
whereas an average of all measurements, ignoring mixture differences, 
yields \HFAGtauball.

%% ====================================================================
\mysubsubsection{\Bd and \Bu lifetimes}
%% ====================================================================
\labs{taubd}
\labs{taubu}
\labs{lifetime_ratio}

In LEP and CDF experiments, the most precise measurements of the
\Bd and \Bu lifetimes
have originated from two classes of partially reconstructed decays.
In the first class the decay 
\particle{B\to \bar{D}^{(*)} \ell^+ \nu_{\ell} \mbox{$X$}}
is used in which the charge of the charmed meson distinguishes between
neutral and charged \B mesons; the ratio \Bu/\Bd 
lifetime ratio is usually also extracted directly, as 
the \Bd and \Bu lifetime measurements are correlated.
In the second class the charge attached to
the \b-decay vertex is used to achieve this separation.

With the benefit from very large statistics, the asymmetric \B-factory
experiments, \babar and \belle, now provide more precise lifetime measurements 
using exclusively reconstructed decays, as well as partially reconstructed
decays.
With increased data sample, CDF also provides improved measurements
using large samples of exclusive
\particle{\Bd\to J/\psi K^{(*)0}} and \particle{\Bu\to J\psi K^+} decays.

% While individual lifetimes are often of interest to experiments, \eg\ in
% extraction of CKM matrix elements, the ratios of the lifetimes are more
% interesting from the theoretical perspective as they are predicted more
% precisely.

\begin{table}[tp]
\caption{Measurements of the \Bd lifetime.}
\labt{lifebd}
\begin{center}
\begin{tabular}{lcccl} \hline
Experiment & Method                    & Data set      & $\tau(\Bd)$ (ps)           & Ref.  \\  
\hline
ALEPH  & \particle{D^{(*)} \ell}       & 91--95 & $1.518\pm 0.053\pm 0.034$       & \cite{ALEB01}\\
ALEPH  & Exclusive                     & 91--94 & $1.25^{+0.15}_{-0.13} \pm 0.05$ & \cite{ALEB0} \\
ALEPH  & Partial rec.\ $\pi^+\pi^-$    & 91--94 & $1.49^{+0.17+0.08}_{-0.15-0.06}$& \cite{ALEB0} \\
CDF    & \particle{D^{(*)} \ell}       & 92--95 & $1.474\pm 0.039^{+0.052}_{-0.051}$ & \cite{CDFB02}\\
CDF    & Excl. \particle{J/\psi K}     & 92--95 & $1.497\pm 0.073\pm 0.032$       & \cite{CDFB01}\\
CDF    & Excl. \particle{J/\psi K}     & 02--03 & $1.49\pm 0.06\pm0.062$          & \cite{CDFB03}$^p$\\
%% the above and what's in the reference (EPS HEP 2003, Aachen) somehow
%% differ.  I don't worry all that much, b/c this will hopefully get
%% superseeded before the final edition of the paper
DELPHI & \particle{D^{(*)} \ell}       & 91--93 & $1.61^{+0.14}_{-0.13} \pm 0.08$ & \cite{DELB01}\\
DELPHI & Charge sec.\ vtx              & 91--93 & $1.63 \pm 0.14 \pm 0.13$        & \cite{DELB02}\\
DELPHI & Inclusive \particle{D^* \ell} & 91--93 & $1.532\pm 0.041\pm 0.040$       & \cite{DELB03}\\
DELPHI & Charge sec.\ vtx              & 94--95 & $1.531 \pm 0.021\pm0.031$       & \cite{DELB04}\\
L3     & Charge sec.\ vtx              & 94--95 & $1.52 \pm 0.06 \pm 0.04$        & \cite{L3B01} \\
OPAL   & \particle{D^{(*)} \ell}       & 91--93 & $1.53 \pm 0.12 \pm 0.08$        & \cite{OPAB0} \\
OPAL   & Charge sec.\ vtx              & 93--95 & $1.523\pm 0.057\pm 0.053$       & \cite{OPAB1} \\
OPAL   & Inclusive \particle{D^* \ell} & 91--00 & $1.541\pm 0.028\pm 0.023$       & \cite{OPAB2} \\
SLD    & Charge sec.\ vtx $\ell$       & 93--95 & $1.56^{+0.14}_{-0.13} \pm 0.10$ & \cite{SLDB01}$^a$\\
SLD    & Charge sec.\ vtx              & 93--95 & $1.66 \pm 0.08 \pm 0.08$        & \cite{SLDB01}$^a$\\
\babar & Exclusive                     & 99--00 & $1.546\pm 0.032\pm 0.022$       & \cite{BABAR1}\\
\babar & Inclusive \particle{D^* \ell} & 99--01 & $1.529\pm 0.012\pm 0.029$       & \cite{BABAR2}\\
\babar & Exclusive \particle{D^* \ell} & 99--02 & $1.523^{+0.024}_{-0.023}\pm 0.022$ & \cite{BABAR3}\\
\babar & Incl.\ \particle{D^*\pi}, \particle{D^*\rho} 
                                       & 99--01 & $1.533\pm 0.034 \pm 0.038$      & \cite{BABAR4} \\
\belle & Exclusive                     & 00--01 & $1.554\pm 0.030 \pm 0.019$      & \cite{BELLE1} \\ 
\hline
Average&                               &        & $1.534 \pm 0.013 $              & \\   
\hline\hline
\multicolumn{5}{c}{Recent measurements not yet included in the
average}\\
\hline
CDF    & Excl. \particle{J/\psi K}     & 02--04 & $1.539\pm 0.051\pm 0.008$       & \cite{CDFBS2}$^{b,p}$\\
\babar & Inclusive \particle{D^* \ell} & 99--04 & $1.501\pm 0.008\pm 0.030$ & \cite{BABAR5}\\ %% 99 in this line needs to be verified
               %- 04 also. 81/fb, i.e. by summer02 (to be confirmed by David) 
\belle & Exclusive                     & 00--03 & $1.534\pm 0.008\pm 0.010$
& \cite{BELLE2}$^c$\\  % Belle   not use 99 data, for 140/fb by summer03  
\hline\hline           %                     though reported in summer04
\multicolumn{5}{l}{$^a$ \footnotesize The combined SLD result 
quoted in \cite{SLDB01} is 1.64 $\pm$ 0.08 $\pm$ 0.08 ps.}\\[-0.5ex]
\multicolumn{5}{l}{$^b$ {\footnotesize To replace~\cite{CDFB03}.}
              ~~   $^c$ {\footnotesize To replace~\cite{BELLE1}.}
              ~~   $^p$ {\footnotesize Preliminary.}}
\end{tabular}
\end{center}
\end{table}

The averaging is summarized in \Tablesss{lifebd}{lifebu}{liferatioBuBd}.
The following sources of
correlated systematic uncertainties have been considered:
% (central values and errors scaled accordingly):
\particle{D^{**}} branching ratio uncertainties~\cite{LEPHFS},
momentum estimation of \B mesons from \particle{Z^0} decays
(\b-quark fragmentation parameter
$\langle X_E \rangle = 0.702 \pm 0.008$~\cite{LEPHFS}),
\Bs and \b-baryon lifetimes (see \Secss{taubs}{taulb}),
and \b-hadron fractions at high energy (see \Table{fractions}).  
The world averages are:
\begin{eqnarray}
\tau(\Bd) & = & \HFAGtauBd \,, \\
\tau(\Bu) & = & \HFAGtauBu \,, \\
\tau(\Bu)/\tau(\Bd) & = & \HFAGtauBuBd \,.
\end{eqnarray}
As indicated in the tables, these averages do not include
recent CDF~\cite{CDFBS2}, \dzero~\cite{D0B01}, \babar~\cite{BABAR5}, 
and \belle~\cite{BELLE2} results; it is planned to incoporate them 
in the next update of this report. 

%% ref CDFBS2 needs to be updated (i.e. this should become a separate
%% public CDF note)
%% CDF Collaboration, CDF Note 6551
%% http://www-cdf.fnal.gov/physics/new/bottom/040428.blessed-lft/blessed-lft.ps
%% I am working on making this happen

\begin{table}[tp]
\caption{Measurements of the \Bu lifetime.}
%           a) The combined ALEPH result quoted in \cite{ALEB0} is 1.58 $\pm$ 0.09 $\pm$ 0.03 ps.\\            
% p) preliminary\\            
\labt{lifebu}
\begin{center}
\begin{tabular}{lcccl} \hline
Experiment & Method                 & Data set      & $\tau(\Bu)$ (ps)           & Ref.         \\
\hline
ALEPH  & \particle{D^{(*)} \ell}    & 91--95 & $1.648\pm 0.049 \pm 0.035$        & \cite{ALEB01}\\
ALEPH  & Exclusive                  & 91--94 & $1.58^{+0.21+0.04}_{-0.18-0.03}$  & \cite{ALEB0} \\
CDF    & \particle{D^{(*)} \ell}    & 92--95 & $1.637\pm 0.058^{+0.045}_{-0.043}$& \cite{CDFB02}\\
CDF    & Excl.\ \particle{J/\psi K} & 92--95 & $1.636 \pm 0.058 \pm 0.025$       & \cite{CDFB01}\\
CDF    & Excl. ($J/\psi K$)         & 02--03 & $1.64 \pm 0.05 \pm 0.02$          & \cite{CDFB03}$^p$\\
DELPHI & \particle{D^{(*)} \ell}    & 91--93 & $1.61 \pm 0.16 \pm 0.12$          & \cite{DELB01}$^a$ \\
DELPHI & Charge sec.\ vtx           & 91--93 & $1.72 \pm 0.08 \pm 0.06$          & \cite{DELB02}$^a$ \\
DELPHI & Charge sec.\ vtx           & 94--95 & $1.624 \pm 0.014 \pm 0.018$       & \cite{DELB04}\\
L3     & Charge sec.\ vtx           & 94--95 & $1.66\pm  0.06 \pm 0.03$          & \cite{L3B01} \\
OPAL   & \particle{D^{(*)} \ell}    & 91--93 & $1.52 \pm 0.14 \pm 0.09$          & \cite{OPAB0} \\
OPAL   & Charge sec.\ vtx           & 93--95 & $1.643\pm 0.037 \pm 0.025$        & \cite{OPAB1} \\
SLD    & Charge sec.\ vtx $\ell$    & 93--95 & $1.61^{+0.13}_{-0.12} \pm 0.07$   & \cite{SLDB01}$^b$ \\
SLD    & Charge sec.\ vtx           & 93--95 & $1.67\pm 0.07 \pm 0.06$           & \cite{SLDB01}$^b$ \\
\babar & Exclusive                  & 99--00 & $1.673\pm 0.032 \pm 0.023$        & \cite{BABAR1}\\
\belle & Exclusive                  & 00--01 & $1.695\pm 0.026 \pm 0.015$        & \cite{BELLE1}\\
\hline
Average &                           &        & $1.652 \pm 0.014$                 & \\
\hline\hline
\multicolumn{5}{c}{Recent measurements not yet included in the
average}\\
\hline
CDF    & Excl. \particle{J/\psi K}     & 02--04 & $1.662\pm 0.033\pm 0.008$       & \cite{CDFBS2}$^{c,p}$\\
\belle & Exclusive                     & 00--03 & $1.635\pm 0.011\pm 0.011$       & \cite{BELLE2}$^d$\\
\hline\hline
\multicolumn{5}{l}{$^a$ \footnotesize The combined DELPHI result quoted 
in~\cite{DELB02} is $1.70 \pm 0.09$ ps.} \\[-0.5ex]
\multicolumn{5}{l}{$^b$ \footnotesize The combined SLD result 
quoted in~\cite{SLDB01} is $1.66 \pm 0.06 \pm 0.05$ ps.}\\[-0.5ex]
\multicolumn{5}{l}{$^c$ {\footnotesize To replace~\cite{CDFB03}.}
              ~~   $^d$ {\footnotesize To replace~\cite{BELLE1}.}
              ~~   $^p$ {\footnotesize Preliminary.}}
\end{tabular}
\end{center}
\end{table}

\begin{table}[tb]
\caption{Measurements of the ratio $\tau(\Bu)/\tau(\Bd)$.}
\labt{liferatioBuBd}
\begin{center}
\begin{tabular}{lcccl} 
\hline
Experiment & Method & Data set & Ratio $\tau(\Bu)/\tau(\Bd)$ & Ref. \\
\hline
ALEPH  & \particle{D^{(*)} \ell} & 91--95 & $1.085 \pm 0.059 \pm 0.018$      & \cite{ALEB01}\\
ALEPH  & Exclusive               & 91--94 & $1.27^{+0.23+0.03}_{-0.19-0.02}$ & \cite{ALEB0} \\
CDF    & \particle{D^{(*)} \ell} & 92--95 & $1.110\pm 0.056^{+0.033}_{-0.030}$ & \cite{CDFB02}\\
CDF    & Excl.\ \particle{J/\psi K} & 92--95 & $1.093 \pm 0.066 \pm 0.028$   & \cite{CDFB01}\\
DELPHI & \particle{D^{(*)} \ell} & 91--93 & $1.00^{+0.17}_{-0.15} \pm 0.10$  & \cite{DELB01}\\
DELPHI & Charge sec.\ vtx        & 91--93 & $1.06^{+0.13}_{-0.11} \pm 0.10$  & \cite{DELB02}\\
DELPHI & Charge sec.\ vtx        & 94--95 & $1.060 \pm 0.021 \pm 0.024$      & \cite{DELB04}\\
L3     & Charge sec.\ vtx        & 94--95 & $1.09  \pm 0.07  \pm 0.03$       & \cite{L3B01} \\
OPAL   & \particle{D^{(*)} \ell} & 91--93 & $0.99  \pm 0.14^{+0.05}_{-0.04}$ & \cite{OPAB0} \\
OPAL   & Charge sec.\ vtx        & 93--95 & $1.079 \pm 0.064 \pm 0.041$      & \cite{OPAB1} \\
SLD    & Charge sec.\ vtx $\ell$ & 93--95 & $1.03^{+0.16}_{-0.14} \pm 0.09$  & \cite{SLDB01}$^a$\\
SLD    & Charge sec.\ vtx        & 93--95 & $1.01^{+0.09}_{-0.08} \pm 0.05$  & \cite{SLDB01}$^a$\\
\babar & Exclusive               & 99--00 & $1.082 \pm 0.026 \pm 0.012$      & \cite{BABAR1}\\
\belle & Exclusive               & 00--01 & $1.091 \pm 0.023 \pm 0.014$      & \cite{BELLE1}\\
\hline
Average &                        &        & $1.081 \pm 0.015$                & \\   
\hline\hline
\multicolumn{5}{c}{Recent measurements not yet included in the average}\\
\hline
\dzero & \particle{D^{*+} \mu} \particle{D^0 \mu} ratio
	                             & 02--04 & $1.080\pm 0.016 \pm 0.014$   & \cite{D0B01} \\ 
CDF    & Excl.\ \particle{J/\psi K} & 02--04 & $1.080\pm 0.042$              & \cite{CDFBS2}$^p$\\
\belle & Exclusive               & 00--03 & $1.066 \pm 0.008 \pm 0.008$      & \cite{BELLE2}$^b$\\
\hline\hline
\multicolumn{5}{l}{$^a$ \footnotesize The combined SLD result quoted
	   in~\cite{SLDB01} is $1.01 \pm 0.07 \pm 0.06$.} \\[-0.5ex] 
\multicolumn{5}{l}{$^b$ {\footnotesize To replace~\cite{BELLE1}.}
              ~~   $^p$ {\footnotesize Preliminary.}}
\end{tabular}
\end{center}
\end{table}

  % from Konstantin
% Bs lifetime, Bc lifetime, Lambda_b and b-baryon lifetimes, theoretical predictions
%%%%%%%%%%%%%%%%%%%%%%%%%%%%%%%%%%%%%%%%%%%%%%%%
%
% This is file life_mix_tau2.tex containing
% the second part of the chapter on the b-hadron lifetimes: 
% Bs, Bc, lambda_b and b-baryon lifetimes
% as well as theroretical predictions for all b-hadron lifetimes.
%
%%%%%%%%%%%%%%%%%%%%%%%%%%%%%%%%%%%%%%%%%%%%%%%
%

\mysubsubsection{\Bs lifetime}
\labs{taubs}

Some of the more precise measurements of the \Bs lifetime
originate from partially reconstructed decays in which
a \particle{D^-_s} meson has been completely reconstructed.

The following correlated systematic errors were considered:
average \B lifetime used in backgrounds,
\Bs decay multiplicity, and branching ratios used to determine 
backgrounds (\eg\ \BR{B\to D_s D}).

A knowledge of the multiplicity of \Bs decays is important for
measurements that partially reconstruct the final state such as 
\particle{\B\to D_s \mbox{$X$}} (where $X$ is not a lepton). 
The boost deduced from Monte Carlo simulation depends on the multiplicity used.
Since this is not well known, the multiplicity in the simulation is
varied and this range of values observed is taken to be a systematic.

Similarly not all the branching ratios for the potential background
processes are measured. Where they are available, the PDG values are
used for the error estimate. Where no measurements are available
estimates can usually be made by using measured branching ratios of
related processes and using some reasonable extrapolation.

The inputs used to form the average \Bs lifetime
are given in \Table{lifebs}.

% , and the world average is:
% \begin{equation}
% \tau(\Bs) = \HFAGtauBs \,.
% \end{equation}

\begin{table}[tb]
\caption{Measurements of the \Bs lifetime.}
% Excluding the CDF J/$\psi \phi$ measurement the average is : $ \tau_{{\rm B}_s} = 1.471\pm 0.059$.}
% To extract the \Bs lifetime,
% a mean \b hadron lifetime of 1.4 ps was assumed in the papers flagged with an (a).
% In this case the numbers quoted in the Table have been scaled
% to the current average $B$ hadron lifetime of 1.55 ps.
% a) The combined DELPHI result quoted
%    in \cite{DELBS2} is 1.67 $\pm$ 0.14 ps.}
\labt{lifebs}
\begin{center}
\begin{tabular}{lcccl} \hline
Experiment & Method           & Data set & $\tau(\Bs)$ (ps)               & Ref. \\
\hline
ALEPH  & \particle{D_s \ell}  & 91--95 & $1.54^{+0.14}_{-0.13}\pm 0.04$   & \cite{ALEBS1}          \\
CDF    & \particle{D_s \ell}  & 92--96 & $1.36\pm 0.09 ^{+0.06}_{-0.05}$  & \cite{CDFBS}           \\
DELPHI & \particle{D_s \ell}  & 91--95 & $1.42^{+0.14}_{-0.13}\pm 0.03$   & \cite{DELBS0}          \\
OPAL   & \particle{D_s \ell}  & 90--95 & $1.50^{+0.16}_{-0.15}\pm 0.04$   & \cite{OPABS1_OPALAM2}  \\ \hline
\multicolumn{3}{l}{Average of \particle{D_s \ell} measurements} &  $1.442\pm 0.066$ & \\  % \HFAGtauBsfs
\hline
ALEPH  & \particle{D_s h}     & 91--95 & $1.47\pm 0.14\pm 0.08$           & \cite{ALEBS2}          \\
DELPHI & \particle{D_s h}     & 91--95 & $1.53^{+0.16}_{-0.15}\pm 0.07$   & \cite{DELBS1_dms_excl} \\
DELPHI & \particle{D_s} incl. & 91--94 & $1.60\pm 0.26^{+0.13}_{-0.15}$   & \cite{DELBS2}          \\
OPAL   & \particle{D_s} incl. & 90--95 & $1.72^{+0.20+0.18}_{-0.19-0.17}$ & \cite{OPABS2}          \\ 
\hline
\multicolumn{3}{l}{Average of all above \particle{D_s} measurements} &  $1.469 \pm 0.059$ & \\ 
\hline\hline
CDF      & \particle{J/\psi\phi} & 92--95  & $1.34^{+0.23}_{-0.19}    \pm 0.05$ & \cite{CDFIN_BS1} \\
CDF      & \particle{J/\psi\phi} & 02--04  & $1.369 \pm 0.100 ^{+0.008}_{-0.010}$ & \cite{CDFBS2}$^p$ \\
\dzero   & \particle{J/\psi\phi} & 02--04  & $1.444^{+0.098}_{-0.090} \pm 0.02$ & \cite{D0BS1}  \\ \hline 
\multicolumn{3}{l}{Average of \particle{J/\psi \phi} measurements} &  $1.404 \pm 0.066$ & \\ 
\hline
\multicolumn{5}{l}{$^p$ \footnotesize Preliminary.}
\end{tabular}
\end{center}
\end{table}

It is important to note that
similar to the kaon system, neutral \B mesons contain
short- and long-lived components, since the two
mass eigenstates $B_L$ and $B_H$ differ not only
in their masses, but also in their widths with
$\Delta\Gamma = \Gamma_L - \Gamma_H$. In the 
Standard Model for the \Bs system, 
this difference can be large, \ie,
$\DGs/\Gs = 0.12 \pm 0.06$~\cite{delta_gams}.
Specific measurements of this parameter are explained
in more detail in \Sec{DGs}.

Flavor-specific decays, such as semileptonic
$\particle{B_s} \to \particle{D_s \ell \nu}$ will
have equal contributions of $\tau_H = 1/\Gamma_H$ and
$\tau_L = 1/\Gamma_L$. For this reason, lifetime measurements
via these decays are broken out separately
as given in \Table{lifebs}, and their world average is:
\begin{equation}
\tau(\Bs)_{\particle{D_s\ell}} = \HFAGtauBsfs \,.
\end{equation}

Any final state can be decomposed into its
\CP-even and \CP-odd component, and in the Standard
Model, ${\DGs}_{\CP} = \DGs$. Fully exclusive decays of \Bs into
\particle{\Bs\to J/\psi\phi} are expected to be
dominated by the \CP-even state and its lifetime.
First measurements of the \CP mix for this decay mode
are outlined in \Sec{DGs}.
CDF and \dzero measurements from this particular mode
\particle{\Bs\to J/\psi\phi} are combined into an
average
given in \Table{lifebs}.  There are no correlations
between the measurements for this fully exclusive
channel, and the world average for this 
specific decay is:
\begin{equation}
\tau(\Bs)_{\particle{J/\psi \phi}} = \HFAGtauBsjpsi \,.
\end{equation}

Finally, the remaining measurements are variations
of lifetimes determined more inclusively 
via \particle{D_s} plus
hadrons, and hence into a more unknown mixture
of flavors and/or \CP-states.  A lifetime
weighted this way can still be a useful input
for analyses examining such an inclusive sample.
These are separated in \Table{lifebs} and combined
with the semileptonic lifetime to obtain:
\begin{equation}
\tau(\Bs)_{\particle{D_s {\rm X}}} = \HFAGtauBs \,.
\end{equation}

\mysubsubsection{\Bc lifetime}
\labs{taubc}

There are currently two measurements of the lifetime of the \Bc meson
from CDF~\cite{CDFBC1} and \dzero~\cite{D0BC1} using the semileptonic decay
mode \particle{\Bc \to J/\psi \ell} and fitting
simultaneously to the mass and lifetime using the vertex formed
from the leptons from the decay of the \particle{J/\psi} and
the third lepton. Correction factors
to estimate the boost due to the missing neutrino are used.
Mass values of
$6.40 \pm 0.39 \pm 0.13$~GeV/$c^2$ and 
$5.95^{+0.14}_{-0.13} \pm 0.34$~GeV/$c^2$, respectively, are
found by fitting
to the tri-lepton invariant mass spectrum. These mass measurements
are consistent to within errors, and no adjustments to the lifetimes
are made.  Correlated systematic errors include the impact
of the uncertainty of the \Bc $p_T$ spectrum on the correction
factors, the level of feed-down from $\psi(2S)$, 
MC modeling of the decay model varying from phase space
to the ISGW model, and mass variations.
Inputs are given in \Table{lifebc} and the world average is
determined to be:
\begin{equation}
\tau(\Bc) = \HFAGtauBc \,.
\end{equation}

\begin{table}[tb]
\caption{Measurements of the \Bc lifetime.}
\labt{lifebc}
\begin{center}
\begin{tabular}{lcccl} \hline
Experiment & Method                    & Data set  & $\tau(\Bc)$ (ps)
      & Ref.\\   \hline
CDF        & \particle{J/\psi \ell} & 92--95  & 0.46$^{+0.18}_{-0.16} \pm$
 0.03   & \cite{CDFBC1}  \\ 
 \dzero & \particle{J/\psi \mu} & 02--04  & $0.448^{+0.123}_{-0.096} 
\pm  0.121$   & \cite{D0BC1}$^p$  \\ \hline
  \multicolumn{2}{l}{Average} &   &  0.45 $\pm$ 0.12
                 &    \\   \hline
\multicolumn{5}{l}{$^p$ \footnotesize Preliminary.}
\end{tabular}
\end{center}
\end{table}

\mysubsubsection{\Lb and \b-baryon lifetimes}
\labs{taulb}

The most precise measurements of the \b-baryon lifetime
originate from two classes of partially reconstructed decays.
In the first class, decays with an exclusively 
reconstructed \Lc baryon
and a lepton of opposite charge are used. These products are
more likely to occur in the decay of \Lb baryons.
In the second class, more inclusive final states with a baryon
(\particle{p}, \particle{\bar{p}}, $\Lambda$, or $\bar{\Lambda}$) 
and a lepton have been used, and these final states can generally
arise from any \b baryon.

The following sources of correlated systematic uncertainties have 
been considered:
experimental time resolution within a given experiment, \b-quark
fragmentation distribution into weakly decaying \b baryons,
\Lb polarization, decay model,
and evaluation of the \b-baryon purity in the selected event samples.
In computing the averages
the central values of the masses are scaled to 
$M(\Lb) = 5624 \pm 9\MeVcc$~\cite{PDGmass} and
$M(\mbox{\b-baryon}) = 5670 \pm 100\MeVcc$.

The meaning of decay model and the correlations are not always clear.
Uncertainties related to the decay model are dominated by
assumptions on the fraction of $n$-body decays.
To be conservative it is assumed
that it is correlated whenever given as an error.
DELPHI varies the fraction of 4-body decays from 0.0 to 0.3. 
In computing the average, the DELPHI
result is corrected for $0.2 \pm 0.2$.

Furthermore, in computing the average,
the semileptonic decay results are corrected for a polarization of 
$-0.45^{+0.19}_{-0.17}$~\cite{LEPHFS} and a 
\Lb fragmentation parameter
$\langle X_E \rangle =0.70\pm 0.03$~\cite{LBFRAG}.

%The ALEPH result for $\Lambda_b$ polarisation is -0.23 $pm$ 0.25
%(CERN-PPE/95-156) while the others use -0.47 +- 0.47.
%The corresponding results and error are corrected for the ALEPH measurement.

%\par Considering only the measurements obtained with $\Lambda_c \ell$ correlations
%     and $\Lambda \ell^- \ell^+$ the average is :
% $$ \tau_{\Lambda_b} = 1.24^{+0.08}_{-0.08}~ps$$

%     Considering the measurements obtained with $\Lambda$-lepton correlation
%     and with $p \mu$ correlation (b-baryons Admixture)
%     the average is :
% $$ \tau_{\Lambda_b} = 1.15^{+0.08}_{-0.08}~ps$$

Inputs to the averages are given in \Table{lifelb}.
The world average lifetime of \b baryons is then:
\begin{equation}
\langle\tau(\mbox{\b-baryon})\rangle = \HFAGtaubbaryon \,.
\end{equation}
Keeping only \particle{\Lambda^{\pm}_c \ell^{\mp}}
and $\Lambda \ell^- \ell^+$ final states, as representative of
the \Lb baryon, the following lifetime is obtained:
\begin{equation}
\tau(\Lb) = \HFAGtauLb \,. 
\end{equation}

Averaging the measurements based on the $\Xi^{\mp} \ell^{\mp}$
final states~\cite{ALEPH_fxi,DELPHI_fxi} gives
a lifetime value for a sample of events
containing $\Xib^0$ and $\Xib^-$ baryons:
\begin{equation}
\langle\tau(\Xib)\rangle = \HFAGtauXib \,.
\end{equation}

\begin{table}[t]
\caption{Measurements of the \b-baryon lifetimes.
%Measurements of the \b-baryon and \Lb lifetime.
%The DELPHI and ALEPH $\Xi \ell$ results are not included 
%in the quoted average since the selected data samples
%contain mostly \Xib while 
%the data samples in the other measurements contain mostly \Lb.
}
\labt{lifelb}
\begin{center}
\begin{tabular}{lcccl} 
\hline
Experiment&Method                &Data set& Lifetime (ps) & Ref. \\\hline
ALEPH  &$\Lc\ell$             & 91--95 &$1.18^{+0.13}_{-0.12} \pm 0.03$ & \cite{ALEPH_fla}\\
ALEPH  &$\Lambda\ell^-\ell^+$ & 91--95 &$1.30^{+0.26}_{-0.21} \pm 0.04$ & \cite{ALEPH_fla}\\
CDF    &$\Lc\ell$             & 91--95 &$1.32 \pm 0.15        \pm 0.06$ & \cite{CDFLAM}\\
CDF    &$J/\psi \Lambda$      & 02--03 &$1.25 \pm 0.26 \pm 0.10$        & \cite{CDFLAM2}$^p$ \\
\dzero &$J/\psi \Lambda$      & 02--04 &$1.22^{+0.22}_{-0.18} \pm 0.04$ & \cite{D0LAMB} \\
DELPHI &$\Lc\ell$             & 91--94 &$1.11^{+0.19}_{-0.18} \pm 0.05$ & \cite{DELLAM0}$^a$\\
OPAL   &$\Lc\ell$, $\Lambda\ell^-\ell^+$ 
                                 & 90--95 & $1.29^{+0.24}_{-0.22} \pm 0.06$ & \cite{OPABS1_OPALAM2}\\ 
\hline
\multicolumn{3}{l}{Average of above 7 (\Lb lifetime)} & $1.232 \pm 0.072$ & \\
\hline
ALEPH  &$\Lambda\ell$         & 91--95 &$1.20^{+0.08}_{-0.08} \pm 0.06$ & \cite{ALEPH_fla}\\
DELPHI &$\Lambda\ell\pi$ vtx  & 91--94 &$1.16 \pm 0.20 \pm 0.08$        & \cite{DELLAM0}$^a$\\
DELPHI &$\Lambda\mu$ i.p.     & 91--94 &$1.10^{+0.19}_{-0.17} \pm 0.09$ & \cite{DELLAM1}$^a$ \\
DELPHI &\particle{p\ell}      & 91--94 &$1.19 \pm 0.14 \pm 0.07$        & \cite{DELLAM0}$^a$\\
OPAL   &$\Lambda\ell$ i.p.    & 90--94 &$1.21^{+0.15}_{-0.13} \pm 0.10$ & \cite{OPALAM1}$^b$  \\
OPAL   &$\Lambda\ell$ vtx     & 90--94 &$1.15 \pm 0.12 \pm 0.06$        & \cite{OPALAM1}$^b$ \\ 
\hline
\multicolumn{3}{l}{Average of above 13 ($b$-baryon lifetime)} & $1.210 \pm 0.048$ & \\  
\hline\hline
ALEPH  &$\Xi\ell$             & 90--95 &$1.35^{+0.37+0.15}_{-0.28-0.17}$ & \cite{ALEPH_fxi}\\
DELPHI &$\Xi\ell$             & 91--93 &$1.5 ^{+0.7}_{-0.4} \pm 0.3$     & \cite{DELPHI_fxi} \\
\hline
\multicolumn{3}{l}{Average of above 2 (\Xib lifetime)} & $1.39^{+0.34}_{-0.28}$ & \\
\hline
\multicolumn{5}{l}{$^a$ \footnotesize The combined DELPHI result quoted 
in \cite{DELLAM0} is $1.14 \pm 0.08 \pm 0.04$ ps.} \\[-0.5ex]
\multicolumn{5}{l}{$^b$ \footnotesize The combined OPAL result quoted 
in \cite{OPALAM1} is $1.16 \pm 0.11 \pm 0.06$ ps.} \\[-0.5ex]
\multicolumn{5}{l}{$^p$ \footnotesize Preliminary.}
\end{tabular}
\end{center}
\end{table}

\mysubsubsection{Summary and comparison to theoretical predictions}
\labs{lifesummary}

Averages of lifetimes of specific \b hadron species are collected
below in \Table{sumlife}.

\begin{table}[t]
\caption{Summary of lifetimes of different \b hadron species.}
\labt{sumlife}
\begin{center}
\begin{tabular}{lc} \hline
\b hadron species & Measured lifetime \\ \hline
\Bu                         & \HFAGtauBu      \\
\Bd                         & \HFAGtauBd      \\
\Bs ($\to$ flavor specific) & \HFAGtauBsfs    \\
\Bs ($\to J/\psi\phi$)      & \HFAGtauBsjpsi  \\
\Bc                         & \HFAGtauBc      \\ 
\Lb                         & \HFAGtauLb      \\
\Xib mixture                & \HFAGtauXib     \\
\b-baryon mixture           & \HFAGtaubbaryon \\
\b-hadron mixture           & \HFAGtauball    \\
\hline
\end{tabular}
\end{center}
%\end{table}
%\begin{table}[t]
\caption{Ratios of \b-hadron lifetimes relative to
the \Bd lifetime and theoretical ranges predicted
by theory~\cite{Gabbiani_et_al}.}
\labt{liferatio}
\begin{center}
\begin{tabular}{lcc} \hline
Lifetime ratio & Measured value & Predicted range \\ \hline
$\tau(\Bu)/\tau(\Bd)$ & \HFAGtauBuBd & 1.04 -- 1.08 \\
$\tau(\Bs)/\tau(\Bd)^a$ & \HFAGtauBsfsBd & 0.99 -- 1.01 \\
$\tau(\Lb)/\tau(\Bd)$ & \HFAGtauLbBd & 0.81 -- 0.91    \\
$\tau(\mbox{\b-baryon})/\tau(\Bd)$  & \HFAGtaubbaryonBd & 0.81 -- 0.91 \\
\hline
\multicolumn{3}{l}{$^a$ \footnotesize Using the 
$\Bs \rightarrow$~flavor specific lifetime for definiteness.}
\end{tabular}
\end{center}
\end{table}

As described in \Sec{lifetimes},
Heavy Quark Effective Theory
can be employed to explain the hierarchy of
$\tau(\Bc) \ll \tau(\Lb) < \tau(\Bs) \approx \tau(\Bd) < \tau(\Bu)$,
and used to predict the ratios between lifetimes.
A recent prediction of the ratio between the \Bu and \Bd lifetimes,
is $1.06 \pm 0.02$~\cite{tarantino}.
The ratio $\tau(\Lb)/\tau(\Bd)$ has particularly
been the source of theoretical
scrutiny since earlier calculations~\cite{lblife_early}
predicted a value greater than 0.90, almost two sigma higher
than the world average at the time. Recent calculations
of this ratio that include higher order effects predict a
ratio
between the 
\Lb and \Bd lifetimes of $0.86 \pm 0.05$~\cite{Gabbiani_et_al}
and reduces this difference.
\Ref{Gabbiani_et_al} presents probability density functions
of its predictions with variation of theoretical inputs, and the
indicated errors (and ranges in \Table{liferatio} below) 
are the RMS of the distributions.
Measured lifetime ratios compared to predicted ranges
are given in \Table{liferatio}.

%{\em Here is the list of lifetime averages:}
%\begin{eqnarray}
%\tau(\Bd) & =& \HFAGtauBd \\
%\tau(\Bu) & =& \HFAGtauBu \\
%\tau(\Bu)/\tau({\Bd}) & =& \HFAGtauBuBd \\
%\tau(\Bs) & =& \HFAGtauBs \\
%\tau(\Bc) & =& \HFAGtauBc \\
%\tau(\Lb) & =& \HFAGtauLb \\
%\langle\tau(\Xib)\rangle & =& \HFAGtauXib \\
%\langle\tau(\mbox{\b-baryon})\rangle & =& \HFAGtaubbaryon \\
%\langle\tau(\mbox{\b-hadron})\rangle & =& \HFAGtaubhadron
%\end{eqnarray}
  % from Rick

%------------------------------------------------
\mysubsection{Neutral \B-meson mixing}
%------------------------------------------------
\labs{mixing}

There are two neutral $\B-\bar{\B}$ systems, $\Bd-\Bdbar$ and $\Bs-\Bsbar$, which 
both exhibit the phenomenon of particle-antiparticle mixing. For each of these systems, 
there are two mass eigenstates which are linear combinations of the two flavour states,
\B or $\bar{\B}$. 
%%% \marginpar{xxx}
%%% \marginpar{Define heavy and light ?}
We consider the case where a neutral \B meson is produced and 
detected in a flavour state, through its decay to a flavour-specific final state. 
There are four different time-dependent probabilities; if \CPT is conserved (which  
will be assumed throughout), they can be written as 
\begin{equation}
\left\{
\begin{array}{rcl}
{\cal P}(\B\to\B) & = &  \frac{e^{-\Gamma t}}{2} 
\left[ \cosh\!\left(\frac{\Delta\Gamma}{2}t\right) + \cos\!\left(\Delta m t\right)\right]  \\
{\cal P}(\B\to\bar{\B}) & = &  \frac{e^{-\Gamma t}}{2} 
\left[ \cosh\!\left(\frac{\Delta\Gamma}{2}t\right) - \cos\!\left(\Delta m t\right)\right] 
\left|\frac{q}{p}\right|^2 \\
{\cal P}(\bar{\B}\to\B) & = &  \frac{e^{-\Gamma t}}{2} 
\left[ \cosh\!\left(\frac{\Delta\Gamma}{2}t\right) - \cos\!\left(\Delta m t\right)\right] 
\left|\frac{p}{q}\right|^2 \\
{\cal P}(\bar{\B}\to\bar{\B}) & = &  \frac{e^{-\Gamma t}}{2} 
\left[ \cosh\!\left(\frac{\Delta\Gamma}{2}t\right) + \cos\!\left(\Delta m t\right)\right] 
\end{array} \right. \,,
\labe{oscillations}
\end{equation}
where $t$ is the proper time of the system (\ie\ the time interval between the production 
and the decay in the rest frame of the \B meson) and $\Gamma = 1/\tau(\B)$ 
is the average decay width.
At the \B factories, only the proper-time difference $\Delta t$ between the decays
of the two neutral \B mesons from the \Ups can be determined, but, 
because the two \B mesons evolve coherently (keeping opposite flavours as long as
none of them has decayed), the 
above formulae remain valid 
if $t$ is replaced with $\Delta t$ and the production flavour is replaced by the flavour 
at the time of the decay of the accompanying \B meson in a flavour specific state.
As can be seen in the above expressions,
the mixing probabilities 
depend on the following three observables: the mass difference $\Delta m$ and the decay 
width difference $\Delta\Gamma$ between the two mass eigenstates, and the parameter 
$|q/p|^2$ which signals \CP violation in the mixing if $|q/p|^2 \ne 1$.

In the following sections we review in turn the experimental knowledge
on these three parameters, separately 
for the \Bd meson (\dmd, \DGd, $|q/p|_{\particle{d}}$) 
and the \Bs meson (\dms, \DGs, $|q/p|_{\particle{s}}$). 

%------------------------------------------------
\mysubsubsection{\Bd mixing parameters}
%------------------------------------------------

%---------------------------------------------------------------
\subsubsubsection{\boldmath \CP violation parameter $|q/p|_{\particle{d}}$}
%---------------------------------------------------------------
\labs{qpd}

Evidence for \CP violation in \Bd mixing
%, which is predicted to be very small in the Standard Model,
has been searched for,
both with flavor-specific and inclusive \Bd decays, 
in samples where the initial 
flavor state is tagged. In the case of semileptonic 
(or other flavor-specific) decays, 
where the final state tag is 
also available, the following asymmetry
\begin{equation} 
 {\cal A}_{\rm SL} = 
\frac{
N(\hbox{\Bdbar}(t) \to \ell^+      \nu_{\ell} X) -
N(\hbox{\Bd}(t)    \to \ell^- \bar{\nu}_{\ell} X) }{
N(\hbox{\Bdbar}(t) \to \ell^+      \nu_{\ell} X) +
N(\hbox{\Bd}(t)    \to \ell^- \bar{\nu}_{\ell} X) } 
= \frac{|p/q|_{\particle{d}}^2 - |q/p|_{\particle{d}}^2}%
{|p/q|_{\particle{d}}^2 + |q/p|_{\particle{d}}^2}
% \simeq 1 - |q/p|^2_{\particle{d}} 
\labe{ASL}
\end{equation} 
has been measured, either in time-integrated analyses at 
CLEO~\cite{CLEO_chid_CP,CLEO_chid_CP_y,CLEO_CP_semi} 
and CDF~\cite{CDF_CP_semi}, or in time-dependent analyses at 
OPAL~\cite{OPAL_CP_semi}, ALEPH~\cite{ALEPH_CP}, 
\babar~\cite{BABAR_DGd_qp,BABAR_CP_semi} and 
\belle~\cite{BELLE_CP_preliminary}.
In the inclusive case, also investigated and published
% at LEP~\cite{DELPHI_CP,ALEPH_CP,OPAL_CP_incl},
at ALEPH~\cite{ALEPH_CP} and OPAL~\cite{OPAL_CP_incl},
no final state tag is used, and the asymmetry~\cite{incl_asym}
\begin{equation} 
\frac{
N(\hbox{\Bd}(t) \to {\rm all}) -
N(\hbox{\Bdbar}(t) \to {\rm all}) }{
N(\hbox{\Bd}(t) \to {\rm all}) +
N(\hbox{\Bdbar}(t) \to {\rm all}) } 
\simeq
{\cal A}_{\rm SL} \left[ \frac{\dmd}{2\Gd} \sin(\dmd \,t) - 
\sin^2\left(\frac{\dmd \,t}{2}\right)\right] 
\labe{ASLincl}
\end{equation} 
must be measured as a function of the proper time to extract information 
on \CP violation.
In all cases asymmetries compatible with zero have been found,  
with a precision limited by the available statistics. A simple 
average of all published 
results for the \Bd 
meson~\cite{CLEO_chid_CP_y,CLEO_CP_semi,OPAL_CP_semi,ALEPH_CP,BABAR_DGd_qp,BABAR_CP_semi,OPAL_CP_incl}
and of the preliminary \belle result~\cite{BELLE_CP_preliminary}
yields 
\begin{equation}
{\cal A}_{\rm SL} = \HFAGASL 
\end{equation}
or, equivalently through \Eq{ASL},
\begin{equation}
|q/p|_{\particle{d}} = \HFAGqp \,.
\end{equation}
This result\footnote{Early analyses and (perhaps hence) the PDG use the complex
parameter $\epsilon_{\B} = (p-q)/(p+q)$; if \CP violation in the mixing in small, 
${\cal A}_{\rm SL} \cong 4 {\rm Re}(\epsilon_{\B})/(1+|\epsilon_{\B}|^2)$ and our 
current world average  % for the \Bd system 
is ${\rm Re}(\epsilon_{\B})/(1+|\epsilon_{\B}|^2)=\HFAGreb$.}, 
summarized in \Table{qoverp},
is compatible 
with no \CP violation in the mixing, an assumption we make for the rest 
of this section.
% and does not yet constrain the Standard Model.

\begin{table}
\caption{Measurements of \CP violation in \Bd mixing and their average in terms of 
both ${\cal A}_{\rm SL}$ and $|q/p|_{\particle{d}}$.
The individual results are listed as quoted in the original publications, 
or converted\addtocounter{footnote}{-1}\protect\footnotemark\
to an ${\cal A}_{\rm SL}$ value.
When two errors are quoted, the first one is statistical and the second one systematic.
}
\labt{qoverp}
\begin{center}
\begin{tabular}{@{}rcl@{$\,\pm$}l@{$\pm$}ll@{$\,\pm$}l@{$\pm$}l@{}}
\hline
%Experiment & {Method} & \multicolumn{3}{c}{${\cal A}_{\rm SL}$} 
%                      & \multicolumn{3}{c}{$|q/p|_{\particle{d}}$} \\
%Exp.\ \& Ref. & Method & ~~~${\cal A}_{\rm SL}$ & stat & syst
%                       & $|q/p|_{\particle{d}}$ & stat & syst \\
Exp.\ \& Ref. & Method & \multicolumn{3}{c}{Measured ${\cal A}_{\rm SL}$} 
                       & \multicolumn{3}{c}{Measured $|q/p|_{\particle{d}}$} \\
\hline
% CLEO   \cite{CLEO_chid_CP_y} & $D^{*\pm}\pi^{\mp}$, $D^{*\pm}\rho^{\mp}$ (part.\ rec.) 
CLEO   \cite{CLEO_chid_CP_y} & partial hadronic rec. 
                             & $+0.017$ & 0.070 & 0.014 
                             & \multicolumn{3}{c}{} \\
CLEO   \cite{CLEO_CP_semi}   & dileptons 
                             & $+0.013$ & 0.050 & 0.005 
                             & \multicolumn{3}{c}{} \\
CLEO   \cite{CLEO_CP_semi}   & average of above two 
                             & $+0.014$ & 0.041 & 0.006 
                             & \multicolumn{3}{c}{} \\
OPAL   \cite{OPAL_CP_semi}   & leptons     
                             & $+0.008$ & 0.028 & 0.012 
                             & \multicolumn{3}{c}{} \\
OPAL   \cite{OPAL_CP_incl}   & inclusive (\Eq{ASLincl}) 
                             & $+0.005$ & 0.055 & 0.013 
                             & \multicolumn{3}{c}{} \\
ALEPH  \cite{ALEPH_CP}       & leptons 
                             & $-0.037$ & 0.032 & 0.007 
                             & \multicolumn{3}{c}{} \\
ALEPH  \cite{ALEPH_CP}       & inclusive (\Eq{ASLincl}) 
                             & $+0.016$ & 0.034 & 0.009 
                             & \multicolumn{3}{c}{} \\
ALEPH  \cite{ALEPH_CP}       & average of above two 
                             & $-0.013$ & \multicolumn{2}{l}{0.026 (tot)} 
                             & \multicolumn{3}{c}{} \\
\babar \cite{BABAR_CP_semi}  & dileptons
                             & $+0.005$ & 0.012 & 0.014 
                             & 0.998 & 0.006 & 0.007 \\ 
\babar \cite{BABAR_DGd_qp}   & full hadronic rec. 
                             & \multicolumn{3}{c}{}  
                             & $1.029$ & 0.013 & 0.011  \\
\belle \cite{BELLE_CP_preliminary} & dileptons (prel.) 
                             & $-0.0013$ & 0.0060 & 0.0056 
                             & 1.0006 & 0.0030 & 0.0028 \\
\hline
& Average of all above       & \multicolumn{3}{l}{\HFAGASL\ (tot)} 
                             & \multicolumn{3}{l}{\HFAGqp\  (tot)} \\ 
\hline
\end{tabular}
\end{center}
\end{table}

%--------------------------------------------------------------------------
\subsubsubsection{\boldmath Mass and decay width differences \dmd and \DGd}
%--------------------------------------------------------------------------
\labs{dmd}
\labs{DGd}

\begin{table}
\caption{Time-dependent measurements included in the \dmd average.
All these measurements have been adjusted to a common set of physics
parameters before being combined. The CDF2 and D0 measurements are preliminary.
The new \babar~\cite{BABAR5} %-{BABAR_dmd_preliminary} 
and \belle~\cite{BELLE2}  %-{BELLE_dmd_tau_3D}  
measurements are not included here, but listed in \Table{dmd2D}.}
\labt{dmd}
\begin{center}
\begin{tabular}{@{}rc@{}cc@{}c@{}cc@{}c@{}c@{}}
\hline
Experiment & \multicolumn{2}{c}{Method} & \multicolumn{3}{l}{Published value}   
                                        & \multicolumn{3}{l}{Adjusted value}     \\
and Ref.   &  rec. & tag                & \multicolumn{3}{l}{of \dmd in\invps} 
                                        & \multicolumn{3}{l}{of \dmd in\invps} \\
\hline
%    0.404 +-0.045 +-0.027      from ALEPH LEPTON/QJET PUBLISHED 1 l/Qjet (91-94)
%    0.452 +-0.039 +-0.044      from ALEPH LEPTON/LEPTON PUBLISHED 1 l/l (91-94)
 ALEPH~\cite{ALEPH_dmd}  & \particle{ \ell  } & \particle{ \Qjet  } & $  0.404 $ & $ \pm  0.045 $ & $ \pm  0.027 $ & & & \\
 ALEPH~\cite{ALEPH_dmd}  & \particle{ \ell  } & \particle{ \ell  } & $  0.452 $ & $ \pm  0.039 $ & $ \pm  0.044 $ & & & \\
 ALEPH~\cite{ALEPH_dmd}  & \multicolumn{2}{c}{above two combined} & $  0.422 $ & $ \pm  0.032 $ & $ \pm  0.026 $ & $  0.441 $ & $ \pm  0.032 $ & $ \pm  0.021 $ \\
 ALEPH~\cite{ALEPH_dmd}  & \particle{ D^*  } & \particle{ \ell,\Qjet  } & $  0.482 $ & $ \pm  0.044 $ & $ \pm  0.024 $ & $  0.482 $ & $ \pm  0.044 $ & $ \pm  0.024 $ \\
 DELPHI~\cite{DELPHI_dmd}  & \particle{ \ell  } & \particle{ \Qjet  } & $  0.493 $ & $ \pm  0.042 $ & $ \pm  0.027 $ & $  0.504 $ & $ \pm  0.042 $ & $ \pm  0.025 $ \\
 DELPHI~\cite{DELPHI_dmd}  & \particle{ \pi^*\ell  } & \particle{ \Qjet  } & $  0.499 $ & $ \pm  0.053 $ & $ \pm  0.015 $ & $  0.501 $ & $ \pm  0.053 $ & $ \pm  0.015 $ \\
 DELPHI~\cite{DELPHI_dmd}  & \particle{ \ell  } & \particle{ \ell  } & $  0.480 $ & $ \pm  0.040 $ & $ \pm  0.051 $ & $  0.487 $ & $ \pm  0.040 $ & $ ^{+  0.049 }_{-  0.048 } $ \\
 DELPHI~\cite{DELPHI_dmd}  & \particle{ D^*  } & \particle{ \Qjet  } & $  0.523 $ & $ \pm  0.072 $ & $ \pm  0.043 $ & $  0.517 $ & $ \pm  0.072 $ & $ \pm  0.043 $ \\
 DELPHI~\cite{DELPHI_dmd_dms_vtx}  & \particle{ \mbox{vtx}  } & \particle{ \mbox{comb}  } & $  0.531 $ & $ \pm  0.025 $ & $ \pm  0.007 $ & $  0.530 $ & $ \pm  0.025 $ & $ \pm  0.006 $ \\
 L3~\cite{L3_dmd}  & \particle{ \ell  } & \particle{ \ell  } & $  0.458 $ & $ \pm  0.046 $ & $ \pm  0.032 $ & $  0.469 $ & $ \pm  0.046 $ & $ \pm  0.029 $ \\
 L3~\cite{L3_dmd}  & \particle{ \ell  } & \particle{ \Qjet  } & $  0.427 $ & $ \pm  0.044 $ & $ \pm  0.044 $ & $  0.436 $ & $ \pm  0.044 $ & $ \pm  0.042 $ \\
 L3~\cite{L3_dmd}  & \particle{ \ell  } & \particle{ \ell\mbox{(IP)}  } & $  0.462 $ & $ \pm  0.063 $ & $ \pm  0.053 $ & $  0.480 $ & $ \pm  0.063 $ & $ \pm  0.047 $ \\
 OPAL~\cite{OPAL_dmd_dilepton}  & \particle{ \ell  } & \particle{ \ell  } & $  0.430 $ & $ \pm  0.043 $ & $ ^{+  0.028 }_{-  0.030 } $ & $  0.462 $ & $ \pm  0.043 $ & $ ^{+  0.018 }_{-  0.017 } $ \\
 OPAL~\cite{OPAL_dmd_lepton}  & \particle{ \ell  } & \particle{ \Qjet  } & $  0.444 $ & $ \pm  0.029 $ & $ ^{+  0.020 }_{-  0.017 } $ & $  0.465 $ & $ \pm  0.029 $ & $ ^{+  0.015 }_{-  0.014 } $ \\
 OPAL~\cite{OPAL_dmd_dstar}  & \particle{ D^*\ell  } & \particle{ \Qjet  } & $  0.539 $ & $ \pm  0.060 $ & $ \pm  0.024 $ & $  0.545 $ & $ \pm  0.060 $ & $ \pm  0.023 $ \\
 OPAL~\cite{OPAL_dmd_dstar}  & \particle{ D^*  } & \particle{ \ell  } & $  0.567 $ & $ \pm  0.089 $ & $ ^{+  0.029 }_{-  0.023 } $ & $  0.571 $ & $ \pm  0.089 $ & $ ^{+  0.028 }_{-  0.022 } $ \\
 OPAL~\cite{OPAL_dmd_slowpion}  & \particle{ \pi^*\ell  } & \particle{ \Qjet  } & $  0.497 $ & $ \pm  0.024 $ & $ \pm  0.025 $ & $  0.496 $ & $ \pm  0.024 $ & $ \pm  0.025 $ \\
 CDF1~\cite{CDF1_dmd_dlepton_SST}  & \particle{ D\ell  } & \particle{ \mbox{SST}  } & $  0.471 $ & $ ^{+  0.078 }_{-  0.068 } $ & $ ^{+  0.033 }_{-  0.034 } $ & $  0.471 $ & $ ^{+  0.078 }_{-  0.068 } $ & $ ^{+  0.033 }_{-  0.034 } $ \\
 CDF1~\cite{CDF1_dmd_dimuon}  & \particle{ \mu  } & \particle{ \mu  } & $  0.503 $ & $ \pm  0.064 $ & $ \pm  0.071 $ & $  0.514 $ & $ \pm  0.064 $ & $ \pm  0.070 $ \\
 CDF1~\cite{CDF1_dmd_lepton}  & \particle{ \ell  } & \particle{ \ell,\Qjet  } & $  0.500 $ & $ \pm  0.052 $ & $ \pm  0.043 $ & $  0.536 $ & $ \pm  0.052 $ & $ \pm  0.037 $ \\
 CDF1~\cite{CDF1_dmd_dstarlepton}  & \particle{ D^*\ell  } & \particle{ \ell  } & $  0.516 $ & $ \pm  0.099 $ & $ ^{+  0.029 }_{-  0.035 } $ & $  0.523 $ & $ \pm  0.099 $ & $ ^{+  0.029 }_{-  0.035 } $ \\
 CDF2~\cite{CDF2_dmd_dlepton_preliminary}  & \particle{ D^{(*)}\ell  } & \particle{ \mbox{SST,SMT,JQT}  } & $  0.536 $ & $ \pm  0.037 $ & $ \pm  0.017 $ & $  0.536 $ & $ \pm  0.037 $ & $ \pm  0.017 $ \\
 CDF2~\cite{CDF2_dmd_exclusive_preliminary}  & \particle{ \Bd  } & \particle{ \mbox{SST}  } & $  0.526 $ & $ \pm  0.056 $ & $ \pm  0.005 $ & $  0.526 $ & $ \pm  0.056 $ & $ \pm  0.005 $ \\
 \dzero~\cite{D0_dmd_preliminary}  & \particle{ D^*\mu  } & \particle{ \mbox{comb}  } & $  0.456 $ & $ \pm  0.034 $ & $ \pm  0.025 $ & $  0.456 $ & $ \pm  0.034 $ & $ \pm  0.025 $ \\
 \babar~\cite{BABAR_dmd_full}  & \particle{ \Bd  } & \particle{ \ell,K,\mbox{NN}  } & $  0.516 $ & $ \pm  0.016 $ & $ \pm  0.010 $ & $  0.518 $ & $ \pm  0.016 $ & $ \pm  0.008 $ \\
 \babar~\cite{BABAR_dmd_dilepton}  & \particle{ \ell  } & \particle{ \ell  } & $  0.493 $ & $ \pm  0.012 $ & $ \pm  0.009 $ & $  0.489 $ & $ \pm  0.012 $ & $ \pm  0.007 $ \\
 \babar~\cite{BABAR3}  & \particle{ D^*\ell\nu  } & \particle{ \ell,K,\mbox{NN}  } & $  0.492 $ & $ \pm  0.017 $ & $ \pm  0.014 $ & $  0.490 $ & $ \pm  0.017 $ & $ \pm  0.013 $ \\
 \belle~\cite{BELLE_dmd_full}  & \particle{ \Bd  } & \particle{ \mbox{comb}  } & $  0.528 $ & $ \pm  0.017 $ & $ \pm  0.011 $ & $  0.529 $ & $ \pm  0.017 $ & $ \pm  0.011 $ \\
 \belle~\cite{BELLE_dmd_dstarlnu}  & \particle{ D^*\ell\nu  } & \particle{ \mbox{comb}  } & $  0.494 $ & $ \pm  0.012 $ & $ \pm  0.015 $ & $  0.496 $ & $ \pm  0.012 $ & $ \pm  0.014 $ \\
 \belle~\cite{BELLE_dmd_dstarpi_partial}  & \particle{ D^*\pi\mbox{(part)}  } & \particle{ \ell  } & $  0.509 $ & $ \pm  0.017 $ & $ \pm  0.020 $ & $  0.511 $ & $ \pm  0.017 $ & $ \pm  0.019 $ \\
 \belle~\cite{BELLE_dmd_dilepton}  & \particle{ \ell  } & \particle{ \ell  } & $  0.503 $ & $ \pm  0.008 $ & $ \pm  0.010 $ & $  0.504 $ & $ \pm  0.008 $ & $ \pm  0.009 $ \\
 \hline \\[-2.0ex]
 \multicolumn{6}{l}{World average (all above measurements included):} & $  0.502 $ & $ \pm  0.004 $ & $ \pm  0.005 $ \\
 % from nov20 directory on lxplus
%\hline  \\[-2.0ex]
%\multicolumn{6}{l}{World average (all above measurements included):}                    
%    & $0.502$ & $\pm 0.004$ & $ \pm0.005$ \\
\\[-2.0ex]
\multicolumn{6}{l}{~~~ -- ALEPH, DELPHI, L3, OPAL and CDF1 only:}
     & $0.496$ & $\pm 0.010$ & $ \pm0.009$ \\
\multicolumn{6}{l}{~~~ -- Above measurements of \babar and \belle only:}
     & $0.503$ & $\pm 0.005$ & $ \pm0.005$ \\
\hline
\end{tabular}
\end{center}
\end{table}

Many time-dependent \Bd--\Bdbar oscillation analyses have been published by the 
ALEPH, \babar, \belle, CDF, DELPHI, L3 and OPAL collaborations. 
The corresponding measurements of \dmd are summarized in \Tabless{dmd}{dmd2D}, 
where only the most recent results
are listed (\ie\ measurements superseded by more recent ones have been omitted). 
Although a variety of different techniques have been used, the 
individual \dmd
results obtained at high-energy colliders have remarkably similar precision.
Their average is compatible with the recent and more precise measurements 
from the asymmetric \B factories.
The systematic uncertainties are not negligible; 
they are often dominated by sample composition, mistag probability,
or \b-hadron lifetime contributions.
Before being combined, the measurements are adjusted on the basis of a 
common set of input values, including the averages of the 
\b-hadron fractions and lifetimes given in this report 
(see \Secss{fractions}{lifetimes}).
Some measurements are statistically correlated. 
Systematic correlations arise both from common physics sources 
(fractions, lifetimes, branching ratios of \b hadrons), and from purely 
experimental or algorithmic effects (efficiency, resolution, flavour tagging, 
background description). Combining all published measurements
listed in \Table{dmd}
% ~%\cite{DELPHI_dmd_dms_vtx,CDF_dmd,ALEPH_dmd,BABAR_dmd,Belle_dmd,DELPHI_dmd,L3_dmd,OPAL_dmd} 
and accounting for all identified correlations
as described in~\cite{LEPHFS} yields $\dmd = \HFAGdmdtimeall$.

On the other hand, ARGUS and CLEO have published 
measurements of the time-integrated mixing probability 
\chid~\cite{ARGUS_chid,CLEO_chid_CP,CLEO_chid_CP_y}, 
which average to $\chid =\HFAGchidint$.
Following \Ref{CLEO_chid_CP_y}, 
the width difference \DGd could 
in principle be extracted from the
measured value of $\Gd=1/\tau(\Bd)$ and the above averages for 
\dmd and \chid 
(provided that \DGd has a negligible impact on 
the \dmd analyses that have assumed $\DGd=0$), 
using the relation
\begin{equation}
\chid = \frac{\xd^2+\yd^2}{2(\xd^2+1)} ~~~ \mbox{with} ~~ \xd=\frac{\dmd}{\Gd} 
~~~ \mbox{and} ~~ \yd=\frac{\DGd}{2\Gd} \,.
\labe{chid_definition}
\end{equation}
However, direct time-dependent studies yield stronger constraints: 
DELPHI published the result
$|\DGd|/\Gd < 18\%$ at \CL{95}~\cite{DELPHI_dmd_dms_vtx}, 
while \babar recently obtained 
$-8.4\% < {\rm sign}({\rm Re} \lambda_{\CP}) \DGGd < 6.8\%$
at \CL{90}~\cite{BABAR_DGd_qp}.

Assuming $\DGd=0$ 
% and no \CP violation in mixing, and using the measured \Bd lifetime,
and using $1/\Gd=\tau(\Bd)=\HFAGtauBd$,
the \dmd and \chid results are combined through \Eq{chid_definition} 
to yield the 
world average
\begin{equation} 
\dmd = \HFAGdmd \,,
\labe{dmd}
\end{equation} 
or, equivalently,
\begin{equation} 
\xd= \HFAGxd ~~~ \mbox{and} ~~~ \chid=\HFAGchid \,.  
\labe{chid}
\end{equation}
\Figure{dmd} compares the \dmd values obtained by the different experiments.

\begin{figure}
\begin{center}
\epsfig{figure=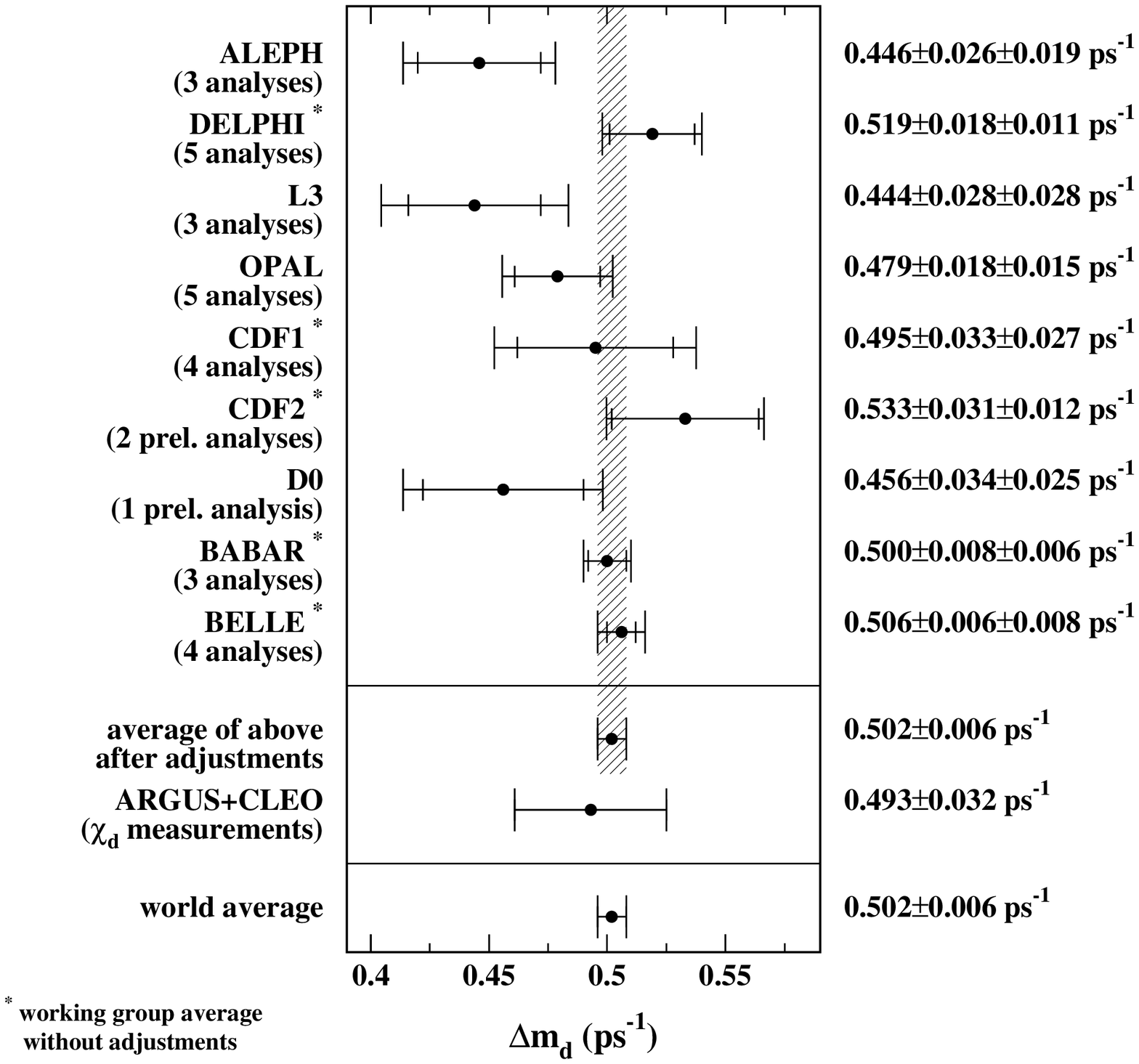,width=\textwidth}
\caption{The \Bd--\Bdbar oscillation frequency \dmd as measured by the different experiments. 
The averages quoted for ALEPH, L3 and OPAL are taken from the original publications, while the 
ones for DELPHI, CDF, \babar, and \belle have been computed from the individual results 
listed in \Table{dmd} without performing any adjustments. The time-integrated measurements 
of \chid from the symmetric \B factory experiments ARGUS and CLEO have been converted 
to a \dmd value using $\tau(\Bd)=\HFAGtauBd$. The two global averages have been obtained 
after adjustments of all the individual \dmd results of \Table{dmd} (see text).}
\labf{dmd}
\end{center}
\end{figure}

The \Bd mixing averages given in \Eqss{dmd}{chid}
and the \b-hadron fractions of \Table{fractions} have been obtained in a fully 
consistent way, taking into account the fact that the fractions are computed using 
the \chid value of \Eq{chid} and that many individual measurements of \dmd
at high energy depend on the assumed values for the \b-hadron fractions.
Furthermore, this set of averages is consistent with the lifetime averages 
of \Sec{lifetimes}.

It should be noted that the above average of \dmd, \Table{dmd} and \Figure{dmd} do not include
two precise measurements that have been released for the Summer 2004 conferences
by \babar and \belle. The new \babar
analysis~\cite{BABAR5},  %-{BABAR_dmd_preliminary}, 
based on partially reconstructed $\Bd \to D^*\ell\nu$ decays, 
extracts simultaneously \dmd and $\tau(\Bd)$ (similarly to the \babar published 
result based on fully reconstructed $\Bd\to D^* \ell\nu$ 
decays~\cite{BABAR3} %-{BABAR_dmd_dstarlnu}),
while the new \belle analysis~\cite{BELLE2}  %-{BELLE_dmd_tau_3D}, 
based on fully reconstructed hadronic \Bd decays and $\Bd \to D^*\ell\nu$ decays, 
extracts simultaneously \dmd, $\tau(\Bd)$ and $\tau(\Bu)$.

\begin{table}
\caption{Simultaneous measurements of \dmd and $\tau(\Bd)$, with the statistical correlations 
$\rho_{\rm stat}$ between them. The \belle analysis also 
measures $\tau(\Bu)$ at the same time, but it is converted here into a two-dimensional measurement 
of \dmd and $\tau(\Bd)$, for an assumed value of $\tau(\Bu)$. 
The first error on \dmd and $\tau(\Bd)$ is statistical and the second one systematic; the 
latter includes a contribution obtained from the variation of $\tau(\Bu)$ or 
$\tau(\Bu)/\tau(\Bd)$ in the indicated range. Units are\invps\ for \dmd
and\unit{ps} for the lifetimes. The second \babar result is still preliminary.}
\labt{dmd2D}
\begin{center}
\begin{tabular}{@{}r@{~~}c@{~~}c@{~~}l@{~}c@{}}
\hline
Exp.\ \& Ref. & Measured \dmd & Measured $\tau(\Bd)$ & $~\rho_{\rm stat}$ & Assumed $\tau(\Bu)$ \\
\hline
\babar \cite{BABAR3}  %{BABAR_dmd_dstarlnu}
                                    & $0.492 \pm 0.018 \pm 0.013$ 
                                    & $1.523 \pm 0.024 \pm 0.022$ 
                                    & $-0.22$ & $(1.083 \pm 0.017)\tau(\Bd)$ \\  
\babar \cite{BABAR5}  %{BABAR_dmd_preliminary}
                                    & $0.523 \pm 0.004 \pm 0.007$ 
                                    & $1.501 \pm 0.008 \pm 0.030$
                                    & $-0.012$ & $1.671 \pm 0.018$ \\  
\belle \cite{BELLE2}  %{BELLE_dmd_tau_3D}
                                    & $0.511 \pm 0.005 \pm 0.006$
                                    & $1.534 \pm 0.008 \pm 0.010$
                                    & $-0.27$ & $1.635 \pm 0.011$ \\  
\hline
\end{tabular}
\end{center}
\end{table}

The results of the three \B-factory analyses extracting \dmd and $\tau(\Bd)$ 
at the same time are listed in \Table{dmd2D}.
Performing a two-dimensional combination of these results 
needs a careful assessment of the statistical and systematic correlations 
between \dmd and $\tau(\Bd)$ in each analysis and between the measurements 
in the different analyses. This will be completed for a next version of this report.

%------------------------------------------------
\mysubsubsection{\Bs mixing parameters}
%------------------------------------------------

%---------------------------------------------------------------
\subsubsubsection{\boldmath \CP violation parameter $|q/p|_{\particle{s}}$}
%---------------------------------------------------------------
\labs{qps}

No measurement or experimental limit exists on $|q/p|_{\particle{s}}$, except in the 
form of a relatively weak constraint from CDF 
on a combination of $|q/p|_{\particle{d}}$ and $|q/p|_{\particle{s}}$,
$f'_{\particle{d}} \,\chid(1-|q/p|^2_{\particle{d}})+
 f'_{\particle{s}} \,\chis(1-|q/p|^2_{\particle{s}})= 0.006\pm 0.017$~\cite{CDF_CP_semi},
using inclusive semileptonic decays of \b hadrons. 
The result is compatible with no \CP violation in the 
mixing, an assumption made in all results described below. 

%--------------------------------------
\subsubsubsection{\boldmath Mass difference \dms}
%--------------------------------------
\labs{dms}

The time-integrated measurements of \chibar (see \Sec{chibar}), when compared to our knowledge
of \chid and the \b-hadron fractions, indicate that \Bs mixing is large, with a value of 
\chis close to its maximal possible value of $1/2$.
However, the time dependence of this mixing (called \Bs oscillations) has not been 
observed yet, mainly because the period of these oscillations turns out to be so small 
that it can't be resolved with the proper-time resolutions achieved so far. 

The statistical significance ${\cal S}$ of a \Bs oscillation signal can be
approximated as~\cite{amplitude}
\begin{equation}
{\cal S} \approx \sqrt{\frac{N}{2}} \,f_{\rm sig}\, (1-2w)\,
\exp{\left(-\frac{1}{2}\left(\frac{\dms}{\sigma_t}\right)^2\right)}\,,
% e^{-(\dms\sigma_t)^2/2}\,, 
\labe{significance}
\end{equation}
where $N$ and $f_{\rm sig}$
are the number of \Bs candidates and the fraction of \Bs signal
in the selected sample, $w$ is the total mistag probability, 
and $\sigma_t$ is the resolution on proper time.
As can be seen, the quantity ${\cal S}$ decreases very quickly as 
\dms increases: this dependence is controlled by $\sigma_t$, 
which is therefore the most critical parameter for \dms analyses. 
The method widely used for \Bs oscillation searches
consists of measuring a \Bs oscillation amplitude ${\cal A}$
at several different test values of \dms, 
using a maximum likelihood fit based on the functions 
of \Eq{oscillations} where the cosine terms have been multiplied 
by ${\cal A}$.
One expects ${\cal A}=1$ at the true 
value of \dms and to ${\cal A}=0$ at a test value of \dms 
(far) below the true value.
To a good approximation, the statistical uncertainty on ${\cal A}$
is Gaussian and equal to $1/{\cal S}$~\cite{amplitude}.

\begin{figure}
\begin{center}
\epsfig{figure=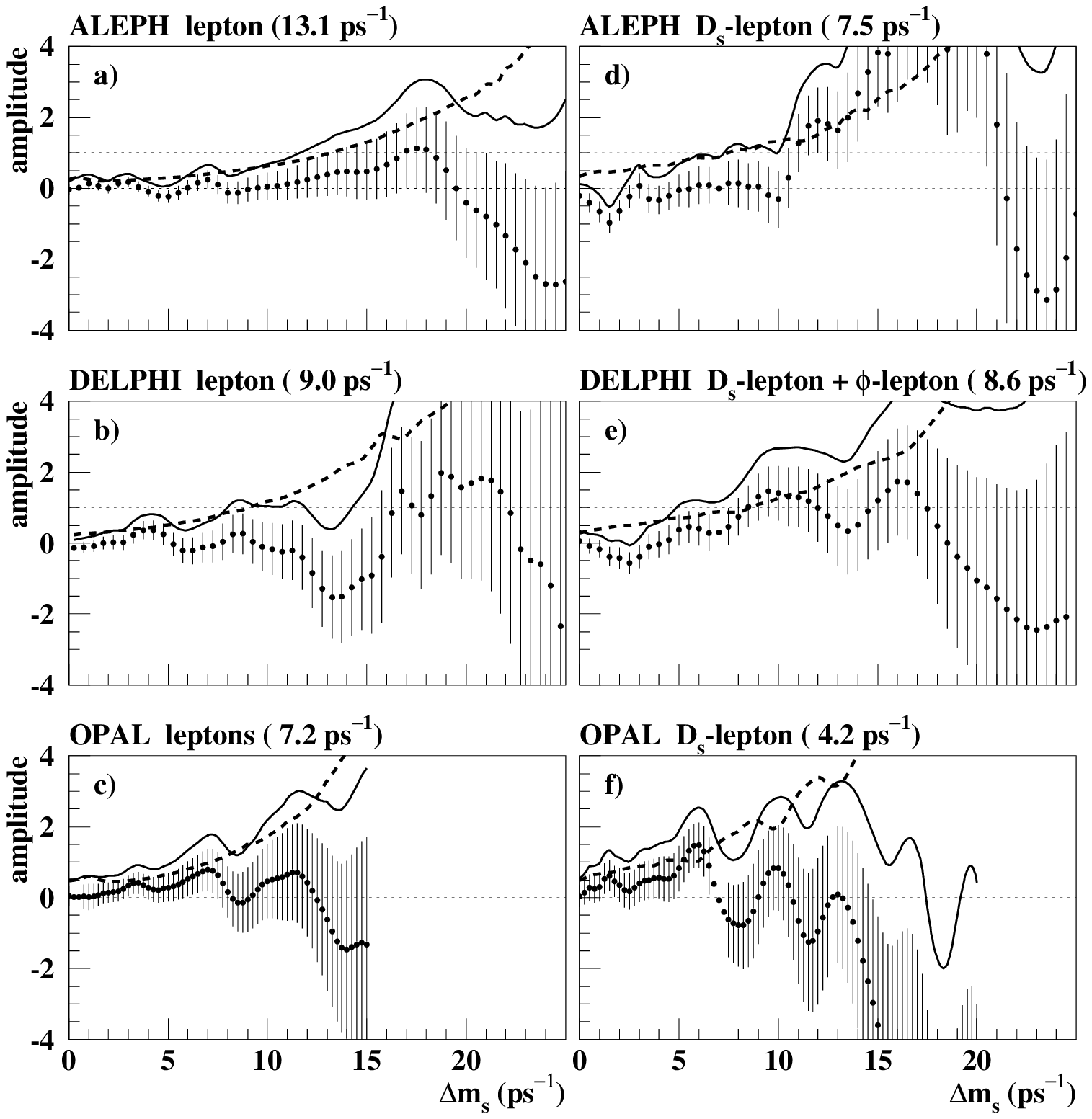,width=\textwidth,%
bbllx=55,bblly=55,bburx=490,bbury=490}
\caption{\Bs-oscillation amplitude spectra, displayed separately for each 
\Bs oscillation analysis. 
The points and error bars represent the measurements of the amplitude ${\cal A}$ and 
their total uncertainties $\sigma_{\cal A}$, adjusted to a set of physics parameters
common to all analyses (including $\fBs=\HFAGfBs$).
Values of \dms where the solid curve 
(${\cal A}+1.645\,\sigma_{\cal A}$) is below 1 are excluded at \CL{95}. 
The dashed curve shows $1.645\,\sigma_{\cal A}$; the number in parenthesis indicates where 
this curve is equal to 1, and is a measure of the sensitivity of the analysis. 
a) ALEPH inclusive lepton~\cite{ALEPH_dms},
b) DELPHI inclusive lepton~\cite{DELPHI_dms_last}, 
c) OPAL inclusive lepton and dilepton~\cite{OPAL_dms_l},
d) ALEPH \particle{D_s}-$\ell$~\cite{ALEPH_dms}, 
e) DELPHI \particle{D_s}-$\ell$~\cite{DELPHI_dms_last} and $\phi$-$\ell$~\cite{DELPHI_dms_dgs},
f) OPAL \particle{D_s}-$\ell$~\cite{OPAL_dms_dsl} (continued on \Fig{individual_amplitudes_2}).}
\labf{individual_amplitudes_1}
\end{center}
\end{figure}

\begin{figure}
\begin{center}
\epsfig{figure=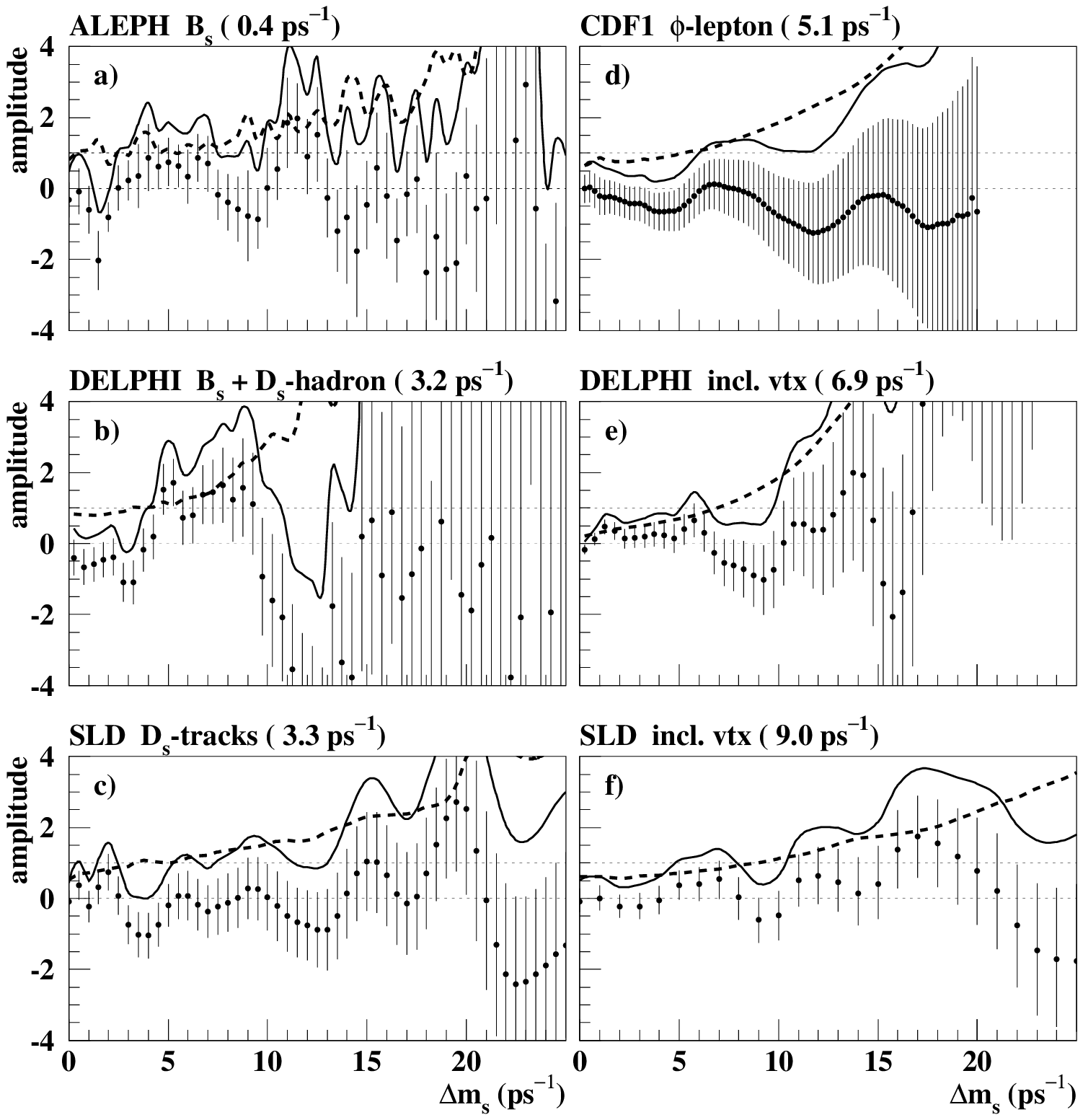,width=\textwidth,%
bbllx=55,bblly=55,bburx=490,bbury=490}
\caption{(continuation of \Fig{individual_amplitudes_1})
\Bs-oscillation amplitude spectra, displayed separately for each 
\Bs oscillation analysis, in the same manner as in \Fig{individual_amplitudes_1}. 
a) ALEPH fully reconstructed \Bs~\cite{ALEPH_dms}, 
b) DELPHI fully reconstructed \Bs and \particle{D_s}-hadron~\cite{DELBS1_dms_excl},
c) SLD \particle{D_s}+tracks~\cite{SLD_dms_ds},
d) CDF $\phi$-$\ell$~\cite{CDF_dms}, 
e) DELPHI inclusive vertex~\cite{DELPHI_dmd_dms_vtx}, 
f) SLD inclusive vertex dipole~\cite{SLD_dms_dipole}.}
\labf{individual_amplitudes_2}
\end{center}
\end{figure}

\begin{figure}
\begin{center}
\epsfig{figure=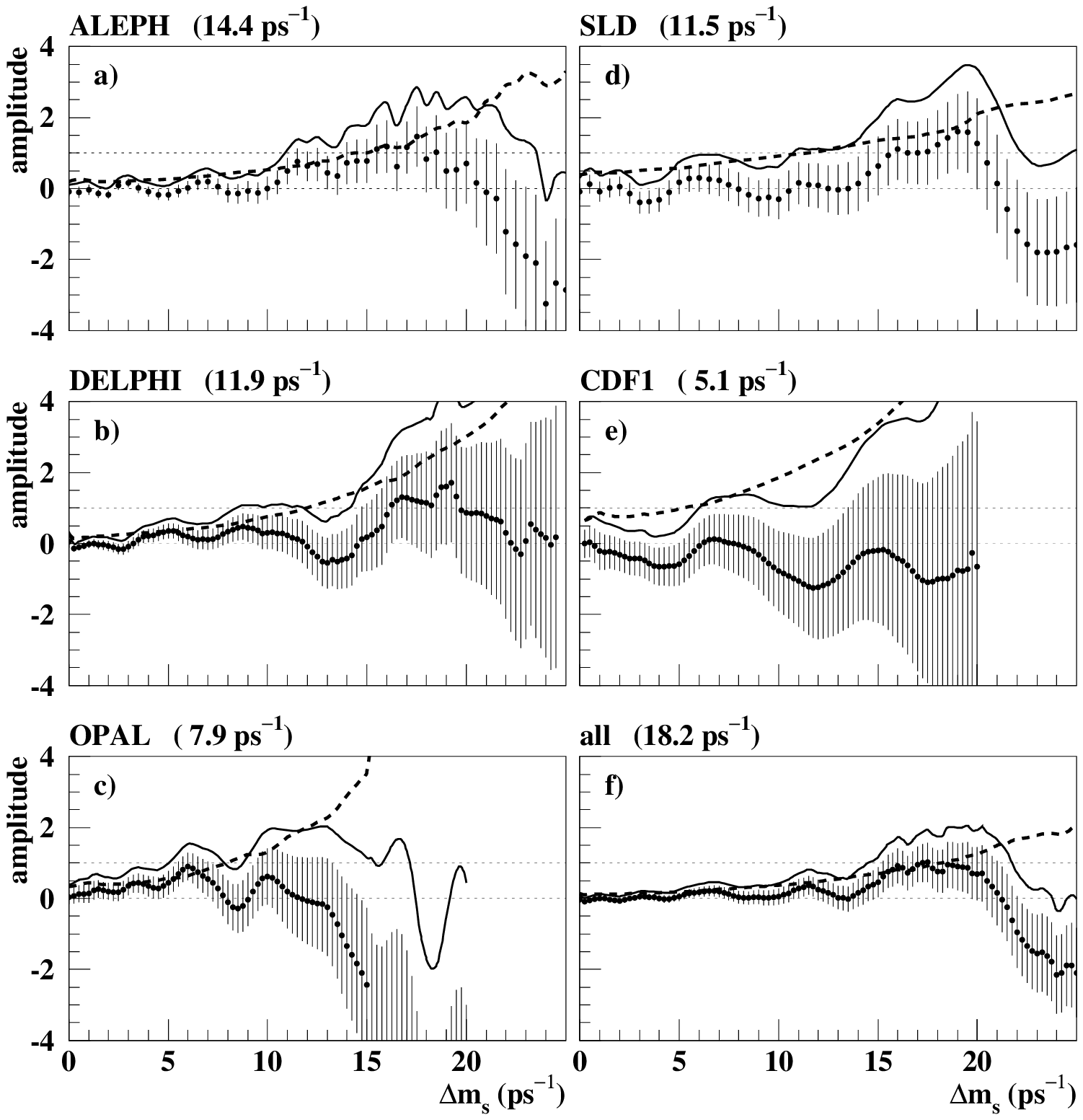,width=\textwidth,%
bbllx=55,bblly=55,bburx=490,bbury=490}
\caption{Combined \Bs-oscillation amplitude spectra, displayed separately for each 
experiment, in the same manner as in \Fig{individual_amplitudes_1}. 
a) ALEPH~\cite{ALEPH_dms}, 
b) DELPHI~\cite{DELBS1_dms_excl,DELPHI_dms_dgs,DELPHI_dmd_dms_vtx,DELPHI_dms_last},
c) OPAL\cite{OPAL_dms_l,OPAL_dms_dsl}, 
d) SLD~\cite{SLD_dms_ds,SLD_dms_dipole},
e) CDF~\cite{CDF_dms}, 
f) all experiments together.
}
\labf{individual_amplitudes_3}
\end{center}
\end{figure}

\Figuress{individual_amplitudes_1}{individual_amplitudes_2} show the
amplitude spectra published by ALEPH~\cite{ALEPH_dms}, CDF~\cite{CDF_dms}, 
DELPHI~\cite{DELPHI_dmd_dms_vtx,DELPHI_dms_dgs,DELBS1_dms_excl,DELPHI_dms_last},
OPAL~\cite{OPAL_dms_l,OPAL_dms_dsl} and 
SLD~\cite{SLD_dms_dipole,SLD_dms_ds}.\footnote{An unpublished analysis 
from SLD~\cite{SLD_dms_leptDvtx_unpublished}, 
based on an inclusive reconstruction from a 
lepton and a topologically reconstructed \particle{D} meson, 
is not included in the plots or
combined results quoted in this section. However, nothing is known 
to be wrong about this analysis, and including it would increase the 
combined \dms limit of \Eq{dmslimit} by less than $0.1\invps$ and the combined 
sensitivity by 0.9\invps.}
In each analysis, a particular value of \dms
can be excluded at \CL{95} if ${\cal A}+ 1.645\,\sigma_{\cal A} < 1$, 
where $\sigma_{\cal A}$ is the total uncertainty on ${\cal A}$.
Because of the proper time resolution, the quantity $\sigma_{\cal A}(\dms)$
is an increasing function of \dms (see \Eq{significance} which merely models  
$1/\sigma_{\cal A}(\dms)$ since all results are limited 
by the available statistics). Therefore, 
if the true value of \dms were infinitely large, one 
expects to be able to exclude all values of \dms up to $\dms^{\rm sens}$, 
where $\dms^{\rm sens}$, called here the
sensitivity of the analysis, is defined by
$1.645\,\sigma_{\cal A}(\dms^{\rm sens}) = 1$. 
The most sensitive analyses appear to be the ones based 
on inclusive lepton samples at LEP, where reasonable statistics is available. 
Because of their better proper time resolution, the small data samples 
analyzed inclusively at SLD, as well as the few fully reconstructed \Bs decays 
at LEP, turn out to be also very useful to explore the high \dms region.

\begin{figure}
\begin{center}
\epsfig{figure=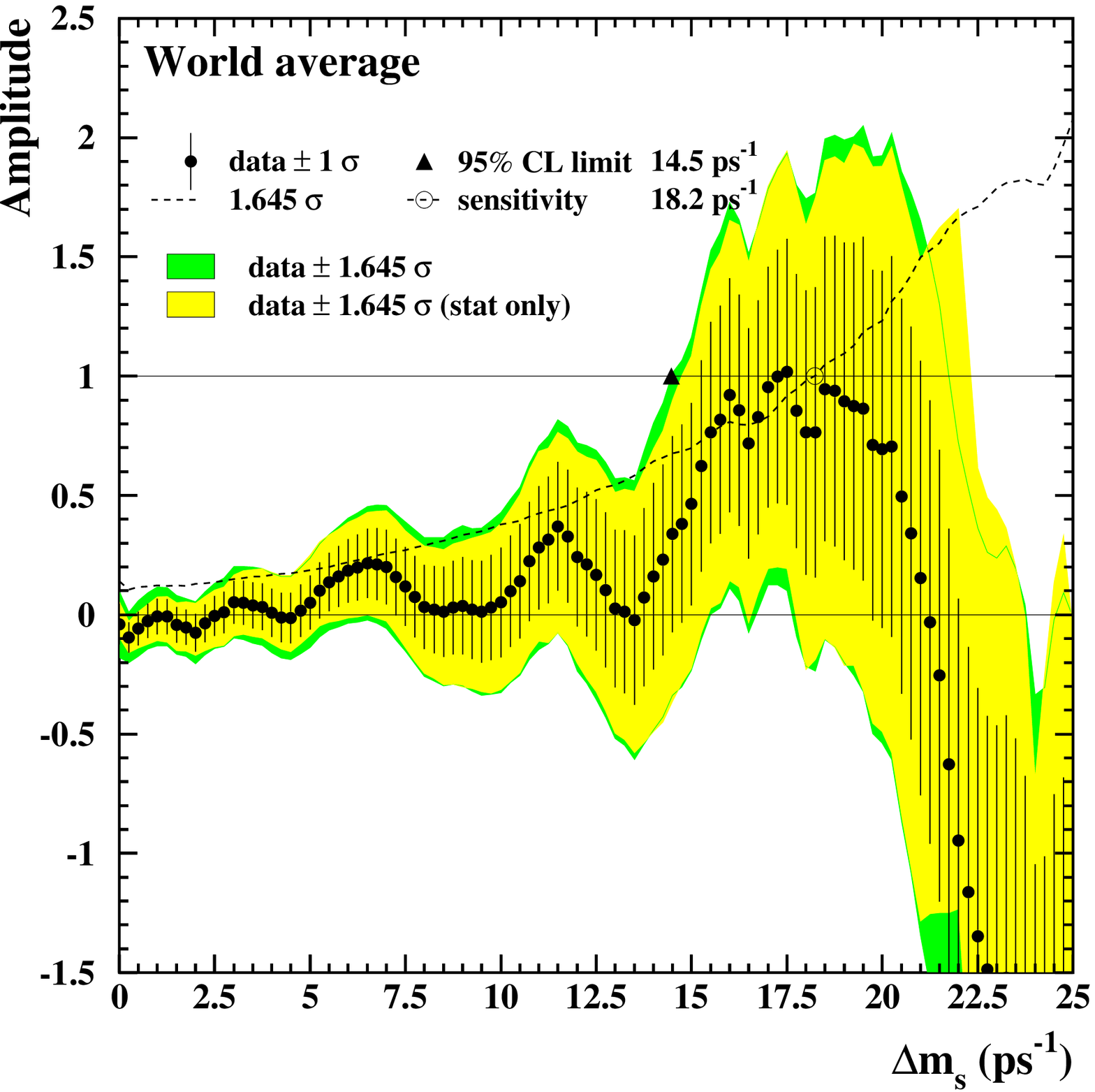,width=\textwidth}
\caption{Combined measurements of the \Bs oscillation amplitude as a 
function of \dms, including all results published 
by Summer 2004~% 
\cite{SLD_dms_dipole,DELPHI_dmd_dms_vtx,DELPHI_dms_dgs,DELBS1_dms_excl,ALEPH_dms,%
CDF_dms,DELPHI_dms_last,OPAL_dms_l,OPAL_dms_dsl,SLD_dms_ds}.
The measurements are dominated by statistical uncertainties. 
Neighboring points are statistically correlated.}
\labf{amplitude}
\end{center}
\end{figure}

These oscillation searches can easily be combined 
by averaging the measured amplitudes ${\cal A}$ at each test value 
of \dms. The combined amplitude spectra for the individual experiments are 
displayed in \Fig{individual_amplitudes_3}, and the world average spectrum is 
displayed in \Fig{amplitude}.
The individual results have been adjusted to common physics inputs, 
and all known correlations have been accounted for; 
in the case of the inclusive analyses, the sensitivities (\ie\ 
the statistical uncertainties on ${\cal A}$), which depend directly 
through \Eq{significance} on the assumed fraction $f_{\rm sig}\sim\fBs$
of \Bs mesons in an unbiased sample of weakly-decaying \b hadrons, 
have also been rescaled to a common average of $\fBs = \HFAGfBs$.
The combined sensitivity for \CL{95} exclusion of \dms values is found 
to be \HFAGdmssens.
All values of \dms below \HFAGdmslim\ are excluded at \CL{95},
which we express as
\begin{equation}
\dms > \HFAGdmslimCL \,.
\labe{dmslimit}
\end{equation}
The values between \HFAGdmslim\ and \HFAGdmslimax\ cannot be excluded, because 
the data is compatible with a signal in this region. However,
no deviation from ${\cal A}=0$ is seen in \Fig{amplitude} that would
indicate the observation of a signal.

It should be noted that most \dms analyses assume no decay-width difference in the \Bs system.
Due to the presence of the $\cosh$ terms in \Eq{oscillations}, a non-zero value of 
\DGs would reduce the oscillation amplitude with a small time-dependent factor that would be 
very difficult to distinguish from time resolution effects.

Convoluting the average \Bs lifetime, \HFAGtauBs, with the limit of \Eq{dmslimit} yields
\begin{equation}
\xs = \dms\,\tau(\Bs) >\HFAGxslimCL\,. 
\labe{xs}
\end{equation}
Under the assumption $\DGs=0$, \ie\ $\ys=\DGs/(2\Gs)=0$ 
(and no \CP violation in the mixing), this is 
equivalent to
\begin{equation}
\chis = \frac{\xs^2+\ys^2}{2(\xs^2+1)} > \HFAGchislimCL\,.
\labe{chis}
\end{equation}

%%---------------------------------------------
%\subsubsubsection{Decay width difference \DGs}
%%---------------------------------------------

%%%%%%%%%%%%%%%%%%%%%%%%%%%%%%%%%%%%%%%%%%%%%%%%
%
% This is file life_mix_dgs.tex containing
% the end of the B mixing section about the 
% decay width difference in the Bs system
%
%%%%%%%%%%%%%%%%%%%%%%%%%%%%%%%%%%%%%%%%%%%%%%%
%

%---------------------------------------------
\subsubsubsection{Decay width difference \DGs}
%---------------------------------------------
\labs{DGs}

% The following text written by Olivier

Information on \DGs can be obtained by studying the proper time 
distribution of untagged data samples enriched in 
\Bs mesons~\cite{Hartkorn_Moser}.
In the case of an inclusive \Bs selection~\cite{L3B01} or a semileptonic 
\Bs decay selection~\cite{DELPHI_dms_dgs,CDFBS}, 
both the short- and long-lived
components are present, and the proper time distribution is a superposition 
of two exponentials with decay constants 
$\Gs\pm\DGs/2$.
In principle, this provides sensitivity to both \Gs and 
$(\DGGs)^2$. Ignoring \DGs and fitting for 
a single exponential leads to an estimate of \Gs with a 
relative bias proportional to $(\DGGs)^2$. 
An alternative approach, which is directly sensitive to first order in \DGGs, 
is to determine the lifetime of \Bs candidates decaying to \CP
eigenstates; measurements exist for 
\particle{\Bs\to J/\psi\phi}~\cite{CDFIN_BS1,CDFBS2,D0BS1} and 
\particle{\Bs\to D_s^{(*)+} D_s^{(*)-}}~\cite{ALEPH_DGs}, which are 
mostly \CP-even states~\cite{Aleksan}. 
However, a time-dependent angular analysis of \particle{\Bs\to J/\psi\phi} 
allows the simultaneous extraction of \DGGs and the \CP-even and \CP-odd 
amplitudes~\cite{CDF2_DGs}. 
An estimate of \DGGs
has also been obtained directly from a measurement of the 
\particle{\Bs\to D_s^{(*)+} D_s^{(*)-}} branching ratio~\cite{ALEPH_DGs}, 
under the assumption that 
these decays account for all the \CP-even final states 
(however, no systematic uncertainty due to this assumption is given, so 
the average quoted below will not include this estimate).

Present published data is not precise enough to efficiently constrain 
both \Gs and \DGGs; since the \Bs and \Bd 
lifetimes are predicted to be equal within a 
percent~\cite{equal_lifetimes,Gabbiani_et_al}, an expectation compatible with 
the current experimental data (see \Table{liferatio}),
the constraint $\Gs = \Gd$ can also be used to improve the
extraction of \DGGs.
Applying the combination procedure of \Ref{LEPHFS} 
on the published results~%
\cite{DELPHI_dms_dgs,CDFBS,CDFIN_BS1,ALEPH_DGs,ALEBS1,OPABS1_OPALAM2}
yields
\begin{equation}
|\DGs|/\Gs < 0.54 ~~~ \hbox{at \CL{95}}
\end{equation}
without external constraint, or
\begin{equation}
|\DGs|/\Gs < 0.29 ~~~ \mbox{at \CL{95}}  
\end{equation}
when constraining $1/\Gs$ to the measured \Bd lifetime.
This can be compared with the recent preliminary measurement of CDF~\cite{CDF2_DGs}
obtained from the time-dependent angular analysis of \particle{\Bs\to J/\psi\phi} decays:
\begin{equation}
\DGGs = 0.65 ^{+0.25}_{-0.33} \pm 0.01 \,. 
\end{equation}
These results are not yet precise enough to test the Standard Model predictions.

%%%%%%%%%%%%%%%%%%%%%%%%%%%%%%%%%%%%%%%%%%%%
%%%%%%%%%%%%%%%%%%%%%%%%%%%%%%%%%%%%%%%%%%%%
%%%%%%%%%%%%%%%%%%%%%%%%%%%%%%%%%%%%%%%%%%%%

% The following text provided by Donatella is based on previos DGs notes
% This is commented out for the time being, until Donatella provides something 
% more final. 

\newcommand{\comment}[1]{}\comment{

{\em This text below taken by Donatella from previous \DGs notes \ldots}

\newcommand{\Bh}{B^{\rm heavy}_{d,s}}
\newcommand{\Bl}{B^{\rm light}_{d,s}}
\newcommand{\Mh}{m^{\rm heavy}_{d,s}}
\newcommand{\Ml}{m^{\rm light}_{d,s}}
\newcommand{\Gh}{\Gamma^{\rm heavy}_{d,s}}
\newcommand{\Gl}{\Gamma^{\rm light}_{d,s}}
\newcommand{\Gsho}{\Gamma^{\rm short}_{d,s}}
\newcommand{\Glon}{\Gamma^{\rm long}_{d,s}}
\newcommand{\G}{\Gamma_{d,s}}
\newcommand{\tb}{\tau(B^0_{d,s})}
\newcommand{\dg}{\Delta\Gamma_{d,s}}
\newcommand{\tbssemi}{\tau(B_s)_{\rm semi}}
\newcommand{\Bssh}{B^{\rm short}_s}
\newcommand{\tbsshort}{\tau(\Bssh)}

The Standard Model predicts that the \Bs and \Bd can mix before decay. 
The phenomenology of this interaction can be 
described in terms of a $2\times 2$ effective Hamiltonian matrix, 
$M - i\Gamma /2$.
This results in new states called heavy and light, 
$\Bh$  and $\Bl$, for \Bs and \Bd with masses $\Mh$ ,
$\Ml$. Also the  widths  $\Gh$ and $\Gl$ could be different.

Neglecting \CP violation, the mass eigenstates are also \CP eigenstates, the ``long''  state being 
\CP even and the short one being \CP odd.  For convenience of notation, in the following
we therefore substitute 
$\Gl \equiv \Gsho$ and $\Gh \equiv \Glon$, and 
define $\G=1/\tb=(\Glon+\Gsho)/2$ and 
$\Delta \G = \Gsho-\Glon$ which is positive.

$\dg$ is related to the off-diagonal matrix elements, which have been recently 
calculated at the NLO including NLO QCD correction~\cite{Ciuchini2}. 
The theoretical values are:
\begin{equation}
\DGGs = (7.4\pm 2.4 )\times 10^{-2} \,, \hspace{1truecm} \DGGd = (2.42 \pm 0.59 ) \times 10^{-3} \,.
\end{equation}

In the same work the ratio $\DGd/\DGs$ is evaluated since the uncertainties 
coming from higher orders of QCD and $\Lambda_{\rm QCD}/m_{\b}$ corrections cancel out:
\begin{equation}
\DGd/\DGs = (3.2\pm 0.8)\times10^{-2}
\end{equation}

Experimentally \DGs can be measured fitting the lifetime of the 
light and heavy component of the \Bs.
An alternative method is based on the measurement of the  
branching fraction \particle{\Bs\to D_s^{(*)+}D_s^{(*)-}}.
Methods based on lifetime measurements have two different approaches.
Double exponential lifetime fits to samples containing a mixture of \CP eigenstates like
inclusive or semileptonic \Bs decays or $\Bs\to D_s$-hadron have a quadratic sensitivity 
to \DGs.
Whereas the isolation of a single \CP eigenstate as $\Bs\to\phi\phi$ or 
\particle{\Bs\to J/\psi\phi} to extract the lifetime of the \CP-even or odd state have 
a linear dependence on \DGs and it is more sensitive to \DGs but tend 
to suffer from reduced statistics.
The branching fraction method, exploited by ALEPH~\cite{ALEPH-phiphi}, 
is based on several theoretical assumptions~\cite{theoBR}, and allows to have 
information on \DGs only through the branching fraction measurement:
\begin{equation}
\BR{B_s\to D_s^{(*)+}
D_s^{(*)-}} = \frac{\DGs}{\Gs\left(1+\frac{\DGs}{2\Gs}\right)} \,.
\label{eq:dg_ratio}
\end{equation}

The available results are summarized in \Table{dgammat}. 
The values of the limit on \DGGs quoted in the last column of this 
table have been obtained by the working group.

Details on how these measurements are included in the average can be found  
in the previous summaries~\cite{Pcomb}.

\begin{table}
\caption{Experimental constraints on \DGGs. The upper limits,
which have been obtained by the working group, are quoted at the \CL{95}.}
\labt{dgammat}
\begin{center}
\begin{tabular}{|l|c|c|c|} 
\hline
Experiment & Selection        & Measurement            & $\Delta \Gs/\Gs$ \\ 
\hline
L3~\cite{L3B01}         & inclusive \b-sample              &                               & $<0.67$         \\
DELPHI~\cite{DELBS0}     & $\Bsb\to D_s^+\ell^- \overline{\nu_{\ell}} X$ & $\tbssemi=(1.42^{+0.14}_{-0.13}\pm0.03)$~ps  & $<0.46$ \\
others~\cite{ref:others}& $\Bsb\to D_s^+\ell^-  \overline{\nu_{\ell}} X$  & $\tbssemi=(1.46\pm{0.07})$~ps & $<0.30$ \\
ALEPH~\cite{ALEPH-phiphi}      & $\Bs\to\phi\phi X$      & 
${\rm BR}(\Bssh \to {\rm D}_s^{(*)+} {\rm D}_s^{(*)-}) =(23\pm10^{+19}_{-~9})\%$       & $0.26^{+0.30}_{-0.15}$ \\
ALEPH~\cite{ALEPH-phiphi}      & $\Bs\to\phi\phi X$      & $\tbsshort=(1.27\pm0.33\pm0.07)$~ps           & 
$0.45^{+0.80}_{-0.49}$ \\ 
%CDF~\cite{CDF-Dsl}        & $\Bsb \to D_s^+ \ell^- \overline{\nu_{\ell}}  X$ & $\tbssemi=(1.36\pm0.09^{+0.06}_{-0.05})$~ps  & $<0.83$ \\ 
DELPHI~\cite{DELBS0}$^a$    & $\Bsb \to D_s^+$ hadron    
&  $\tau_{\rm B^{D_s-had.}_s}=(1.53^{+0.16}_{-0.15}\pm0.07)$~ps                          & $<0.69$         \\
CDF~\cite{CDFB01} & $\Bs \to {\rm J}/\psi\phi$        
& $\tau_{\rm B^{{\rm J}/\psi \phi}_s}=(1.34^{+0.23}_{-0.19}\pm0.05)$~ps & $0.33^{+0.45}_{-0.42}$ \\ 
\hline
\multicolumn{4}{l}{$^a$ \footnotesize 
The value quoted for the measured lifetime differs
slightly from the one quoted in \Table{bs} because it} \\[-1ex]
\multicolumn{4}{l}{~~ \footnotesize 
corresponds to the present status of the analysis in which the information
on \DGs has been obtained.}
\end{tabular}
\end{center}
\end{table}

Here only a short description will be given.

L3 and DELPHI use inclusively reconstructed
\Bs and $\Bs\to \particle{D_s} \ell\nu X$ events respectively.
If those sample are fitted assuming a single exponential lifetime then,
assuming \DGGs is small, the measured lifetime is given by:
\begin{equation}
\tau(\Bs)_{\rm incl.} = \frac{1}{\Gs} \frac{1}{1-\left(\frac{\DGs}{2\Gs}\right)^2}
\quad \quad ; \quad \quad
\tau(\Bs)_{\rm semi.} = \frac{1}{\Gs} 
\frac{{1+\left(\frac{\DGs}{2\Gs}\right)^2}}{{1-\left(\frac{\DGs}{2\Gs}\right)^2}}.     
\end{equation}

The single lifetime fit is thus more sensitive to the effects of 
\DGs in the semileptonic case than in the fully inclusive case.

The same method is used for the \Bs world average lifetime
(recomputed without the DELPHI measurement) obtained
by using only the semileptonic decays and referenced in \Table{dgammat} as {\it others}.

The technique of reconstructing only decays at defined \CP
has been exploited by ALEPH, DELPHI and CDF.

ALEPH reconstructs the decay
$\Bs\to \particle{D_s^{(*)+}D_s^{(*)-}} \to \phi\phi X$
which is predominantly \CP even.
The proper time dependence of the \Bs component is a simple
exponential and the lifetime is related to \DGs via
\begin{equation}
\frac{\Delta \Gs}{\Gs}=2(\frac{1}{\Gs~\tbsshort}-1).  
\end{equation} 
 The same data have been used by ALEPH to exploit the branching fraction method.

DELPHI uses a sample of $\Bs\to D_s$-hadron,
which is expected to have an increased \CP-even component as the contribution
due to \particle{D_s^{(*)+}D_s^{(*)-}} events is enhanced by
selection criteria.

CDF reconstructs \particle{\Bs\to J/\psi\phi} with
\particle{J/\psi\to\mu^+\mu^-} and \particle{\phi\to K^+K^-}
where the \CP-even component is equal to $0.84\pm 0.16$ obtained by
combining CLEO~\cite{cleo} measurement of \CP-even fraction in
\particle{\Bd\to J/\psi K^{*0}} and possible SU(3) symmetry
correction.

In order to combine all the measurements~\footnote{L3 is not 
included since the likelihood for the results
was not available} the two-dimensional log-likelihood in the ($1/\Gs$, \DGGs) 
plane is summed and normalized with respect to its minimum.
The 68\%, 95\% and \CL{99} contours of the combined negative 
log-likelihood are shown in \Fig{dgplot} (left)
The corresponding limit on \DGGs is:
\begin{eqnarray}
\DGGs & = & 0.16^{+0.15}_{-0.16}  \,, \\
\DGGs & < & 0.54~\mbox{at \CL{95}} \,. 
\end{eqnarray}

\begin{figure}
\begin{center}
\epsfig{figure=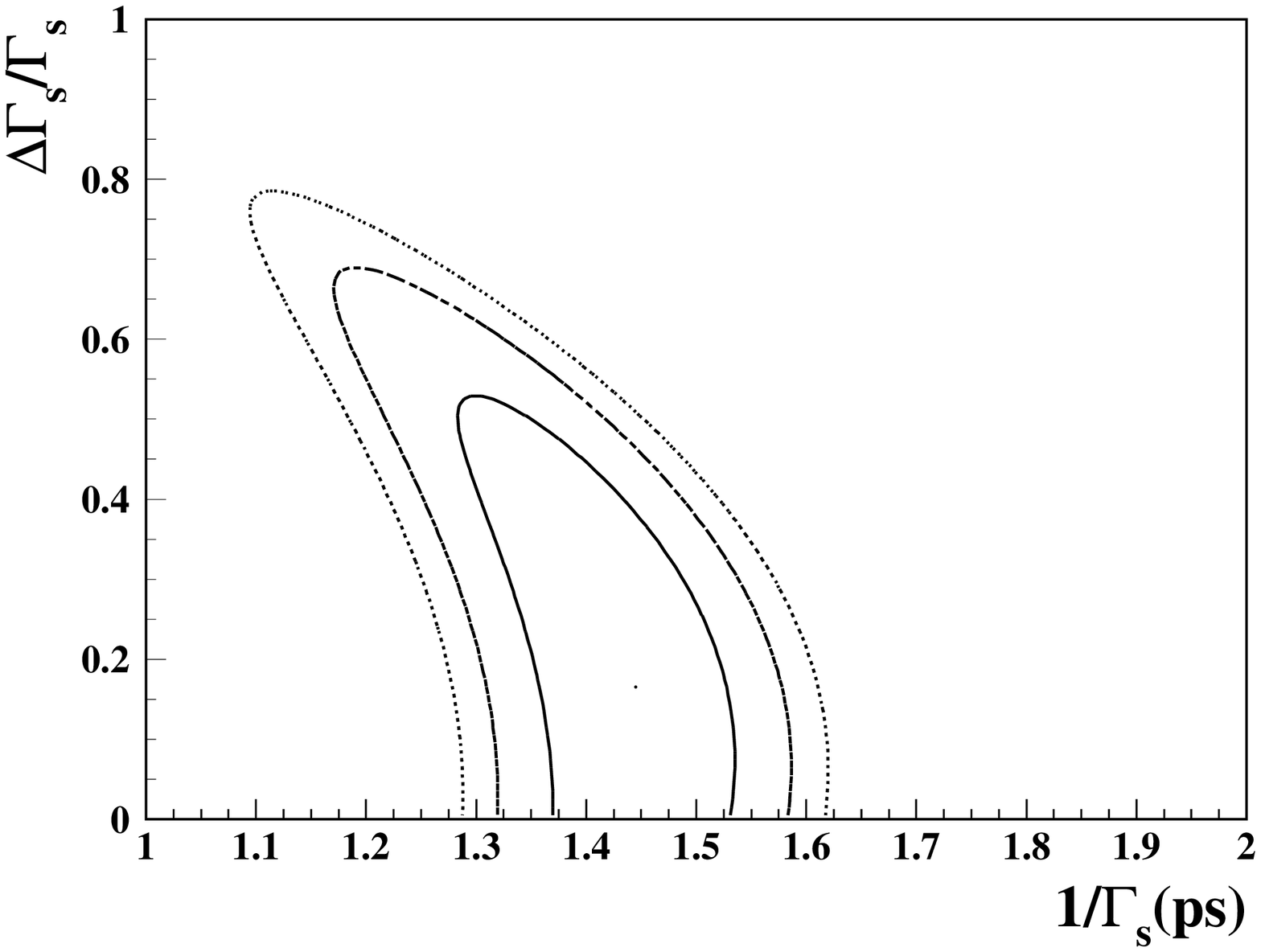,width=\textwidth}
\epsfig{figure=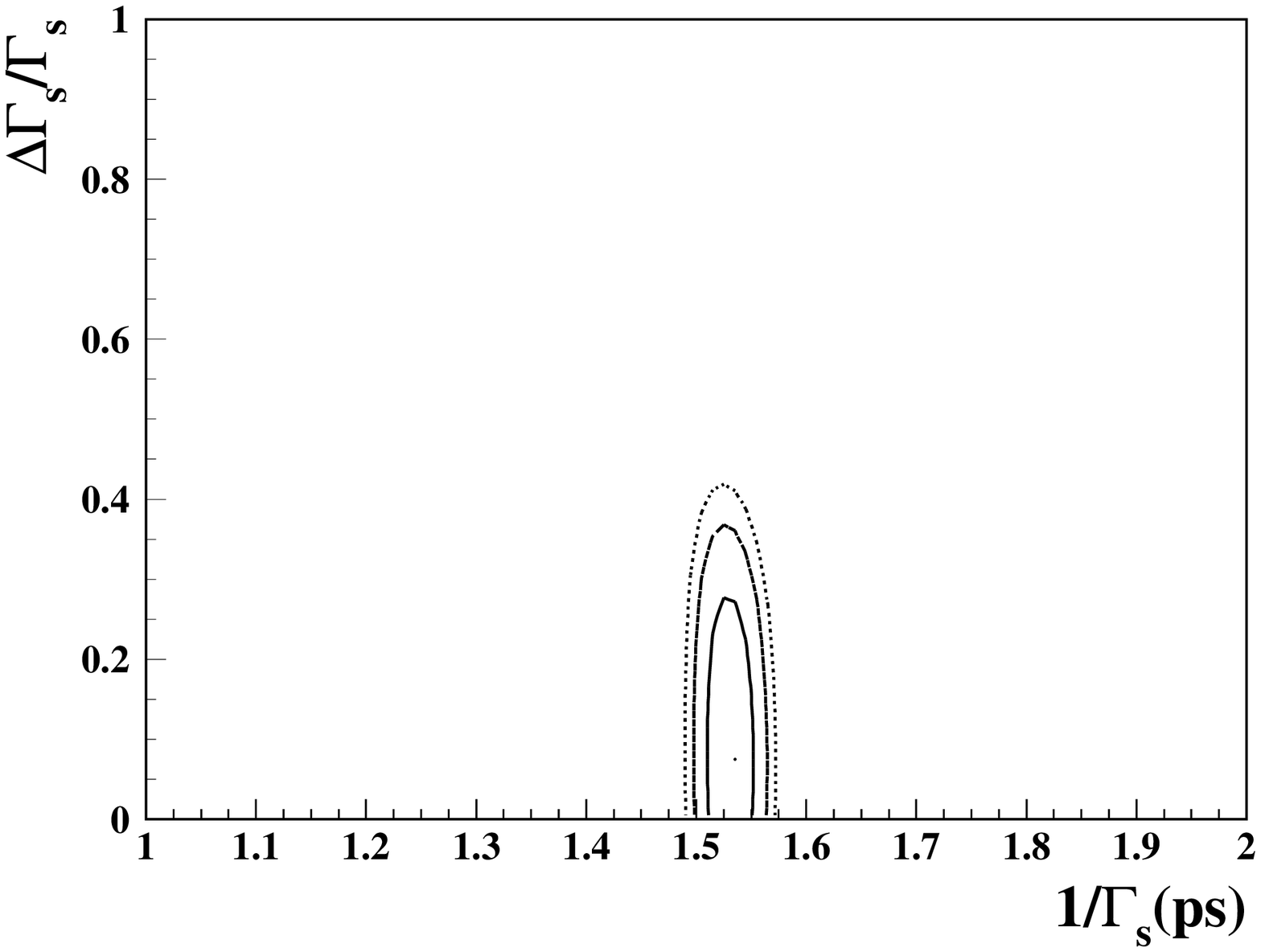,width=\textwidth}
\end{center}
\caption{Top: 68\%, 95\% and \CL{99} contours of the negative log-likelihood 
distribution in the plane ($1/\Gs$, \DGGs).
Bottom: Same, but with the constraint $1/\Gs \equiv\tau_{\Bd}$} 
\labf{dgplot}
\end{figure}

\begin{figure}
\begin{center}
\epsfig{figure=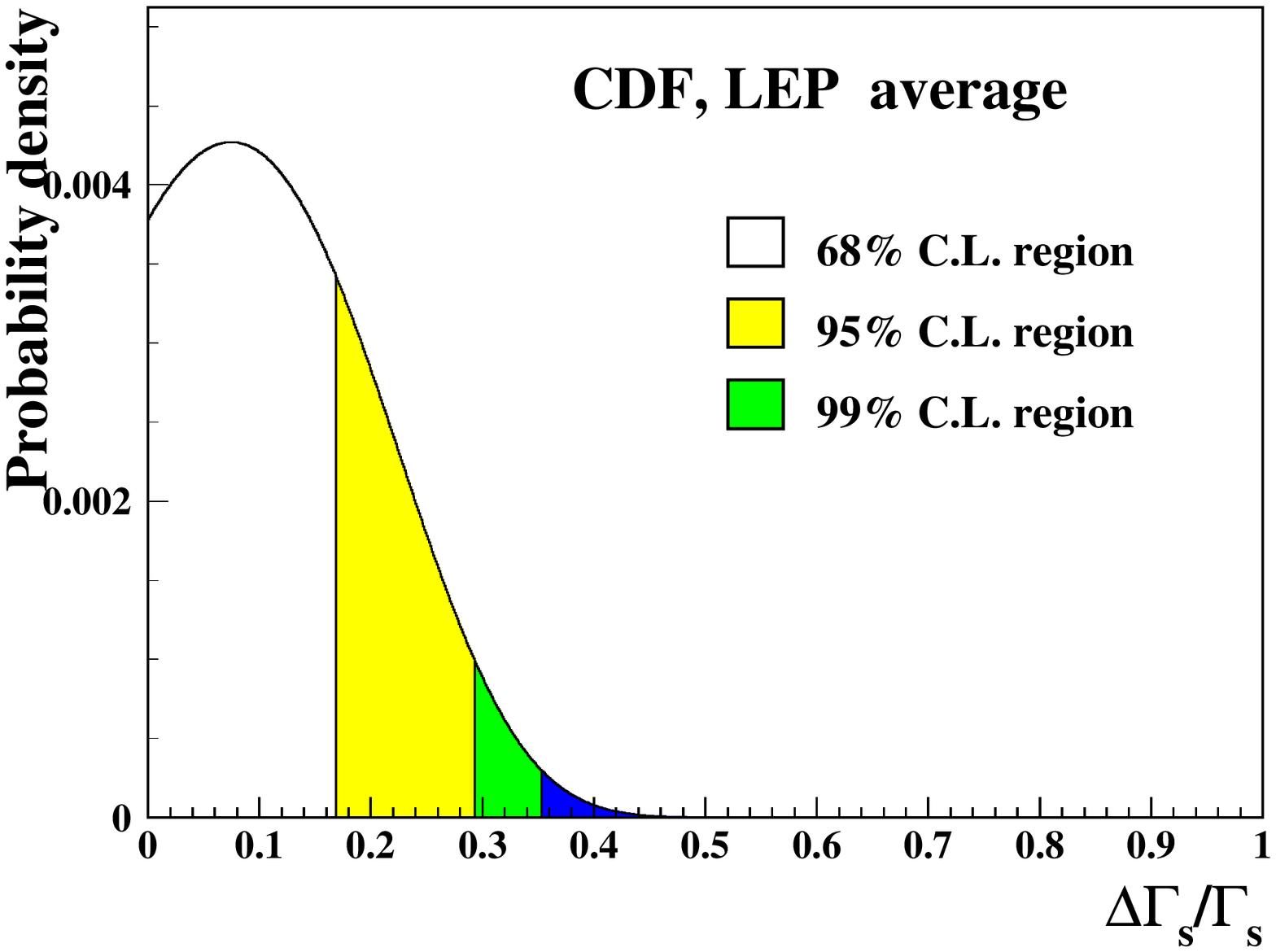,width=\textwidth}
\end{center}
\caption{Probability density distribution for \DGGs after applying the constraint; 
the three shaded regions show the limits at the 68\%, 95\% and \CL{99} respectively.} 
\labf{dgprobplot}
\end{figure}

An improved limit on \DGGs can be obtained by applying the $\tau_{\Bd}=\HFAGtauBd$ constraint.
The world average \Bs lifetime is not used, as its meaning 
is not clear if $\Delta \Gs$ is non-zero.
This is well motivated theoretically, as 
the total widths of the \Bs and \Bd mesons
are expected to be 
equal within less than one percent~\cite{bigilife}, \cite{Beneke}
and \DGd is expected to be small. 
 
The two-dimensional log-likelihood obtained, after including the constraint is shown in 
\Fig{dgplot} (right). The resulting probability density distribution for \DGGs is 
shown in \Fig{dgprobplot}. The corresponding limit on \DGGs is:
\begin{eqnarray}
\DGGs & = & 0.07^{+0.09}_{-0.07} \,, \\
\DGGs & < & 0.29~~\mbox{at \CL{95}} \,.
\end{eqnarray}

} % end comment
  % from Donatella

%\section{ Averages of Semileptonic B-decay Parameters and 
%                      $|V_{cb}|$, $|V_{ub}|$ }
% -- \include{slbdecays.tex}
% ======================================================================

%\cleardoublepage

\section{Semileptonic $B$ decays}
\label{slbdecays}

The original charge of the ``Semileptonic $B$ decays'' working group
consisted in 
\begin{itemize}
\item
the determination  of the best  values of the inclusive  and exclusive
semileptonic $B$  decay branching ratios from the  combined data, both
for Cabibbo-favored  and Cabibbo-suppressed decays,  fully taking into
account correlations between the experiments;

\item
determining  the best  possible  values of  $F(1)\vcb$ from  exclusive
semileptonic $B$ decays and of \vcb\ from the inclusive semileptonic $B$
decay rate;

\item 
a detailed understanding of  the correlated theoretical errors and the
encouragement of  a uniform and consistent error  estimation among the
active experiments.
\end{itemize}

\noindent Currently,  this program is implemented  to varying degrees.
The work  of the ``LEP Heavy  Flavor Steering Group''  has resulted in
averages  for  Cabibbo-favored  exclusive and  inclusive  semileptonic
decays (see,  \eg, Ref.~\cite{maeb}).  Recently,  the determination of
\vcb\ from inclusive decays has entered a new level precision with the
inclusion of data  on the moments of invariant  mass and lepton energy
distribution  in semileptonic $B$  decays.  This  approach is  not yet
implemented in HFAG.

In  this  edition  of the  HFAG  updates,  we  present an  average  of
inclusive determinations of \vub.  Work  is in progress to achieve the
same  for  exclusive  decays  and  eventually  for  a  combination  of
inclusive  and   exclusive  results.   See   Ref.~\cite{mblg}  for  an
alternative best estimate of \vub.

In  the  following  a  detailed  description  of  all  parameters  and
published  analyses (including preliminary  results) relevant  for the
determination of the combined results is provided.  The description is
based on the information available on the web-page at

\medskip
 \centerline{\tt http://www.slac.stanford.edu/xorg/hfag/semi/summer04/summer04.shtml}
\medskip

In the  combination of the  published results, the central  values and
errors are rescaled to a common set of input parameters, summarized in
Table~\ref{tab:common.param}   and   provided   in   the   file   {\tt
common.param} (accessible from the  web-page).  All measurements with a
dependency  on any  of these  parameters are  rescaled to  the central
values  given  in  Table~\ref{tab:common.param},  and their  error  is
recalculated based on the  error provided in the column ``Excursion''.
The  detailed dependency for  each measurement  is contained  in files
(provided by the experiments) accessible from the web-page.

\begin{table}[!htb]
\caption{Common input  parameters for the  combination of semileptonic
$B$    decays.   Most    of    the   parameters    are   taken    from
Ref.~\cite{Eidelman:2004wy}.  This  table  is  encoded in  the  file  {\tt
common.param}. The units are picoseconds for lifetimes and percentage
for branching fractions.}
\begin{center}
\begin{tabular}{|l|c|c|l|}\hline
Parameter    &Assumed Value  &Excursion             &Description\\
\hline\hline 
rb           &$21.646$       &$\pm0.065$            &$R_b$\\
bdst         &$1.27$         &$\pm0.021$            &$\cbf(\Bb\to \Dstar\tau\nub)$\\
bdsd         &$1.62$         &$\pm0.040$            &$\cbf(\Bb\to \Dstar D)$\\
bdst2        &$0.65$         &$\pm0.013$            &$\cbf(b\to \Dstar \tau)$ (OPAL incl) \\
bdsd2        &$4.2$          &$\pm1.5$              &$\cbf(b\to \Dstar D)$ (OPAL incl)\\
bdsd3        &$0.87$         &${}^{+0.23}_{-0.19}$  &$\cbf(b\to \Dstar D)$ (DELPHI incl)\\
xe           &$0.702$        &$\pm0.008$            &$B$ fragmentation:
$\langle E_B\rangle/E_{\rm beam}$\\
bdsi         &$17.3$         &$\pm2.0$              &$\cbf(b\to \Dstarp\ \mathrm{incl})$\\
cdsi         &$22.6$         &$\pm1.4$              &$\cbf(c\to \Dstarp\ \mathrm{incl})$\\
\hline
tb0          &$1.534$        &$\pm0.013$            &$\tau(\Bz)$\\
tbplus       &$1.653$        &$\pm0.014$            &$\tau(\Bp)$\\
tbps         &$1.442$        &$\pm0.066$            &$\tau(\Bs)$\\
\hline
fbd          &$39.8$         &$\pm1.0$              &$\Bz$ fraction at $\sqrt{s} = m_{Z^0}$\\
fbs          &$10.5$         &$\pm1.5$              &$\Bs$ fraction at $\sqrt{s} = m_{Z^0}$\\
fbar         &$9.9$          &$\pm1.7$              &Baryon fraction at $\sqrt{s} = m_{Z^0}$\\
\hline
dst          &$67.7$         &$\pm0.5$              &$\cbf(\Dstarp\to \Dz\pi^+)$\\
dkpp         &$9.20$         &$\pm0.6$              &$\cbf(\Dp \to K^-\pi^+\pi^+)$\\
dkp          &$3.80$         &$\pm0.09$             &$\cbf(\Dz \to K^-\pi^+)$\\
dkpzp        &$13.0$         &$\pm0.8$              &$\cbf(\Dz \to K^-\pi^+\piz)$\\
dkppp        &$7.46$         &$\pm0.31$             &$\cbf(\Dz \to K^-\pi^+\pi^+\pi^-)$\\
dkzpp        &$2.99$         &$\pm0.18$             &$\cbf(\Dz \to K^0\pi^+\pi^-)$\\
dkln         &$7.0$          &$\pm0.4$              &$\cbf(\Dz \to K^-\ell^+\nu)$\\
dkk          &$4.3$          &$\pm0.2$              &$\cbf(\Dz \to K^-K^+)$\\
dkx          &$1.100$        &$\pm0.025$            &$K^-\pi^+X$ rates\\
dkox         &$0.42$         &$\pm0.05$             &$\cbf(\Dz \to K^0X)$\\
dnlx         &$6.87$         &$\pm0.28$             &$\cbf(\Dz \to X\ell\nub)$\\
dkpcl        &$61.2$         &$\pm2.9$              &$\cbf(\Dstarz \to \Dz\piz)$\\
dssR         &$0.64$         &$\pm0.11$             &$\cbf(b\to D^{**}\ell\nub)\times\cbf(D^{**}\to D^{*+}X)$ \\
\hline
fb0          &$49.4$         &$\pm0.8$              &$f^{00} = \cbf(\FourS\to \Bz\Bzb)$\\
chid         &$0.186$        &$\pm0.004$            &$\chi_d$, time-integrated probability for \Bz\ mixing\\
chi          &$0.092$        &$\pm0.002$            &$\chi = \chi_d\times (f^{00}/100)$\\
\hline
\end{tabular}
\end{center}
\label{tab:common.param}
\end{table}

%% \clearpage

% -- \include{b2cexcl.tex}
% ======================================================================
\subsection{Exclusive Cabibbo-favored decays}
\label{slbdecays_b2cexcl}
% -------------------------------------------

Aspects of  the phenomenology of exclusive  Cabibbo-favored $B$ decays
and their  use in the determination  of \vcb\ in the  context of Heavy
Quark Effective  Theory (HQET) are  described in many places,  \eg, in
Ref.~\cite{maeb} and will not be repeated here.

Averages are provided for both the branching ratios $\cbf(\BzbDstarlnu)$
plus $\cbf(\BzbDplnu)$ and the CKM matrix element \vcb\ multiplied by the
form factor  at zero  recoil of the  decay $\BzbDstarlnu$  and $\BzbDplnu$,
respectively.

% ----------------------------------------------------------------------
\subsubsection{\BzbDstarlnu}
\label{slbdecays_dstarlnu}
% ------------------------

The    measurements    included    in    the   average,    shown    in
Table~\ref{tab:dstarlnu}  are  scaled to  a  consistent  set of  input
parameters  and   their  errors.    Therefore  some  of   the  (older)
measurements are subject to considerable adjustments.

\begin{itemize}
 
  \item  In  order  to  reduce  the dependence  on  theoretical  error
 estimates, the central  values and errors for the  form factors $R_1$
 and     $R_2$     are    taken     from     the    measurement     by
 CLEO~\cite{Duboscq:1995mv}. However, all  experiments (except for the
 CLEO~\cite{Adam:2002uw}, the recent DELPHI~\cite{Abdallah:2004rz} and
 the   \babar~\cite{Aubert:2004bw}   measurements)   quote  in   their
 abstracts  $F(1)\vcb$ based  on  form factors  (and their  respective
 errors) from  theory.  \belle provides a second  result evaluated with
 the  CLEO  form factors.   All  other  experiments have  recalculated
 $F(1)\vcb$  to rely  on  the CLEO  form  factors.  In  the future,  a
 substantial improvement in the  error associated with form factors is
 expected by including form factor measurements at the $B$ factories.
 
  \item  Updates in  the branching  fractions  of $D$  mesons and  the
 production of  $D^{**}$ mesons in  $B$ decays have generally  lead to
 increased rates for $\cbf(\BzbDstarlnu)$.
 
  \item The  average \Bz\ lifetime has changed  considerably since the
 ALEPH  measurement,   especially  with  the   much  higher  precision
 available at  the $B$ factories. This  effect is less visible  in the
 other measurements as they are more recent.
 
  \item The  production of \Bz\ mesons  at $\sqrt{s} =  m_{Z^0}$ has a
 direct  impact on  all LEP  measurements.  Adjusting  results  on the
 \FourS\ for the branching  fraction of $\cbf(\FourS\to \BzBzb) \equiv
 f^{00}$ yields  an increase compared  to the assumption of  $f^{00} =
 0.5$.
 
  \item  Many input  parameters are  now known  with a  much increased
 precision---this  decreases  some of  the  systematic  errors of  the
 rescaled results with respect to the original publication.
 
\end{itemize}

The largest correlated errors are the fraction of \Bz\ mesons (fbd and
fb0,  respectively), the form  factors $R_1(1)$  and $R_2(1)$  at zero
recoil, \Bz\  meson lifetime, branching  fractions of $D$  mesons, and
the details of $D^{**}$ modeling.

\begin{figure}[!ht]
 \begin{center}
  \unitlength1.0cm % coordinates in cm
  \begin{picture}(14.,8.0)  %ys(25.,6.0)
   \put( -0.5,  0.0){\includegraphics[width=7.5cm]{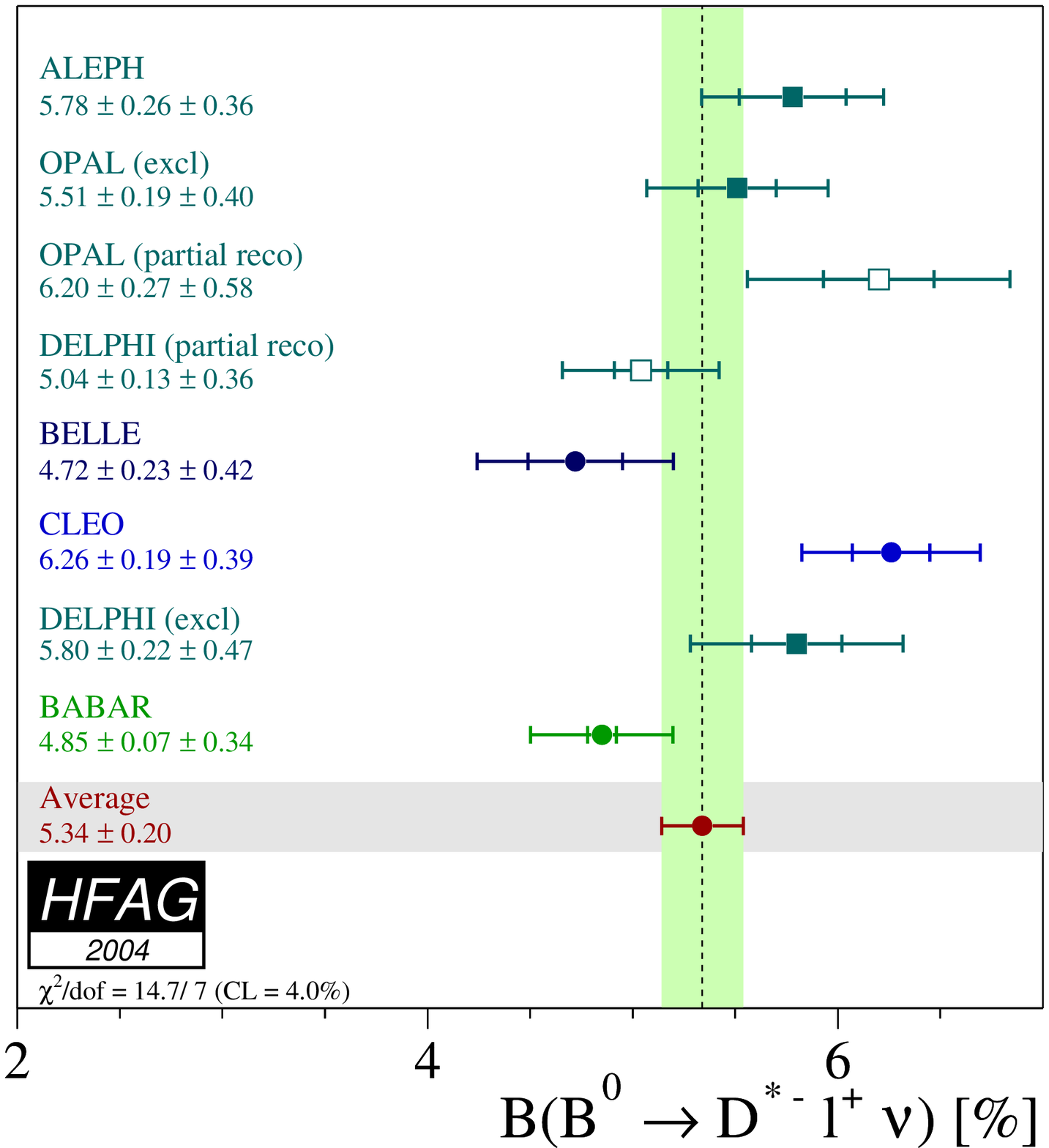}}
    %ys {\includegraphics[width=0.50\textwidth]{figures/slb/br_dsl.eps}}

   \put(  8.0,  0.0){\includegraphics[width=7.5cm]{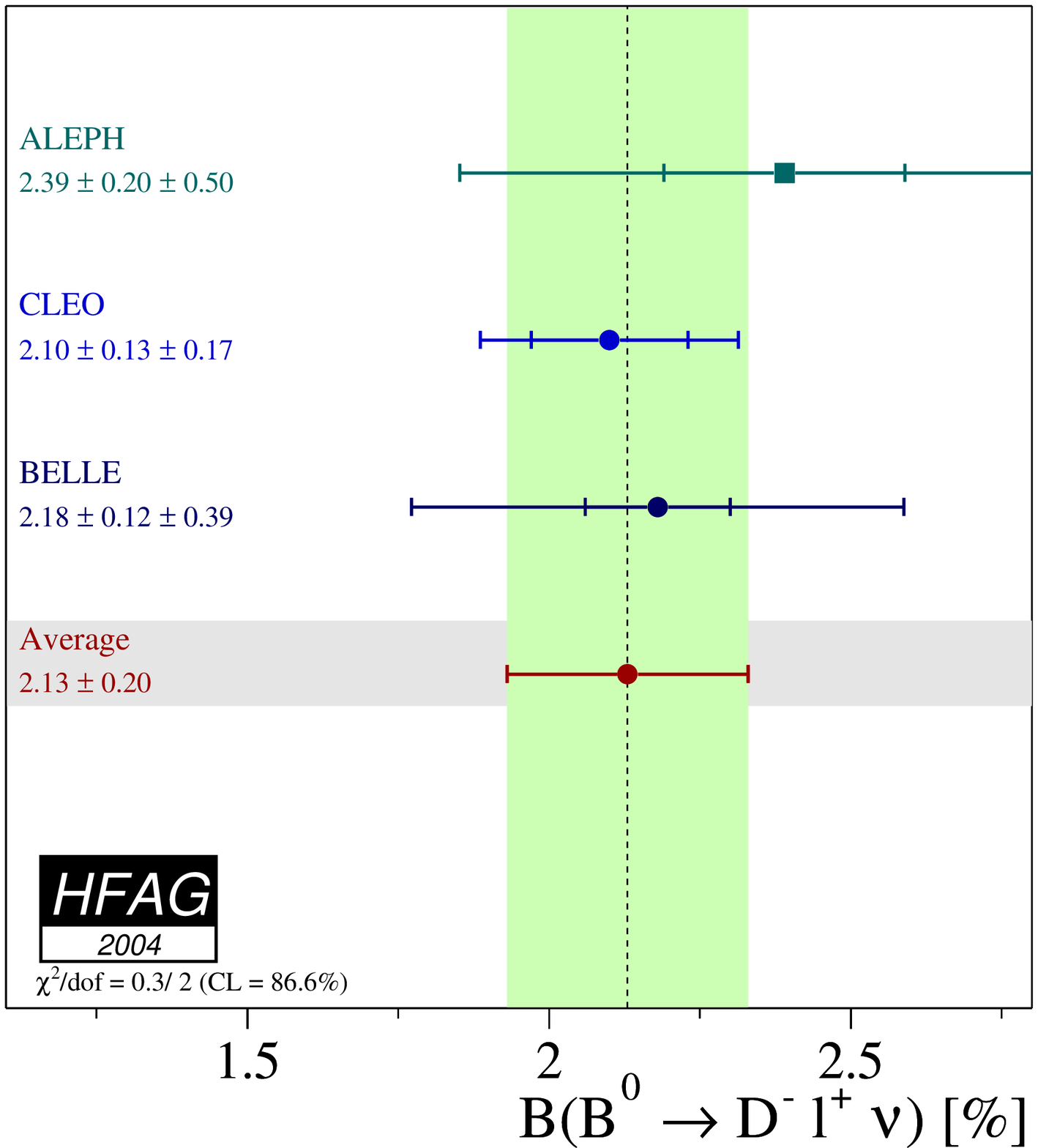}}
    %ys {\includegraphics[width=0.50\textwidth]{figures/slb/br_dl.eps}}
    
   \put(  5.5,  6.8){{\large\bf a)}}
   \put( 14.0,  6.8){{\large\bf b)}}
  \end{picture}
  \caption{Average branching ratio of  exclusive semileptonic $B$ decays. (a) \BzbDstarlnu\ and (b) \BzbDplnu.}
  \label{fig:brdsl}
 \end{center}
\end{figure}

At LEP, the  measurements of \BzbDstarlnu\ decays have  been done both
with ``inclusive''  analyses based on a partial  reconstruction of the
\BzbDstarlnu\ decay and a full  reconstruction of the exclusive decay. 
The average  branching ratio  $\cbf(\BzbDstarlnu)$ is determined  in a
one-dimensional    fit    from    the   measurements    provided    in
Table~\ref{tab:dstarlnu}.   The  statistical  correlation between  two
analyses from  the same experiment (DELPHI and  OPAL, respectively) is
taken   into  account.    Figure~\ref{fig:brdsl}(a)   illustrates  the
measurements and the resulting average.  The $\chi^2/\dof = 14.7/7$ is
slightly  above the  level where  the  PDG starts  to introduce  scale
factors.  The measurement by CLEO provides the largest contribution to
the $\chi^2$.

% ----------------------------------------------------------------------
\begin{table}[!htb]
\caption{Average branching ratio $\cbf(\BzbDstarlnu)$ and individual
  results. See the description in the text for explanations why the
  published results are lower than the rescaled results. }
\begin{center}
\begin{tabular}{|l|c|c|}\hline
Experiment    &$\cbf(\BzbDstarlnu) [\%]$    (rescaled)     &$\cbf(\BzbDstarlnu) [\%]$  (published) \\
\hline\hline 
ALEPH (excl)~\hfill\cite{Buskulic:1996yq}  &$5.78\pm 0.26_{\rm stat} \pm 0.36_{\rm syst}$ &$5.53\pm 0.26_{\rm stat} \pm 0.52_{\rm syst}$ \\
OPAL (excl)~\hfill\cite{Abbiendi:2000hk}   &$5.51\pm 0.19_{\rm stat} \pm 0.40_{\rm syst}$ &$5.11\pm 0.19_{\rm stat} \pm 0.49_{\rm syst}$ \\
OPAL (incl)~\hfill\cite{Abbiendi:2000hk}   &$6.20\pm 0.27_{\rm stat} \pm 0.58_{\rm syst}$ &$5.92\pm 0.27_{\rm stat} \pm 0.68_{\rm syst}$ \\
DELPHI (incl)~\hfill\cite{Abreu:2001ic}    &$5.04\pm 0.13_{\rm stat} \pm 0.36_{\rm syst}$ &$4.70\pm 0.13_{\rm stat} \ {}^{+0.36}_{-0.31}\ {}_{\rm syst}$ \\
\belle  (excl)~\hfill\cite{Abe:2001cs}      &$4.72\pm 0.23_{\rm stat} \pm 0.42_{\rm syst}$ &$4.60\pm 0.23_{\rm stat} \pm 0.40_{\rm syst}$ \\
CLEO  (excl)~\hfill\cite{Adam:2002uw}      &$6.26\pm 0.19_{\rm stat} \pm 0.39_{\rm syst}$ &$6.09\pm 0.19_{\rm stat} \pm 0.40_{\rm syst}$ \\
DELPHI (excl)~\hfill\cite{Abdallah:2004rz} &$5.80\pm 0.22_{\rm stat} \pm 0.47_{\rm syst}$ &$5.90\pm 0.22_{\rm stat} \pm 0.50_{\rm syst}$ \\
\babar\ (excl)~\hfill\cite{Aubert:2004bw}  &$4.85\pm 0.07_{\rm stat} \pm 0.34_{\rm syst}$ &$4.90\pm 0.07_{\rm stat} \pm 0.36_{\rm syst}$ \\
\hline 
{\bf Average}                              &\mathversion{bold}$5.34\pm0.20$           &\mathversion{bold}$\chi^2/\dof = 14.7/7$ \\
\hline 
\end{tabular}
\end{center}
\label{tab:dstarlnu}
\end{table}
% ----------------------------------------------------------------------

\begin{figure}[!ht]
 \begin{center}
  \unitlength1.0cm % coordinates in cm
  \begin{picture}(14.,8.0)  %ys(25.,6.)
   \put( -0.5,  0.0){\includegraphics[width=7.5cm]{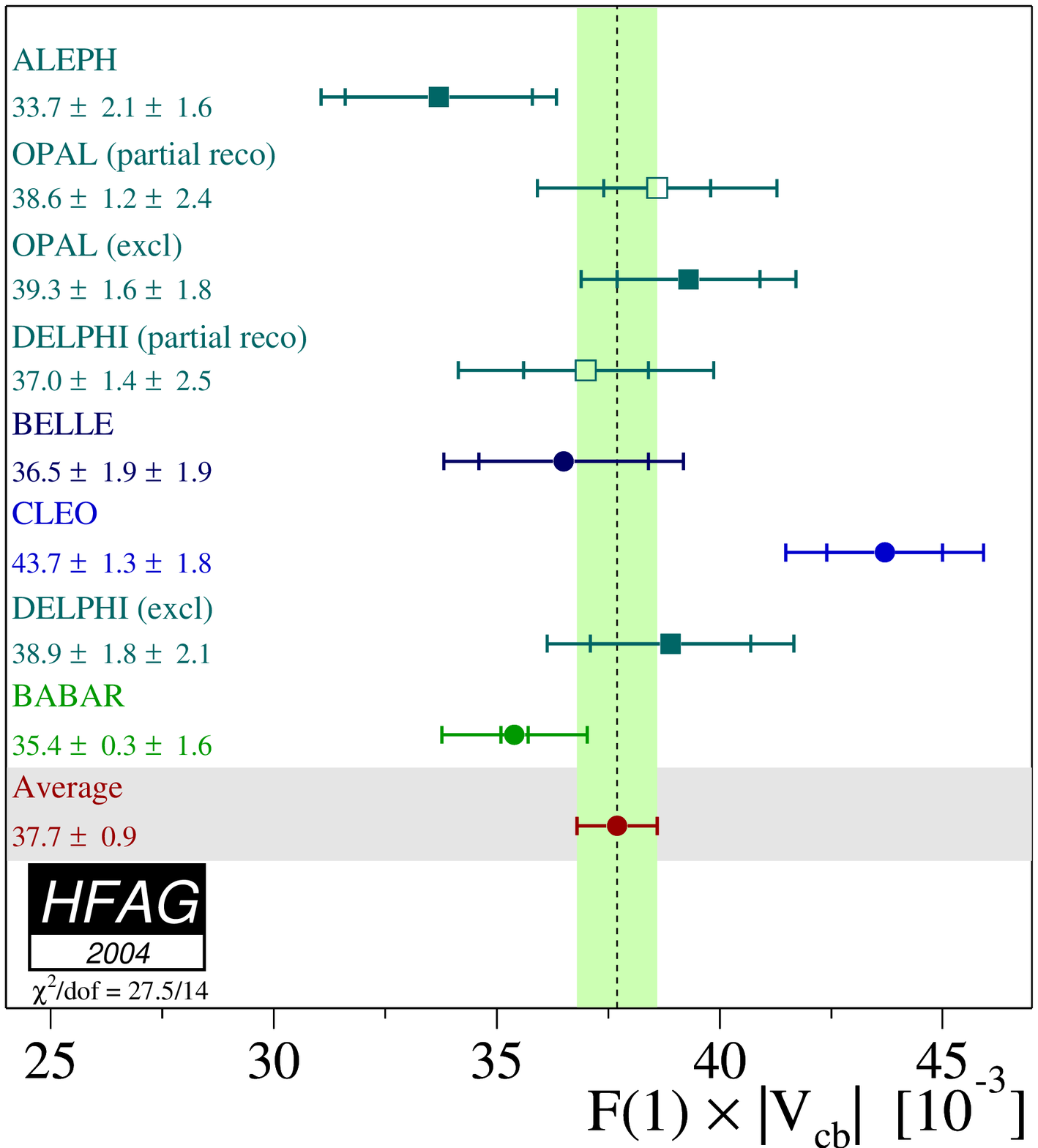}}
   %ys{\includegraphics[width=0.49\textwidth]{figures/slb/vcbf1.eps}}
   \put(  8.0, -0.2){\includegraphics[width=8.0cm]{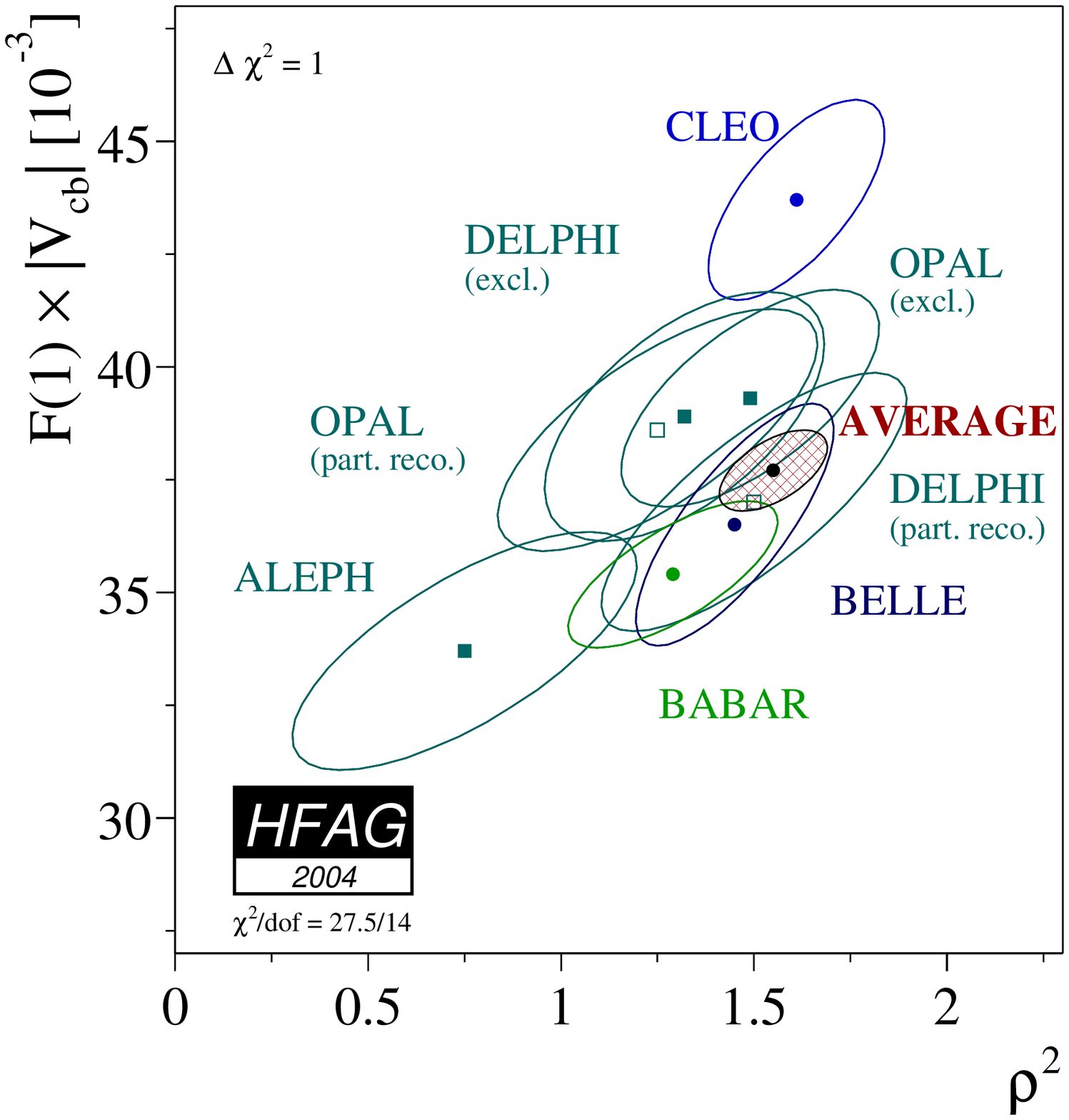}}
   %ys{\includegraphics[width=0.51\textwidth]{figures/slb/vcbf1.vs.rho2.eps}}
   \put(  5.5,  6.8){{\large\bf a)}}  
   \put( 14.4,  6.8){{\large\bf b)}}
   \end{picture} \caption{(a)  Illustration of the  average $F(1)\vcb$
   and   rescaled  measurements   of   exclusive  \BzbDstarlnu\   decays
   determined   in  a  two-dimensional   fit.   (b)   Illustration  of
   $F(1)\vcb$   vs.  $\rho^2$.  The   error  ellipses   correspond  to
   $\Delta\chi^2 = 1$.  }  \label{fig:vcbf1} \end{center}
\end{figure}

The  average  for  $F(1)\vcb$  is determined  by  the  two-dimensional
combination  of the results  provided in  Table~\ref{tab:vcbf1}.  This
allows  to fully  incorporate the  correlation between  $F(1)\vcb$ and
$\rho^2$. Figure~\ref{fig:vcbf1}(b)  illustrates the average $F(1)\vcb$
and the measurements included in the average. Figure~\ref{fig:vcbf1}(a)
provides a one-dimensional  projection for illustrative purposes.  The
$\chi^2/\dof  = 27.5/14$ is  below the  level where  the PDG  starts to
introduce scale factors.  The measurement by CLEO provides the largest
contribution to the $\chi^2$.

% ----------------------------------------------------------------------
\begin{table}[!htb]

\caption{Average of $F(1)\vcb$ determined in the decay \BzbDstarlnu\ and
individual  results.   The  fit  for  the average  has  $\chi^2/\dof  =
27.5/14$.  The total  correlation between  the average  $F(1)\vcb$ and
$\rho^2$ is 0.57.}

\begin{center}
\begin{tabular}{|l|c|c|}\hline
Experiment                                 &$F(1)\vcb [10^{-3}]$ (rescaled) &$\rho^2$ (rescaled) \\
                                           &$F(1)\vcb [10^{-3}]$ (published) &$\rho^2$ (published) \\
\hline\hline 
ALEPH (excl)~\hfill\cite{Buskulic:1996yq}  &$33.7\pm 2.1_{\rm stat} \pm 1.6_{\rm syst}$   &$0.75\pm 0.25_{\rm stat} \pm 0.37_{\rm syst}$ \\
                                           &$31.9\pm 1.8_{\rm stat} \pm 1.9_{\rm syst}$   &$0.37\pm 0.26_{\rm stat} \pm 0.14_{\rm syst}$ \\
\hline
OPAL (incl)~\hfill\cite{Abbiendi:2000hk}   &$38.6\pm 1.2_{\rm stat} \pm 2.4_{\rm syst}$   &$1.25\pm 0.14_{\rm stat} \pm 0.39_{\rm syst}$ \\
                                           &$37.5\pm 1.2_{\rm stat} \pm 2.5_{\rm syst}$   &$1.12\pm 0.14_{\rm stat} \pm 0.29_{\rm syst}$ \\
\hline
OPAL (excl)~\hfill\cite{Abbiendi:2000hk}   &$39.3\pm 1.6_{\rm stat} \pm 1.8_{\rm syst}$   &$1.49\pm 0.21_{\rm stat} \pm 0.26_{\rm syst}$ \\
                                           &$36.8\pm 1.6_{\rm stat} \pm 2.0_{\rm syst}$   &$1.31\pm 0.21_{\rm stat} \pm 0.16_{\rm syst}$ \\
\hline
DELPHI (incl)~\hfill\cite{Abreu:2001ic}    &$37.0\pm 1.4_{\rm stat} \pm 2.5_{\rm syst}$   &$1.50\pm 0.14_{\rm stat} \pm 0.37_{\rm syst}$ \\
                                  &$35.5\pm1.4_{\rm stat}\ {}^{+2.3}_{-2.4}{}_{\rm syst}$ &$1.34\pm0.14_{\rm stat}\ {}^{+0.24}_{-0.22}{}_{\rm syst}$\\
\hline
\belle  (excl)~\hfill\cite{Abe:2001cs}     &$36.5\pm 1.9_{\rm stat} \pm 1.9_{\rm syst}$   &$1.45\pm 0.16_{\rm stat} \pm 0.20_{\rm syst}$ \\
                                           &$35.8\pm 1.9_{\rm stat} \pm 1.9_{\rm syst}$   &$1.45\pm 0.16_{\rm stat} \pm 0.20_{\rm syst}$ \\
\hline
CLEO  (excl)~\hfill\cite{Adam:2002uw}      &$43.7\pm 1.3_{\rm stat} \pm 1.8_{\rm syst}$   &$1.61\pm 0.09_{\rm stat} \pm 0.21_{\rm syst}$ \\
                                           &$43.1\pm 1.3_{\rm stat} \pm 1.8_{\rm syst}$   &$1.61\pm 0.09_{\rm stat} \pm 0.21_{\rm syst}$ \\
\hline
DELPHI (excl)~\hfill\cite{Abdallah:2004rz} &$38.9\pm 1.8_{\rm stat} \pm 2.1_{\rm syst}$   &$1.32\pm 0.15_{\rm stat} \pm 0.33_{\rm syst}$ \\
                                           &$39.2\pm 1.8_{\rm stat} \pm 2.3_{\rm syst}$   &$1.32\pm 0.15_{\rm stat} \pm 0.33_{\rm syst}$ \\
\hline
\babar\ (excl)~\hfill\cite{Aubert:2004bw}  &$35.4\pm 0.3_{\rm stat} \pm 1.6_{\rm syst}$   &$1.29\pm 0.03_{\rm stat} \pm 0.27_{\rm syst}$ \\
                                           &$35.5\pm 0.3_{\rm stat} \pm 1.6_{\rm syst}$   &$1.29\pm 0.03_{\rm stat} \pm 0.27_{\rm syst}$ \\
\hline 
{\bf Average }                             &\mathversion{bold}$37.7\pm0.9$        &\mathversion{bold}$1.55\pm0.14$ \\
\hline 
\end{tabular}
\end{center}
\label{tab:vcbf1}
\end{table}
% ----------------------------------------------------------------------

For  a determination  of \vcb,  the form factor  at zero  recoil $F(1)$
needs   to   be   computed.     A   possible   choice   is   $F(1)   =
0.91\pm0.04_{\rm theo}$~\cite{Battaglia:2003in}, resulting in

\begin{displaymath}
\vcb = (41.4 \pm 1.0_{\rm exp} \pm 1.8_{\rm theo}) \times 10^{-3}.
\end{displaymath}

\noindent
% The value for $F(1)$ and its error  is based on a comparison
% of  results  by  QCD  sum  rules  and  (quenched)  HQET-based  lattice
% calculations (see Ref.~\cite{Battaglia:2003in} for more details).
%- try to include Elizabetta's suggestion (YS)
The value for $F(1)$ and its error  is based on a comparison
of estimates % in a quark model\cite{neubert-vcb}\cite{neubert-vcb2}, 
using OPE sum rules  %\cite{uraltsev-vcb}, 
and with an HQET based lattice gauge calculation %\cite{kronfeld}. 
(see Ref.~\cite{Battaglia:2003in} for more details).

% ----------------------------------------------------------------------

% ----------------------------------------------------------------------

% ----------------------------------------------------------------------
\subsubsection{\BzbDplnu}
\label{slbdecays_dlnu}
% --------------------

The  average branching  ratio $\cbf(\BzbDplnu)$  is determined  by the
combination  of  the results  provided  in Table~\ref{tab:dlnu}.   The
error    sources    here    are    the   same    as    discussed    in
Sec.~\ref{slbdecays_dstarlnu}, but generally  at a higher level due
to larger background levels, less stringent kinematic constraints, and
larger     kinematic     suppression      at     the     endpoint.     
Figure~\ref{fig:brdsl}(b)   illustrates   the   measurements  and   the
resulting average.

% ----------------------------------------------------------------------0
\begin{table}[!htb]
\caption{Average of the branching ratio $\cbf(\BzbDplnu)$ and individual
results. }
\begin{center}
\begin{tabular}{|l|c|c|}\hline
Experiment                                 &$\cbf(\BzbDplnu) [\%]$ (rescaled) &$\cbf(\BzbDplnu) [\%]$ (published) \\
\hline\hline 
ALEPH ~\hfill\cite{Buskulic:1996yq}        &$2.39 \pm0.20_{\rm stat} \pm0.50_{\rm syst}$  &$2.35 \pm0.20_{\rm stat} \pm0.44_{\rm syst}$ \\
CLEO  ~\hfill\cite{Bartelt:1998dq}         &$2.10 \pm0.13_{\rm stat} \pm0.17_{\rm syst}$  &$2.20 \pm0.16_{\rm stat} \pm0.19_{\rm syst}$ \\
\belle  ~\hfill\cite{Abe:2001yf}           &$2.18 \pm0.12_{\rm stat} \pm0.39_{\rm syst}$  &$2.13 \pm0.12_{\rm stat} \pm0.39_{\rm syst}$ \\
\hline 
{\bf Average}                              &\mathversion{bold}$2.13 \pm0.20$      &\mathversion{bold}$\chi^2/\dof = 0.3/2$ \\
\hline 
\end{tabular}
\end{center}
\label{tab:dlnu}
\end{table}
% ----------------------------------------------------------------------

\begin{figure}[!ht]
 \begin{center}
  \unitlength1.0cm % coordinates in cm
  \begin{picture}(14.,8.) %ys(25.,6.)
   \put( -0.5,  0.0){\includegraphics[width=7.5cm]{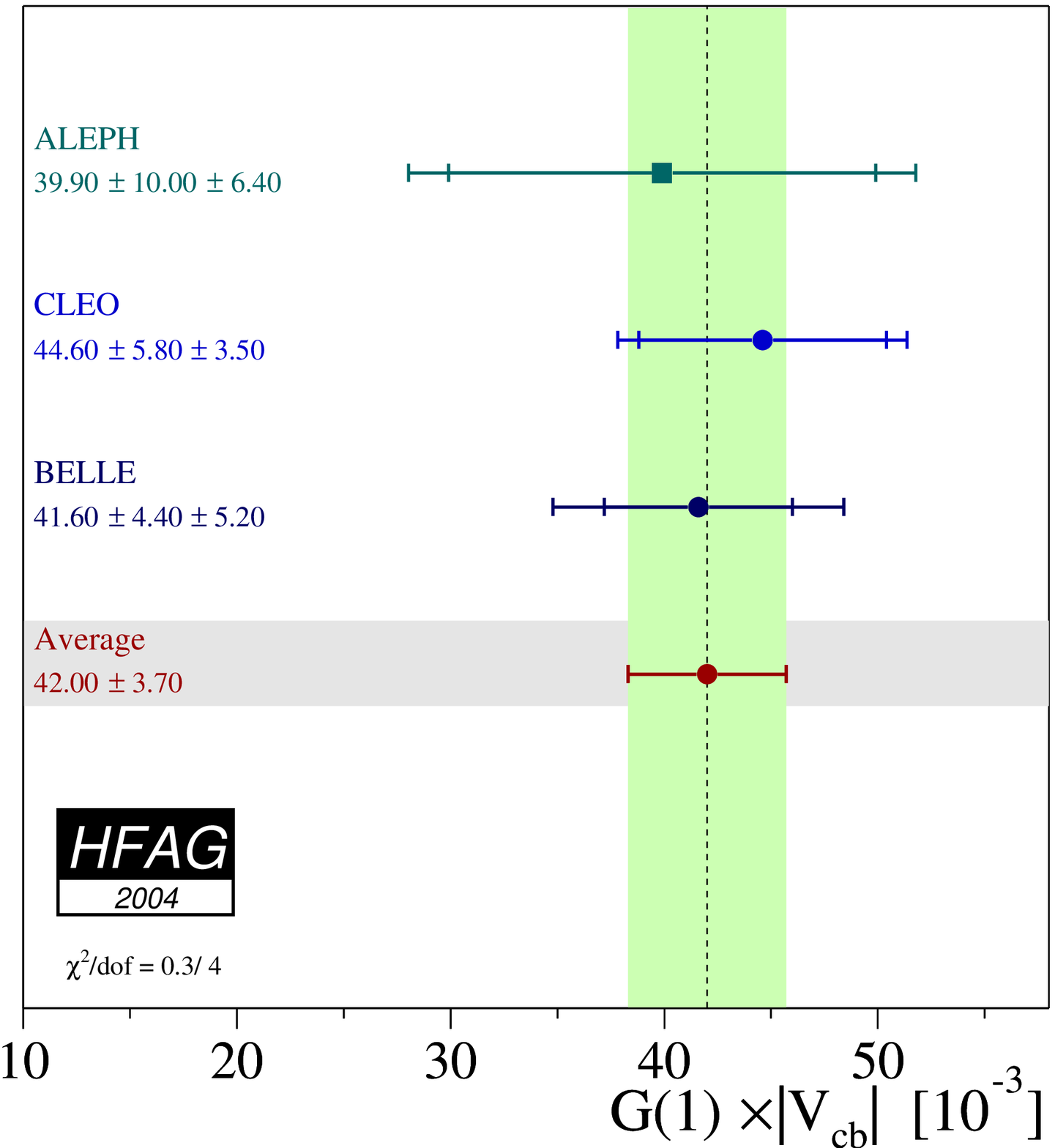}}
    %ys{\includegraphics[width=0.49\textwidth]{figures/slb/vcbg1.eps}}
   \put(  8.0, -0.2){\includegraphics[width=8.0cm]{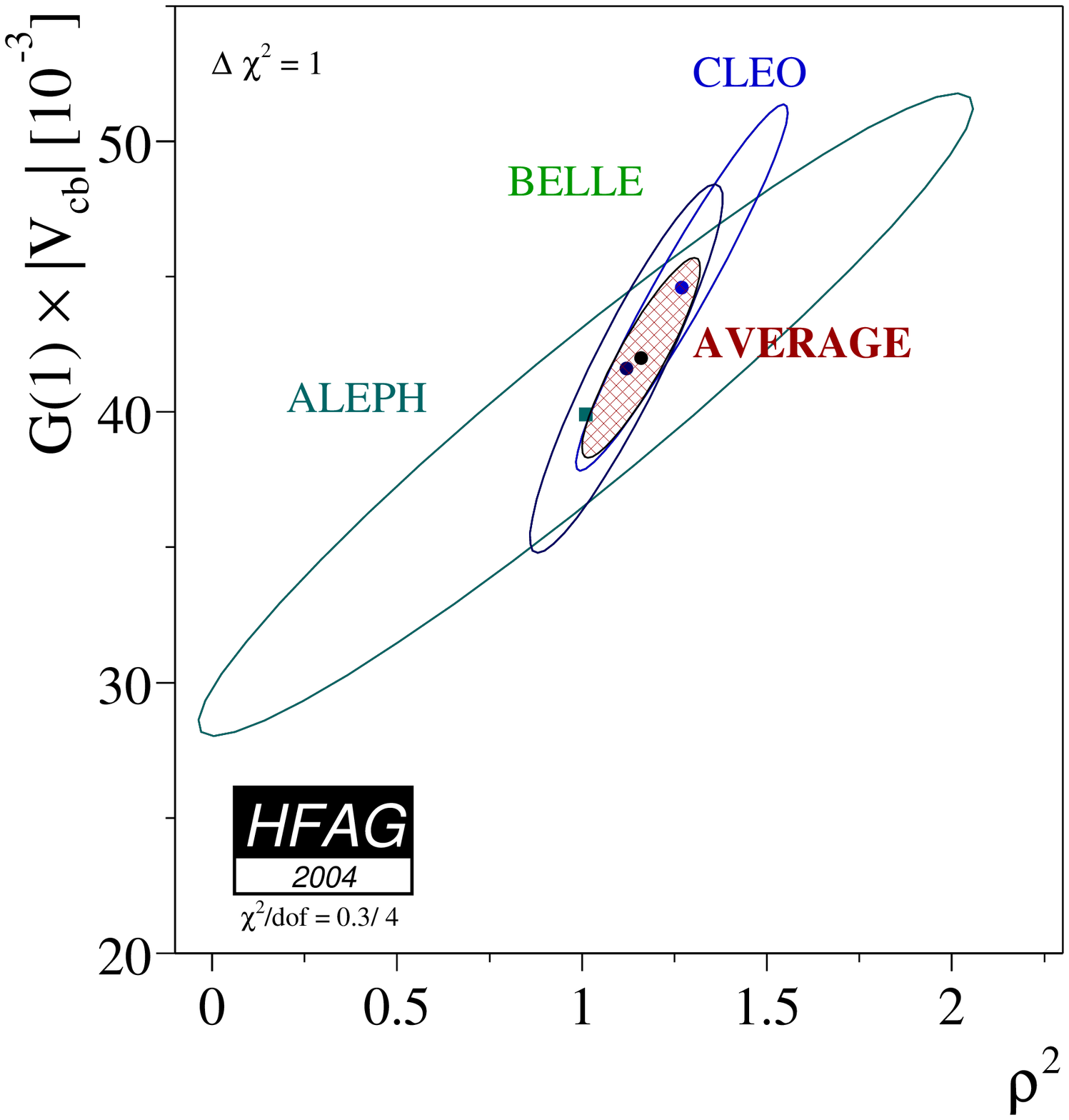}}
   %ys{\includegraphics[width=0.51\textwidth]{figures/slb/vcbg1.vs.rho2.eps}}
   \put(  5.5,  6.8){{\large\bf a)}}
   \put( 14.4,  6.8){{\large\bf b)}}
  \end{picture}
  \caption{(a)  Illustration of the  average $G(1)\vcb$
   and   rescaled  measurements   of   exclusive  \BzbDplnu\   decays
   determined   in  a  two-dimensional   fit.   (b)   Illustration  of
   $F(1)\vcb$   vs.  $\rho^2$.  The   error  ellipses   correspond  to
   $\Delta\chi^2 = 1$. }
  \label{fig:vcbg1}
 \end{center}
\end{figure}

The  average  for  $G(1)\vcb$  is determined  by  the  two-dimensional
combination  of   the  results  provided   in  Table~\ref{tab:vcbg1}.  
Figure~\ref{fig:vcbg1}(b)  illustrates the  average $F(1)\vcb$  and the
measurements   included  in   the   average.  Figure~\ref{fig:vcbg1}(a)
provides a one-dimensional projection for illustrative purposes.

\begin{table}[!htb]
\caption{Average  of $G(1)\vcb$  determined  in the  decay \BzbDplnu\  and
individual  results. The  fit  for the  average  has $\chi^2/\dof  =
0.3/4$.}
\begin{center}
\begin{tabular}{|l|c|c|}\hline
Experiment &$G(1)\vcb [10^{-3}]$ (rescaled)  &$\rho^2$ (rescaled) \\ 
           &$G(1)\vcb [10^{-3}]$ (published) &$\rho^2$ (published) \\
\hline\hline 
ALEPH~\hfill\cite{Buskulic:1996yq}  &$39.9 \pm10.0_{\rm stat} \pm6.4_{\rm syst}$   &$1.01 \pm0.98_{\rm stat} \pm0.37_{\rm syst}$ \\
                                    &$31.1 \pm9.9_{\rm stat}  \pm8.6_{\rm syst}$   &$0.20 \pm0.98_{\rm stat} \pm0.50_{\rm syst}$ \\
\hline
CLEO ~\hfill\cite{Bartelt:1998dq}   &$44.9 \pm5.8_{\rm stat} \pm3.5_{\rm syst}$    &$1.27 \pm0.25_{\rm stat} \pm0.14_{\rm syst}$ \\
                                    &$44.8 \pm6.1_{\rm stat} \pm3.7_{\rm syst}$  &$1.30 \pm0.27_{\rm stat} \pm0.14_{\rm syst}$ \\
\hline
\belle~\hfill\cite{Abe:2001yf}       &$41.6 \pm4.4_{\rm stat} \pm5.2_{\rm syst}$    &$1.12 \pm0.22_{\rm stat} \pm0.14_{\rm syst}$ \\
                                    &$41.1 \pm4.4_{\rm stat} \pm5.1_{\rm syst}$    &$1.12 \pm0.22_{\rm stat} \pm0.14_{\rm syst}$ \\
\hline 
{\bf Average }                      &\mathversion{bold}$42.1 \pm3.7$       &\mathversion{bold}$1.15 \pm0.16$      \\
\hline 
\end{tabular}
\end{center}
\label{tab:vcbg1}
\end{table}

For  a determination  of \vcb,  the form factor  at zero  recoil $G(1)$
needs   to    be   computed.   A   possible   choice    is   $G(1)   =
1.04\pm0.06_{\rm theo}$~\cite{Battaglia:2003in}, resulting in

\begin{displaymath}
\vcb = (40.4 \pm 3.6_{\rm exp} \pm 2.3_{\rm theo}) \times 10^{-3}.
\end{displaymath}

%% \clearpage

% -- \include{b2cincl.tex}
% ======================================================================
\subsection{Inclusive Cabibbo-favored decays}
\label{slbdecays_b2cincl}
% -------------------------------------------

Aspects of  the theory and phenomenology  of inclusive Cabibbo-favored
$B$ decays and their use in  the determination of \vcb\ in the context
of  the Heavy  Quark Expansion  (HQE), an  Operator  Product Expansion
based   on   HQET,  are   described   in   many   places  (see,   \eg,
Ref.~\cite{b2xintros} and references therein).

Averages  are  provided for  the  total  semileptonic branching  ratio
$\cbf(\Bb\to  X\ell\nub)$  and   the  partial  semileptonic  branching
fraction $\cbf(\Bb\to X\ell\nub; E_\ell>0.6\,\gev)$.

The measurements of the  total semileptonic branching ratio $\cbf(b\to
X\ell\nub)$   at   LEP   (see,  \eg,   Ref~\cite{Eidelman:2004wy}   or
Ref.~\cite{lepewwg}) represent a different  analysis class with a more
explicit model  dependence than  the (lepton-)tagged analyses  used at
the  \FourS.  Therefore  the  LEP  measurements are  not  used in  the
averages computed here.

% ----------------------------------------------------------------------
\subsubsection{Total semileptonic branching fraction}
\label{slbdecays_b2cinclrate}
% ---------------------------------------------------

The average for the total branching ratio $\cbf(\Bb\to X \ell\nub)$ is
determined    by   the   combination of the   results    provided   in
Table~\ref{tab:brisltot}.  In  this average, the  extrapolation of the
measured rate to the total decay rate is performed by each experiment,
usually with a fit of several components to the experimental spectrum.

\begin{table}[!htb]
\caption{Average of the total semileptonic branching fractions $\cbf(\Bb\to X \ell\nub)$
determined in tagged measurements on the \FourS.}
\begin{center}
\begin{tabular}{|l|c|c|}\hline
Experiment                                   &$\cbf_{tot}(\Bb\to X \ell\nub) [\%]$ (rescaled)    &$\cbf_{tot}(\Bb\to X \ell\nub) [\%]$ (published) \\
\hline\hline 
ARGUS ($\ell$-tag)~\hfill\cite{Albrecht:1993pu}      &$9.76  \pm0.50 \pm0.39$            &$9.70  \pm0.50 \pm0.60$ \\
\babar (\breco-tag)~\hfill\cite{Langenegger:2001qp}  &$10.40 \pm0.50 \pm0.46$            &$10.40 \pm0.50 \pm0.46$ \\
\belle ($\ell$-tag)~\hfill\cite{Abe:2002du}           &$10.97 \pm0.12 \pm0.49$            &$10.90 \pm0.12 \pm0.49$ \\
\babar ($e$-tag)~\hfill\cite{Aubert:2002uf}          &$10.92 \pm0.18 \pm0.29$            &$10.87 \pm0.18 \pm0.30$ \\
\belle (\breco-tag)~\hfill\cite{belle:breco}          &$11.19 \pm0.20 \pm0.31$            &$11.19 \pm0.20 \pm0.31$ \\
CLEO  ($\ell$-tag)~\hfill\cite{Mahmood:2004kq}       &$10.91 \pm0.08 \pm0.30$            &$10.88 \pm0.08 \pm0.33$ \\
\hline 
{\bf Average}                                        &\mathversion{bold}$10.90\pm0.23$   &\mathversion{bold}$\chi^2/\dof = 5.0/5$ \\
\hline 
\end{tabular}
\end{center}
\label{tab:brisltot}
\end{table}

It  is  possible  to  determine  \vcb\  from  the  total  semileptonic
branching  fraction and  the lifetime  of $B$  mesons.   However, this
approach has  limitations so that  it is no longer  applied. Recently,
several  groups have published  global fits  to the  partial branching
fraction plus  the moments of  the hadronic mass distribution  and the
lepton energy  has been done  by different groups~\cite{Aubert:2004aw,
  Bauer:2004ve}.

% ----------------------------------------------------------------------
\subsubsection{Partial semileptonic branching fraction}
\label{slbdecays_b2cinclprate}
% ---------------------------------------------------

The average  for the visible branching ratio  $\cbf(\Bb\to X \ell\nub;
p_\ell > 0.6\,\gevc)$ is determined  by the combination of the results
provided in Table~\ref{tab:brisl}.  For the determination of \vcb, the
extrapolation to the full spectrum is not necessary, as the HQE allows
the direct determination of \vcb\ from a partial rate.

\begin{table}[!htb]
\caption{Average of $\cbf(\Bb\to X \ell\nub; p_\ell > 0.6\,\gevc)$
determined in (model-independent) lepton-tagged measurements on the
\FourS.}
\begin{center}
\begin{tabular}{|l|c|c|}\hline
Experiment    &$\cbf_{vis}(\Bb\to X \ell\nub) [\%]$ (rescaled) &$\cbf_{vis}(\Bb\to X \ell\nub) [\%]$ (published) \\
\hline\hline 
\belle ($\ell$-tag)~\hfill\cite{Abe:2002du}                &$10.31 \pm0.11 \pm0.47$ &$10.24 \pm0.11 \pm0.46$ \\
BABAR ($e$-tag)~\hfill\cite{Aubert:2002uf}                &$10.37 \pm0.06 \pm0.23$ &$10.36 \pm0.06 \pm0.23$ \\
CLEO  ($\ell$-tag)~\hfill\cite{Mahmood:2004kq,stepaniak}  &$10.23 \pm0.08 \pm0.22$ &$10.21 \pm0.08 \pm0.22$ \\

\hline 
{\bf Average}                              &\mathversion{bold}$10.29\pm0.18$       &\mathversion{bold}$\chi^2/\dof = 0.22/2$ \\
\hline 
\end{tabular}
\end{center}
\label{tab:brisl}
\end{table}

% ----------------------------------------------------------------------
\subsubsection{Determination of \vcb}
\label{slbdecays_b2cinclvcb}
% -----------------------------------

The  determination  of  \vcb\  directly  from  the  visible  branching
fraction in combination with moments of the hadronic mass distribution
and    the    lepton   energy    has    been    done   by    different
groups~\cite{Battaglia:2002tm,  Aubert:2004aw,  Bauer:2004ve}.  It  is
foreseen to use a similar approach here as well in the future.

%% \clearpage

% -- \include{b2uexcl.tex}
% ======================================================================
\subsection{Exclusive Cabibbo-suppressed decays}
\label{slbdecays_b2uexcl}
% ----------------------------------------------

Here we list  results on exclusive determinations of  \vub. An average
of (these)  exclusive $b\to u\ell\nub$  results is envisioned  for the
future. The measurements  are separated into two classes:  a first one
averaging     over    the    entire     $\qq$    range     (shown    in
Table~\ref{tab:xslvuballqq}), and a second  class, where the decay rate
is  measured   differentially  in  (few)   bins  of  $\qq$   (shown  in
Table~\ref{tab:xslvublimqq}).

% ----------------------------------------------------------------------
\begin{table}[!htb]
\caption{Summary of exclusive determinations of $\cbf(\Bb\to  X
\ell\nub)$ and \vub\ using the entire $\qq$ range. The  errors quoted
on \vub\ correspond to  statistical,
experimental systematic and theoretical systematic, respectively.}
\begin{center}
\begin{small}
\begin{tabular}{|l|l|c|c|}\hline
Experiment             &Mode                       &$\cbf [10^{-4}]$           &\vub\ $[10^{-3}]$ (rescaled) \\
\hline\hline 
CLEO  ~\hfill\cite{Behrens:1999vv}  &$\Bz\to \rho^-\ell^+\nu$   &$2.69 \pm0.41 \ {}^{+0.35}_{-0.40} \pm0.50$   
                                                                               &$3.24  \pm0.25 \ {}^{+0.21}_{-0.24}\pm0.58$ \\
\babar ~\hfill\cite{Aubert:2003zd}  &$\Bz\to \rho^-\ell^+\nu$   &$3.29 \pm0.42 \pm0.47            \pm0.60$     
                                                                               &$3.59  \pm0.23 \pm0.26 \pm0.66$ \\
\belle ~\hfill\cite{Schwanda:2004fa} &$\Bz\to \omega\ell^+\nu$   &$1.3  \pm0.4  \pm0.2             \pm0.3$      
                                                                               &$3.1   \pm0.2  \pm0.2  \pm0.6$ \\
\hline 
\end{tabular}
\end{small}
\end{center}
\label{tab:xslvuballqq}
\end{table}
% ----------------------------------------------------------------------

% ----------------------------------------------------------------------
\begin{table}[!htb]
\caption{Summary of exclusive determinations of $\cbf(\Bb\to  X
\ell\nub)$ and \vub\ binned in $\qq$. The  errors quoted on \vub\ correspond to  statistical,
experimental systematic, theoretical systematic, and signal form-factor shape, respectively.}
\begin{center}
\begin{small}
\begin{tabular}{|l|l|c|c|}\hline
Experiment             &Mode                     &$\cbf [10^{-4}]$  &\vub\ $[10^{-3}]$ (rescaled) \\
\hline\hline 
CLEO  ~\hfill\cite{Athar:2003yg}   &$\Bz\to \pi^-\ell^+\nu$  &$1.33 \pm0.18 \pm0.11            \pm0.01 \pm0.07$   
                                                                    &$2.88  \pm0.55 \pm0.30 \ {}^{+0.45}_{-0.35} \pm0.18$ \\
CLEO  ~\hfill\cite{Athar:2003yg}   &$\Bz\to \rho^-\ell^+\nu$ &$2.17 \pm0.34\ {}^{+0.47}_{-0.54} \pm0.41 \pm0.01$   
                                                                    &$3.34  \pm0.32\ {}^{+0.27}_{-0.36}\ {}^{+0.50}_{-0.40}$ \\
\belle ~\hfill\cite{Abe:2004zm}     &$\Bz\to \pi^-\ell^+\nu$ &$1.76 \pm0.28\ \pm0.20 \pm0.03$   
                                                                    &$3.90  \pm0.71 \pm0.23\ {}^{+0.62}_{-0.48}$ \\
\belle ~\hfill\cite{Abe:2004zm}     &$\Bz\to \rho^-\ell^+\nu$ &$2.54 \pm0.78\ \pm0.85 \pm0.30$ & \\
\babar~\hfill\cite{danieledelre}   &$\Bz\to \pi^-\ell^+\nu$ &$1.46 \pm0.27\ \pm0.28$ & \\
\hline 
\end{tabular}
\end{small}
\end{center}
\label{tab:xslvublimqq}
\end{table}
% ----------------------------------------------------------------------

%% \clearpage

% -- \include{b2uincl.tex}
% ======================================================================
\subsection{Inclusive Cabibbo-suppressed decays}
\label{slbdecays_b2uincl}
% ----------------------------------------------

A recent discussion of the  theoretical issues and errors is provided,
\eg, in Refs.~\cite{Luke:ga} and~\cite{Ligeti:2003hp}.  An independent
estimate of the size of  the theoretical uncertainties can be found in
Ref.~\cite{mblg}. The following description is focusing on a technical
description of the relevant details of the averaging procedure.

% ----------------------------------------------------------------------
\subsubsection{Determination of \vub}
\label{slbdecays_b2uinclvub}
% -----------------------------------

Inclusive determinations of \vub\  based on charmless semileptonic $B$
decays have been presented  by all four LEP experiments~\cite{LEP:vub}
and, more recently,  by the experiments at the  \FourS.  In all cases,
the  experimental approach  is to  measure the  charmless semileptonic
rate in a  restricted region of phase space  where the background from
$B\to  X_c \ell\nub$  is  suppressed.  Theoretical  input  is used  to
extrapolate  to   the  full  rate  and/or  to   determine  \vub.   The
determination of \vub\ from  the full rate $\cbf(\Bb\to X_u \ell\nub)$
is accomplished with the following formula:

\begin{displaymath}
\vub = 0.00424\cdot\sqrt{\frac{\cbf(\Bb\to X_u \ell\nub)}{0.002}
  \frac{1.604\ps}{\tau_B}
}\times(1.000 \pm 0.028_{\rm pert} \pm 0.039_{1/m_b^3}),
\end{displaymath}

\noindent where $\tau_B$ is the average lifetime of \Bz\ and \Bp\
mesons     and     the    theoretical     error     is    based     on
Refs.~\cite{Hoang:1998hm,Uraltsev:1999rr}   using   the   moments
measurements of the \babar\ collaboration~\cite{Aubert:2004aw}.

Here we combine results of  the \babar, \belle, and CLEO collaborations
as they  present measurements in well-defined  phase-space regions and
are based on the same theoretical description of the decay in terms of
an  OPE. Due  to the  large background  from  Cabibbo-favored \Bxclnu\ 
decays,  the measurements  are sensitive  to \Bxulnu\  decays  only in
restricted  regions of  phase-space.  The  convergence of  the  OPE is
affected by these restrictions and complicate the extrapolation to the
full rate and thus the determination of \vub.

In the theoretical description, the motion of the $b$ quark inside the
$B$  meson is  described with  a  ``shape function'',  usually in  the
two-parameter   ``exponential  form''.    These   parameters  can   be
determined from  a measurement  of the photon  energy spectrum  in the
rare  decay  $b\to  s\gamma$.   This  decay has  a  smaller  branching
fraction than  the decay \Bxulnu, which limits  the precision possible
in this  approach.  Recently, there  has been theoretical  interest to
use  moments   measured  in  semileptonic  \Bxclnu\   decays  for  the
determination of the shape function parameters.

In the average \vub\ presented  here, all measurements are scaled to a
common set of  the shape function parameters and  their uncertainties. 
Both  are taken  from the  photon  energy spectrum  in $B\to  s\gamma$
decays as measured  by the \belle collaboration~\cite{Limosani:2004jk}. 
The (large) errors  of the shape function parameters  are dominated by
statistical  uncertainties in  the  measurement of  the photon  energy
spectrum. 

For the average,  the systematic error is divided  into the categories
shown in  Table~\ref{tab:vuberrors}.  In the  averaging procedure, the
errors belonging to a specific  category are taken as 100\% correlated
between the different measurements.  The numerical values of the errors
for each measurement are given in a spreadsheet~\cite{spreadsheet}.

% ----------------------------------------------------------------------
\begin{table}[!htb]
\caption{Errors in  the average of \vub. The  numerical values for
 all       measurements       can       be      found       in       a
 spreadsheet~\cite{spreadsheet}.  The entries  in the  ``Name'' column
 correspond to the names in the spreadsheet. }
\begin{center} 
 \begin{small}
 \begin{tabular}{|l|l|} 
  \hline
  Name             &Explanation                                                 \\
  \hline     
  \multicolumn{2}{|l|}{{\bf Uncorr}    \hskip 1.3cm {\bf Quadratic sum of statistical and experimental errors}}    \\
  \hline     
  Statistical      &Statistical                                                 \\
  Exp              &Experimental detector systematics (uncorrelated)            \\
  \hline
  \multicolumn{2}{|l|}{{\bf Corr}    \hskip 1.7cm {\bf Quadratic sum of correlated errors}}                 \\
  \hline     
  B2C(bg)          &Modeling of \Bxclnu\ (Branching fractions, form factors, etc)                          \\
  B2U(eff)         &Modeling of \Bxulnu\ (Branching fractions, $\varepsilon_{vis}$, \ssbar\ popping, etc) \\
  BF$\to$Vub       &Theoretical error $\Gamma\to\vub$ ($m_b$, $\alpha_s$)       \\
  fu(extrapol)     &Extrapolation from visible range                            \\
  \hline 
  \multicolumn{2}{|l|}{{\bf Additional errors not provided by the experiments}} \\
  \hline     
  SSF              &Subleading shape functions (power corrections)              \\
  WA               &Weak annihilation                                           \\
  QHD              &(Local) Quark-Hadron Duality                                \\
  \hline 
 \end{tabular}
 \end{small}
\end{center}
\label{tab:vuberrors}
\end{table}
% ----------------------------------------------------------------------

In   Table~\ref{tab:vuberrors},   the   largest  and   most   critical
uncertainties are  the extrapolation ``fu'' and  the additional errors
not provided by the experiments:

\begin{itemize}
 
  \item fu: This error  estimates the uncertainty in the extrapolation
 from the observed phase-space to the full phase-space.  For this, all
 measurements    use    the    triple-differential   calculation    of
 Ref.~\cite{DeFazio:1999sv}.  The shape  function is parametrized with
 the ``exponential form''. All measurements are rescaled to correspond
 to the  shape function parameters  determined from the  photon energy
 spectrum  measurement of  \belle~\cite{Limosani:2004jk}.   In previous
 update (Winter 2004), the  data of  CLEO~\cite{Chen:2001fj} had
 been  used for  this purpose.  The change  substantially  reduces the
 errors and leads to a  shift in the central value.  The extrapolation
 errors for each measurement are  determined by the uncertainty of the
 shape function parameters~\cite{Gibbons:2004dg}, largely dominated by
 statistics.   Recent theoretical  work~\cite{Bosch:2004th}  may allow
 the   extraction  of   the   shape  function   parameters  from   the
 high-precision  measurements  of  moments in  semileptonic  \Bxclnu\ 
 decays.
 
  \item  SSF: Uncertainties  due to  power corrections  are summarized
 into the  term labeled  ``subleading shape function''  effects.  They
 have        been       calculated       for        the       endpoint
 measurements~\cite{Neubert:2002yx}; the errors of the corrections are
 used as estimate for the  error (without scaling the central values). 
 These  corrections strongly  depend  on the  fraction of  phase-space
 covered in a measurement. They are much larger for the CLEO and \belle
 endpoint analyses  with the higher  lepton energy threshold.   We use
 the  uncertainty for the  case $E_\ell>2.2\gev$  as estimate  for the
 \mx\ analysis of \babar.  This is roughly consistent with theoretical
 expectations~\cite{Burrell:2003cf}.   For  the  \mxqq\  results,  the
 error is  expected to be significantly  smaller; here we  use half of
 the    \mx\   error.     The   SSF    errors   are    summarized   in
 Table~\ref{tab:vubscale}.

\item  WA: The error  for weak  annihilation estimates  the uncertainty
from  parametrically  enhanced  nonperturbative effects  predominantly
expected at high-$\qq$.  The estimates in Ref.~\cite{Bauer:2001rc} are
used  for the  \mxqq\  analysis  and scaled  to  the other  phase-space
regions  as   shown  in  Table~\ref{tab:vubscale}.    Because  of  the
concentration  at high-$\qq$,  the  effects of  weak annihilation  are
diluted as more phase-space is covered.

 \item QHD: The error due  to the violation of quark-hadron duality is
estimated for the \mxqq\ analysis in Ref.~\cite{mblg}.  It is found to
be substantially smaller than the uncertainties from weak annihilation
and power corrections. Therefore, we neglect it here.

\end{itemize}

% ----------------------------------------------------------------------
\begin{table}[!htb]
\caption{Scaling  of   theoretical  errors  in   different  regions  of
phase-space.  The \mx\  analysis requires  for the  hadronic  system in
\Bxulnu \ decays  $\mx < 1.55\,\gevcc$, the \mxqq\  analysis selects the
phase-space $\mx<1.7\,\gevcc, \qq>8\,\gev^2$.}
\begin{center} 
 \begin{small}
 \begin{tabular}{|l|c|c|c|c|c|} 
  \hline
  Component          &\mx  &\mxqq  &$E_\ell>2.0\gev$ &$E_\ell>2.2\gev$   &$E_\ell>2.3\gev$ \\
  \hline                                                                     
  SSF                &4\%    &2\%    &2\%               &4\%                 &6\% \\
  WA                 &2\%    &4\%    &4\%               &6\%                 &8\% \\
  \hline 
 \end{tabular}
 \end{small}
\end{center}
\label{tab:vubscale}
\end{table}
% ----------------------------------------------------------------------

The  average of  \vub\  is  based on  the  measurements summarized  in
Table~\ref{tab:islvub}. The \babar\ and \belle collaborations provided
these central  values. The average  $B$ meson lifetime used  varies by
less  than $0.5\%$  between the  \babar\ and  Belle  collaboration; no
rescaling to a  common lifetime value is applied  because of its small
effect.  No  re-evaluation of the endpoint results  was available from
the CLEO  collaboration.  Therefore  their previous result  was scaled
with $(2.23/1.72)^{1/2}$, the ratio of branching fractions for $E_\ell
> 2\gev$ as determined with the  \belle and CLEO $b\to s\gamma$ photon
energy  spectra,  as   tabulated  in  Ref.~\cite{Aubert:2004bv}.   The
relative extrapolation  error in Ref.~\cite{Aubert:2004bv}  is applied
for the  CLEO measurement.  After  this adjustment, the  average \vub\ 
amounts to

\begin{displaymath}
\vub = (4.70 \pm 0.44) \times 10^{-3}.
\end{displaymath}

This error contains contributions  from the perturbative expansion and
the  uncertainty of the  $b$ quark  mass.  The  error of  this average
contains  a relative contribution  of $4.8\%$  for the  translation of
$\Gamma(b\to  u\ell\nub)$ to \vub,  which has  been derived  using the
semileptonic      moments     measurements     of      the     \babar\ 
collaboration~\cite{Aubert:2004aw}.

The current extrapolation  error on \vub\ is based  on the approach of
Ref.~\cite{DeFazio:1999sv}  for  the shape  function  effect, where  the
shape function  parameter errors are  based only on the  photon energy
spectrum. Recent reports~\cite{Bosch:2004th, Bauer:2003pi} studied the
radiative corrections  to the shape  function and the  prescription of
Ref.~\cite{DeFazio:1999sv}  must  be   modified.   In  particular  Bosch
\etal~\cite{Bosch:2004th}   showed  that   the   sensitivity  of   the
extrapolation to the shape function is reduced for the \mx\ and \mxqq\
analyses, but becomes larger  for the endpoint analyses.  Furthermore,
it may become  possible to substantially improve the  precision of the
shape function parameters (and  hence reduce the extrapolation errors)
when using the moments measurements in semileptonic $B$ decays.

%For example the f_u error for the CLEO's ellipse becomes ~ a few % for
%Mx<1.7GeV andq^2>8GeV^2, ~19% for Mx<1.55 GeV.

% ----------------------------------------------------------------------
\begin{table}[!htb]

\caption{Average of  \vub\ determined in inclusive  measurements on the
\FourS. The  errors quoted for the measurements correspond to  uncorrelated and correlated,
respectively. The correlated error of the measurements do not include the common and
constant error of $2.8\% \oplus 3.9\%$ from the full rate to \vub;
however, this error is included in the average \vub.}
\begin{center}
\begin{tabular}{|l|l|c|c|}\hline
Experiment    &Method ($[\gev^{(2)}]$)                      &\vub\
$[10^{-3}]$ (rescaled)             &\vub\ $[10^{-3}]$ (published) \\
\hline\hline 
CLEO  ~\hfill\cite{Bornheim:2002du}     &Endpoint, $p_\ell>2.2$       &$4.69 \pm0.23 \pm0.63$    &$4.08  \pm0.20 \pm0.62$ \\
\belle ~\hfill\cite{Kakuno:2003fk}       &$(\mx < 1.70, \qq > 8)$      &$4.75 \pm0.46 \pm0.46$    &$4.66  \pm0.45 \pm0.61$ \\
\belle ~\hfill\cite{belle:endpoint}      &Endpoint, $p_\ell>2.3$       &$4.46 \pm0.23 \pm0.61$    &$3.99  \pm0.20 \pm0.61$ \\
\babar ~\hfill\cite{Aubert:2004bv}      &Endpoint, $p_e>2.0$          &$4.40 \pm0.15 \pm0.44$    &$4.40  \pm0.15 \pm0.44$ \\
\babar ~\hfill\cite{Aubert:2004bq}      &$\mx < 1.55$                 &$5.22 \pm0.30 \pm0.43$    &$5.22  \pm0.30 \pm0.43$ \\
\babar ~\hfill\cite{Aubert:2004bq}      &$(\mx < 1.70, \qq > 8)$      &$5.18 \pm0.52 \pm0.42$    &$5.18  \pm0.52 \pm0.42$ \\
\babar ~\hfill\cite{Aubert:2004tw}      &$(E_\ell, \qq)$              &$4.99 \pm0.34 \pm0.51$    &$4.99  \pm0.34 \pm0.51$ \\
\belle ~\hfill\cite{Abe:2004sc}          &$(\mx < 1.70, \qq > 8)$      &$5.54 \pm0.65 \pm0.54$    &$5.54  \pm0.65 \pm0.54$ \\
\hline 
{\bf Average}                           &  &\mathversion{bold}$4.70\pm0.44$    &\mathversion{bold}$\chi^2/\dof = 6.7/7$ \\
\hline 
\end{tabular}
\end{center}
\label{tab:islvub}
\end{table}
% ----------------------------------------------------------------------

\begin{figure}[!htb]
 \begin{center}
  \unitlength1.0cm % coordinates in cm
  \begin{picture}(14.,8.)  %ys(25.,6.)
   \put( -0.5,  0.0){\includegraphics[width=7.5cm]{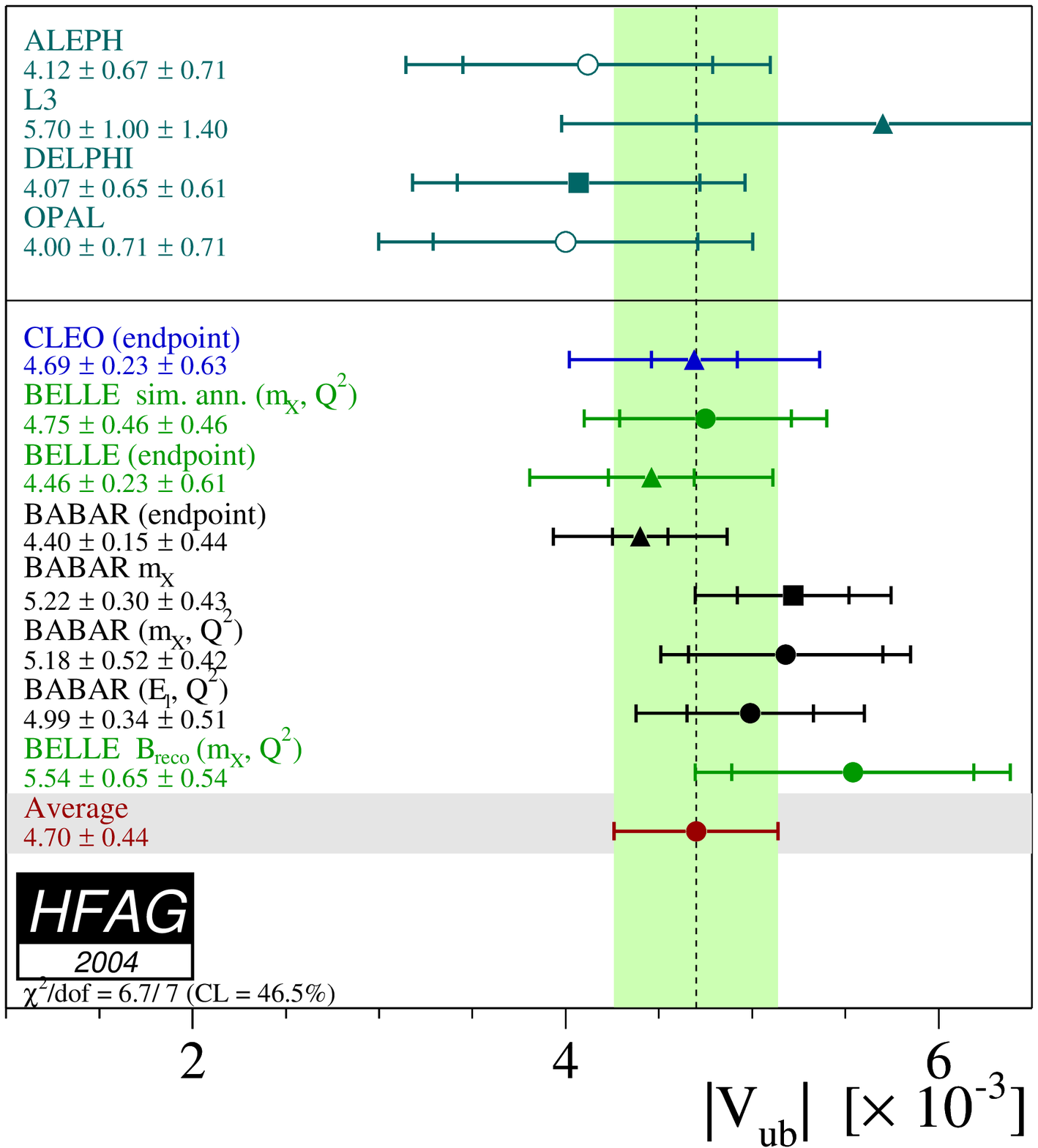}}
    %ys{\includegraphics[width=0.49\textwidth]{figures/slb/vub_isl.eps}}
   \put(  8.0,  0.0){\includegraphics[width=7.5cm]{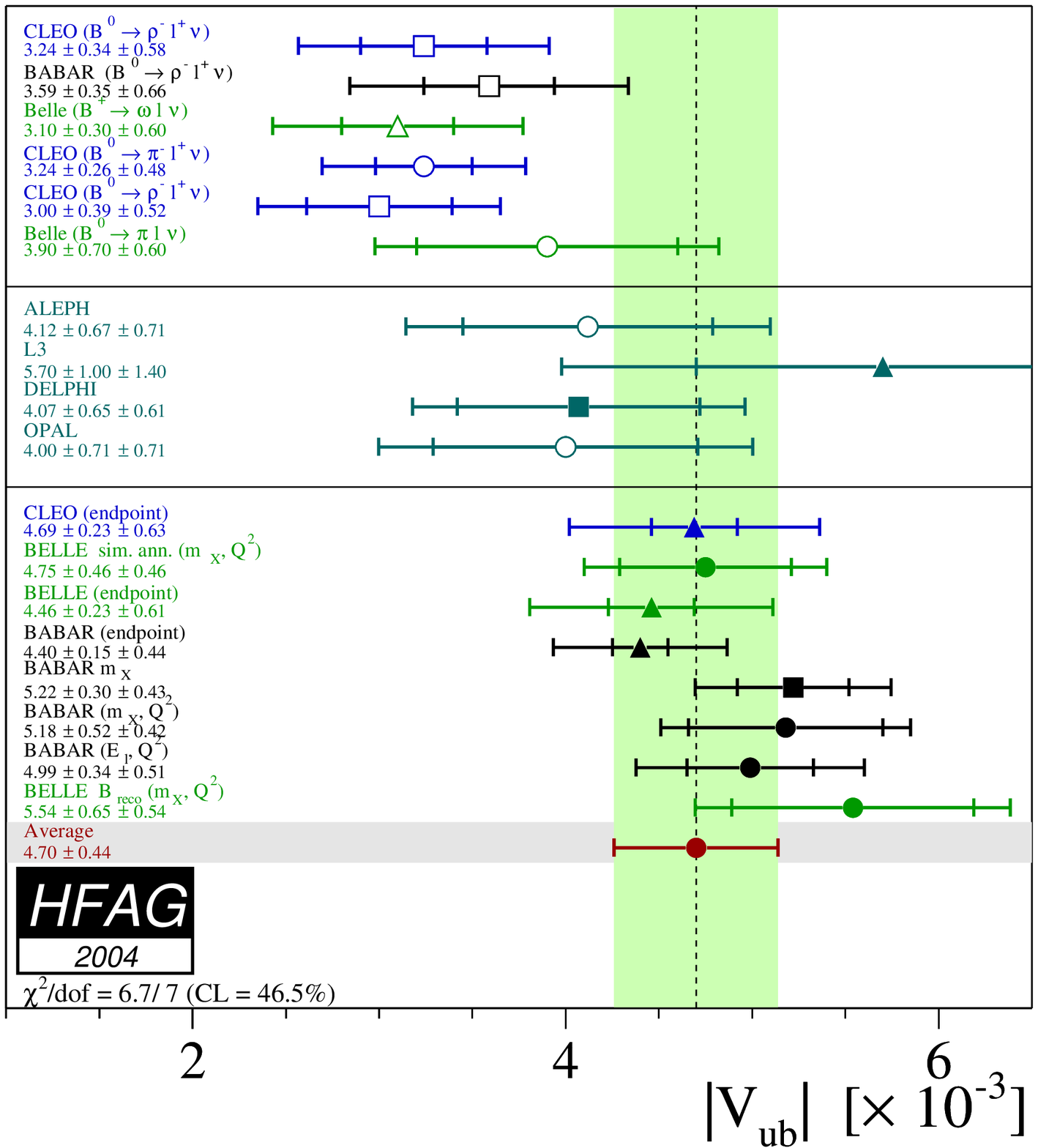}}
    %ys{\includegraphics[width=0.49\textwidth]{figures/slb/vub.eps}}
   \put(  5.5,  6.8){{\large\bf a)}}
   \put( 14.0,  6.8){{\large\bf b)}}
  \end{picture}
  \caption{(a)  Illustration  of  the  average  \vub\  from  inclusive
  semileptonic $B$ decays measured  at the \FourS. The measurements
  from LEP~\cite{LEP:vub} are shown only for illustration, they are not included in
  the average. (b) Illustration of additional exclusive \vub\  measurements
  together   with  the  average  \vub\  from
  inclusive semileptonic $B$ decays measured at the \FourS. }
  \label{fig:brisl}
 \end{center}
\end{figure}

% ----------------------------------------------------------------------
\subsubsection{Discussion}
\label{slbdecays_discussion}
% --------------------------

In  the future,  improvements  to  this average  will  be studied  and
implemented.   The following is  a starting  point, not  an exhaustive
list, for possible improvements.

\begin{itemize}

\item  The  paper by  de Fazio  and Neubert~\cite{DeFazio:1999sv}
  with a shape function parametrization provides a triple differential
  decay rate for \Bxulnu\ decays that is used in all measurements over
  the entire  phase-space.  Recent  theoretical work seems  to indicate
  that  it can  be improved  even at  its nominal  order  ($1/m_B$ and
  $\alpha_s$)~\cite{Bosch:2004th, Bauer:2003pi}.

\item Every  experiment should provide the  result in an  unfolded rate so
  that theoretical improvements can be applied in the future. The idea
   is to separate  the ``unfoldings'' of a rate  or distribution from the
   ``extrapolation''   (the   former   is   experimental,   the   latter
   theoretical).

\item All three experiments have a ``hybrid'' MC model for signal decays. 
  The  differential  spectra  should  be  compared.  Its applicability
   should be understood (probably good for unfolding, but questionable
   for extrapolation).

\item There should be agreement on one specific recipe to evaluate the common
  errors. This can include items like 

  \begin{itemize}
  \item \Bxclnu\ modeling:

    \begin{itemize}  

      \item[-] Variation  of inclusive and  exclusive branching fractions
      for $B$ and $D$ decays  (full list and range) 

      \item[-] Variation of form factors describing \Bxclnu\  decays 

     \end{itemize}

  \item \Bxulnu\  modeling and theoretical error

    \begin{itemize}
     \item[-] Variation of exclusive branching fraction (full list and range)
  
     \item[-] Comparison with inclusive-only  signal model
       and comparison with exclusive-only signal model.
     
     \item[-] \ssbar\ popping (if applicable in the analysis)

     \item[-] Central values of shape function parameters

     \item[-] Variation of shape function  parameters (this is linked to recent
       theoretical work)
    \end{itemize}

  \end{itemize}

\end{itemize}

\noindent  A  close  collaboration  between  the  experiments  will  be
necessary to  obtain an even  more consistent treatment of  errors not
only for  the most  critical ones  (as shown here),  but also  for the
remaining smaller errors.

%\clearpage
%
% Averages of measurements related to the Unitarity Triangle
%
% \documentclass[12pt]{article}
% \usepackage[lmargin=20mm,rmargin=25mm,vmargin=25mm]{geometry}
% \usepackage{graphicx}
% \include{symbols}
% \begin{document}
% \begin{flushright}
%   \today
% \end{flushright}

\mysection{Measurements related to Unitarity Triangle angles
}
\label{sec:cp_uta}

The charge of the ``$\CP(t)$ and Unitarity Triangle angles'' group
is to provide averages of measurements related (mostly) to the 
angles of the Unitarity Triangle (UT).
To date, most of the measurements that can be used to 
obtain model-independent information on the UT angles
come from time-dependent $\CP$ asymmetry analyses.
% only one measurement included in the present averages does not 
% fall into this category.
In cases where considerable theoretical input is required to 
extract the fundamental quantities, no attempt is made to do so at 
this stage. However, straightforward interpretations of the averages 
are given, where possible.

In Sec.~\ref{sec:cp_uta:introduction} 
a brief introduction to the relevant phenomenology is given.
In Sec.~\ref{sec:cp_uta:notations}
an attempt is made to clarify the various different notations in use.
In Sec.~\ref{sec:cp_uta:common_inputs}
the common inputs to which experimental results are rescaled in the
averaging procedure are listed. We also briefly introduce the treatment
of experimental errors. 
In the remainder of this section,
the experimental results and their averages are given,
divided into subsections based on the underlying quark-level decays.

\mysubsection{Introduction
}
\label{sec:cp_uta:introduction}

The Standard Model Cabibbo-Kobayashi-Maskawa (CKM) quark mixing matrix $\VCKM$ 
must be unitary. A $3 \times 3$ unitary matrix has four free parameters,\footnote{
  In the general case there are nine free parameters,
  but five of these are absorbed into unobservable quark phases.}
and these are conventionally written by the product
of three (complex) rotation matrices~\cite{ref:cp_uta:chau}, where the rotations are 
characterized by the Euler angles $\theta_{12}$, $\theta_{13}$ 
and $\theta_{23}$, which are the mixing angles
between the generations, and one overall phase $\delta$,
\begin{equation}
\label{eq:ckmPdg}
\VCKM =
        \left(
          \begin{array}{ccc}
            V_{ud} & V_{us} & V_{ub} \\
            V_{cd} & V_{cs} & V_{cb} \\
            V_{td} & V_{ts} & V_{tb} \\
          \end{array}
        \right)
        =
        \left(
        \begin{array}{ccc}
        c_{12}c_{13}    
                &    s_{12}c_{13}   
                        &   s_{13}e^{-i\delta}  \\
        -s_{12}c_{23}-c_{12}s_{23}s_{13}e^{i\delta} 
                &  c_{12}c_{23}-s_{12}s_{23}s_{13}e^{i\delta} 
                        & s_{23}c_{13} \\
        s_{12}s_{23}-c_{12}c_{23}s_{13}e^{i\delta}  
                &  -c_{12}s_{23}-s_{12}c_{23}s_{13}e^{i\delta} 
                        & c_{23}c_{13} 
        \end{array}
        \right)
\end{equation}
where $c_{ij}=\cos\theta_{ij}$, $s_{ij}=\sin\theta_{ij}$ for 
$i<j=1,2,3$. 

Following the observation of a hierarchy between the different
matrix elements, the Wolfenstein parameterization~\cite{ref:cp_uta:wolfenstein}
is an expansion of $\VCKM$ in terms of the four real parameters $\lambda$
(the expansion parameter), $A$, $\rho$ and $\eta$. Defining to 
all orders in $\lambda$~\cite{ref:cp_uta:buras}
\begin{eqnarray}
  \label{eq:burasdef}
  s_{12}             &\equiv& \lambda,\nonumber \\ 
  s_{23}             &\equiv& A\lambda^2, \\
  s_{13}e^{-i\delta} &\equiv& A\lambda^3(\rho -i\eta),\nonumber
\end{eqnarray}
and inserting these into the representation of Eq.~(\ref{eq:ckmPdg}), 
unitarity of the CKM matrix is achieved to all orders.
A Taylor expansion of $\VCKM$ leads to the familiar approximation
\begin{equation}
  \label{eq:cp_uta:ckm}
  \VCKM
  = 
  \left(
    \begin{array}{ccc}
      1 - \lambda^2/2 & \lambda & A \lambda^3 ( \rho - i \eta ) \\
      - \lambda & 1 - \lambda^2/2 & A \lambda^2 \\
      A \lambda^3 ( 1 - \rho - i \eta ) & - A \lambda^2 & 1 \\
    \end{array}
  \right) + {\cal O}\left( \lambda^4 \right).
\end{equation}
The non-zero imaginary part of the CKM matrix,
which is the origin of $\CP$ violation in the Standard Model,
is encapsulated in a non-zero value of $\eta$.

The unitarity relation $\VCKM^\dagger\VCKM = {\mathit 1}$
results in nine expressions which can be written
$\sum_{i=u,c,t} V^*_{ij}V_{ik} = \delta_{jk}$,
where $\delta_{jk}$ is the Kronecker symbol.
Of the off-diagonal expressions ($j \neq k$),
three can be trivially transformed into the other three 
(under $j \leftrightarrow k$),
leaving six relations, in which three complex numbers sum to zero,
which therefore can be expressed as triangles in the complex plane.

\begin{figure}[t]
  \begin{center}
    \resizebox{0.55\textwidth}{!}{\includegraphics{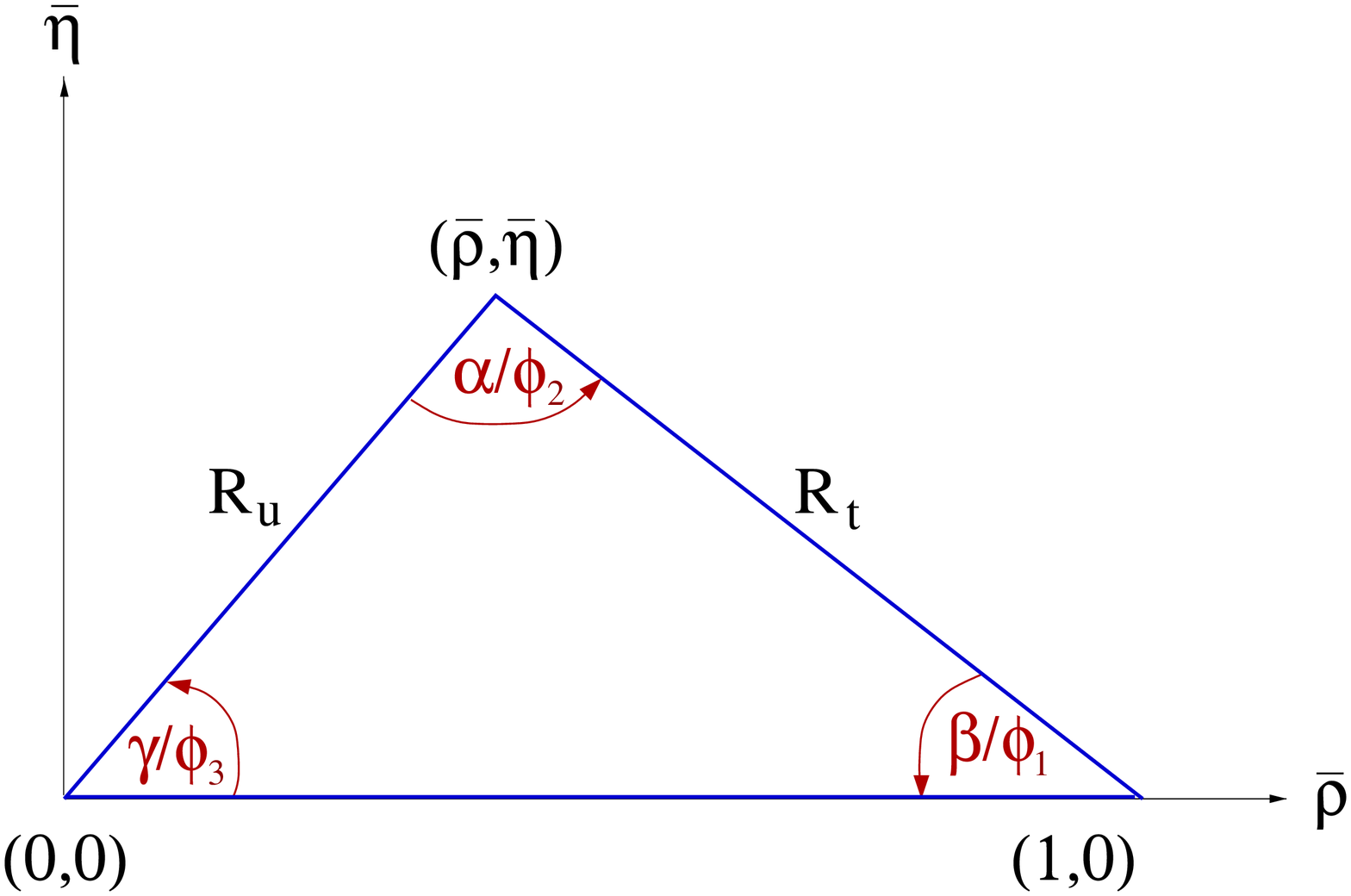}}
    \caption{The Unitarity Triangle.}
    \label{fig:cp_uta:ut}
  \end{center}
\end{figure}

One of these,
\begin{equation}
  \label{eq:cp_uta:ut}
  V_{ud}V^*_{ub} + V_{cd}V^*_{cb} + V_{td}V^*_{tb} = 0,
\end{equation}
is specifically related to $\B$ decays.
The three terms in Eq.~(\ref{eq:cp_uta:ut}) are of the same order 
(${\cal O}\left( \lambda^3 \right)$),
and this relation is commonly known as the Unitarity Triangle.
For presentational purposes,
it is convenient to rescale the triangle by $(V_{cd}V^*_{cb})^{-1}$,
as shown in Fig.~\ref{fig:cp_uta:ut}.

Two popular naming conventions for the UT angles exist in the literature:
\begin{equation}
  \label{eq:cp_uta:abc}
  \alpha  \equiv  \phi_2  = 
  \arg\left[ - \frac{V_{td}V_{tb}^*}{V_{ud}V_{ub}^*} \right],
  \hspace{0.5cm}
  \beta   \equiv   \phi_1 =  
  \arg\left[ - \frac{V_{cd}V_{cb}^*}{V_{td}V_{tb}^*} \right],
  \hspace{0.5cm}
  \gamma  \equiv   \phi_3  =  
  \arg\left[ - \frac{V_{ud}V_{ub}^*}{V_{cd}V_{cb}^*} \right].
  \nonumber
\end{equation}
In this document the $\left( \alpha, \beta, \gamma \right)$ set is used.

The apex of the Unitarity Triangle is given by the following 
definition~\cite{ref:cp_uta:buras}  
%% to all orders in $\lambda$~\cite{ref:cp_uta:buras} 
\begin{eqnarray}
  \label{eq:rhoetabar}
  \rhobar + i\etabar
  \;\equiv\;-\frac{V_{ud}V_{ub}^*}{V_{cd}V_{cb}^*}
  = (\rho + i\eta)
       (1 - \frac{1}{2}\lambda^{2}) + {\cal O}(\lambda^4).
\end{eqnarray}
The sides $R_u$ and $R_t$ of the Unitarity Triangle 
(the third side being normalized to unity) 
are given by
%% read to all orders 
\begin{eqnarray}
  \label{eq:ru}
  R_u &=& 
  \left|\frac{V_{ud}V_{ub}^*}{V_{cd}V_{cb}^*} \right|
  \;=\; \sqrt{\rhobar^2+\etabar^2}, \\
  \label{eq:rt}
  R_t &=& 
  \left|\frac{V_{td}V_{tb}^*}{V_{cd}V_{cb}^*}\right| 
  \;=\; \sqrt{(1-\rhobar)^2+\etabar^2}.
\end{eqnarray} 

\mysubsection{Notations
}
\label{sec:cp_uta:notations}

Several different notations for $\CP$ violation parameters
are commonly used.
This section reviews those found in the experimental literature,
in the hope of reducing the potential for confusion, 
and to define the frame that is used for the averages.

In some cases, when $\B$ mesons decay into 
multibody final states via broad resonances ($\rho$, $\Kstar$, \etc),
the experiments ignore interference effects in the analyses.
%% DP is only for 3body, but Q2B is also true for \rho\rho \etc 
% in the underlying multidimensional Dalitz plots.
%%%% --> that was meant by 'multidimensional' 
This is referred to as the quasi-two-body (Q2B) approximation
in the following.

\mysubsubsection{$\CP$ asymmetries
}
\label{sec:cp_uta:notations:pra}

The $\CP$ asymmetry is defined as the difference between the rate 
involving a $b$ quark and that involving a $\bar b$ quark, divided 
by the sum. For example, the partial rate (or charge) asymmetry for 
a charged $\B$ decay would be given as 
\begin{equation}
  \label{eq:cp_uta:pra}
  \Acp_{f} \;\equiv\; 
  \frac{\Gamma(\Bm \to f)-\Gamma(\Bp \to \bar{f})}{\Gamma(\Bm \to f)+\Gamma(\Bp \to \bar{f})}.
\end{equation}

\mysubsubsection{Time-dependent \CP asymmetries in decays to $\CP$ eigenstates
}
\label{sec:cp_uta:notations:cp_eigenstate}

If the amplitudes for $\Bz$ and $\Bzb$ to decay to a final state $f$, 
which is a $\CP$ eigenstate with eigenvalue $\etacpf$,
are given by $\Af$ and $\Abarf$, respectively, 
then the decay distributions for neutral $\B$ mesons, 
with known flavour at time $\Delta t =0$,
are given by
\begin{eqnarray}
%%   \label{eq:cp_uta:td_cp_asp}
  \Gamma_{\Bzb \to f} (\Delta t) & = &
  \frac{e^{-| \Delta t | / \tau(\Bz)}}{4\tau(\Bz)}
  \left[ 
    1 +
%%    \left\{ 
      \frac{2\, \Im(\lambda_f)}{1 + |\lambda_f|^2} \sin(\Delta m \Delta t) -
      \frac{1 - |\lambda_f|^2}{1 + |\lambda_f|^2} \cos(\Delta m \Delta t)
%%    \right\}  
  \right], \\
  \Gamma_{\Bz \to f} (\Delta t) & = &
  \frac{e^{-| \Delta t | / \tau(\Bz)}}{4\tau(\Bz)}
  \left[ 
    1 -
%%    \left\{ 
      \frac{2\, \Im(\lambda_f)}{1 + |\lambda_f|^2} \sin(\Delta m \Delta t) +
      \frac{1 - |\lambda_f|^2}{1 + |\lambda_f|^2} \cos(\Delta m \Delta t)
%%    \right\}  
  \right].
\end{eqnarray}
Here $\lambda_f = \frac{q}{p} \frac{\Abarf}{\Af}$ 
contains terms related to $\Bz$-$\Bzb$ mixing and to the decay amplitude
(the eigenstates of the effective Hamiltonian in the $\BzBzb$ system 
are $\left| B_\pm \right> = p \left| \Bz \right> \pm q \left| \Bzb \right>$).
This formulation assumes $\CPT$ invariance, 
and neglects possible lifetime differences in the neutral $\B$ meson system.
The time-dependent $\CP$ asymmetry,
again defined as the difference between the rate 
involving a $b$ quark and that involving a $\bar b$ quark,
is then given by
%% don't use {\cal A}_f since this will be used for A (cosine term)
\begin{equation}
  \label{eq:cp_uta:td_cp_asp}
  \Acp_{f} \left(\Delta t\right) \; \equiv \;
  \frac{
    \Gamma_{\Bzb \to f} (\Delta t) - \Gamma_{\Bz \to f} (\Delta t)
  }{
    \Gamma_{\Bzb \to f} (\Delta t) + \Gamma_{\Bz \to f} (\Delta t)
  } \; = \;
  \frac{2\, \Im(\lambda_f)}{1 + |\lambda_f|^2} \sin(\Delta m \Delta t) -
  \frac{1 - |\lambda_f|^2}{1 + |\lambda_f|^2} \cos(\Delta m \Delta t).
\end{equation}

While the coefficient of the $\sin(\Delta m \Delta t)$ term in 
Eq.~(\ref{eq:cp_uta:td_cp_asp}) is everywhere\footnote
{
%  Actually, not quite everywhere.  
  Occasionally one also finds Eq.~(\ref{eq:cp_uta:td_cp_asp}) written as
  $\Acp_{f} \left(\Delta t\right) = 
  {\cal A}^{\rm mix}_f \sin(\Delta m \Delta t) + {\cal A}^{\rm dir}_f \cos(\Delta m \Delta t)$,
  or similar.
} denoted $S_f$:
\begin{equation}
  \label{eq:cp_uta:s_def}
  S_f \;\equiv\; \frac{2\, \Im(\lambda_f)}{1 + \left|\lambda_f\right|^2},
\end{equation}
different notations are in use for the
coefficient of the $\cos(\Delta m \Delta t)$ term:
%% cannot use A_f here as this is already used for the amplitude
%% use {\cal A}_f and clarify in the text
\begin{equation}
  \label{eq:cp_uta:c_def}
  C_f \;\equiv\; - A_f \;\equiv\; \frac{1 - \left|\lambda_f\right|^2}{1 + \left|\lambda_f\right|^2}.
\end{equation}
The $C$ notation is used by the \babar\  collaboration 
(see \eg~\cite{ref:cp_uta:ccs:babar}), 
and also in this document.
The $A$ notation is used by the \belle\ collaboration
(see \eg~\cite{BELLE2}).

Neglecting effects due to $\CP$ violation in mixing 
(\ie, taking $|q/p| = 1$),
if the decay amplitude contains terms with a single weak phase
then $\left|\lambda_f\right| = 1$ and one finds
$S_f = -\etacpf \sin(\phi_{\rm mix} + \phi_{\rm dec})$, $C_f = 0$,
where $\phi_{\rm mix}=\arg(q/p)$ and $\phi_{\rm dec}=\arg(\Abarf/\Af)$.
Note that $\phi_{\rm mix}\approx2\beta $ 
in the Standard Model (in the usual phase convention). 
If amplitudes with different weak phases contribute to the decay, 
no clean interpretation of $S_f$ is possible. If the decay amplitudes
have in addition different $\CP$ conserving strong phases,
then $\left| \lambda_f \right| \neq 1$ and no clean interpretation is possible.
The coefficient of the cosine term becomes non-zero,
indicating direct $\CP$ violation.
The sign of $A_f$ as defined above is consistent with that of $\Acp_{f}$ in 
Eq.~(\ref{eq:cp_uta:pra}).

\mysubsubsection{Time-dependent \CP asymmetries in decays to vector-vector final states
}
\label{sec:cp_uta:notations:vv}

Consider \B decays to states consisting of two vector particles,
such as $\jpsi K^{*0}(\to\KS\piz)$, $D^{*+}D^{*-}$ and $\rho^+\rho^-$,
which are eigenstates of charge conjugation but not of parity.\footnote{
  \noindent
  This is not true of all vector-vector final states,
  \eg, $D^{*\pm}\rho^{\mp}$ is clearly not an eigenstate of 
  charge conjugation.
}
In fact, for such a system, there are three possible final states;
in the helicity basis these can be written $h_{-1}, h_0, h_{+1}$.
The $h_0$ state is an eigenstate of parity, and hence of $\CP$;
however, $\CP$ transforms $h_{+1} \leftrightarrow h_{-1}$ (up to 
an unobservable phase). In the transversity basis, these states 
are transformed into  $h_\parallel =  (h_{+1} + h_{-1})/2$ and 
$h_\perp = (h_{+1} - h_{-1})/2$.
In this basis all three states are $\CP$ eigenstates, 
and $h_\perp$ has the opposite $\CP$ to the others.

The amplitudes to these states are usually given by $A_{0,\perp,\parallel}$
(here we use a normalization such that 
% $\left| A_0 \right|^2 + \left| A_\perp \right|^2 + \left| A_\parallel \right|^2 = 1$).
$| A_0 |^2 + | A_\perp |^2 + | A_\parallel |^2 = 1$).
Then the effective $\CP$ of the vector-vector state is known if 
% $\left| A_\perp \right|^2$ is measured.
$| A_\perp |^2$ is measured.
An alternative strategy is to measure just the longitudinally polarized 
% component,  $\left| A_0 \right|^2$
component,  $| A_0 |^2$
(sometimes denoted by $f_{\rm long}$), 
which allows a limit to be set on the effective $\CP$ since
% $\left| A_\perp \right|^2 \leq \left| A_\perp \right|^2 + \left| A_\parallel \right|^2
% = 1 - \left| A_0 \right|^2$.
$| A_\perp |^2 \leq | A_\perp |^2 + | A_\parallel |^2 = 1 - | A_0 |^2$.
The most complete treatment for 
neutral $\B$ decays to vector-vector final states
is time-dependent angular analysis 
(also known as time-dependent transversity analysis).
In such an analysis, 
the interference between the $\CP$ even and $\CP$ odd states 
provides additional sensitivity to the weak and strong phases involved.

\mysubsubsection{Time-dependent \CP asymmetries in decays to non-$\CP$ eigenstates
}
\label{sec:cp_uta:notations:non_cp}

Consider a non-$\CP$ eigenstate $f$, and its conjugate $\bar{f}$. 
For neutral $\B$ decays to these final states,
there are four amplitudes to consider:
those for $\Bz$ to decay to $f$ and $\bar{f}$
($\Af$ and $\Afbar$, respectively),
and the equivalents for $\Bzb$
($\Abarf$ and $\Abarfbar$).
% $\CP$ invariance in the decay requires 
If $\CP$ is conserved in the decay, then
$\Af = \Abarfbar$ and $\Afbar = \Abarf$.

%% make definition Bbar - B for f then fbar
%% define so that asymmetry is C cos DmDt - S sin DmDt for both f and fbar

The time-dependent decay distributions can be written in many different ways.
Here, we follow Sec.~\ref{sec:cp_uta:notations:cp_eigenstate}
and define $\lambda_f = \frac{q}{p}\frac{\Abarf}{\Af}$ and
$\lambda_{\bar f} = \frac{q}{p}\frac{\Abarfbar}{\Afbar}$.
The time-dependent \CP asymmetries then follow Eq.~(\ref{eq:cp_uta:td_cp_asp}):
\begin{eqnarray}
\label{eq:cp_uta:non-cp-obs}
  {\cal A}_f (\Delta t) \; \equiv \;
  \frac{
    \Gamma_{\Bzb \to f} (\Delta t) - \Gamma_{\Bz \to f} (\Delta t)
  }{
    \Gamma_{\Bzb \to f} (\Delta t) + \Gamma_{\Bz \to f} (\Delta t)
  } & = & S_f \sin(\Delta m \Delta t) - C_f \cos(\Delta m \Delta t), \\
  {\cal A}_{\bar{f}} (\Delta t) \; \equiv \;
  \frac{
    \Gamma_{\Bzb \to \bar{f}} (\Delta t) - \Gamma_{\Bz \to \bar{f}} (\Delta t)
  }{
    \Gamma_{\Bzb \to \bar{f}} (\Delta t) + \Gamma_{\Bz \to \bar{f}} (\Delta t)
  } & = & S_{\bar{f}} \sin(\Delta m \Delta t) - C_{\bar{f}} \cos(\Delta m \Delta t),
\end{eqnarray}
with the definitions of the parameters 
$C_f$, $S_f$, $C_{\bar{f}}$ and $S_{\bar{f}}$,
following Eqs.~(\ref{eq:cp_uta:s_def}) and~(\ref{eq:cp_uta:c_def}).

The time-dependent decay rates are given by
\begin{eqnarray}
  \Gamma_{\Bzb \to f} (\Delta t) & = &
  \frac{e^{-\left| \Delta t \right| / \tau(\Bz)}}{8\tau(\Bz)} 
  ( 1 + \Adirnoncp ) 
  \left\{ 
    1 + S_f \sin(\Delta m \Delta t) - C_f \cos(\Delta m \Delta t) 
  \right\},
  \\
  \Gamma_{\Bz \to f} (\Delta t) & = &
  \frac{e^{-\left| \Delta t \right| / \tau(\Bz)}}{8\tau(\Bz)} 
  ( 1 + \Adirnoncp ) 
  \left\{ 
    1 - S_f \sin(\Delta m \Delta t) + C_f \cos(\Delta m \Delta t) 
  \right\},
  \\
  \Gamma_{\Bzb \to \bar{f}} (\Delta t) & = &
  \frac{e^{-\left| \Delta t \right| / \tau(\Bz)}}{8\tau(\Bz)} 
  ( 1 - \Adirnoncp ) 
  \left\{ 
    1 + S_{\bar{f}} \sin(\Delta m \Delta t) - C_{\bar{f}} \cos(\Delta m \Delta t) 
  \right\},
  \\
  \Gamma_{\Bz \to \bar{f}} (\Delta t) & = &
    \frac{e^{-\left| \Delta t \right| / \tau(\Bz)}}{8\tau(\Bz)} 
  ( 1 - \Adirnoncp ) 
  \left\{ 
    1 - S_{\bar{f}} \sin(\Delta m \Delta t) + C_{\bar{f}} \cos(\Delta m \Delta t) 
  \right\},
\end{eqnarray}
where the time-independent parameter \Adirnoncp
represents an overall asymmetry in the production of the 
$f$ and $\bar{f}$ final states,\footnote{
  This parameter is often denoted ${\cal A}_f$ (or ${\cal A}_{\CP}$),
  but here we avoid this notation to prevent confusion with the
  time-dependent $\CP$ asymmetry.
}
\begin{equation}
  \Adirnoncp = 
  \frac{
    \left( 
      \left| \Af \right|^2 + \left| \Abarf \right|^2
    \right) - 
    \left( 
      \left| \Afbar \right|^2 + \left| \Abarfbar \right|^2
    \right)
  }{
    \left( 
      \left| \Af \right|^2 + \left| \Abarf \right|^2
    \right) +
    \left( 
      \left| \Afbar \right|^2 + \left| \Abarfbar \right|^2
    \right)
  }.
\end{equation}
Assuming $|q/p| = 1$,
the parameters $C_f$ and $C_{\bar{f}}$
can also be written in terms of the decay amplitudes as follows:
\begin{equation}
  C_f = 
  \frac{
    \left| \Af \right|^2 - \left| \Abarf \right|^2 
  }{
    \left| \Af \right|^2 + \left| \Abarf \right|^2
  }
  \hspace{5mm}
  {\rm and}
  \hspace{5mm}
  C_{\bar{f}} = 
  \frac{
    \left| \Afbar \right|^2 - \left| \Abarfbar \right|^2
  }{
    \left| \Afbar \right|^2 + \left| \Abarfbar \right|^2
  },
\end{equation}
giving asymmetries in the decay amplitudes of $\Bz$ and $\Bzb$
to the final states $f$ and $\bar{f}$ respectively.
In this notation, the direct $\CP$ invariance conditions are
$\Adirnoncp = 0$ and $C_f = - C_{\bar{f}}$.
Note that $C_f$ and $C_{\bar{f}}$ are typically non-zero;
\eg, for a flavour-specific final state, 
$\Abarf = \Afbar = 0$ ($\Af = \Abarfbar = 0$), they take the values
$C_f = - C_{\bar{f}} = 1$ ($C_f = - C_{\bar{f}} = -1$).
%\footnote
%{
%  	A formal derivation for the flavour-specific case
%  	should take care to avoid division by zero in the expressions for 
%  	$\lambda_f$ and $\lambda_{\bar f}$.
%}

The coefficients of the sine terms
% \begin{equation}
%   S_f = 
%   \frac{ 
%     \left| \Af \right|^2
%   }{
%     \left| \Af \right|^2 + \left| \Abarf \right|^2
%   } 2\, \Im \left( \frac{q}{p}\frac{\Abarf}{\Af} \right) 
%   \hspace{5mm}
%   {\rm and}
%   \hspace{5mm}
%   S_{\bar{f}} = 
%   \frac{
%     \left| \Afbar \right|^2
%   }{
%     \left| \Afbar \right|^2 + \left| \Abarfbar \right|^2
%   } 2\, \Im \left( \frac{q}{p}\frac{\Abarfbar}{\Afbar} \right)
% \end{equation}
contain information about the weak phase. 
% Assuming $\left| \frac{q}{p} \right| = 1$, then 
In the case that each decay amplitude contains only a single weak phase
(\ie, no direct $\CP$ violation),
these terms can be written
%% should include angular momentum factor here (hep-ph/0304027)
%% or just absorb it into the strong phase difference
\begin{equation}
  S_f = 
  \frac{ 
    - 2 \left| \Af \right| \left| \Abarf \right| 
    \sin( \phi_{\rm mix} + \phi_{\rm dec} - \delta_f )
  }{
    \left| \Af \right|^2 + \left| \Abarf \right|^2
  } 
  \hspace{5mm}
  {\rm and}
  \hspace{5mm}
  S_{\bar{f}} = 
  \frac{
    - 2 \left| \Afbar \right| \left| \Abarfbar \right| 
    \sin( \phi_{\rm mix} + \phi_{\rm dec} + \delta_f )
  }{
    \left| \Afbar \right|^2 + \left| \Abarfbar \right|^2
  },
\end{equation}
where $\delta_f$ is the strong phase difference between the decay amplitudes.
If there is no $\CP$ violation, the condition $S_f = - S_{\bar{f}}$ holds.
If amplitudes with different weak and strong phases contribute,
no clean interpretation of $S_f$ and $S_{\bar{f}}$ is possible.

Since two of the $\CP$ invariance conditions are 
$C_f = - C_{\bar{f}}$ and $S_f = - S_{\bar{f}}$,
there is motivation for a rotation of the parameters:
\begin{equation}
\label{eq:cp_uta:non-cp-s_and_deltas}
  S_{f\bar{f}} = \frac{S_{f} + S_{\bar{f}}}{2},
  \hspace{4mm}
  \Delta S_{f\bar{f}} = \frac{S_{f} - S_{\bar{f}}}{2},
  \hspace{4mm}
  C_{f\bar{f}} = \frac{C_{f} + C_{\bar{f}}}{2},
  \hspace{4mm}
  \Delta C_{f\bar{f}} = \frac{C_{f} - C_{\bar{f}}}{2}.
\end{equation}
With these parameters, the $\CP$ invariance conditions become
$S_{f\bar{f}} = 0$ and $C_{f\bar{f}} = 0$. 
The parameter $\Delta C_{f\bar{f}}$ gives a measure of the ``flavour-specificity''
of the decay:
$\Delta C_{f\bar{f}}=\pm1$ corresponds to a completely flavour-specific decay,
in which no interference between decays with and without mixing can occur,
while $\Delta C_{f\bar{f}} = 0$ results in 
maximum sensitivity to mixing-induced $\CP$ violation.
% describes the ``flavour-eigenstateness'' 
% of  the decay: maximum sensitivity to mixing-induced $\CP$ violation is 
% achieved for $\Delta C_{f\bar{f}}=0$, while for $\Delta C_{f\bar{f}}=\pm1$
% (maximum dilution) no interference between decays with and without 
% mixing can occur. 
The parameter $\Delta S_{f\bar{f}}$ is related to the strong phase difference 
between the decay amplitudes of $\Bz$ to $f$ and to $\bar f$. 
We note that the observables of Eq.~(\ref{eq:cp_uta:non-cp-s_and_deltas})
exhibit experimental correlations 
(typically of $\sim 20\%$, depending on the tagging purity, and other effects)
between $S_{f\bar{f}}$ and  $\Delta S_{f\bar{f}}$, 
and between $C_{f\bar{f}}$ and $\Delta C_{f\bar{f}}$. 
%% TJG try to clarify
% This is not the case for the final state
% specific observables~(\ref{eq:cp_uta:non-cp-obs}). 
On the other hand, 
the final state specific observables of Eq.~(\ref{eq:cp_uta:non-cp-obs})
tend to have low correlations.
% since they are obtained from essentially independent data.
%% TJG not sure this is helpful
% Since the transformation is linear,
% both sets of observables are approximately Gaussian distributed.

Alternatively, if we recall that the $\CP$ invariance
conditions at the amplitude level are
$\Af = \Abarfbar$ and $\Afbar = \Abarf$,
we are led to consider the parameters~\cite{ref:cp_uta:uud:charles}
\begin{equation}
  \label{eq:cp_uta:non-cp-directcp}
  {\cal A}_{f\bar{f}} = 
  \frac{
    \left| \Abarfbar \right|^2 - \left| \Af \right|^2 
  }{
    \left| \Abarfbar \right|^2 + \left| \Af \right|^2
  }
  \hspace{5mm}
  {\rm and}
  \hspace{5mm}
  {\cal A}_{\bar{f}f} = 
  \frac{
    \left| \Abarf \right|^2 - \left| \Afbar \right|^2
  }{
    \left| \Abarf \right|^2 + \left| \Afbar \right|^2
  }.
\end{equation}
These are sometimes considered more physically intuitive parameters
since they characterize direct $\CP$ violation 
in decays with particular topologies.
For example, in the case of $\Bz \to \rho^\pm\pi^\mp$
(choosing $f =  \rho^+\pi^-$ and $\bar{f} = \rho^-\pi^+$),
${\cal A}_{f\bar{f}}$ (also denoted ${\cal A}^{+-}_{\rho\pi}$)
parameterizes direct $\CP$ violation
in decays in which the produced $\rho$ meson does not contain the 
spectator quark,
while ${\cal A}_{\bar{f}f}$ (also denoted ${\cal A}^{-+}_{\rho\pi}$)
parameterizes direct $\CP$ violation 
in decays in which it does.
Note that we have again followed the sign convention that the asymmetry 
is the difference between the rate involving a $b$ quark and that
involving a $\bar{b}$ quark, \cf\ Eq.~(\ref{eq:cp_uta:pra}). 
Of course, these parameters are not independent of the 
other sets of parameters given above, and can be written
\begin{equation}
  {\cal A}_{f\bar{f}} =
  - \frac{
    \Adirnoncp + C_{f\bar{f}} + \Adirnoncp \Delta C_{f\bar{f}} 
  }{
    1 + \Delta C_{f\bar{f}} + \Adirnoncp C_{f\bar{f}} 
  }
  \hspace{5mm}
  {\rm and}
  \hspace{5mm}
  {\cal A}_{\bar{f}f} =
  \frac{
    - \Adirnoncp + C_{f\bar{f}} + \Adirnoncp \Delta C_{f\bar{f}} 
  }{
    - 1 + \Delta C_{f\bar{f}} + \Adirnoncp C_{f\bar{f}}  
  }.
\end{equation}
They usually exhibit strong correlations.

We now consider the various notations which have been used 
in experimental studies of
time-dependent $\CP$ asymmetries in decays to non-$\CP$ eigenstates.

\mysubsubsubsection{$\Bz \to D^{*\pm}D^\mp$
}
\label{sec:cp_uta:notations:non_cp:dstard}

The above set of parameters 
($\Adirnoncp$, $C_f$, $S_f$, $C_{\bar{f}}$, $S_{\bar{f}}$),
has been used by both
\babar~\cite{ref:cp_uta:ccd:babar:dstard} and
\belle~\cite{ref:cp_uta:ccd:belle:dstard} in the $D^{*\pm}D^{\mp}$ system
($f = D^{*+}D^-$, $\bar{f} = D^{*-}D^+$).
However, slightly different names for the parameters are used:
\babar\ uses 
(${\cal A}$, $C_{+-}$, $S_{+-}$, $C_{-+}$, $S_{-+}$);
\belle\ uses
(${\cal A}$, $C_{+}$,  $S_{+}$,  $C_{-}$,  $S_{-}$).
In this document, we follow the notation used by \babar.

\mysubsubsubsection{$\Bz \to \rho^{\pm}\pi^\mp$
}
\label{sec:cp_uta:notations:non_cp:rhopi}

In the $\rho^\pm\pi^\mp$ system, the 
($\Adirnoncp$, $C_{f\bar{f}}$, $S_{f\bar{f}}$, $\Delta C_{f\bar{f}}$, 
$\Delta S_{f\bar{f}}$)
set of parameters has been used 
originally by 
\babar~\cite{ref:cp_uta:uud:babar:rhopi_old}, and more recently by
\belle~\cite{ref:cp_uta:uud:belle:rhopi}, in the Q2B approximation;
the exact names\footnote{
  \babar\ has used the notations
  $A_{\CP}^{\rho\pi}$~\cite{ref:cp_uta:uud:babar:rhopi_old} and 
  ${\cal A}_{\rho\pi}$~\cite{ref:cp_uta:uud:babar:rhopi}
  in place of ${\cal A}_{\CP}^{\rho\pi}$.
}
used in this case are
$\left( 
  {\cal A}_{\CP}^{\rho\pi}, C_{\rho\pi}, S_{\rho\pi}, \Delta C_{\rho\pi}, \Delta S_{\rho\pi}
\right)$,
and these names are also used in this document.

Since $\rho^\pm\pi^\mp$ is reconstructed in the final state $\pi^+\pi^-\pi^0$,
the interference between the $\rho$ resonances
can provide additional information about the phases.
\babar~\cite{ref:cp_uta:uud:babar:rhopi} has performed 
a time-dependent Dalitz plot analysis, 
from which the weak phase $\alpha$ is directly extracted.
In such an analysis, the measured Q2B parameters are 
also naturally corrected for interference effects.

\mysubsubsubsection{$\Bz \to D^{\pm}\pi^{\mp}, D^{*\pm}\pi^{\mp}, D^{\pm}\rho^{\mp}$
}
\label{sec:cp_uta:notations:non_cp:dstarpi}

Time-dependent $\CP$ analyses have also been performed for the
final states $D^{\pm}\pi^{\mp}$, $D^{*\pm}\pi^{\mp}$ and $D^{\pm}\rho^{\mp}$.
In these theoretically clean cases, no penguin contributions are possible,
so there is no direct $\CP$ violation.
Furthermore, due to the smallness of the ratio of the magnitudes of the 
suppressed ($b \to u$) and favoured ($b \to c$) amplitudes (denoted $R_f$),
to a very good approximation, $C_f = - C_{\bar{f}} = 1$
(using $f = D^{(*)-}h^+$, $\bar{f} = D^{(*)+}h^-$ $h = \pi,\rho$),
and the coefficients of the sine terms are given by
\begin{equation}
  S_f = - 2 R_f \sin( \phi_{\rm mix} + \phi_{\rm dec} - \delta_f )
  \hspace{5mm}
  {\rm and}
  \hspace{5mm}
  S_{\bar{f}} = - 2 R_f \sin( \phi_{\rm mix} + \phi_{\rm dec} + \delta_f ).
\end{equation}
Thus weak phase information can be cleanly obtained from measurements
of $S_f$ and $S_{\bar{f}}$, 
although external information on at least one of $R_f$ or $\delta_f$ is necessary.
(Note that $\phi_{\rm mix} + \phi_{\rm dec} = 2\beta + \gamma$ for all the decay modes 
in question, while $R_f$ and $\delta_f$ depend on the decay mode.)

Again, different notations have been used in the literature.
\babar\xspace\cite{ref:cp_uta:cud:babar:full,ref:cp_uta:cud:babar:partial}
defines the time-dependent probability function by
\begin{equation}
  f^\pm (\eta, \Delta t) = \frac{e^{-|\Delta t|/\tau}}{4\tau} 
  \left[  
    1 \mp S_\zeta \sin (\Delta m \Delta t) \mp \eta C_\zeta \cos(\Delta m \Delta t) 
  \right],
\end{equation} 
where the upper (lower) sign corresponds to 
the tagging meson being a $\Bz$ ($\Bzb$). 
[Note here that a tagging $\Bz$ ($\Bzb$) corresponds to $-S_\xi$ ($+S_\xi$).]
The parameters $\eta$ and $\zeta$ take the values $+1$ and $+$ ($-1$ and $-$) 
when the final state is, \eg, $D^-\pi^+$ ($D^+\pi^-$). 
However, in the fit, the substitutions $C_\zeta = 1$ and 
$S_\zeta = a \mp \eta b_i - \eta c_i$ are made. 
[Note that, neglecting $b$ terms, $S_+ = a - c$ and $S_- = a + c$, 
so that $a = (S_+ + S_-)/2$, $c = (S_- - S_+)/2$, in analogy to 
the parameters of Eq.~(\ref{eq:cp_uta:non-cp-s_and_deltas}).] 
The subscript $i$ denotes the tagging category. 
These are motivated by the possibility of 
$\CP$ violation on the tag side~\cite{ref:cp_uta:cud:tagside}, 
which is absent for semileptonic $\B$ decays (mostly lepton tags). 
The parameter $a$ is not affected by tag side $\CP$ violation. 

The parameters used by \belle\ in the analysis using 
partially reconstructed $\B$ decays~\cite{ref:cp_uta:cud:belle:partial}, 
are similar to the $S_\zeta$ parameters defined above. 
However, in the \belle\ convention, 
a tagging $\Bz$ corresponds to a $+$ sign in front of the sine coefficient; 
furthermore the correspondence between the super/subscript 
and the final state is opposite, so that $S_\pm$ (\babar) = $- S^\mp$ (\belle). 
In this analysis, only lepton tags are used, 
so there is no effect from tag side $\CP$ violation. 
In the \belle\ analysis using 
fully reconstructed $\B$ decays~\cite{ref:cp_uta:cud:belle:full}, 
this effect is measured and taken into account using $\Dstar l \nu$ decays; 
in neither \belle\ analysis are the $a$, $b$ and $c$ parameters used. 
In the latter case, the measured parameters are 
$2 R_{D^{(*)}\pi} \sin( 2\phi_1 + \phi_3 \pm \delta_{D^{(*)}\pi} )$; 
the definition is such that 
$S^\pm$ (\belle) = $- 2 R_{\Dstar \pi} \sin( 2\phi_1 + \phi_3 \pm \delta_{\Dstar \pi} )$. 
However, the definition includes an 
angular momentum factor $(-1)^L$~\cite{ref:cp_uta:cud:fleischer}, 
and so for the results in the $D\pi$ system, 
there is an additional factor of $-1$ in the conversion.

Explicitly, the conversion then reads as given in 
Table~\ref{tab:cp_uta:notations:non_cp:dstarpi}, 
where we have neglected the $b_i$ terms used by \babar.
For the averages in this document,
we use the $a$ and $c$ parameters,
and give the explicit translations used in 
Table~\ref{tab:cp_uta:notations:non_cp:dstarpi2}.
It is to be fervently hoped that the experiments will
converge on a common notation in future.

\begin{table}
  \begin{center} 
    \caption{
      Conversion between the various notations used for 
      $\CP$ violation parameters in the 
      $D^{\pm}\pi^{\mp}$, $D^{*\pm}\pi^{\mp}$ and $D^{\pm}\rho^{\mp}$ systems.
      The $b_i$ terms used by \babar\ have been neglected.
    }
    \vspace{0.2cm}
    \setlength{\tabcolsep}{0.0pc}
    \begin{tabular*}{\textwidth}{@{\extracolsep{\fill}}cccc} \hline 
      & \babar\ & \belle\ partial rec. & \belle\ full rec. \\
      \hline
      $S_{D^+\pi^-}$    & $- S_- = - (a + c_i)$ &  N/A  &
      $2 R_{D\pi} \sin( 2\phi_1 + \phi_3 + \delta_{D\pi} )$ \\
      $S_{D^-\pi^+}$    & $- S_+ = - (a - c_i)$ &  N/A  &
      $2 R_{D\pi} \sin( 2\phi_1 + \phi_3 - \delta_{D\pi} )$ \\
      $S_{D^{*+}\pi^-}$ & $- S_- = - (a + c_i)$ & $S^+$ &   
      $- 2 R_{\Dstar \pi} \sin( 2\phi_1 + \phi_3 + \delta_{\Dstar \pi} )$ \\
      $S_{D^{*-}\pi^+}$ & $- S_+ = - (a - c_i)$ & $S^-$ &
      $- 2 R_{\Dstar \pi} \sin( 2\phi_1 + \phi_3 - \delta_{\Dstar \pi} )$ \\
      $S_{D^+\rho^-}$    & $- S_- = - (a + c_i)$ &  N/A  &  N/A  \\
      $S_{D^-\rho^+}$    & $- S_+ = - (a - c_i)$ &  N/A  &  N/A  \\
      \hline 
    \end{tabular*}
    \label{tab:cp_uta:notations:non_cp:dstarpi}
  \end{center}
\end{table}
   
\begin{table}
  \begin{center} 
    \caption{
      Translations used to convert the parameters measured by \belle
      to the parameters used for averaging in this document.
      The angular momentum factor $L$ is $-1$ for $\Dstar\pi$ and $+1$ for $D\pi$.
    }
    \vspace{0.2cm}
    \setlength{\tabcolsep}{0.0pc}
    \begin{tabular*}{\textwidth}{@{\extracolsep{\fill}}ccc} \hline 
        & $\Dstar\pi$ partial rec. & $D^{(*)}\pi$ full rec. \\
        \hline
        $a$ & $- (S^+ + S^-)$ &
        $\frac{1}{2} (-1)^{L+1}
        \left(
          2 R_{D^{(*)}\pi} \sin( 2\phi_1 + \phi_3 + \delta_{D^{(*)}\pi} ) + 
          2 R_{D^{(*)}\pi} \sin( 2\phi_1 + \phi_3 - \delta_{D^{(*)}\pi} )
        \right)$ \\
        $c$ & $- (S^+ - S^-)$ & 
        $\frac{1}{2} (-1)^{L+1}
        \left(
          2 R_{D^{(*)}\pi} \sin( 2\phi_1 + \phi_3 + \delta_{D^{(*)}\pi} ) -
          2 R_{D^{(*)}\pi} \sin( 2\phi_1 + \phi_3 - \delta_{D^{(*)}\pi} )
        \right)$ \\
        \hline 
      \end{tabular*}
    \label{tab:cp_uta:notations:non_cp:dstarpi2}
  \end{center}
\end{table}

\mysubsubsubsection{Time-dependent asymmetries in radiative $\B$ decays
}
\label{sec:cp_uta:notations:non_cp:radiative}

As a special case of decays to non-$\CP$ eigenstates,
let us consider radiative $\B$ decays.
Here, the emitted photon has a distinct helicity,
which is in principle observable, but in practice is not usually measured.
Thus the measured time-dependent decay rates 
are given by~\cite{ref:cp_uta:bsg:ags,ref:cp_uta:bsg:aghs}
\begin{eqnarray}
  \Gamma_{\Bzb \to X \gamma} (\Delta t) & = &
  \Gamma_{\Bzb \to X \gamma_L} (\Delta t) + \Gamma_{\Bzb \to X \gamma_R} (\Delta t) \\ \nonumber
  & = &
  \frac{e^{-\left| \Delta t \right| / \tau(\Bz)}}{4\tau(\Bz)} 
  \left\{ 
    1 + 
    \left( S_L + S_R \right) \sin(\Delta m \Delta t) - 
    \left( C_L + C_R \right) \cos(\Delta m \Delta t) 
  \right\},
  \\
  \Gamma_{\Bz \to X \gamma} (\Delta t) & = & 
  \Gamma_{\Bz \to X \gamma_L} (\Delta t) + \Gamma_{\Bz \to X \gamma_R} (\Delta t) \\ \nonumber 
  & = &
  \frac{e^{-\left| \Delta t \right| / \tau(\Bz)}}{8\tau(\Bz)} 
  \left\{ 
    1 - 
    \left( S_L + S_R \right) \sin(\Delta m \Delta t) + 
    \left( C_L + C_R \right) \cos(\Delta m \Delta t) 
  \right\},
\end{eqnarray}
where in place of the subscripts $f$ and $\bar{f}$ we have used $L$ and $R$
to indicate the photon helicity.
In order for interference between decays with and without $\Bz$-$\Bzb$ mixing
to occur, the $X$ system must not be flavour-specific,
\eg, in case of $\Bz \to K^{*0}\gamma$, the final state must be $\KS \pi^0 \gamma$.
The sign of the sine term depends on the $C$ eigenvalue of the $X$ system.
The photons from $b \to q \gamma$ ($\bar{b} \to \bar{q} \gamma$) are predominantly
left (right) polarized, with corrections of order of $m_q/m_b$,
thus interference effects are suppressed.
% In the case of $b \to s \gamma$, a single weak phase dominates,
% so one expects that $C_L + C_R \approx 0$ and
% $\left| S_L + S_R \right| \lesssim \frac{2 m_s}{m_b} \sin \left( 2\beta \right)$.
% In the case of $b \to d \gamma$, more than one weak phase is possible,
% so direct $\CP$ violation can occur, 
% but the $S$ term should be vanishingly small.
The predicted smallness of the $S$ terms in the Standard Model
results in sensitivity to new physics contributions.

\mysubsubsection{Asymmetries in $\B \to \DorDstar K^{(*)}$ decays
}
\label{sec:cp_uta:notations:cus}

$\CP$ asymmetries in $\B \to \DorDstar K^{(*)}$ decays are sensitive to $\gamma$.
The neutral $D^{(*)}$ meson produced 
% in the decay $\Bm \to \DorDstar K^{(*)-}$
is an admixture of $\DorDstarz$ (produced by a $b \to c$ transition) and 
$\DorDstarzb$ (produced by a colour-suppressed $b \to u$ transition) states.
If the final state is chosen so that both $\DorDstarz$ and $\DorDstarzb$ 
can contribute, the two amplitudes interfere,
and the resulting observables are sensitive to $\gamma$, 
the relative weak phase between the two $\B$ decay amplitudes.
% \footnote{
%   The same is true for $\Bzb \to \DorDstar \bar{K}^{(*)0}$ decays,
%   where both $b \to c$ and $b \to u$ transitions are colour-suppressed.
% }
Various methods have been proposed to exploit this interference,
including those where the neutral $D$ meson is reconstructed 
as a $\CP$ eigenstate (GLW)~\cite{ref:cp_uta:cus:glw},
in a suppressed final state (ADS)~\cite{ref:cp_uta:cus:ads},
or in a self-conjugate three-body final state, 
such as $\KS \pi^+\pi^-$ (Dalitz)~\cite{ref:cp_uta:cus:dalitz}.
% Each of these approaches, while theoretically clean,
% has some difficulty for $\gamma$ extraction due to intrinsic ambiguities
% or model dependence;
% these can be overcome by combining the results from different techniques,
% and including other modes, such as $\Bmp \to \Dstar \Kmp$ and $\Bmp \to D \Kstarmp$.

Consider the case of $\Bmp \to D \Kmp$,
with $D$ decaying to a final state $f$,
which is accessible to both $\Dz$ and $\Dzb$.
We can write the decay rates for $\Bm$ and $\Bp$ ($\Gamma_\mp$), 
the charge averaged rate ($\Gamma = (\Gamma_- + \Gamma_+)/2$)
and the charge asymmetry 
(${\cal A} = (\Gamma_- - \Gamma_+)/(\Gamma_- + \Gamma_+)$, see Eq.~(\ref{eq:cp_uta:pra})) as 
\begin{eqnarray}
  \label{eq:cp_uta:dk:rate_def}
  \Gamma_\mp  & \propto & 
  r_B^2 + r_D^2 + 2 r_B r_D \cos \left( \delta_B + \delta_D \mp \gamma \right), \\
  \label{eq:cp_uta:dk:av_rate_def}
  \Gamma & \propto &  
  r_B^2 + r_D^2 + 2 r_B r_D \cos \left( \delta_B + \delta_D \right) \cos \left( \gamma \right), \\
  \label{eq:cp_uta:dk:acp_def}
  {\cal A} & = & 
  \frac{
    2 r_B r_D \sin \left( \delta_B + \delta_D \right) \sin \left( \gamma \right)
  }{
    r_B^2 + r_D^2 + 2 r_B r_D \cos \left( \delta_B + \delta_D \right) \cos \left( \gamma \right),  
  }
\end{eqnarray}
where the ratio of $\B$ decay amplitudes\footnote{
  Note that here we use the notation $r_B$ to denote the ratio
  of $\B$ decay amplitudes, 
  whereas in Sec.~\ref{sec:cp_uta:notations:non_cp:dstarpi} 
  we used, \eg, $R_{D\pi}$, for a rather similar quantity.
  The reason is that here we need to be concerned also with 
  $D$ decay amplitudes,
  and so it is convenient to use the subscript to denote the decaying particle.
  Hopefully, using $r$ in place of $R$ will help reduce potential confusion.
} 
is usually defined to be less than one,
\begin{equation}
  \label{eq:cp_uta:dk:rb_def}
  r_B = 
  \frac{
    \left| A\left( \Bm \to \Dzb K^- \right) \right|
  }{
    \left| A\left( \Bm \to \Dz  K^- \right) \right|
  },
\end{equation}
and the ratio of $D$ decay amplitudes is correspondingly defined by
\begin{equation}
  \label{eq:cp_uta:dk:rd_def}
  r_D = 
  \frac{
    \left| A\left( \Dz  \to f \right) \right|
  }{
    \left| A\left( \Dzb \to f \right) \right|
  }.
\end{equation}
The strong phase differences between the $\B$ and $D$ decay amplitudes 
are given by $\delta_B$ and $\delta_D$, respectively.
% Note that $r_B$ and $\delta_B$ take different values for different $\B$ decays;
% the values for $\Bm \to D \Km$ and $\Bm \to \Dstar \Km$ are not the same.
% On the other hand, the value of $r_{D^{(*)}}$ depends only on the final state of
% the $D$ decay, since the amplitudes for $D^{*0}$ and $\bar{D}^{*0}$ decays
% to $D^{0}$ and $\bar{D}^{0}$, respectively,
% via emission of either a pion or a photon, will cancel in the ratio.
The values of $r_D$ and $\delta_D$ depend on the final state $f$:
for the GLW analysis, $r_D = 1$ and $\delta_D$ is trivial (either zero or $\pi$),
in the Dalitz plot analysis $r_D$ and $\delta_D$ vary across the Dalitz plot,
and depend on the $D$ decay model used,
for the ADS analysis, the values of $r_D$ and $\delta_D$ are not trivial.

In the GLW analysis, the measured quantities are the 
partial rate asymmetry, and the charge averaged rate,
which are measured both for $\CP$ even and $\CP$ odd $D$ decays.
For the latter, it is experimentally convenient to measure a double ratio,
\begin{equation}
  \label{eq:cp_uta:dk:double_ratio}
  R_{\CP} = 
  \frac{
    \Gamma\left( \Bm \to D_{\CP} \Km  \right) \, / \, \Gamma\left( \Bm \to \Dz \Km \right)
  }{
    \Gamma\left( \Bm \to D_{\CP} \pim \right) \, / \, \Gamma\left( \Bm \to \Dz \pim \right)
  }
\end{equation}
that is normalized both to the rate for the favoured $\Dz \to \Km\pip$ decay, 
and to the equivalent quantities for $\Bm \to D\pim$ decays
(charge conjugate modes are implicitly 
included in Eq.~(\ref{eq:cp_uta:dk:double_ratio})).
In this way the constant of proportionality drops out of 
Eq.~(\ref{eq:cp_uta:dk:av_rate_def}).

For the ADS analysis, using a suppressed $D \to f$ decay,
the measured quantities are again the partial rate asymmetry, 
and the charge averaged rate.
In this case it is sufficient to measure the rate in a single ratio
(normalized to the favoured $D \to \bar{f}$ decay)
since detection systematics cancel naturally.
In the ADS analysis, there are an additional two unknowns ($r_D$ and $\delta_D$)
compared to the GLW case.  
However, the value of $r_D$ can be measured using 
decays of $D$ mesons of known flavour.

The relations between the measured quantities and the
underlying parameters are summarized in Table~\ref{tab:cp_uta:notations:dk}.
Note carefully that the hadronic factors $r_B$ and $\delta_B$ 
are different, in general, for each $\B$ decay mode.
In the Dalitz plot analysis,
once a model is assumed for the $D$ decay, 
which gives the values of $r_D$ and $\delta_D$ across the Dalitz plot,
the values of ($\gamma$, $r_B$, $\delta_B$) can be directly extracted from
a simultaneous fit to the $\Bm$ and $\Bp$ data.

\begin{table}
  \begin{center} 
    \caption{
      Summary of relations between measured and physical parameters 
      in GLW and ADS analyses of $\B \to \DorDstar K^{(*)}$.
    }
    \vspace{0.2cm}
    \setlength{\tabcolsep}{1.0pc}
    \begin{tabular}{cc} \hline 
      \multicolumn{2}{c}{GLW analysis} \\
      $R_{\CP\pm}$ & $1 + r_B^2 \pm 2 r_B \cos \left( \delta_B \right) \cos \left( \gamma \right)$ \\
      $A_{\CP\pm}$ & $\pm 2 r_B \sin \left( \delta_B \right) \sin \left( \gamma \right) / R_{\CP\pm}$ \\
      \hline
      \multicolumn{2}{c}{ADS analysis} \\
      $R_{ADS}$ & $r_B^2 + r_D^2 + 2 r_B r_D \cos \left( \delta_B + \delta_D \right) \cos \left( \gamma \right)$ \\
      $A_{ADS}$ & $2 r_B r_D \sin \left( \delta_B + \delta_D \right) \sin \left( \gamma \right) / R_{ADS}$ \\
      \hline
    \end{tabular}
    \label{tab:cp_uta:notations:dk}
  \end{center}
\end{table}

\mysubsection{Common inputs and error treatment
}
\label{sec:cp_uta:common_inputs}

The common inputs used for rescaling are listed in 
Table~\ref{tab:cp_uta:common_inputs}.
The $\Bz$ lifetime ($\tau(\Bz)$) and mixing parameter ($\Delta m_d$)
averages are provided by the HFAG Lifetimes and Oscillations 
subgroup (Sec.~\ref{sec:life_mix}).
The fraction of the perpendicularly polarized component 
($\left| A_{\perp} \right|^2$) in $\B \to \jpsi \Kstar(892)$ decays,
which determines the $\CP$ composition, 
is averaged from results by 
\babar~\cite{ref:cp_uta:ccs:babar:psi_kstar} and
\belle~\cite{ref:cp_uta:ccs:belle:psi_kstar}.

At present, we only rescale to a common set of input parameters
for modes with reasonably small statistical errors
($b \to c\bar{c}s$ and $b \to q\bar{q}s$ transitions).
Correlated systematic errors are taken into account
in these modes as well.
For all other modes, the effect of such a procedure is 
currently negligible.

\begin{table}
  \begin{center}
    \caption{
      Common inputs used in calculating the averages.
    }
    \vspace{0.2cm}
    \setlength{\tabcolsep}{1.0pc}
    \begin{tabular}{cc} \hline 
      $\tau(\Bz)$ $({\rm ps})$ 	& $1.536 \pm 0.014$  \\
      $\Delta m_d$ $({\rm ps}^{-1})$ & $0.502 \pm 0.007$ \\
      $\left| A_{\perp} \right|^2 (\jpsi \Kstar)$ & $0.211 \pm 0.011$ \\
      \hline
    \end{tabular}
    \label{tab:cp_uta:common_inputs}
  \end{center}
\end{table}

As explained in Sec.~\ref{sec:intro},
we do not apply a rescaling factor on the error of an average
that has $\chi^2/\dof > 1$ 
(unlike the procedure currently used by the PDG~\cite{Eidelman:2004wy}).
We provide a confidence level of the fit so that
one can know the consistency of the measurements included in the average,
and attach comments in case some care needs to be taken in the interpretation.
Note that, in general, results obtained from data samples with low statistics
will exhibit some non-Gaussian behaviour.
For measurements where one error is given, 
it represents the total error, 
where statistical and systematic uncertainties have been added in quadrature.
If two errors are given, the first is statistical and the second systematic.
If more than two errors are given,
the origin of the third will be explained in the text.

Averages are computed by maximizing a log-likelihood function 
${\cal L}$ assuming Gaussian statistical and systematic errors.
When observables are correlated (\eg, sine and cosine coefficients 
in time-dependent \CP asymmetries), a combined minimization is 
performed, taking into account the correlations. Asymmetric errors
are treated by defining an asymmetric log-likelihood
function: ${\cal L}_i = (x - x_i)^2/(2\sigma_{i}^2)$,
where $\sigma_i=\sigma_{i,+}$ ($\sigma_i=\sigma_{i,-}$) if 
$x>x_i$ ($x<x_i$), and where $x_i$ is the $i$th measurement of
the observable $x$ that is averaged. This example assumes no 
correlations between observables. The correlated case is a 
straightforward extension to this.

\mysubsection{Time-dependent $\CP$ asymmetries in $b \to c\bar{c}s$ transitions
}
\label{sec:cp_uta:ccs}

\begin{figure}
  \begin{center}
    \resizebox{0.65\textwidth}{!}{\includegraphics{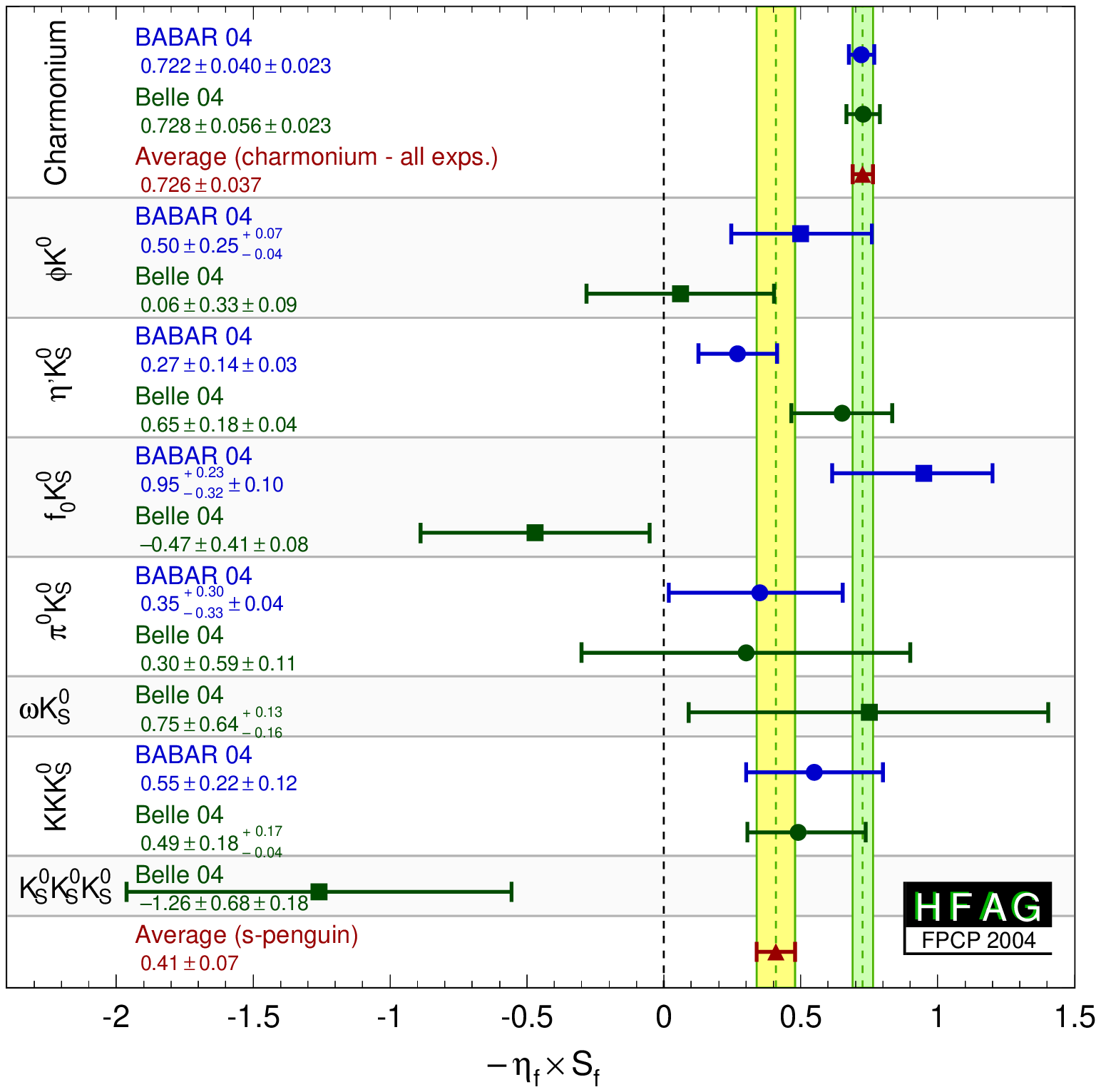}}
  \end{center}
  \vspace{-0.8cm}
  \caption{
    $S_{b \to c\bar c s}$ and $S_{b \to q\bar q s}$.
  }
  \label{fig:cp_uta:s_qqs}
\end{figure}

\begin{figure}
  \begin{center}
    \begin{tabular}{cc}
      \resizebox{0.46\textwidth}{!}{\includegraphics{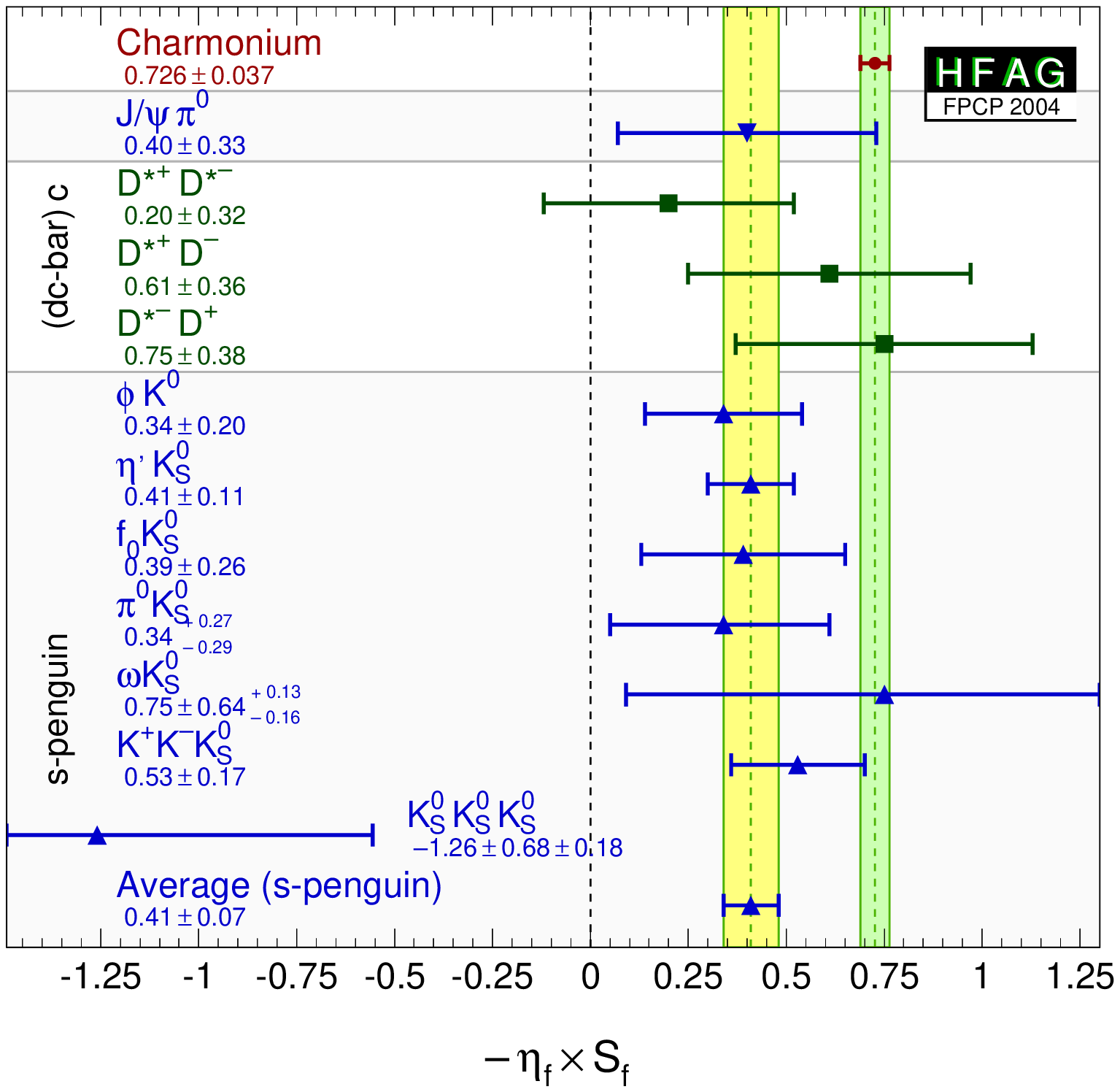}}
      &
      \resizebox{0.46\textwidth}{!}{\includegraphics{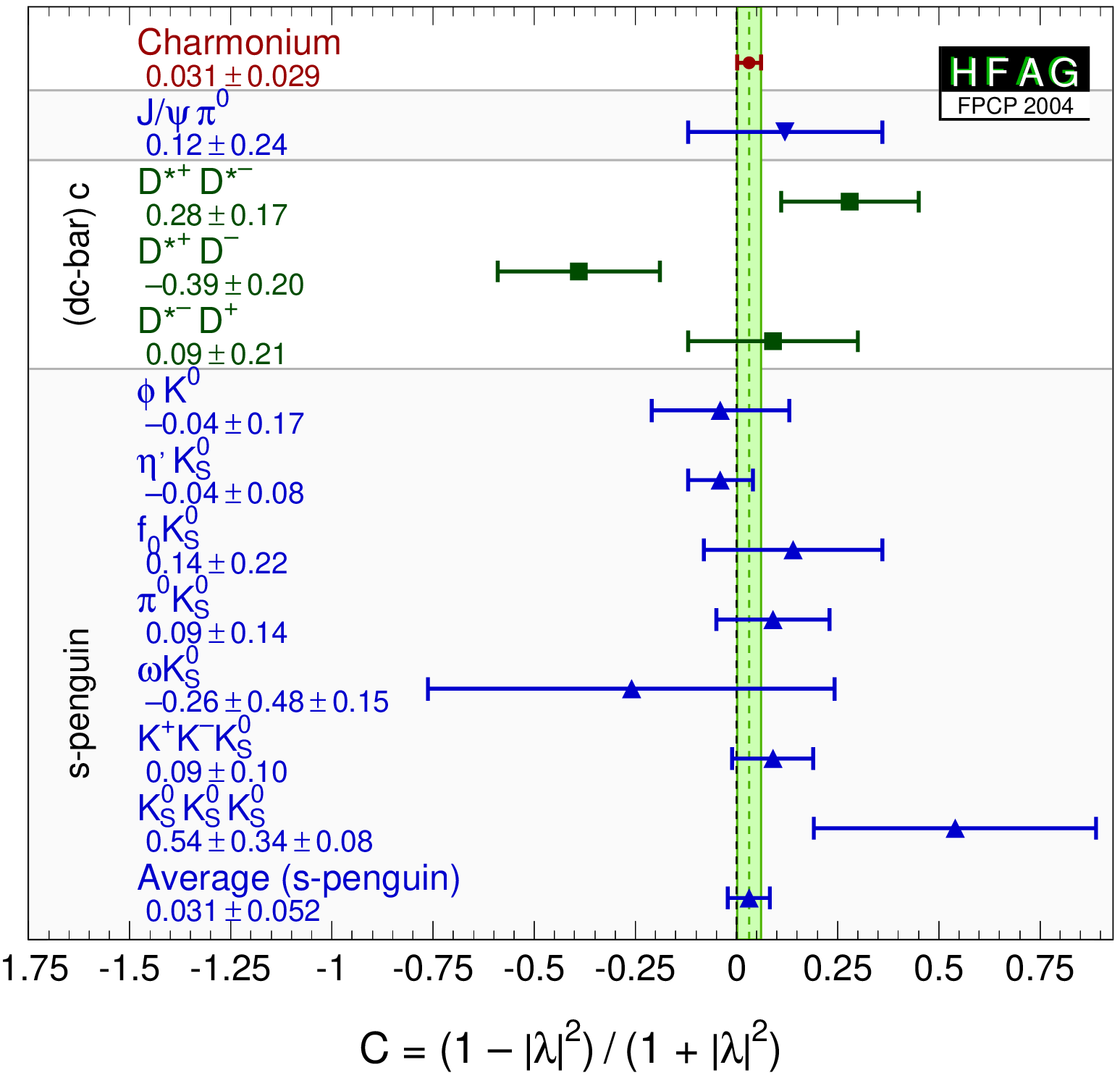}}
    \end{tabular}
  \end{center}
  \vspace{-0.8cm}
  \caption{
    Comparisons of the averages of (left)
    $S_{b \to c\bar c s}$, $S_{b \to c\bar c d}$ and $S_{b \to q\bar q s}$,
    and (right)
    $C_{b \to c\bar c s}$, $C_{b \to c\bar c d}$ and $C_{b \to q\bar q s}$.
  }
  \label{fig:cp_uta:qqs_ccd}
\end{figure}

In the Standard Model, the time-dependent parameters for
$b \to c\bar c s$ transitions are predicted to be: 
$S_{b \to c\bar c s} = - \etacp \sin(2\beta)$,
$C_{b \to c\bar c s} = 0$ to very good accuracy.
%% Due to the cleanliness of this prediction,
%% averages of $S_{b \to c\bar c s}$ are provided with the constraint
%% $C_{b \to c\bar c s} = 0$ in Table~\ref{tab:cp_uta:s_ccs},
%% and averages for $C_{b \to c\bar c s}$ are provided separately
%% in Table~\ref{tab:cp_uta:c_ccs}.
The averages for $S_{b \to c\bar c s}$ and $C_{b \to c\bar c s}$
are provided in Table~\ref{tab:cp_uta:ccs}.
The averages for $S_{b \to c\bar c s}$ are shown in Fig.~\ref{fig:cp_uta:s_qqs};
averages for $C_{b \to c\bar c s}$ are included in Fig.~\ref{fig:cp_uta:qqs_ccd}.

Both \babar\  and \belle\ use the $\etacp = -1$ modes
$\jpsi \KS$, $\psi(2S) \KS$, $\chi_{c1} \KS$ and $\eta_c \KS$, 
as well as $\jpsi \KL$, which has $\etacp = +1$
and $\jpsi K^{*0}(892)$, which is found to have $\etacp$ close to $+1$
based on the measurement of $\left| A_\perp \right|$ 
(see Sec.~\ref{sec:cp_uta:common_inputs}).
ALEPH, OPAL and CDF use only the $\jpsi \KS$ final state.

\begin{table}
  \begin{center}
    \caption{
      $S_{b \to c\bar c s}$ and $C_{b \to c\bar c s}$.
    }
    \vspace{0.2cm}
    \setlength{\tabcolsep}{0.0pc}
    \begin{tabular*}{\textwidth}{@{\extracolsep{\fill}}lrcc} \hline 
      \multicolumn{2}{l}{Experiment} & 
      $- \etacp S_{b \to c\bar c s}$ & $C_{b \to c\bar c s}$ \\
      \hline
      \babar & \cite{ref:cp_uta:ccs:babar} & 
      $0.722 \pm 0.040 \pm 0.023$ & $\ph{-}0.051 \pm 0.033 \pm 0.014$ \\
      \belle & \cite{BELLE2} & 
      $0.728 \pm 0.056 \pm 0.023$ & $-0.007 \pm 0.041 \pm 0.033$ \\
      \hline
      \multicolumn{2}{l}{\bf \boldmath $\B$ factory average} & 
      $0.725 \pm 0.037$ & $0.031 \pm 0.029$ \\
      \multicolumn{2}{l}{\small Confidence level} & 
      \small $0.91$ & \small $0.30$ \\
      \hline
      ALEPH & \cite{ref:cp_uta:ccs:aleph} & $0.84 \, ^{+0.82}_{-1.04} \pm 0.16$ \\
      OPAL  & \cite{ref:cp_uta:ccs:opal}  & $3.2 \, ^{+1.8}_{-2.0} \pm 0.5$ \\
      CDF   & \cite{ref:cp_uta:ccs:cdf}   & $0.79 \, ^{+0.41}_{-0.44}$ \\
      \hline
      \multicolumn{2}{l}{\bf Average} & 
      $0.726 \pm 0.037$ & $0.031 \pm 0.029$ \\
      \hline
    \end{tabular*}
    \label{tab:cp_uta:ccs}
  \end{center}
\end{table}

\begin{figure}[p]
  \begin{center}
    \resizebox{0.80\textwidth}{!}{
      \includegraphics{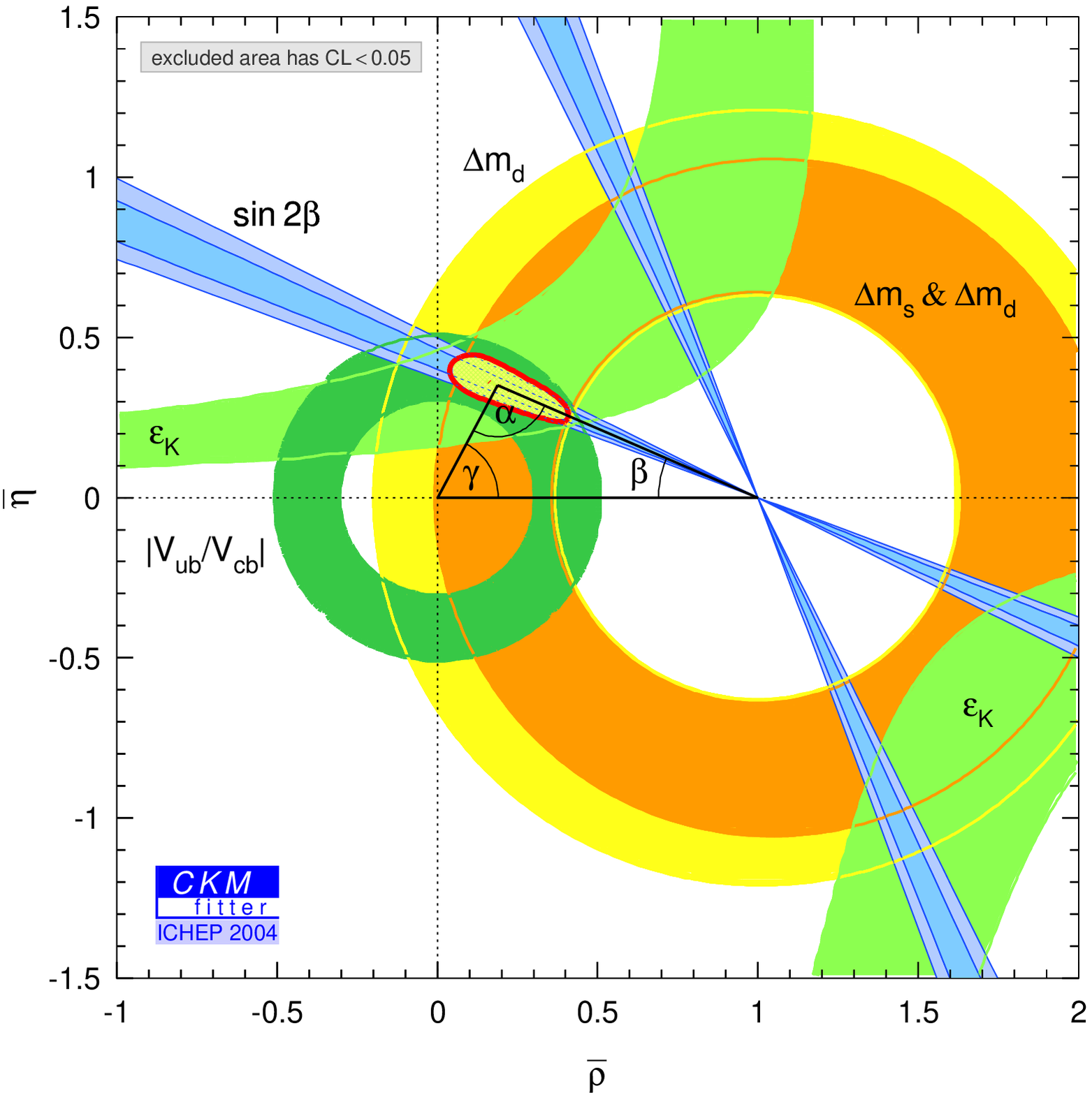}
    }
    \resizebox{0.95\textwidth}{!}{\hspace{1cm}
      \includegraphics{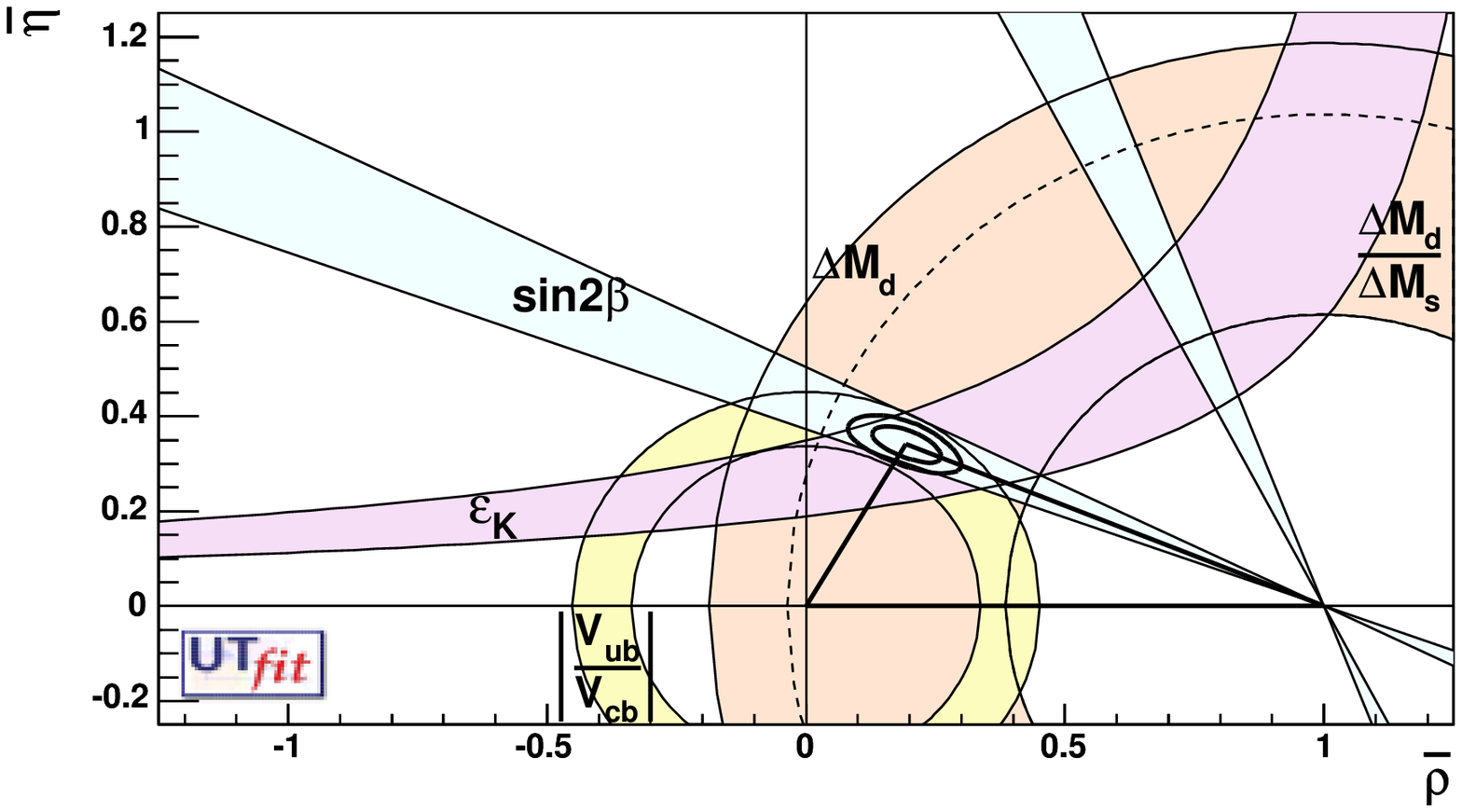}
    }
  \end{center}
  \vspace{-0.5cm}
  \caption{
    Standard Model constraints on the $(\rhobar,\etabar)$ plane,
    from (top)~\cite{ref:cp_uta:ckmfitter} and 
    (bottom)~\cite{ref:cp_uta:utfit}.
  }
  \label{fig:cp_uta:ckmfitter_sin2beta}
\end{figure}

%% straightforward interpretation
These results give a precise constraint on the $(\rhobar,\etabar)$ plane,
in remarkable agreement with other constraints from 
$\CP$ conserving quantities, 
and with $\CP$ violation in the kaon system, in the form of the parameter $\epsilon_K$.
Such comparisons have been performed by various phenomenological groups,
such as CKMfitter~\cite{ref:cp_uta:ckmfitter} 
and UTFit~\cite{ref:cp_uta:utfit}.
Figure~\ref{fig:cp_uta:ckmfitter_sin2beta} displays the constraints 
obtained from these two groups.

\mysubsection{Time-dependent transversity analysis of $\Bz \to J/\psi K^{*0}$
}
\label{sec:cp_uta:ccs_vv}

$\B$ meson decays to the vector-vector final state $J/\psi K^{*0}$
are also mediated by the $b \to c \bar c s$ transition.
When a final state which is not flavour-specific ($K^{*0} \to \KS \pi^0$) is used,
a time-dependent transversity analysis can be performed 
allowing sensitivity to both $\sin(2\beta)$ and $\cos(2\beta)$.
Such analyses have been performed by both $\B$ factory experiments.
% \babar~\cite{ref:cp_uta:ccs:babar:psi_kstar} and
% \belle~\cite{ref:cp_uta:ccs:belle:psi_kstar}.
In principle, the strong phases between the transversity amplitudes
are not uniquely determined by such an analysis, 
leading to a discrete ambiguity in the sign of $\cos(2\beta)$.
The \babar~\cite{ref:cp_uta:ccs:babar:psi_kstar} collaboration resolves 
this ambiguity using the known variation~\cite{ref:cp_uta:ccs:lass}
of the P-wave phase (fast) relative to the S-wave phase (slow) 
with the invariant mass of the $K\pi$ system 
in the vicinity of the $K^*(892)$ resonance. 
The result is in agreement with the prediction from 
$s$ quark helicity conservation,
and corresponds to Solution II defined by Suzuki~\cite{ref:cp_uta:ccs:suzuki}.
We use this phase convention for the averages given in 
Table~\ref{tab:cp_uta:ccs:psi_kstar}.

\begin{table}
  \begin{center}
    \caption{
      Averages from $\Bz \to J/\psi K^{*0}$ transversity analyses.
    }
    \vspace{0.2cm}
    \setlength{\tabcolsep}{0.0pc}
    \begin{tabular*}{\textwidth}{@{\extracolsep{\fill}}lrcc} \hline 
      \multicolumn{2}{l}{Experiment} & $\sin(2\beta)$ & $\cos(2\beta)$ \\
      \hline
      \babar & \cite{ref:cp_uta:ccs:babar:psi_kstar} & 
      $-0.10 \pm 0.57 \pm 0.14$ & $ 3.32 \, ^{+0.76}_{-0.96} \pm 0.27$ \\
      \belle & \cite{ref:cp_uta:ccs:belle:psi_kstar} &
      $\ph{-}0.30 \pm 0.32 \pm 0.02$ & $ 0.31 \pm 0.91 \pm 0.11$ \\
      \hline
      \multicolumn{2}{l}{\bf Average} & $0.21 \pm 0.28$ & $1.69 \pm 0.67 $ \\
      \hline
    \end{tabular*}
    \label{tab:cp_uta:ccs:psi_kstar}
  \end{center}
\end{table}

While the statistical errors are large, 
and exhibit non-Gaussian behaviour, 
$\cos(2\beta)>0$ is preferred 
by the experimental data in $J/\psi \Kstar$.

\mysubsection{Time-dependent $\CP$ asymmetries in $b \to q\bar{q}s$ transitions
}
\label{sec:cp_uta:qqs}

The flavour changing neutral current $b \to s$ penguin
can be mediated by any up-type quark in the loop, 
and hence the amplitude can be written as
\begin{equation}
  \label{eq:cp_uta:b_to_s}
  \begin{array}{ccccc}
    A_{b \to s} & = & 
    \multicolumn{3}{l}{F_u V_{ub}V^*_{us} + F_c V_{cb}V^*_{cs} + F_t V_{tb}V^*_{ts}} \\
    & = & (F_u - F_c) V_{ub}V^*_{us} & + & (F_t - F_c) V_{tb}V^*_{ts} \\
    & = & {\cal O}(\lambda^4) & + & {\cal O}(\lambda^2) \\
  \end{array}
\end{equation}
using the unitarity of the CKM matrix.
Therefore, in the Standard Model, 
this amplitude is dominated by $V_{tb}V^*_{ts}$, 
and to within a few degrees ($\delta\beta\lesssim2^\circ$ for 
$\beta\simeq23.3^\circ$) 
the time-dependent parameters 
can be written\footnote
{
  	The presence of a small (${\cal O}(\lambda^2)$) weak phase in 
	the dominant amplitude of the $s$ penguin decays introduces 
	a phase shift given by
	$S_{b \to q\bar q s} = -\eta\sin(2\beta)\cdot(1 + \Delta)$. 
	Using the CKMfitter results for the Wolfenstein 
	parameters~\cite{ref:cp_uta:uud:charles}, one finds: 
	$\Delta \simeq 0.033$, which corresponds to a shift of 
	$2\beta$ of $+2.1$ degrees. Nonperturbative contributions
      	can alter this result.
}
$S_{b \to q\bar q s} \approx - \etacp \sin(2\beta)$,
$C_{b \to q\bar q s} \approx 0$,
assuming $b \to s$ penguin contributions only ($q = u,d,s$).

Due to the large virtual mass scales occurring in the penguin loops,
additional diagrams from physics beyond the Standard Model,
with heavy particles in the loops, may contribute.
In general, these contributions will affect the values of 
$S_{b \to q\bar q s}$ and $C_{b \to q\bar q s}$.
A discrepancy between the values of 
$S_{b \to c\bar c s}$ and $S_{b \to q\bar q s}$
can therefore provide a clean indication of new physics.

However, there is an additional consideration to take into account.
The above argument assumes only the $b \to s$ penguin contributes
to the $b \to q\bar q s$ transition.
For $q = s$ this is a good assumption, 
which neglects only rescattering effects.
However, for $q = u$ there is a colour-suppressed $b \to u$ tree diagram
(of order ${\cal O}(\lambda^4)$), which has a different weak 
(and possibly strong) phase.
In the case $q = d$, any light neutral meson that is formed from
$d \bar{d}$ also has a $u \bar{u}$ component,
and so again there is ``tree pollution''. The \Bz decays to 
$\piz\KS$ and $\omega\KS$ belong to this category.
The mesons $f_0$ and $\etapr$ are expected to have predominant $s\bar s$ parts,
which reduces the possible tree pollution. If the inclusive 
decay $\Bz\to\Kp\Km\Kz$ (excluding $\phi\Kz$) is dominated by
a non-resonant three-body transition, an OZI-rule suppressed 
tree-level diagram can occur through insertion of an 
$s\sbar$ pair. The corresponding penguin-type transition 
proceeds via insertion of a $u\ubar$ pair, which is expected
to be favored over the $s\sbar$ insertion by fragmentation models.
Neglecting rescattering, the final state $\Kz\Kzb\Kz$ has no 
tree pollution.
%% this is repeated below
% Note that the calculation of an average between different modes
% implicitly neglects contributions with different weak phases 
% to the $b \to s$ penguin amplitude.

The averages for $S_{b \to q\bar q s}$ and $C_{b \to q\bar q s}$
can be found in Table~\ref{tab:cp_uta:qqs}.
The averages for $S_{b \to q\bar q s}$ are shown in Fig.~\ref{fig:cp_uta:s_qqs};
averages for $C_{b \to q\bar q s}$ are included in Fig.~\ref{fig:cp_uta:qqs_ccd}.
Results from both \babar\  and \belle\ are averaged for the modes
$\phi K^0$ (both $\phi\KS$ and $\phi\KL$ are used), 
$\etapr \KS$, $K^+K^-\KS$, $f_0 \KS$ and $\pi^0 \KS$.
In addition, results from \belle\ are taken for the modes
$\omega\KS$ and $\KS\KS\KS$. Of these modes,
$\phi\KS$, $\etapr \KS$, $\pi^0 \KS$ and $\omega\KS$ have $\CP$ 
eigenvalue $\etacp = -1$, while $\phi\KL$, $f_0 \KS$ and $\KS\KS\KS$ 
have $\etacp = +1$.

The final state $K^+K^-\KS$ 
(contributions from $\phi \KS$ are implicitly excluded) 
is not a $\CP$ eigenstate.
However, the $\CP$ composition can be determined using either an 
isospin argument (used by \belle\ to determine a $\CP$ even fraction of 
$1.03 \pm 0.15 \pm 0.05$~\cite{ref:cp_uta:qqs:belle:kkks_cp})
or a moments analysis (used by \babar\ who finds a 
$\CP$ even fraction of $0.89 \pm 0.08 \pm 0.06$~\cite{ref:cp_uta:qqs:babar:kkks}).
The uncertainty in the $\CP$ even fraction leads to an 
asymmetric error on $S_{b \to q\bar q s}$, which is taken to be 
correlated among the experiments.
To combine, we rescale the results to the 
average $\CP$ even fraction of $0.93 \pm 0.09$.

\begin{table}
  \begin{center}
    \caption{
      $S_{b \to q\bar q s}$ and $C_{b \to q\bar q s}$.
    }
    \vspace{0.2cm}
    \setlength{\tabcolsep}{0.0pc}
    \begin{tabular*}{\textwidth}{@{\extracolsep{\fill}}lrcc} \hline 
      \multicolumn{2}{l}{Experiment} & 
      $- \etacp S_{b \to q\bar q s}$ & $C_{b \to q\bar q s}$ \\
      \hline
      \multicolumn{4}{c}{$\phi K^0$} \\
      \babar & \cite{ref:cp_uta:qqs:babar:phiks} & 
      $ 0.50 \pm 0.25 \, ^{+0.07}_{-0.04}$ & $ 0.00 \pm 0.23 \pm 0.05$ \\
      \belle & \cite{ref:cp_uta:qqs:belle} & 
      $ 0.06 \pm 0.33 \pm 0.09$              & $-0.08 \pm 0.22 \pm 0.09$ \\
      \multicolumn{2}{l}{\bf Average} & $0.34 \pm 0.20$ & $-0.04 \pm 0.17$ \\
      \multicolumn{2}{l}{\small Confidence level} & 
      \small $0.30$ & \small $0.81$ \\
      \hline
      \multicolumn{4}{c}{$\etapr \KS$} \\
      \babar & \cite{ref:cp_uta:qqs:babar:etapks} & 
      $ 0.27 \pm 0.14 \pm 0.03$ & $-0.21 \pm 0.10 \pm 0.03$ \\
      \belle & \cite{ref:cp_uta:qqs:belle} & 
      $ 0.65 \pm 0.18 \pm 0.04$ & $ 0.19 \pm 0.11 \pm 0.05$ \\
      \multicolumn{2}{l}{\bf Average} & $0.41 \pm 0.11$ & $-0.04 \pm 0.08$ \\
      \multicolumn{2}{l}{\small Confidence level} & 
      \small $0.10$ & \small $0.01~(2.5\sigma)$ \\
      \hline
      \multicolumn{4}{c}{$f_0 \KS$} \\
      \babar & \cite{ref:cp_uta:qqs:babar:f0ks} &
      $ 0.95 \, ^{+0.23}_{-0.32} \pm 0.10$ & $-0.24 \pm 0.31 \pm 0.15$ \\
      \belle & \cite{ref:cp_uta:qqs:belle} & 
      $-0.47 \pm 0.41 \pm 0.08$ & $ 0.39 \pm 0.27 \pm 0.08$ \\
      \multicolumn{2}{l}{\bf Average} & $ 0.39 \pm 0.26$ & $ 0.14 \pm 0.22$ \\
      \multicolumn{2}{l}{\small Confidence level} & 
      \small $0.008~(2.7\sigma)$ & \small $0.16~(1.4\sigma)$ \\
      \hline
      \multicolumn{4}{c}{$\pi^0 \KS$} \\
      \babar & \cite{ref:cp_uta:qqs:babar:pi0ks} &
      $ 0.35 \, ^{+0.30}_{-0.33} \pm 0.04$ & $ 0.06 \pm 0.18 \pm 0.06$ \\
      \belle & \cite{ref:cp_uta:qqs:belle} & 
      $ 0.30 \pm 0.59 \pm 0.11$ & $ 0.12 \pm 0.20 \pm 0.07$\\
      \multicolumn{2}{l}{\bf Average} & $0.34 \, ^{+0.27}_{-0.29}$ & $0.09 \pm 0.14 $ \\
      \multicolumn{2}{l}{\small Confidence level} & 
      \small $0.94$ & \small $0.83$ \\
      \hline
      \multicolumn{4}{c}{$\omega \KS$} \\
      \belle & \cite{ref:cp_uta:qqs:belle} & 
      $ 0.75 \pm 0.64 \, ^{+0.13}_{-0.16}$ & $-0.26 \pm 0.48 \pm 0.15$ \\
      \hline
      \multicolumn{4}{c}{$K^+K^-\KS$} \\
      \babar  & \cite{ref:cp_uta:qqs:babar:kkks} &
      $ 0.55 \pm 0.22 \pm 0.04 \pm 0.11$ & $ 0.10 \pm 0.14 \pm 0.06$ \\
      \belle & \cite{ref:cp_uta:qqs:belle} & 
      $ 0.49 \pm 0.18 \pm 0.04 \, ^{+0.17}_{-0.00}$ & $ 0.08 \pm 0.12 \pm 0.07$ \\
      \multicolumn{2}{l}{\bf Average} & $ 0.53 \pm 0.17$ & $ 0.09 \pm 0.10$ \\
      \multicolumn{2}{l}{\small Confidence level} & 
      \small $0.72$ & \small $0.92$ \\
      \hline
      \multicolumn{4}{c}{$\KS\KS\KS$} \\
      \belle & \cite{ref:cp_uta:qqs:belle:ksksks} & 
      $-1.26 \pm 0.68 \pm 0.18$ & $ 0.54 \pm 0.34 \pm 0.08$ \\
      \hline 
      \multicolumn{2}{l}{\bf \boldmath Average of all $b \to q\bar q s$} & 
      $0.41 \pm 0.07$ & $0.03 \pm 0.05$ \\
      \multicolumn{2}{l}{\small Confidence level} & 
      \small $0.10~(1.7\sigma)$ & \small $0.32$ \\
      \hline 
      \multicolumn{2}{l}{\bf \boldmath Average including $b \to c\bar c s$} &
      $ 0.665 \pm 0.033$ & $ 0.031 \pm 0.025$ \\
      \multicolumn{2}{l}{\small Confidence level} & 
      \small $1.2\times 10^{-4}~(3.1\sigma)$ & \small $0.40$ \\
      \hline
    \end{tabular*}
    \label{tab:cp_uta:qqs}
  \end{center}
\end{table}

%% straightforward interpretation
If we treat the combined error as a Gaussian quantity,
we note that the average of $- \etacp S_{b \to q\bar q s}$ 
of all $b \to q\bar q s$ dominated modes ($0.41 \pm 0.07$)
is more than $5\sigma$ from zero, 
and hence $\CP$ violation in $b \to q\bar q s$ transitions is established.
Furthermore, the averages of $- \etacp S_{b \to q\bar q s}$ 
for the modes $\etapr\KS$ and $K^+K^-\KS$ are more than $3\sigma$ from zero.

\begin{figure}
  \begin{center}
    \resizebox{0.60\textwidth}{!}{\includegraphics{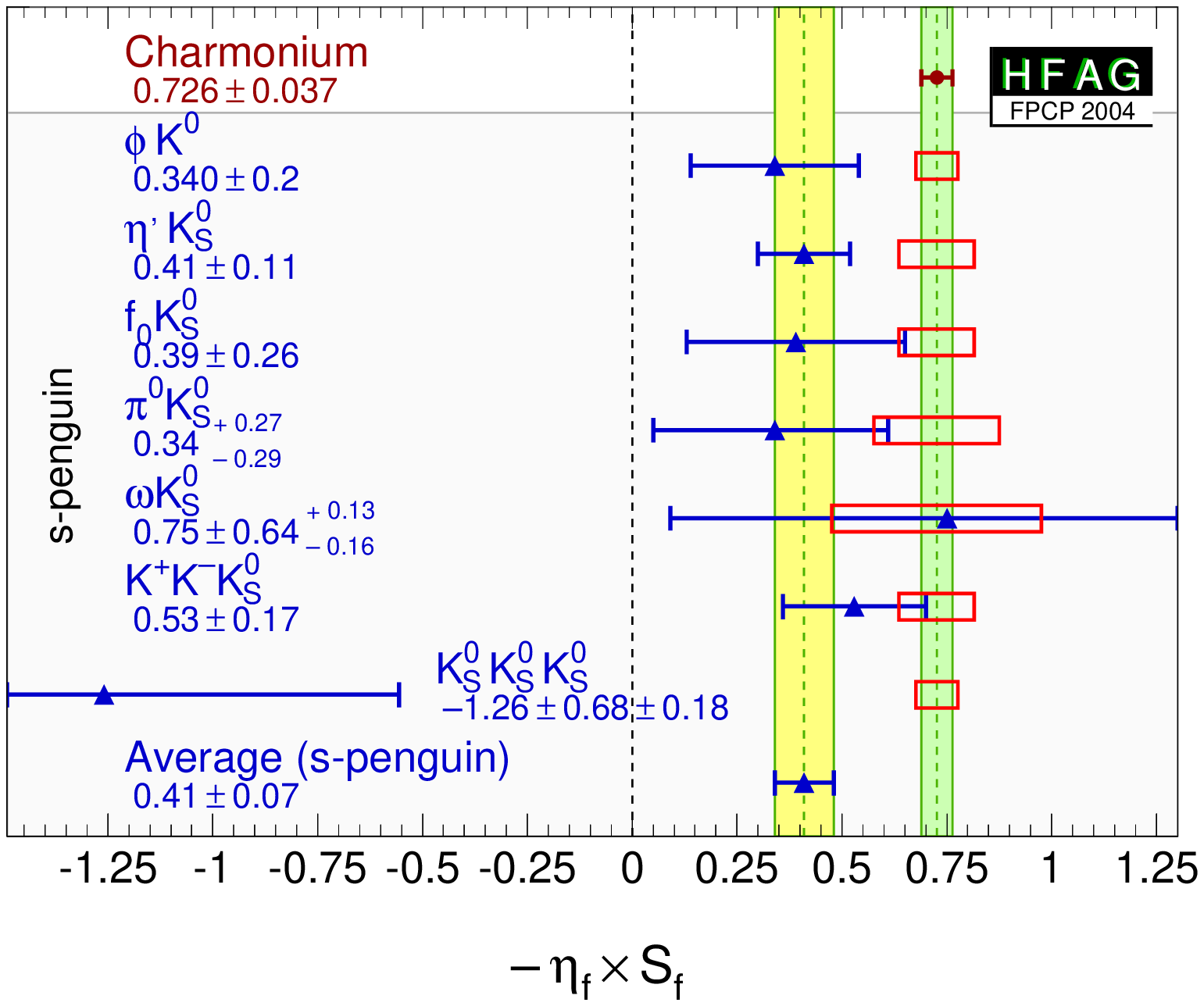}}
  \end{center}
  \vspace{-0.8cm}
  \caption{
    Averages for $S_{b \to q\bar q s}$ compared to $S_{b \to c\bar c s}$,
    with coarse dimensional estimates of associated theoretical uncertainties 
    indicated by the open boxes.
  }
  \label{fig:cp_uta:s_qqs_theo}
\end{figure}

Neglecting theory errors due to suppressed contributions with 
different weak phases,
% which are not well quantified,
the difference between $S_{b \to c\bar c s}$ and $S_{b \to q\bar q s}$
can be calculated.
We find the confidence level (CL) of 
the joint $S_{b \to c\bar c s}$ and $S_{b \to q\bar q s}$ average 
to be $0.00012$, which corresponds to a $3.8\sigma$ discrepancy.
To give an idea of the theoretical uncertainties involved,
Fig.~\ref{fig:cp_uta:s_qqs_theo} shows 
coarse estimates of the theoretical errors
associated with the non-charmonium modes.
These crude estimates are obtained from dimensional arguments only, 
based on the CKM suppression of the $V_{ub}$ penguin, 
and on the naive contribution from tree diagrams. Including these 
estimates according to the procedure defined in Ref.~\cite{ref:cp_uta:uud:charles} 
improves the CL of the joint average to about $3\sigma$.

\mysubsection{Time-dependent $\CP$ asymmetries in $b \to c\bar{c}d$ transitions
}
\label{sec:cp_uta:ccd}

The transition $b \to c\bar c d$ can occur via either a $b \to c$ tree
or a $b \to d$ penguin amplitude.  
Similarly to Eq.~(\ref{eq:cp_uta:b_to_s}), the amplitude for 
the $b \to d$ penguin can be written
\begin{equation}
  \label{eq:cp_uta:b_to_d}
  \begin{array}{ccccc}
    A_{b \to d} & = & 
    \multicolumn{3}{l}{F_u V_{ub}V^*_{ud} + F_c V_{cb}V^*_{cd} + F_t V_{tb}V^*_{td}} \\
    & = & (F_u - F_c) V_{ub}V^*_{ud} & + & (F_t - F_c) V_{tb}V^*_{td} \\
    & = & {\cal O}(\lambda^3) & + & {\cal O}(\lambda^3). \\
  \end{array}
\end{equation}
From this it can be seen that the $b \to d$ penguin amplitude 
does not have a dominant weak phase.

In the above, we have followed Eq.~(\ref{eq:cp_uta:b_to_s}) 
by eliminating the $F_c$ term using unitarity.
However, we could equally well write
\begin{equation}
  \label{eq:cp_uta:b_to_d_alt}
  \begin{array}{ccccc}
    A_{b \to d} 
    & = & (F_u - F_t) V_{ub}V^*_{ud} & + & (F_c - F_t) V_{cb}V^*_{cd}, \\
    & = & (F_c - F_u) V_{cb}V^*_{cd} & + & (F_t - F_u) V_{tb}V^*_{td}. \\
  \end{array}
\end{equation}
Since the $b \to c\bar{c}d$ tree amplitude 
has the weak phase of $V_{cb}V^*_{cd}$,
either of the above expressions allow the penguin to be decomposed into 
parts with weak phases the same and different to the tree amplitude
(the relative weak phase can be chosen to be either $\beta$ or $\gamma$).
However, if the tree amplitude dominates,
there is little sensitivity to any phase 
other than that from $\Bz$-$\Bzb$ mixing.

\begin{figure}
  \begin{center}
    \begin{tabular}{cc}
      \resizebox{0.46\textwidth}{!}{\includegraphics{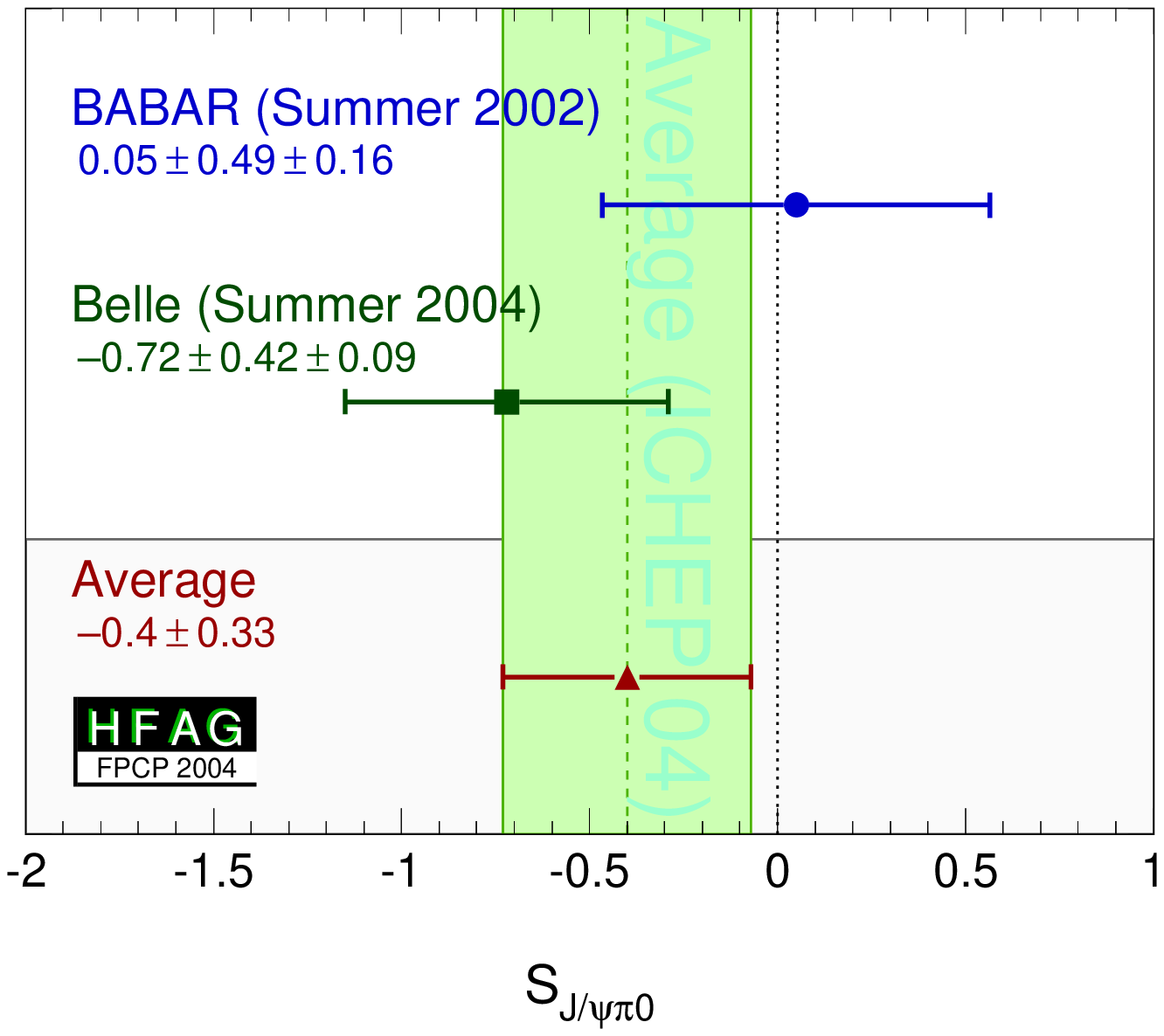}}
      &
      \resizebox{0.46\textwidth}{!}{\includegraphics{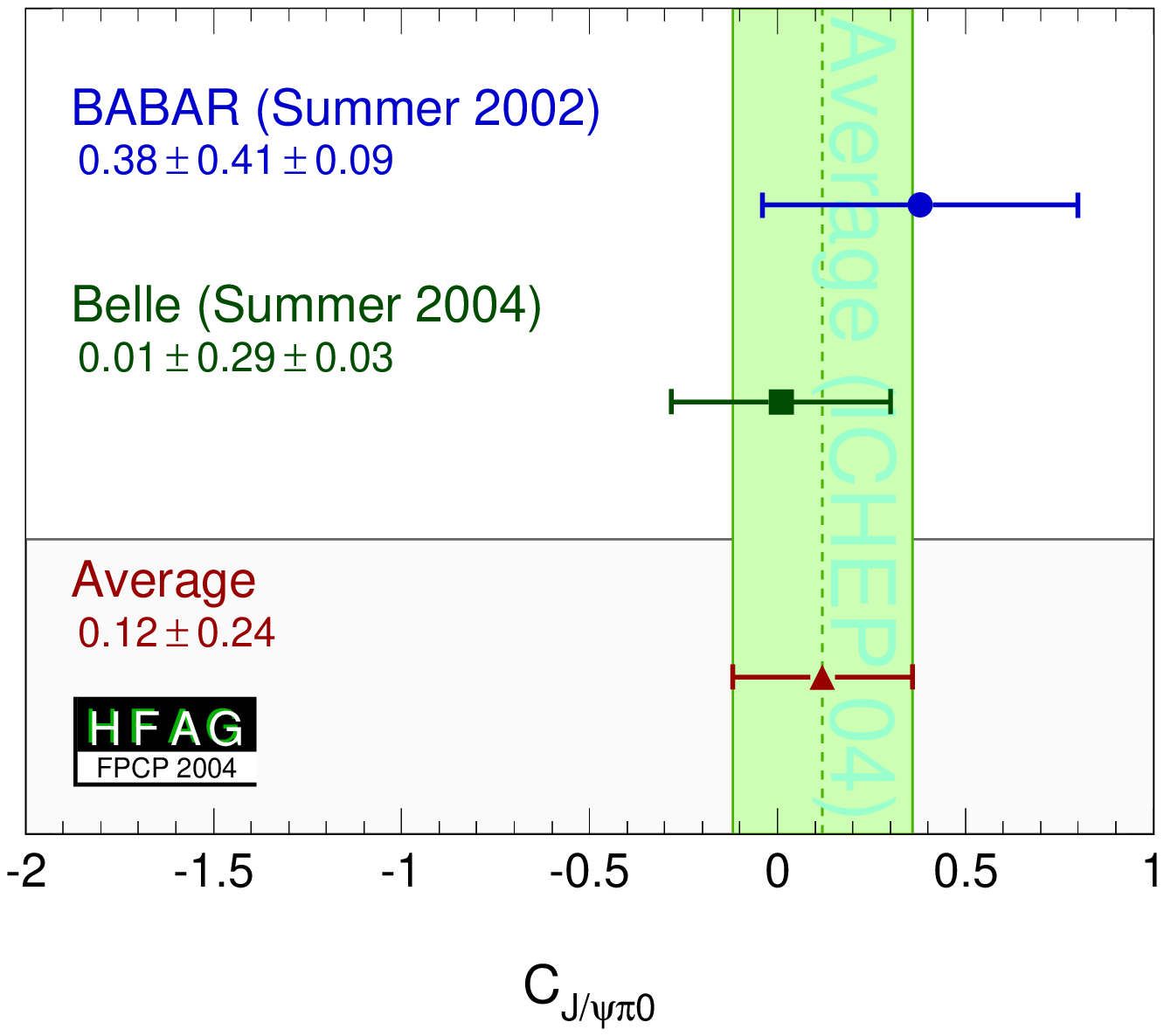}}
    \end{tabular}
  \end{center}
  \vspace{-0.8cm}
  \caption{
    Averages of 
    (left) $S_{b \to c\bar c d}$ and (right) $C_{b \to c\bar c d}$ 
    for the mode $\Bz \to J/ \psi \pi^0$.
  }
  \label{fig:cp_uta:ccd:psipi0}
\end{figure}

The $b \to c\bar{c}d$ transitions can be investigated with studies 
of various different final states. 
Results are available from both \babar\  and \belle\ 
using the final states $\jpsi \pi^0$, $D^{*+}D^{*-}$ and $D^{*\pm}D^{\mp}$;
the averages of these results are given in Table~\ref{tab:cp_uta:ccd}.
The results using the $\CP$ eigenstate ($\etacp = +1$) $\jpsi \pi^0$
are shown in Fig.~\ref{fig:cp_uta:ccd:psipi0}.
The vector-vector mode $D^{*+}D^{*-}$ 
is found to be dominated by the $\CP$ even longitudinally polarized component;
\babar\ measures a $\CP$ odd fraction of 
$0.063 \pm 0.055 \pm 0.009$~\cite{ref:cp_uta:ccd:babar:dstardstar} while
\belle\ measures a $\CP$ odd fraction of 
$0.19  \pm 0.08  \pm 0.01 $~\cite{ref:cp_uta:ccd:belle:dstardstar}
(here we do not average these fractions and rescale the inputs,
% {\sc \bf i think we should, though}
however the average is almost independent of the treatment).
For the non-$\CP$ eigenstate mode $D^{*\pm}D^{\mp}$
\babar\ uses fully reconstructed events while 
\belle\ combines both fully and partially reconstructed samples.

\begin{table}
  \begin{center}
    \caption{
      Averages for $b \to c \bar c d$ modes.
    }
    \vspace{0.2cm}
    \setlength{\tabcolsep}{0.0pc}
    \begin{tabular*}{\textwidth}{@{\extracolsep{\fill}}lrcc} \hline 
      \multicolumn{2}{l}{Experiment} & 
      $S_{b \to c\bar c d}$ & $C_{b \to c\bar c d}$ \\
      \hline
      \multicolumn{4}{c}{$\jpsi \pi^0$} \\
      \babar & \cite{ref:cp_uta:ccd:babar:psipi0} & 
      $\ph{-}0.05 \pm 0.49 \pm 0.16$ & $ 0.38 \pm 0.41 \pm 0.09$ \\
      \belle & \cite{ref:cp_uta:ccd:belle:psipi0} & 
      $-0.72 \pm 0.42 \pm 0.09$ & $ 0.01 \pm 0.29 \pm 0.03$ \\
      \multicolumn{2}{l}{\bf Average} & $-0.40 \pm 0.33$ & $ 0.12 \pm 0.24$ \\
      \multicolumn{2}{l}{\small Confidence level} & 
      \mc{2}{c}{\small combined average: $0.36$} \\
      \hline
      \multicolumn{4}{c}{$D^{*+}D^{*-}$} \\
      \babar & \cite{ref:cp_uta:ccd:babar:dstardstar} &
      $\ph{-}0.06 \pm 0.37 \pm 0.13$ & $ 0.28 \pm 0.23 \pm 0.02$ \\
      \belle & \cite{ref:cp_uta:ccd:belle:dstardstar} &
      $-0.75 \pm 0.56 \pm 0.12$ & $ 0.26 \pm 0.26 \pm 0.04$ \\
      \multicolumn{2}{l}{\bf Average} & $-0.20 \pm 0.32$ & $ 0.28 \pm 0.17$ \\
      \multicolumn{2}{l}{\small Confidence level} & 
      \mc{2}{c}{\small combined average: $0.49$} \\
      \hline
    \end{tabular*}

    \vspace{2ex}

% make this tabular (not tabular*) and resize down to \textwidth
% change @{\extracolsep{\fill}} to @{\extracolsep{2mm}}
    \resizebox{\textwidth}{!}{
      \setlength{\tabcolsep}{0.0pc}
      \begin{tabular}{@{\extracolsep{2mm}}lrccccc} \hline 
        \multicolumn{2}{l}{Experiment} & 
        $S_{+-}$ & $C_{+-}$ & $S_{-+}$ & $C_{-+}$ & $A$ \\
        \hline
        \multicolumn{7}{c}{$D^{*\pm}D^{\mp}$} \\        
        \babar & \cite{ref:cp_uta:ccd:babar:dstard} &
        $-0.82 \pm 0.75 \pm 0.14$ & $-0.47 \pm 0.40 \pm 0.12$ & 
        $-0.24 \pm 0.69 \pm 0.12$ & $-0.22 \pm 0.37 \pm 0.10$ & $-0.03 \pm 0.11 \pm 0.05$ \\
        \belle & \cite{ref:cp_uta:ccd:belle:dstard} &
        $-0.55 \pm 0.39 \pm 0.12$ & $-0.37 \pm 0.22 \pm 0.06$ & 
        $-0.96 \pm 0.43 \pm 0.12$ & $\ph{-}0.23 \pm 0.25 \pm 0.06$ & $\ph{-}0.07 \pm 0.08 \pm 0.04$ \\
        \multicolumn{2}{l}{\bf Average} & 
        $-0.61 \pm 0.36$ & $-0.39 \pm 0.20$ & 
        $-0.75 \pm 0.38$ & $ 0.09 \pm 0.21$ & $ 0.03 \pm 0.07$ \\
        \hline 
      \end{tabular}
    }

    \label{tab:cp_uta:ccd}
  \end{center}
\end{table}

In the absence of the penguin contribution,
the time-dependent parameters would be given by
$S_{b \to c\bar c d} = - \etacp \sin(2\beta)$,
$C_{b \to c\bar c d} = 0$,
$S_{+-} = \sin(2\beta + \delta)$,
$S_{-+} = \sin(2\beta - \delta)$,
$C_{+-} = - C_{-+}$ and 
$A_{+-} = 0$,
where $\delta$ is the strong phase difference between the 
$D^{*+}D^-$ and $D^{*-}D^+$ decay amplitudes.
In the presence of the penguin contribution,
there is no clean interpretation in terms of CKM parameters,
however
direct $\CP$ violation may be observed as any of
$C_{b \to c\bar c d} \neq 0$, $C_{+-} \neq - C_{-+}$ or $A_{+-} \neq 0$.

\begin{figure}
  \begin{center}
    \begin{tabular}{cc}
      \resizebox{0.46\textwidth}{!}{\includegraphics{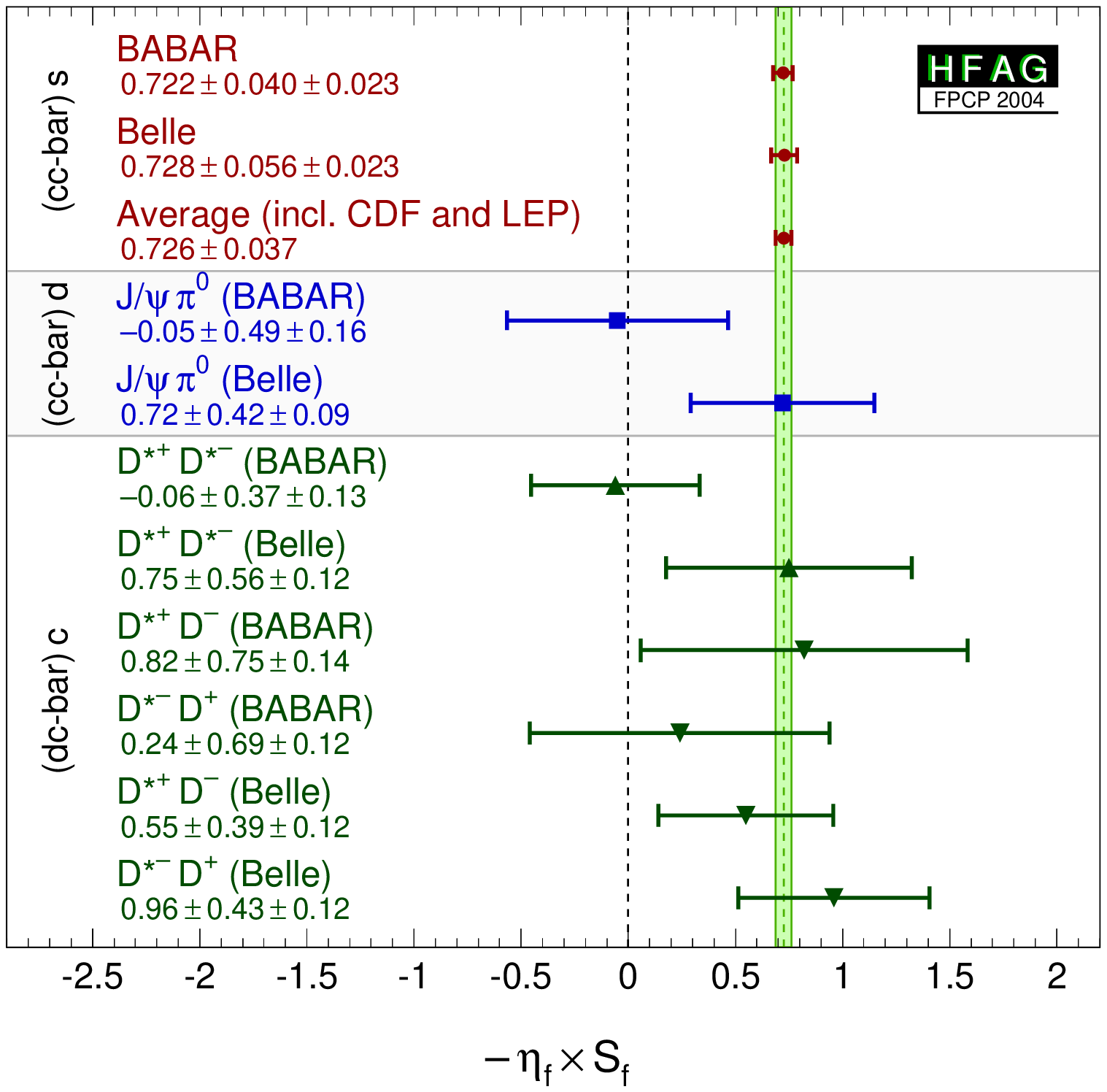}}
      &
      \resizebox{0.46\textwidth}{!}{\includegraphics{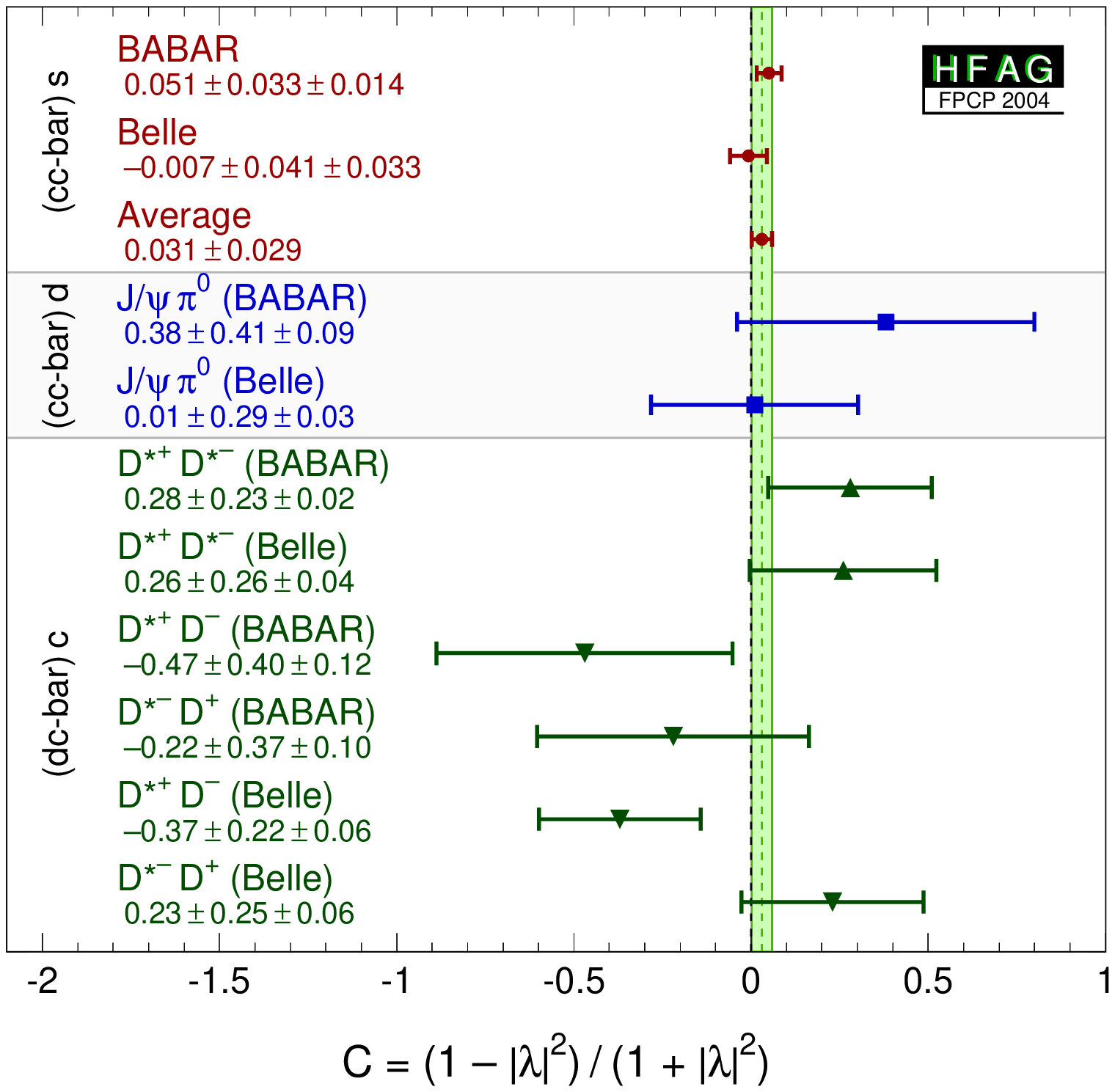}}
    \end{tabular}
  \end{center}
  \vspace{-0.8cm}
  \caption{
    Averages of 
    (left) $S_{b \to c\bar c d}$ and (right) $C_{b \to c\bar c d}$,
    compared to $S_{b \to c\bar c s}$ and $C_{b \to c\bar c s}$, respectively.
  }
  \label{fig:cp_uta:ccd}
\end{figure}

The averages for the $b \to c\bar c d$ modes 
are shown in Fig.~\ref{fig:cp_uta:ccd}.
Comparisons of the results for the $b \to c\bar c d$ modes 
to the $b \to c\bar c s$ and $b \to q\bar q s$ modes,
can be seen in Fig.~\ref{fig:cp_uta:qqs_ccd}.

%% straightforward interpretation

\mysubsection{Time-dependent asymmetries in $b \to s\gamma$ transitions
}
\label{sec:cp_uta:bsg}

The radiative decays $b \to s\gamma$ produce photons 
which are highly polarized in the Standard Model.
%-leftover ?% Since the polarization is, in principle, observable,
The decays $\Bz \to F \gamma$ and $\Bzb \to F \gamma$ 
produce photons with opposite helicities, 
and since the polarization is, in principle, observable,
these final states cannot interfere.
The finite mass of the $s$ quark introduces small corrections,
to the limit of maximum polarization,
but any large mixing induced $\CP$ violation would be a signal for new physics.
Since a single weak phase dominates the $b \to s \gamma$ transition in the 
Standard Model, the cosine term is also expected to be small.

The \babar\ collaboration has studied time-dependent asymmetries in
$b \to s\gamma$ transitions using the decay 
$\Bz \to \Kstar \gamma$, with $\Kstar \to \KS \pi^0$.
The \belle\ collaboration also uses the decay $\Bz \to \KS \pi^0 \gamma$,
but do not restrict themselves to the 
$\Kstar$ mass region~\cite{ref:cp_uta:bsg:aghs}
(the invariant mass range 
$0.60 \ {\rm GeV}/c^2 < M_{\KS\pi^0} < 1.80 \ {\rm GeV}/c^2$ is used).
These results, and their averages, are given in Table~\ref{tab:cp_uta:bsg}.

\begin{table}
  \begin{center}
    \caption{
      Averages for $b \to s \gamma$ modes.
    }
    \vspace{0.2cm}
    \setlength{\tabcolsep}{0.0pc}
    \begin{tabular*}{\textwidth}{@{\extracolsep{\fill}}lrcc} \hline 
      \multicolumn{2}{l}{Experiment} & 
      $S_{b \to s \gamma}$ & $C_{b \to s \gamma}$ \\
      \hline
      \multicolumn{4}{c}{$\KS \pi^0 \gamma$} \\
      \babar & \cite{ref:cp_uta:bsg:babar:kstargamma} & 
      $\ph{-}0.25 \pm 0.63 \pm 0.14$ & $-0.57 \pm 0.32 \pm 0.09$ \\
      \belle & \cite{ref:cp_uta:bsg:belle:kspi0gamma} & 
      $-0.58 \, ^{+0.46}_{-0.38} \pm 0.11$ & $\ph{-}0.03 \pm 0.34 \pm 0.11$ \\
      \multicolumn{2}{l}{\bf Average} & $-0.29 \pm 0.38$ & $-0.32 \pm 0.24$ \\
      \multicolumn{2}{l}{\small Confidence level} & 
      \mc{2}{c}{\small combined average: $0.31~(1.0\sigma)$} \\
      \hline 
    \end{tabular*}
      
    \label{tab:cp_uta:bsg}
  \end{center}
\end{table}

\mysubsection{Time-dependent $\CP$ asymmetries in $b \to u\bar{u}d$ transitions
}
\label{sec:cp_uta:uud}

The $b \to u \bar u d$ transition can be mediated by either 
a $b \to u$ tree amplitude or a $b \to d$ penguin amplitude.
These transitions can be investigated using 
the time dependence of $\Bz$ decays to final states containing light mesons.
Results are available from both \babar\ and \belle\ for the 
$\CP$ eigenstate ($\etacp = +1$) $\pi^+\pi^-$ final state.
\babar\  has also performed an analysis on 
the vector-vector final state $\rho^+\rho^-$, 
which they find to be dominated by the $\CP$ even
longitudinally polarized component
(they measure $f_{\rm long} = 
0.99 \pm 0.03 \, ^{+0.04}_{-0.03}$~\cite{ref:cp_uta:uud:babar:rhorho}).

\begin{figure}
  \begin{center}
    \begin{tabular}{cc}
      \resizebox{0.46\textwidth}{!}{\includegraphics{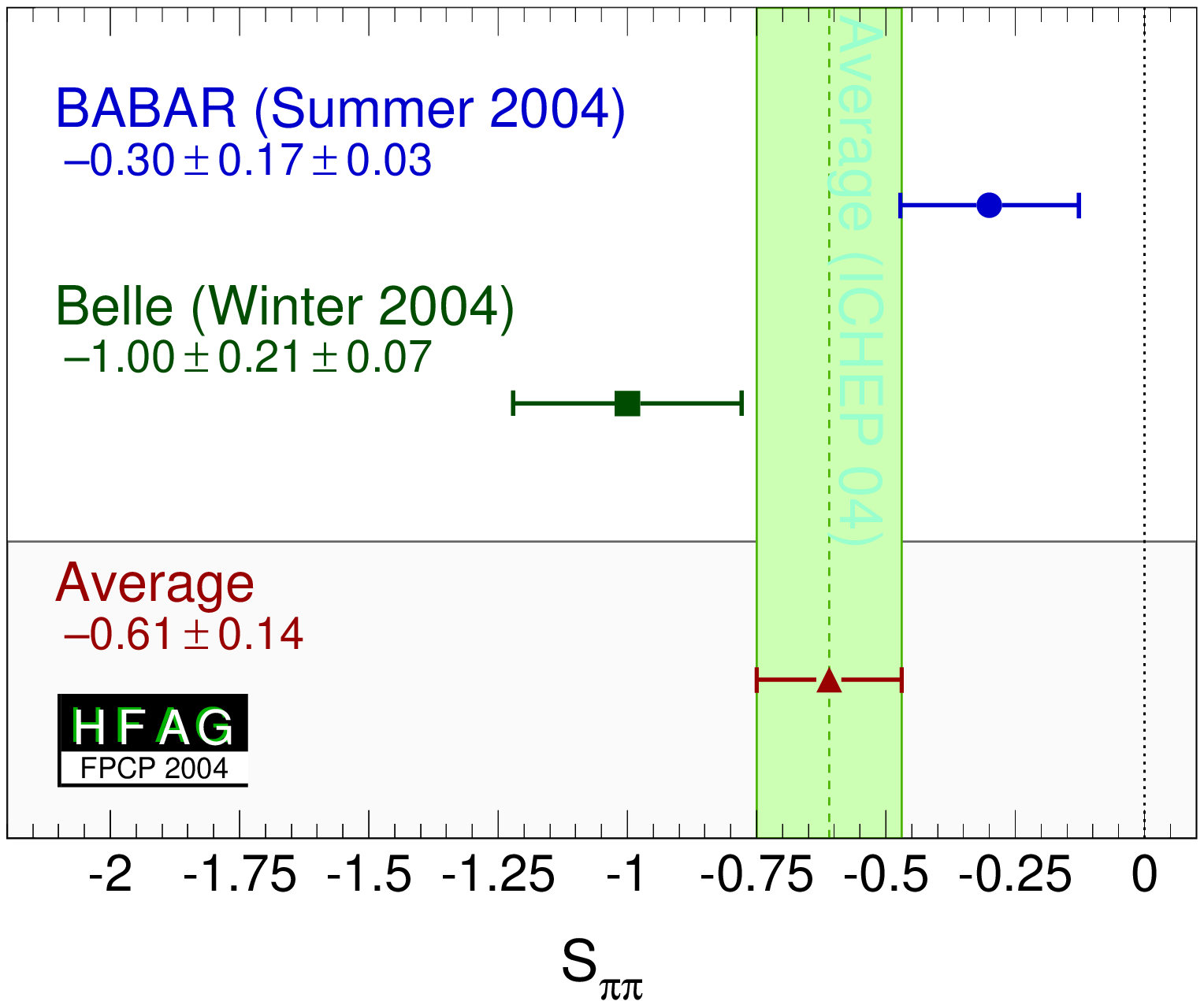}}
      &
      \resizebox{0.46\textwidth}{!}{\includegraphics{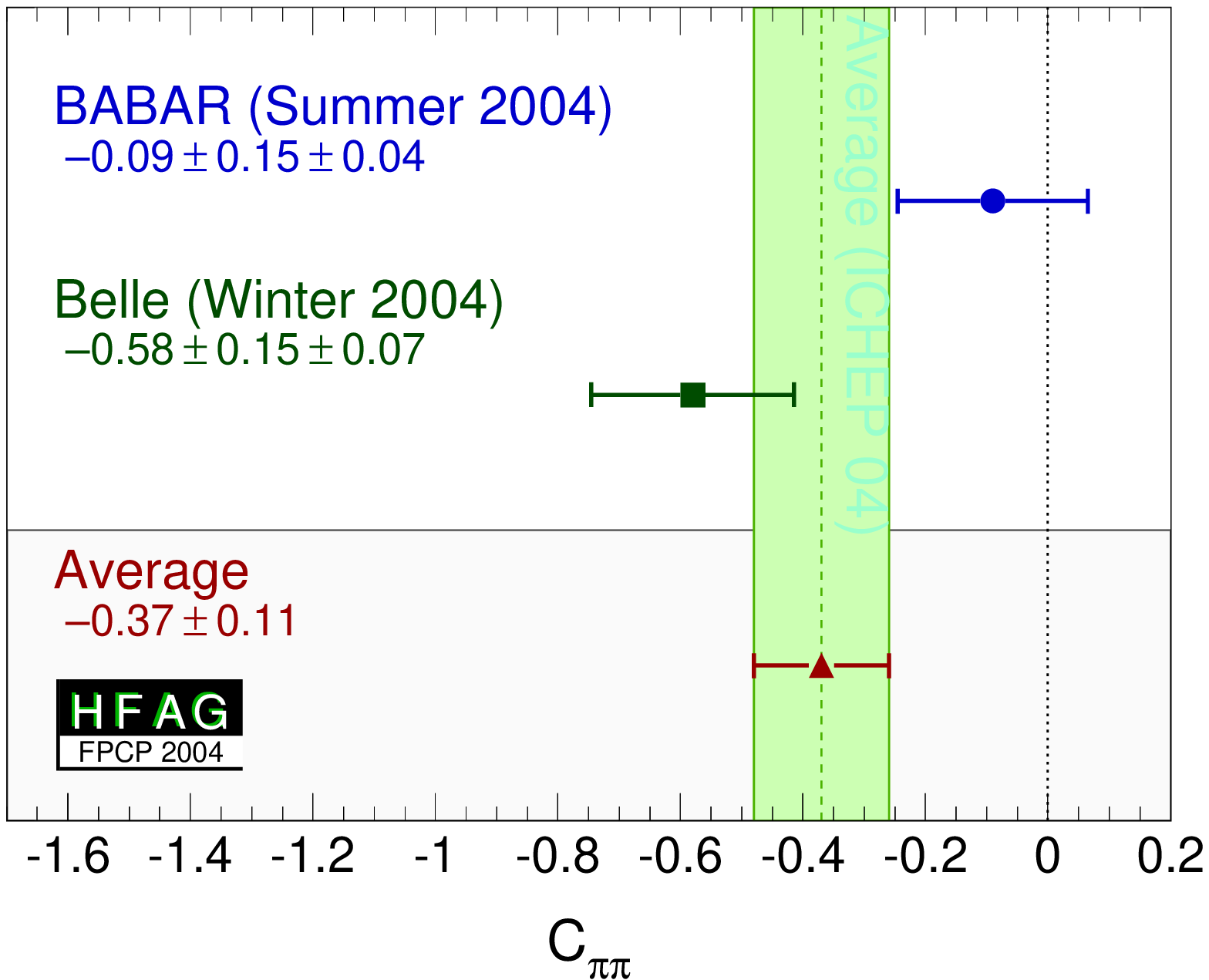}}
    \end{tabular}
  \end{center}
  \vspace{-0.8cm}
  \caption{
    Averages of 
    (left) $S_{b \to u\bar u d}$ and (right) $C_{b \to u\bar u d}$ 
    for the mode $\Bz \to \pi^+\pi^-$.
  }
  \label{fig:cp_uta:uud:pipi}
\end{figure}

For the non-$\CP$ eigenstate $\rho^{\pm}\pi^{\mp}$, 
\belle\ has performed a quasi-two-body analysis,
while  \babar\ performs a time-dependent Dalitz plot (DP) analysis
of the $\pi^+\pi^-\pi^0$ final state~\cite{ref:cp_uta:uud:snyderquinn};
such an analysis allows direct measurements of the phases.
These results, and averages, are listed in Table~\ref{tab:cp_uta:uud}.
The averages for $\pi^+\pi^-$ are shown in Fig.~\ref{fig:cp_uta:uud:pipi}.

\begin{table}
  \begin{center}
    \caption{
      Averages for $b \to u \bar u d$ modes.
    }
    \vspace{0.2cm}
    \setlength{\tabcolsep}{0.0pc}
    \begin{tabular*}{\textwidth}{@{\extracolsep{\fill}}lrcc} \hline 
      \multicolumn{2}{l}{Experiment} & 
      $S_{b \to u\bar u d}$ & $C_{b \to u\bar u d}$ \\
      \hline
      & \multicolumn{3}{c}{$\pi^+\pi^-$} \\
      \babar & ~\cite{ref:cp_uta:uud:babar:pipi} & 
      $-0.30 \pm 0.17 \pm 0.03$ & $-0.09 \pm 0.15 \pm 0.04$ \\
      \belle & ~\cite{ref:cp_uta:uud:belle:pipi} & 
      $-1.00 \pm 0.21 \pm 0.07$ & $-0.58 \pm 0.15 \pm 0.07$ \\
      \multicolumn{2}{l}{\bf Average} & $-0.61 \pm 0.14$ & $-0.37 \pm 0.11$ \\
      \multicolumn{2}{l}{\small Confidence level} & 
      \mc{2}{c}{\small combined average: $1.4\times10^{-3}~(3.2\sigma)$}  \\
      \hline
      & \multicolumn{3}{c}{$\rho^+\rho^-$} \\
      \babar & ~\cite{ref:cp_uta:uud:babar:rhorho} &
      $-0.19 \pm 0.33 \pm 0.11$ & $-0.23 \pm 0.24 \pm 0.14$ \\
%              & ~\cite{ref:cp_uta:uud:babar:rho0rho0} & 
%       \mc{2}{l}{$\alpha = (96 \pm 10 \pm 4 \pm 11)^\circ$,
%                 {\small where the last error is due to }} \\
%              & &
%       \mc{2}{l}{\small the penguin contribution (isospin analysis)} \\
      \hline 
    \end{tabular*}

    \vspace{2ex}

% make this tabular (not tabular*) and resize down to \textwidth
% change @{\extracolsep{\fill}} to @{\extracolsep{2mm}}
    \resizebox{\textwidth}{!}{
      \setlength{\tabcolsep}{0.0pc}
      \begin{tabular}{@{\extracolsep{2mm}}lrccccc} \hline 
        & \multicolumn{6}{c}{$\rho^{\pm}\pi^{\mp}$ Q2B/DP analysis} \\
        \multicolumn{2}{l}{Experiment} & 
        $S_{\rho\pi}$ & $C_{\rho\pi}$ & $\Delta S_{\rho\pi}$ & $\Delta C_{\rho\pi}$ & ${\cal A}_{CP}^{\rho\pi}$ \\
        \hline
        \babar & ~\cite{ref:cp_uta:uud:babar:rhopi} &
        $-0.10 \pm 0.14 \pm 0.04$ & $ 0.34 \pm 0.11 \pm 0.05$ &
        $\ph{-}0.22 \pm 0.15 \pm 0.03$ & $ 0.15 \pm 0.11 \pm 0.03$ & $-0.088 \pm 0.049 \pm 0.013$ \\
        \belle & ~\cite{ref:cp_uta:uud:belle:rhopi} & 
        $-0.28 \pm 0.23 \, ^{+0.10}_{-0.08}$ & $ 0.25 \pm 0.17 \, ^{+0.02}_{-0.06}$ &
        $-0.30 \pm 0.24 \pm 0.09$ & $ 0.38 \pm 0.18 \, ^{+0.02}_{-0.04}$ & $-0.16 \pm 0.10 \pm 0.02$ \\
        \multicolumn{2}{l}{\bf Average} & 
        $-0.13 \pm 0.13$ & $ 0.31 \pm 0.10$ &
        $ 0.09 \pm 0.13$ & $ 0.22 \pm 0.10$ & $-0.102 \pm 0.045$ \\
        \hline
        & & & \multicolumn{2}{c}{${\cal A}^{+-}_{\rho\pi}$} & \multicolumn{2}{c}{${\cal A}^{-+}_{\rho\pi}$} \\
        \hline
        \babar & \cite{ref:cp_uta:uud:babar:rhopi} &
        & \multicolumn{2}{c}{$\ph{-}0.25 \pm 0.17 \, ^{+0.02}_{-0.06}$} & \multicolumn{2}{c}{$-0.47 ^{\,+0.14}_{\,-0.15} \pm 0.06$} \\         
        \belle & \cite{ref:cp_uta:uud:belle:rhopi} & 
        & \multicolumn{2}{c}{$-0.02 \pm 0.16^{\,+0.05}_{\,-0.02}$} & \multicolumn{2}{c}{$-0.53 \pm 0.29^{\,+0.09}_{\,-0.04}$} \\
        \multicolumn{2}{l}{\bf Average} & 
        & \multicolumn{2}{c}{$-0.15 \pm 0.09$} & \multicolumn{2}{c}{$-0.47^{\,+0.13}_{\,-0.14}$} \\
        \hline 
      \end{tabular}
    }

    \vspace{2ex}

    \setlength{\tabcolsep}{0.0pc}
    \begin{tabular*}{\textwidth}{@{\extracolsep{\fill}}lrcc} \hline 
      \multicolumn{4}{c}{$\rho^{\pm}\pi^{\mp}$ DP analysis} \\
      \multicolumn{2}{l}{Experiment} & $\alpha \ (^\circ)$ & $\delta_{+-} \ (^\circ)$ \\
      \hline
      \babar & \cite{ref:cp_uta:uud:babar:rhopi} &
      $113 \, ^{+27}_{-17} \pm 6$ & $-67 \, ^{+28}_{-31} \pm 7$ \\ 
      \hline 
    \end{tabular*}
      
    \label{tab:cp_uta:uud}
  \end{center}
\end{table}

\begin{figure}
  \begin{center}
    \begin{tabular}{cc}
      \resizebox{0.46\textwidth}{!}{\includegraphics{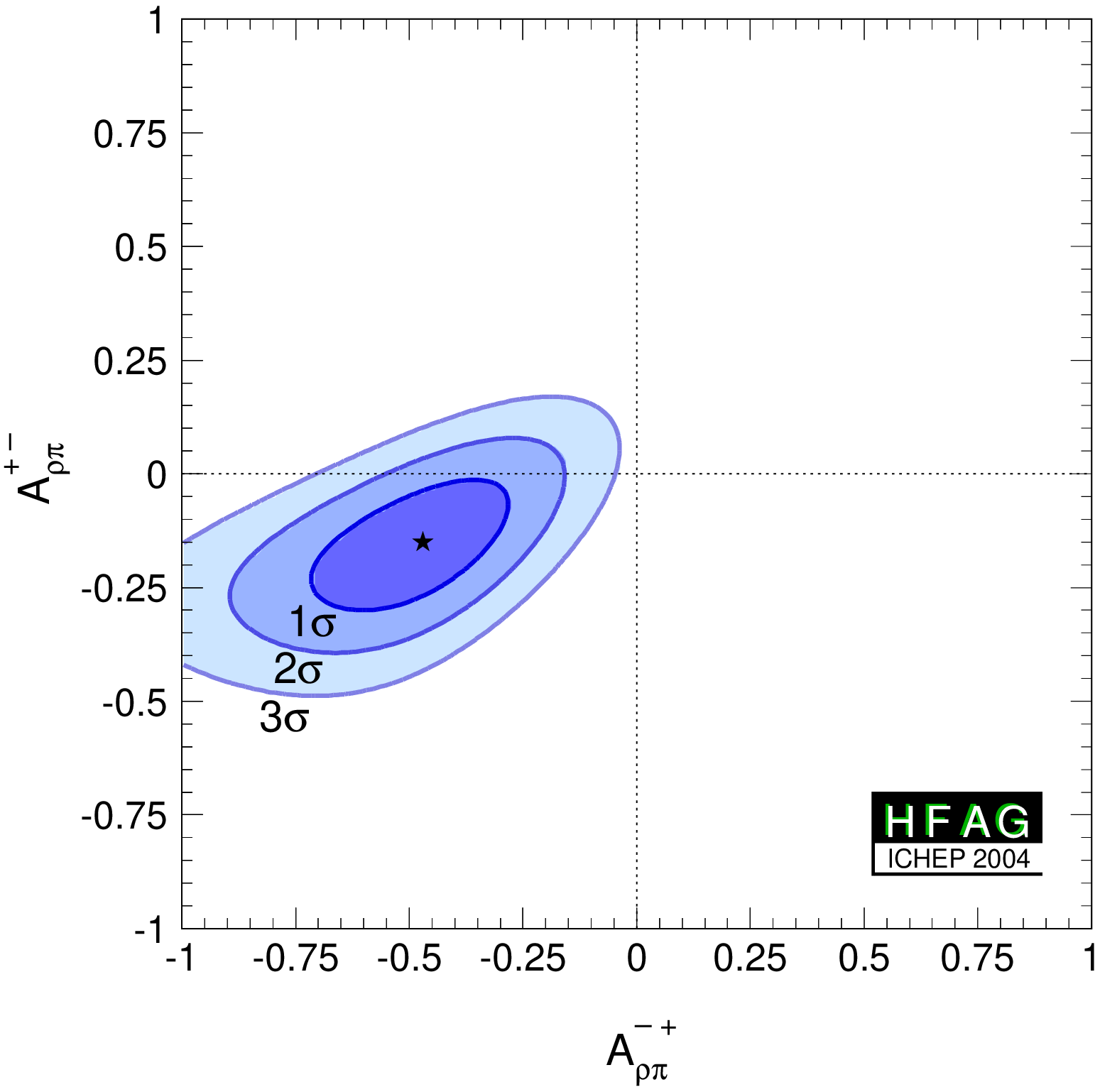}}
    \end{tabular}
  \end{center}
  \vspace{-0.8cm}
  \caption{
    	Direct $\CP$ violation in $\Bz\to\rho^\pm\pi^\mp$. The no-\CP violation
	hypothesis is excluded at the $3.4\sigma$ level.
  }
  \label{fig:cp_uta:uud:rhopi-dircp}
\end{figure}

If the penguin contribution is negligible, 
the time-dependent parameters for $\Bz \to \pi^+\pi^-$ and $\Bz \to \rho^+\rho^-$ 
are given by
$S_{b \to u\bar u d} = \etacp \sin(2\alpha)$ and
$C_{b \to u\bar u d} = 0$.
With the notation described in Sec.~\ref{sec:cp_uta:notations}
(Eq.~(\ref{eq:cp_uta:non-cp-s_and_deltas})), 
the time-dependent parameters for $\Bz \to \rho^\pm\pi^\mp$ are,
neglecting penguin contributions, given by
$S_{\rho\pi}   = \sqrt{1 - (\frac{\Delta C}{2})^2}\sin(2\alpha)\cos(\delta)$,
$\Delta S_{\rho\pi} = \sqrt{1 - (\frac{\Delta C}{2})^2}\cos(2\alpha)\sin(\delta)$ and
$C_{\rho\pi} = {\cal A}_{\CP}^{\rho\pi} = 0$,
where $\delta=\arg(A_{-+}A^*_{+-})$ is the strong phase difference 
between the $\rho^-\pi^+$ and $\rho^+\pi^-$ decay amplitudes.
In the presence of the penguin contribution, there is no straightforward 
interpretation of $\Bz \to \rho^\pm\pi^\mp$ in terms of CKM parameters.
However direct $\CP$ violation may arise,
resulting in either or both of $C_{\rho\pi} \neq 0$ and ${\cal A}_{\CP}^{\rho\pi} \neq 0$.
Equivalently,
direct $\CP$ violation may be seen by either of
the decay-type-specific observables ${\cal A}^{+-}_{\rho\pi}$ 
and ${\cal A}^{-+}_{\rho\pi}$, defined in Eq.~(\ref{eq:cp_uta:non-cp-directcp}), 
deviating from zero.
Results and averages for these parameters
are also given in Table~\ref{tab:cp_uta:uud}.
They exhibit a linear correlation coefficient of $+0.59$.
The significance of observing direct $\CP$ violation 
computed from the difference of the $\chi^2$ obtained in the nominal average, 
compared to setting 
%YS- $C_{\rho\pi}=A_{\rho\pi}=0$,  
%- but it may be  ${\cal A}^{+-}_{\rho\pi}={\cal A}^{-+}_{\rho\pi}=0$ ?
$C_{\rho\pi} = {\cal A}^{\rho\pi}_{\CP} = 0$
is found to be $3.4\sigma$ in this mode. 
The confidence level
contours of ${\cal A}^{+-}_{\rho\pi}$ versus ${\cal A}^{-+}_{\rho\pi}$ 
are shown in Fig.~\ref{fig:cp_uta:uud:rhopi-dircp}.


%% straightforward interpretation
Some difference is seen between the 
\babar\ and \belle\ measurements in the $\pi^+\pi^-$ system.
The confidence level of the average is $0.0014$,
which corresponds to a $3.2\sigma$ discrepancy.  Since there is no
evidence of systematic problems in either analysis,
we do not rescale the errors of the averages.

The precision of the measured $\CP$ violation parameters in
$b \to u\bar{u}d$ transitions allows constraints to be set on the UT angle $\alpha$. 
In addition to the value of $\alpha$ from the \babar\ time-dependent DP analysis,
given in Table~\ref{tab:cp_uta:uud},
constraints have been obtained with various methods:
\begin{itemize}
\item In the \belle\ analysis of $\Bz \to \pi^+\pi^-$~\cite{ref:cp_uta:uud:belle:pipi},  
  constraints on the time-dependent \CP violation parameters are used to 
  obtain $90^\circ < \alpha < 146^\circ$ ($95.5\%$ CL) following the method proposed by 
  Gronau and Rosner~\cite{ref:cp_uta:uud:gronaurosner}.
  This result includes an assumption on the relative size of
  tree and penguin amplitudes.
% \item 
%   \belle~\cite{ref:cp_uta:uud:belle:rhopi} cite the result of 
%   Gronau and Zupan\cite{ref:cp_uta:uud:gronauzupan},
%   who obtain $\alpha = (102 \pm 11 \pm 15)^\circ$ using the \belle\ results 
%   and ``broken-flavour SU(3).''  
%   The first uncertainty is experimental 
%   and the second is due to SU(3) breaking effects.
\item Using the measured time-dependent \CP violation parameters 
  in longitudinally polarized $\Bz \to \rho^+\rho^-$ 
  decays~\cite{ref:cp_uta:uud:babar:rhorho},
  in combination with the upper limit for the 
  $\Bz \to \rho^0\rho^0$ branching fraction~\cite{ref:cp_uta:uud:babar:rho0rho0},
  and the measurement of the branching fraction and longitudinal polarization
  of $\Bp \to \rho^+\rho^0$~\cite{ref:cp_uta:uud:babar:rho0rho+,ref:cp_uta:uud:belle:rho0rho+},
  \babar\ performs an isospin analysis~\cite{ref:cp_uta:uud:gronaulondon}
  and obtains $\alpha = (96 \pm 10 \pm 4 \pm 11)^\circ$,
  where the third error is due to the unknown penguin contribution.
\item The CKMfitter group~\cite{ref:cp_uta:ckmfitter} uses the 
  measurements from \belle\ and \babar\ given in Table~\ref{tab:cp_uta:uud},
  with other branching fractions and \CP asymmetries in 
  $\B\to\pi\pi,~\rho\pi$ and $\rho\rho$ modes, 
  to perform isospin analyses for each system.  They then combine the results
  to obtain $\alpha = (100^{\,+\ph{1}9}_{\,-10})^\circ$.
\end{itemize}
Note that each method suffers from ambiguities in the solutions.
All the above measurements correspond to the choice
that is in agreement with the global CKM fit.

At present we make no attempt to provide an HFAG average for $\alpha$.
More details on procedures to calculate a best fit value for $\alpha$ 
can be found in Refs.~\cite{ref:cp_uta:ckmfitter,ref:cp_uta:utfit}.

% Due to the presence of sizeable penguin 
% contributions, an isospin analysis using the \CP asymmetries 
% and branching fractions of all isospin-related neutral and charged
% final states~\cite{ref:cp_uta:uud:gronaulondon},
% or a Dalitz plot analysis~\cite{ref:cp_uta:uud:snyderquinn} must be 
% performed. This has been done by the \babar\   collaboration for the modes 
% $\B\to\rho\rho$~\cite{ref:cp_uta:uud:babar:rhorho,ref:cp_uta:uud:babar:rho0rho0} and 
% $\Bz\to\pip\pim\piz$ (dominated by $(\rho\pi)^0$)~\cite{ref:cp_uta:uud:babar:rhopi}
% (see results in Table~\ref{tab:cp_uta:uud}). The CKMfitter group combines the 
% likelihoods of the Belle and \babar\ measurements shown in 
% Table~\ref{tab:cp_uta:uud}, including the isospin analysis for the 
% world average \CP asymmetry and branching fraction measurements 
% of $\B\to\pi\pi$, to obtain 
% $\alpha = (100^{\,+\ph{1}9}_{\,-10})^\circ$~\cite{ref:cp_uta:ckmfitter}.
% HFAG has not yet done this work independently.
% Note that the solution for $\alpha$ comes along with multiple
% ambiguities. All results given by the experiments (and also here) 
% correspond to the solution that is in agreement with the global
% CKM fit. More details on the $\alpha$ determination can be found
% in Refs.~\cite{ref:cp_uta:ckmfitter,ref:cp_uta:utfit}.

\mysubsection{Time-dependent $\CP$ asymmetries in $b \to c\bar{u}d / u\bar{c}d$ transitions
}
\label{sec:cp_uta:cud}

Non-$\CP$ eigenstates such as $D^\pm\pi^\mp$, $D^{*\pm}\pi^\mp$ and $D^\pm\rho^\mp$ can be produced 
in decays of $\Bz$ mesons either via Cabibbo favoured ($b \to c$) or
doubly Cabibbo suppressed ($b \to u$) tree amplitudes. 
Since no penguin contribution is possible,
these modes are theoretically clean.
The ratio of the magnitudes of the suppressed and favoured amplitudes, $R$,
is sufficiently small (predicted to be about $0.02$),
that terms of ${\cal O}(R^2)$ can be neglected, 
and the sine terms give sensitivity to the combination of UT angles $2\beta+\gamma$.

\begin{figure}
  \begin{center}
    \resizebox{0.60\textwidth}{!}{\includegraphics{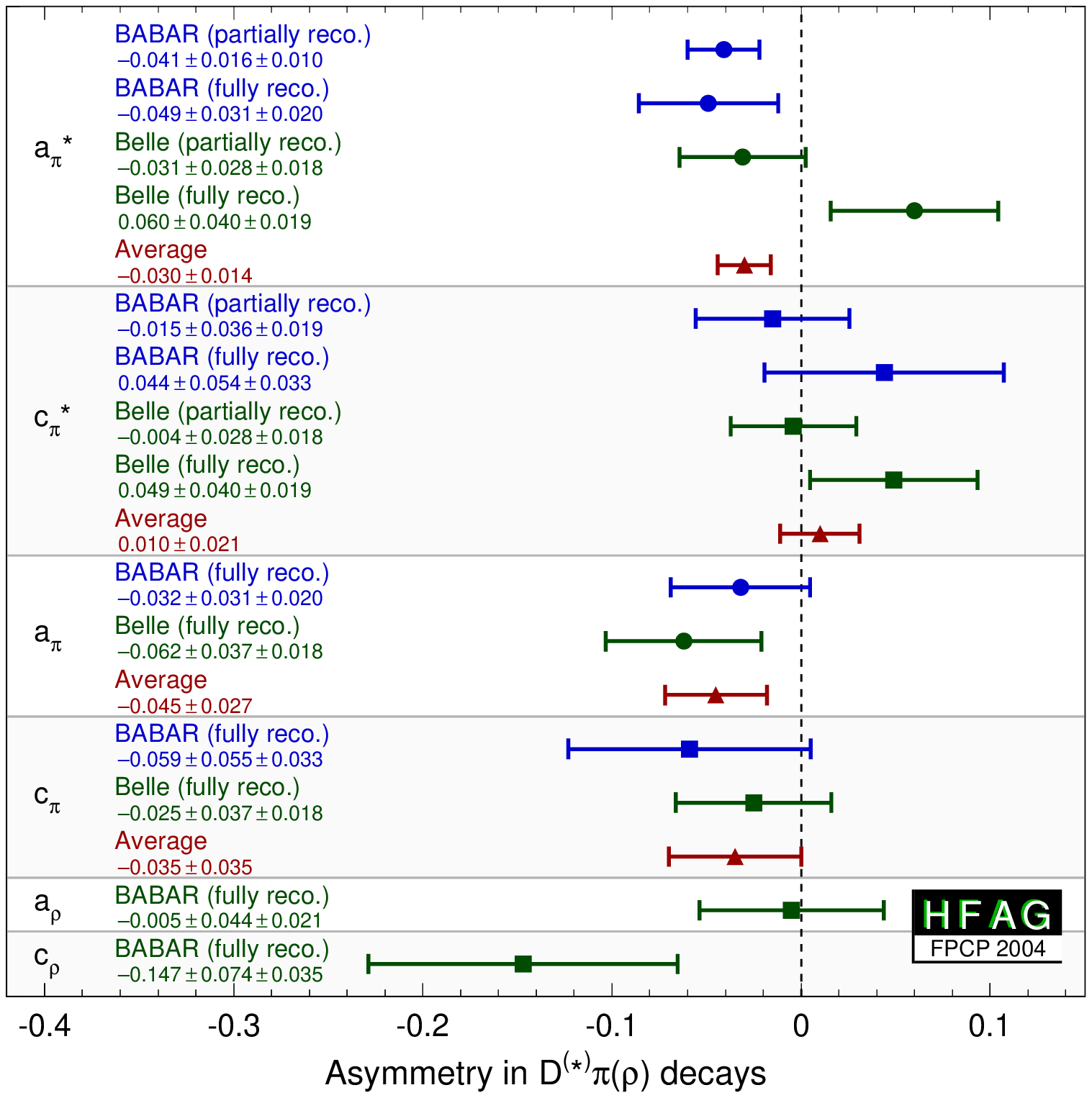}}
  \end{center}
  \vspace{-0.8cm}
  \caption{
    Averages for $b \to c\bar{u}d / u\bar{c}d$ modes.
  }
  \label{fig:cp_uta:cud}
\end{figure}

As described in Sec.~\ref{sec:cp_uta:notations:non_cp:dstarpi},
the averages are given in terms of parameters $a$ and $c$.
$\CP$ violation would appear as $a \neq 0$.
Results are available from both \babar\ and \belle\ in the modes
$D^\pm\pi^\mp$ and $D^{*\pm}\pi^\mp$; for the latter mode both experiments 
have used both full and partial reconstruction techniques.
Results are also available from \babar\ using $D^\pm\rho^\mp$.
These results, and their averages, are listed in Table~\ref{tab:cp_uta:cud},
and are shown in Fig.~\ref{fig:cp_uta:cud}.

\begin{table}
  \begin{center}
    \caption{
      Averages for $b \to c\bar{u}d / u\bar{c}d$ modes.
    }
    \vspace{0.2cm}
    \setlength{\tabcolsep}{0.0pc}
    \begin{tabular*}{\textwidth}{@{\extracolsep{\fill}}lrcc} \hline 
      \multicolumn{2}{l}{Experiment} & $a$ & $c$ \\
      \hline
      \multicolumn{4}{c}{$D^{*\pm}\pi^{\mp}$} \\
      \babar (full rec.) & \cite{ref:cp_uta:cud:babar:full} &
      $-0.049 \pm 0.031 \pm 0.020$ & $\ph{-}0.044 \pm 0.054 \pm 0.033$ \\
      \belle (full rec.) & \cite{ref:cp_uta:cud:belle:full} &
      $\ph{-}0.060 \pm 0.040 \pm 0.019$ & $\ph{-}0.049 \pm 0.040 \pm 0.019$ \\
      \babar (partial rec.) & \cite{ref:cp_uta:cud:babar:partial} &
      $-0.041 \pm 0.016 \pm 0.010$ & $-0.015 \pm 0.036 \pm 0.019 $ \\
      \belle (partial rec.) & \cite{ref:cp_uta:cud:belle:partial} &
      $-0.031 \pm 0.028 \pm 0.018$ & $-0.004 \pm 0.028 \pm 0.018$ \\
      \multicolumn{2}{l}{\bf Average} & $-0.030 \pm 0.014$ & $ 0.010 \pm 0.021$ \\
      \multicolumn{2}{l}{\small Confidence level} & 
      \small $0.19$ & \small $0.66$ \\
      \hline
      \multicolumn{4}{c}{$D^{\pm}\pi^{\mp}$} \\
      \babar (full rec.) & \cite{ref:cp_uta:cud:babar:full} &
      $-0.032 \pm 0.031 \pm 0.020$ & $-0.059 \pm 0.055 \pm 0.033$ \\
      \belle (full rec.) & \cite{ref:cp_uta:cud:belle:full} &
      $-0.062 \pm 0.037 \pm 0.018$ & $-0.025 \pm 0.037 \pm 0.018$ \\
      \multicolumn{2}{l}{\bf Average} & $-0.045 \pm 0.027$ & $-0.035 \pm 0.035$ \\
      \multicolumn{2}{l}{\small Confidence level} & 
      \small $0.59$ & \small $0.66$ \\
      \hline
      \multicolumn{4}{c}{$D^{\pm}\rho^{\mp}$} \\
      \babar (full rec.) & \cite{ref:cp_uta:cud:babar:full} &
      $-0.005 \pm 0.044 \pm 0.021$ & $-0.147 \pm 0.074 \pm 0.035$ \\
      \hline 
    \end{tabular*}
    \label{tab:cp_uta:cud}
  \end{center}
\end{table}

\mysubsection{Rates and asymmetries in $\Bmp \to \DorDstar K^{(*)\mp}$ decays
}
\label{sec:cp_uta:cus}

As explained in Sec.~\ref{sec:cp_uta:notations:cus},
rates and asymmetries in $\Bmp \to \DorDstar K^{(*)\mp}$ decays
are sensitive to $\gamma$.
Various methods using different $\DorDstar$ final states exist.

Results are available from both \babar\ and \belle\ on GLW analyses
in the decay modes $\Bmp \to D\Kmp$, $\Bmp \to \Dstar\Kmp$ and $\Bmp \to D\Kstarmp$.
Both experiments use the 
$\CP$ even $D$ decay final states $K^+K^-$ and $\pi^+\pi^-$ in all three modes; 
both experiments also use only the $\Dstar \to D\pi^0$ decay, 
which gives $\CP(\Dstar) = \CP(D)$. 
For $\CP$ odd $D$ decay final states, 
\belle\ uses $\KS\pi^0$, $\KS\eta$ and $\KS\phi$ in all three analyses, 
and also use $\KS\omega$ in $D\Kmp$ and $\Dstar\Kmp$ analyses. 
\babar\ uses $\KS\pi^0$ only for $D\Kmp$ analysis; 
for $D\Kstarmp$ analysis they also use $\KS\phi$ and $\KS\omega$
(and assign an asymmetric systematic error due to $\CP$ even pollution 
in these $\CP$ odd channels~\cite{ref:cp_uta:cus:babar:dstarcpk}).
The results and averages are given in Table~\ref{tab:cp_uta:cus:glw}.

\begin{table}
  \begin{center}
    \caption{
      Averages from GLW analyses of $b \to c\bar{u}d / u\bar{c}d$ modes.
    }
    \vspace{0.2cm}
% make this tabular (not tabular*) and resize down to \textwidth
% change @{\extracolsep{\fill}} to @{\extracolsep{2mm}}
    \resizebox{\textwidth}{!}{
      \setlength{\tabcolsep}{0.0pc}
      \begin{tabular}{@{\extracolsep{2mm}}lrcccc} \hline 
        \multicolumn{2}{l}{Experiment} & 
        $A_{\CP+}$ & $A_{\CP-}$ & $R_{\CP+}$ & $R_{\CP-}$ \\
        \hline
        \multicolumn{6}{c}{$D_{\CP} K^-$} \\
        \babar & \cite{ref:cp_uta:cus:babar:dcpk} &
        $\ph{-} 0.40 \pm 0.15 \pm 0.08$ & $\ph{-} 0.21 \pm 0.17 \pm 0.07$ & 
        $ 0.87 \pm 0.14 \pm 0.06$ & $ 0.80 \pm 0.14 \pm 0.08$ \\
        \belle & \cite{ref:cp_uta:cus:belle:dcpk} &
        $\ph{-} 0.07 \pm 0.14 \pm 0.06$ & $-0.11 \pm 0.14 \pm 0.05$ & 
        $ 0.98 \pm 0.18 \pm 0.10$ & $ 1.29 \pm 0.16 \pm 0.08$ \\
        \multicolumn{2}{l}{\bf Average} & 
        $ 0.22 \pm 0.11$ & $ 0.02 \pm 0.12$ & $ 0.91 \pm 0.12$ & $ 1.02 \pm 0.12$ \\
        \hline
        \multicolumn{6}{c}{$\Dstar_{\CP} K^-$} \\
        \babar & \cite{ref:cp_uta:cus:babar:dstarcpk} &
        $-0.02 \pm 0.24 \pm 0.05$ & $$                    & 
        $ 1.09 \pm 0.26 \, ^{+0.10}_{-0.08}$ & $$ \\
        \belle & \cite{ref:cp_uta:cus:belle:dcpk} &
        $-0.27 \pm 0.25 \pm 0.04$ & $ 0.26 \pm 0.26 \pm 0.03$ & 
        $ 1.43 \pm 0.28 \pm 0.06$ & $ 0.94 \pm 0.28 \pm 0.06$ \\
        \multicolumn{2}{l}{\bf Average} & 
        $-0.14 \pm 0.18$ & $ 0.26 \pm 0.26$ & $ 1.25 \pm 0.20$ & $ 0.94 \pm 0.29$ \\
        \hline
        \multicolumn{6}{c}{$D_{\CP} K^{*-}$} \\
        \babar & \cite{ref:cp_uta:cus:babar:dcpkstar} &
        $-0.09 \pm 0.20 \pm 0.06$ & $-0.33 \pm 0.34 \pm 0.10 \, ^{+0.00}_{-0.06}$ & 
        $ 1.77 \pm 0.37 \pm 0.12$ & $ 0.76 \pm 0.29 \pm 0.06 \, ^{+0.04}_{-0.14}$ \\
        \belle & \cite{ref:cp_uta:cus:belle:dcpkstar} &
        $-0.02 \pm 0.33 \pm 0.07$ & $ 0.19 \pm 0.50 \pm 0.04$ & $$ & $$ \\
        \multicolumn{2}{l}{\bf Average} & 
        $-0.07 \pm 0.18$ & $-0.16 \pm 0.29$ & $$ & $$ \\
        \hline
      \end{tabular}
    }
    \label{tab:cp_uta:cus:glw}
  \end{center}
\end{table}

For ADS analysis, both \babar\ and \belle\ have studied
the mode $\Bmp \to D\Kmp$,
and \babar\ has also analyzed the $\Bmp \to \Dstar\Kmp$ mode
($\Dstar \to D\pi^0$ and $\Dstar \to D\gamma$ are studied separately).
In all cases the suppressed decay $D \to K^+\pi^-$ has been used.
The results and averages are given in Table~\ref{tab:cp_uta:cus:ads}.
Note that although no clear signals for these modes have yet been seen,
the central values are given in Table~\ref{tab:cp_uta:cus:ads}.
In $\Bm \to \Dstar\Km$ decays there is an effective shift of $\pi$
in the strong phase difference between the cases that the $\Dstar$ is 
reconstructed as $D\pi^0$ and $D\gamma$~\cite{ref:cp_uta:cus:bg}.
As a consequence, the different $D^*$ decay modes are treated separately.

\begin{table}
  \begin{center} 
    \caption{
      Averages from ADS analyses of $b \to c\bar{u}d / u\bar{c}d$ modes.
    }
    \vspace{0.2cm}
    \setlength{\tabcolsep}{0.0pc}
    \begin{tabular*}{\textwidth}{@{\extracolsep{\fill}}lrcc} \hline 
      \multicolumn{2}{l}{Experiment} & $A_{ADS}$ & $R_{ADS}$ \\
      \hline
      \multicolumn{4}{c}{$D K^-$, $D \to K^+\pi^-$} \\
      \babar & \cite{ref:cp_uta:cus:babar:dk_ads} &
      & $ 0.013 \, ^{+0.011}_{-0.009}$ \\ 
      \belle & \cite{ref:cp_uta:cus:belle:dk_ads} &
      $0.49 \, ^{+0.53}_{-0.46} \pm 0.06$ & $0.028 \, ^{+0.015}_{-0.014} \pm 0.010$ \\
      \multicolumn{2}{l}{\bf Average} & 
      $0.49 \, ^{+0.53}_{-0.46}$ & $0.017 \pm 0.009$ \\
      \hline
      \multicolumn{4}{c}{$\Dstar K^-$, $\Dstar \to D\pi^0$, $D \to K^+\pi^-$} \\
      \babar & \cite{ref:cp_uta:cus:babar:dk_ads} &
      & $-0.001 \, ^{+0.010}_{-0.006}$ \\
      \hline
      \multicolumn{4}{c}{$\Dstar K^-$, $\Dstar \to D\gamma$, $D \to K^+\pi^-$} \\
      \babar & \cite{ref:cp_uta:cus:babar:dk_ads} &
      & $ 0.011 \, ^{+0.019}_{-0.013}$ \\
      \hline 
    \end{tabular*}
    \label{tab:cp_uta:cus:ads}
  \end{center}
\end{table}

For the Dalitz plot analysis, both \babar\ and \belle\ have studied
the mode $\Bmp \to D\Kmp$.
Both have also studied the mode $\Bmp \to \Dstar\Kmp$;
\belle\ has used only $\Dstar \to D\pi^0$,
while \babar\ has used both $\Dstar$ decay modes and 
taken the effective shift in the strong phase difference into account.
In all cases the decay $D \to \KS\pi^+\pi^-$ has been used.
The results are given in Table~\ref{tab:cp_uta:cus:dalitz}.
Since the measured values of $r_B$ are positive definite, 
and since the error on $\gamma$ depends on the value of $r_B$, 
some statistical treatment is necessary to correct for bias. 
\belle\ has used a frequentist treatment, 
while \babar\ used a Bayesian approach. 
At present, we make no attempt to average the results.

\begin{table}
  \begin{center} 
    \caption{
      Averages from Dalitz plot analyses of $b \to c\bar{u}d / u\bar{c}d$ modes.
    }
    \vspace{0.2cm}
    \setlength{\tabcolsep}{0.0pc}
    \begin{tabular*}{\textwidth}{@{\extracolsep{\fill}}lrccc} \hline 
      \multicolumn{2}{l}{Experiment} & $\gamma \ (^\circ)$ & $\delta_B \ (^\circ)$ & $r_B$ \\
      \hline
      \multicolumn{5}{c}{$D K^-$, $D \to \KS \pi^+\pi^-$} \\
      \babar & \cite{ref:cp_uta:cus:babar:dk_dalitz} &
      $ 70 \pm 44 \pm 10 \pm 10$ & $ 114 \pm 41 \pm \ph{1}8 \pm 10$ & $<0.19$  @ 90\% CL\\
      \belle & \cite{ref:cp_uta:cus:belle:dk_dalitz} &
      $ 64 \pm 19 \pm 13 \pm 11$ & $ 157 \pm 19 \pm 11 \pm 21$ & $ 0.21 \pm 0.08 \pm 0.03 \pm 0.04$ \\
      \multicolumn{2}{l}{\bf Average} & 
      \multicolumn{3}{c}{\sc in preparation} \\
      \hline
      \multicolumn{5}{c}{$\Dstar K^-$, $\Dstar \to D\pi^0$ or $D\gamma$, $D \to \KS \pi^+\pi^-$} \\
      \babar & \cite{ref:cp_uta:cus:babar:dk_dalitz} &
      $ 73 \pm 35 \pm \ph{1}8 \pm 10$ & $ 303 \pm 34 \pm 14 \pm 10$ & $0.16 \, ^{+0.07}_{-0.08} \pm 0.04 \pm 0.02$ \\
      \belle & \cite{ref:cp_uta:cus:belle:dk_dalitz} &
      $ 75 \pm 57 \pm 11 \pm 11$ & $ 321 \pm 57 \pm 11 \pm 21$ & $ 0.12 \, ^{+0.16}_{-0.11} \pm 0.02 \pm 0.04$ \\
      \multicolumn{2}{l}{\bf Average} & 
      \multicolumn{3}{c}{\sc in preparation} \\
      \hline
      \multicolumn{5}{c}{$D K^-$ and $\Dstar K^-$ combined} \\
      \babar & \cite{ref:cp_uta:cus:babar:dk_dalitz} &
      $ 70 \pm 26 \pm 10 \pm 10$ \\
      \belle & \cite{ref:cp_uta:cus:belle:dk_dalitz} &
      $ 68 \, ^{+14}_{-15} \pm 13 \pm 11$ \\
      \multicolumn{2}{l}{\bf Average} & 
      \multicolumn{3}{c}{\sc in preparation} \\
      \hline 
    \end{tabular*}
    \label{tab:cp_uta:cus:dalitz}
  \end{center}
\end{table}

% \end{document}

\section{Averages of charmless \B-decay branching fractions and
               their asymmetries }

\label{sec:rare}
The aim of this section is to provide the branching fractions and
the partial rate asymmetries ($A_{CP}$) of rare $B$ decays. The asymmetry is
defined as $A_{CP} = \frac{N_{\Bbar} -N_B}{N_{\Bbar} +N_B}$, where $N_{\Bbar}$ 
and $N_B$ are number of $\Bzb/\Bm$ and $\Bz/\Bp$, respectively.
Four different $B$ decay categories are     
considered: charmless mesonic, baryonic, radiative and leptonic. Rare mesonic 
decays with charm are not in our scope but results of charmful baryonic decays
are included. Measurements supported with  written documents are accepted in 
our 
the averages; written documents could be journal papers, 
conference contributed papers, preprints or conference proceedings.  
Results from  $A_{CP}$ measurements  obtained from time dependent analyses 
are  listed and described in Sec.~\ref{sec:cp_uta}.

So far all branching fractions assume equal production of charged and
neutral $B$ pairs.  The best measurements to date show that this is
still a good approximation (see Sec.~\ref{sec:bfraction}).
For branching fractions, we provide either averages or the most stringent
90\% confidence level upper limits.  If one or more experiments have
measurements with $>$4$\sigma$ for a decay channel, 
all available central values
for that channel are used in the averaging.  We also give central values
and errors for cases where the significance of the average value is at
least $3 \sigma$, even if no single measurement is above $4 \sigma$.
For $A_{CP}$ we provide averages in all cases.

Our averaging is performed by maximizing the likelihood,
\begin{eqnarray}
    {\mathcal L} = \prod_i {\mathcal P}_i(x) , 
\end{eqnarray}
where ${\mathcal P_i}$ is the probability density function (PDF) of the
$i$th  measurement, and $x$ is the branching fraction or $A_{CP}$.
The PDF is modeled by an asymmetric Gaussian function with the measured
central value as its mean and the quadratic sum of the statistical
and systematic errors as the standard deviations. The experimental
uncertainties are considered to be uncorrelated with each other when the 
averaging is performed. No error scaling is applied when the fit $\chi^2$ is 
greater than 1 since we believe that tends to overestimate the errors
except in cases of extreme disagreement (we have no such cases).

At present, we have measurements of 230 $B$ decay modes and asymmetry
measurements for 35 of these decays.  These results are reported in 112
separate papers.  Because the number of references is so large, we do
not include them with the tables shown here but the full set of
references is available quickly from active gifs at the ICHEP04 link on 
the rare web page: {\tt http://www.slac.stanford.edu/xorg/hfag/rare/index.html}

\subsection{Mesonic charmless decays}

\begin{table}
\begin{center}
%\vspace*{-1.6cm}
\caption{
$B^+$ branching fractions 
(in units of $10^{-6}$). Upper limits are at 90\% CL.
Values in {\red red} ({\blue blue}) are new {\red published} 
({\blue preliminary}) result since PDG2004  [as of August 25th, 2004].
}
\scriptsize
\begin{tabular}{|lccccccc|} 
\sglinespb
RPP\#   & Mode & PDG2004 Avg. & BABAR & Belle & CLEO & CDF & New Avg. \\
\sglinespb
%SCRIPTMARKER -- DO NOT ALTER THIS LINE
117                                               & %         RPP\#
$K^0\pi^+$                                        & %         Mode
$18.8\pm 2.1$                                     & %         PDG2004 Avg.
\blue{$\err{26.0}{1.3}{1.0}$}                     & %   [1]  BABAR
\red{$\err{22.0}{1.9}{1.1}$}                      & %   [31]  Belle
$\aerrsy{18.8}{3.7}{3.3}{2.1}{1.8}$         & %   [61]  CLEO
\nodata                                           & %   [0]  CDF
$24.1 \pm 1.3$                                    \\

118                                               & %         RPP\#
$K^+\pi^0$                                        & %         Mode
$12.9\pm 1.2$                                     & %         PDG2004 Avg.
\blue{$\err{12.0}{0.7}{0.6}$}                     & %   [13]  BABAR
\red{$\berr{12.0}{1.3}{1.3}{0.9}$}                & %   [31]  Belle
$\aerrsy{12.9}{2.4}{2.2}{1.2}{1.1}$               & %   [61]  CLEO
\nodata                                           & %   [0]  CDF
$12.1 \pm 0.8$                                    \\

119                                               & %         RPP\#
$\etapr K^+$                                      & %         Mode
$78\pm 5$                                         & %         PDG2004 Avg.
$\err{76.9}{3.5}{4.4}$                            & %   [3]  BABAR
\blue{$\err{78}{6}{9}$}                           & %   [32]  Belle
$\aerr{80}{10}{9}{7}$                             & %   [62]  CLEO
\nodata                                           & %   [0]  CDF
$\cerr{77.6}{4.6}{4.5}$                           \\
\end{tabular}
\end{center}

\vspace{-0.4cm} 
\scriptsize
$\dag$Product BF - daughter BF taken to be 100\%; 
~\S $M_{\phi\phi}<2.85$ GeV/$c^2$
% \hspace{-1.cm} ~\S Non-resonant state is not phase-space like.
\end{table}

\large

\begin{table}
\begin{center}
\caption{
$B^0$ branching fractions (in units of $10^{-6}$).
Upper limits are at 90\% CL.
Values in {\red red} ({\blue blue}) are new {\red published} 
({\blue preliminary}) result since PDG2004  [as of August 25th, 2004].
}
%\hspace*{-1.2cm}
\scriptsize
\begin{tabular}{|lccccccc|} 
\sglinespb
RPP\#   & Mode & PDG2004 Avg. & BABAR & Belle & CLEO & CDF & New Avg. \\
\sglinespb
%SCRIPTMARKER -- DO NOT ALTER THIS LINE
123                                               & %         RPP\#
$K^+\pi^-$                                        & %         Mode
$18.5\pm 1.1$                                     & %         PDG2004 Avg.
$\err{17.9}{0.9}{0.7}$                            & %   [11]  BABAR
\red{$\err{18.5}{1.0}{0.7}$}                      & %   [31]  Belle
$\aerrsy{18.0}{2.3}{2.1}{1.2}{0.9}$               & %   [61]  CLEO
\nodata                                           & %   [0]  CDF
$18.2 \pm 0.8$                                    \\

124                                               & %         RPP\#
$K^0\pi^0$                                        & %         Mode
$\cerr{9.5}{2.1}{1.9}$                           & %         PDG2004 Avg.
\blue{$\err{11.4}{0.9}{0.6}$}                     & %   [6]  BABAR
\red{$\berr{11.7}{2.3}{1.2}{1.3}$}                & %   [31]  Belle
$\aerrsy{12.8}{4.0}{3.3}{1.7}{1.4}$               & %   [61]  CLEO
\nodata                                           & %   [0]  CDF
$11.5 \pm 1.0$                                    \\

125                                               & %         RPP\#
$\etapr K^0$                                      & %         Mode
$63\pm 7$                                         & %         PDG2004 Avg.
$\err{60.6}{5.6}{4.6}$                            & %   [3]  BABAR
\blue{$\berr{68}{10}{9}{8}$}                      & %   [32]  Belle
$\aerr{89}{18}{16}{9}$                            & %   [62]  CLEO
\nodata                                           & %   [0]  CDF
$\cerr{65.2}{6.0}{5.9}$                           \\
\end{tabular}
\end{center}
\vspace{-0.4cm}
\scriptsize
\hspace{0.1cm} $\dag$Product BF - daughter BF taken to be 100\%
\end{table}

 \clearpage
{\vspace{-2cm}
\subsection{Radiative and leptonic decays}}

\begin{table}[!htbp]
\begin{center}
\caption{
Compilation of $B^+$ semileptonic and radiative branching fractions
(in units of $10^{-6}$). Upper limits are at 90\% CL.
values in {\red red} ({\blue blue}) are new {\red published} 
({\blue preliminary}) result since PDG2004  [as of August 25th, 2004].
}
\end{center}
\scriptsize
\begin{tabular}{|lcccccc|} \sglinespb
RPP\# &Mode & PDG2004 Avg. & BaBar  & Belle & CLEO
 & New Avg.  \\ 
 \sglinespb
%SCRIPTMARKER -- DO NOT ALTER THIS LINE
%13             &
%$e^+\nu\gamma$ &
%$<200$         &
%               &
%               &
%$<200$         &
%$<200$

%14             &
%$\mu^+\nu\gamma$ &
%$<52$          &
%&              &
%$<520$         &
%$<520$

160                                               & %         RPP\#
$K^*(892)^+\gamma$                                & %         Mode
$38\pm 5$                                         & %         PDG2004 Avg.
{\blue $38.7\pm 2.8 \pm 2.6$}                     & %   [1]  BaBar
{\red $42.5\pm 3.1\pm 2.4$}                       & %   [22]  Belle
$\aerr{37.6}{8.9}{8.3}{2.8} $                     & %   [42]  CLEO
$40.3 \pm 2.6$                                    \\

%         New Avg.

161                                               & %         RPP\#
$K_1(1270)^+\gamma$                               & %         Mode
$<99$                                             & %         PDG2004 Avg.
\nodata                                           & %   [0]  BaBar
{\blue $\err{42.8}{9.4}{4.3}$}                    & %   [32]  Belle
\nodata                                           & %   [0]  CLEO
$42.8 \pm 10.3$                                   \\

%         New Avg.

162                                               & %         RPP\#
$K^+\phi\gamma$                                   & %         Mode
$3.4\pm 1.0$                                      & %         PDG2004 Avg.
\nodata                                           & %   [0]  BaBar
$\err{3.4}{0.9}{0.4}$                             & %   [29]  Belle
\nodata                                           & %   [0]  CLEO
$3.4 \pm 1.0$                                     \\
\end{tabular}
~\S~$M_{K\pi\pi} < 2.4$ GeV/$c^2$ 
~\ddag~Central values are not significant.  
\end{table}

\begin{table}
\begin{center}
\caption{
Compilation of $B^0$ semileptonic and radiative branching fractions
(in units of $10^{-6}$).  Upper limits are at 90\% CL.
values in {\red red} ({\blue blue}) are new {\red published} 
({\blue preliminary}) result since PDG2004  [as of August 25th, 2004].
}
\end{center}
\hspace*{-0.0cm} 
\scriptsize
\begin{tabular}{|lcccccc|} \sglinespb
RPP\# &Mode & PDG2004 Avg. & BaBar  & Belle & CLEO
 & New Avg.  \\ \sglinespb
%SCRIPTMARKER -- DO NOT ALTER THIS LINE
162                                               & %         RPP\#
$K^*(892)^0\gamma$                                & %         Mode
$43\pm 4$                                         & %         PDG2004 Avg.
{\blue $39.2\pm 2.0\pm 2.4$}                      & %   [1]  BaBar
$40.1\pm 2.1\pm 1.7$                       & %   [22]  Belle
$\aerr{45.5}{7.2}{6.8}{3.4}$                      & %   [42]  CLEO
$40.1 \pm 2.0$                                    \\

%         New Avg.

163                                               & %         RPP\#
$K^0\phi\gamma$                                   & %         Mode
$<8.3$                                            & %         PDG2004 Avg.
\nodata                                           & %   [0]  BaBar
$<8.3$                                            & %   [29]  Belle
\nodata                                           & %   [0]  CLEO
{$<8.3$}                                          \\

%         New Avg.

164                                               & %         RPP\#
$K^+\pi^-\gamma$~\dag                             & %         Mode
$4.6\pm 1.4$                                      & %         PDG2004 Avg.
\nodata                                           & %   [0]  BaBar
$\aerrsy{4.6}{1.3}{1.2}{0.5}{0.7}$                & %   [23]  Belle
\nodata                                           & %   [0]  CLEO
$4.6 \pm 1.4$                                     \\
\end{tabular}
\break
\hspace*{0.5cm}~\dag~$1.25$ GeV/$c^2 < M_{K\pi} < 1.6$ GeV/$c^2$ 
~\ddag~ Central values are not significant.
\end{table}

\begin{table}
\begin{center}
\caption{
Compilation of $B$ semileptonic and radiative branching fractions
(in units of $10^{-6}$). Upper limits are at 90\% CL.
values in {\red red} ({\blue blue}) are new {\red published} 
({\blue preliminary}) result since PDG2004  [as of August 25th, 2004].
}
\end{center}
\hspace*{0.7cm}
\scriptsize
\begin{tabular}{|lcccccc|} \sglinespb
RPP\# &Mode & PDG2004 Avg. & BaBar  & Belle & CLEO
 & New Avg.  \\ \sglinespb
%SCRIPTMARKER -- DO NOT ALTER THIS LINE
~60                                                & %         RPP\#
$K_3^*(1780)\gamma $                              & %         Mode
$<3000$                                           & %         PDG2004 Avg.
\nodata                                           & %   [0]  BaBar
{\blue $<34$}                                     & %   [33]  Belle
\nodata                                           & %   [0]  CLEO
{ $<34$}                                          \\

%         New Avg.

~67                                                & %         RPP\#
$s\gamma$                                         & %         Mode
$330\pm 40$                                       & %         PDG2004 Avg.
{\blue $\berr{388}{36}{57}{46}$}                  & %   [6]  BaBar
$\berrsyt{355}{32}{30}{31}{11}{7}$                & %   [27]  Belle
$\berr{321}{43}{32}{29}$                          & %   [47]  CLEO
$\cerr{352}{30}{28}$                              \\

%         New Avg.

$~-$                                               & %         RPP\#
$s\gamma$ with baryons                            & %         Mode
New                                               & %         PDG2004 Avg.
\nodata                                           & %   [0]  BaBar
\nodata                                           & %   [0]  Belle
\red $<38$~\dag                                   & %   [50]  CLEO
 $<38$~\dag                                       \\
\end{tabular}
\break
\hspace*{0.5cm}~\dag~$E_\gamma > 2.0$ GeV; ~\ddag $M(\ell^+\ell^-)>0.2$ GeV/$c^2$
\end{table}

\begin{table}
\begin{center}
\caption{
Compilation of $B$  leptonic  branching fractions
(in units of $10^{-6}$). Upper limits are at 90\% CL.
values in {\red red} ({\blue blue}) are new {\red published} 
({\blue preliminary}) result since PDG2004  [as of August 25th, 2004].
}
\vspace{0.3cm}
\begin{tabular}{|lcccccccc|} \sglinespb
RPP\# &Mode & PDG2004 Avg. & BaBar  & Belle & CLEO &CDF & D0
 & New Avg.  \\ \sglinespb
%SCRIPTMARKER -- DO NOT ALTER THIS LINE
~12                                                & %         RPP\#
$e^+\nu$                                          & %         Mode
$<15$                                             & %         PDG2004 Avg.
\nodata                                           & %   [0]  BaBar
{\blue $<5.4$}                                    & %   [21]  Belle
$<15$                                             & %   [41]  CLEO
\nodata                                           & %   [0]  CDF
\nodata                                           & %   [0]  D0
{ $<5.4$}                                         \\

%         New Avg.

~13                                                & %         RPP\#
$\mu^+\nu$                                        & %         Mode
$<21$                                             & %         PDG2004 Avg.
{\red $<6.6$}                                     & %   [10]  BaBar
{\blue $<2.0$}                                    & %   [36]  Belle
$<21$                                             & %   [41]  CLEO
\nodata                                           & %   [0]  CDF
\nodata                                           & %   [0]  D0
{ $<2.0$}                                         \\

%         New Avg.

~14                                                & %         RPP\#
$\tau^+\nu$                                       & %         Mode
$<570$                                            & %         PDG2004 Avg.
\blue $<330$                                      & %   [9]  BaBar
\blue $<290$                                      & %   [37]  Belle
$<840$                                            & %   [0]  CLEO
\nodata                                           & %   [0]  CDF
\nodata                                           & %   [0]  D0
 $<290$                                           \\
\end{tabular}
\end{center}
\end{table}
%\clearpage

\begin{table}
\begin{center}
\caption{
Compilation of $B_s$  leptonic  branching fractions
(in units of $10^{-6}$). Upper limits are at 90\% CL.
values in {\red red} ({\blue blue}) are new {\red published} 
({\blue preliminary}) result since PDG2004  [as of August 25th, 2004].
}
\vskip 0.25cm
\begin{tabular}{|lccccc|} \sglinespb
RPP\# &Mode & PDG2004 Avg. &  CDF & D0
 & New Avg.  \\ \sglinespb
%SCRIPTMARKER -- DO NOT ALTER THIS LINE
$-$                                               & %         RPP\#
$\mu^+\mu^-$                                      & %         Mode
New                                               & %         PDG2004 Avg.
\blue$<0.58 $                                     & %   [51]  CDF
\blue$<0.41$                                      & %   [54]  D0
$<0.41 $                                          \\

%         New Avg.

%SCRIPTMARKER -- DO NOT ALTER THIS LINE
%\\
\sglinespt
\end{tabular}
\end{center}
\end{table}
\clearpage

\subsection{Baryonic decays}

\begin{table}[!htbp]
\begin{center}
\caption{
Compilation of $B^+$ baryonic branching fractions (in units of $10^{-6}$).
Upper limits are at 90\% CL.
values in {\red red} ({\blue blue}) are new {\red published} 
({\blue preliminary}) result since PDG2004  [as of August 25th, 2004].
}
\end{center}
% \hspace*{0.5cm} \vspace*{-0.5cm}
\vspace*{-0.5cm}
\scriptsize
\begin{tabular}{|lcccccc|} 
\sgline
RPP\#   & Mode & PDG2004 Avg. & BABAR & Belle & CLEO & New Avg. \\
\sgline
%SCRIPTMARKER -- DO NOT ALTER THIS LINE
%-----------------------------------------------------
% ppbar pi
%-----------------------------------------------------
201                                               & %         RPP\#
$p\overline p \pi^+$                              & %         Mode
$<3.7$                                            & %         PDG2004 Avg.
\nodata                                           & %   [0]  BABAR
{\red $\aerr{3.06}{0.73}{0.62}{0.37}$}            & %   [2]  Belle
$<160$                                            & %   [8]  CLEO
$\cerr{3.06}{0.82}{0.72}$                         \\

% Our World Average

%-----------------------------------------------------
% ppK  {\blue$\err{6.7}{0.9}{0.6}$}  previous babar number
%-----------------------------------------------------
204                                               & %         RPP\#
$p\overline p K^+$                                & %         Mode
$\cerr{4.3}{1.2}{1.0}$                            & %         PDG2004 Avg.
{\blue$\err{6.7}{0.9}{0.6}$}                      & %   [12]  BABAR
{\red$\aerr{5.66}{0.67}{0.57}{0.62}$}             & %   [2]  Belle
\nodata                                           & %   [0]  CLEO
$\cerr{6.08}{0.71}{0.68}$                         \\

%-----------------------------------------------------
% ppK^*
%-----------------------------------------------------
$~-$                                              & %         RPP\#
$p\overline p K^{*+}$                             & %         Mode
New                                               & %         PDG2004 Avg.
\nodata                                           & %   [0]  BABAR
{\red$\aerrsy{10.31}{3.62}{2.77}{1.34}{1.65}$}    & %   [2]  Belle
\nodata                                           & %   [0]  CLEO
$\cerr{10.31}{3.86}{3.22}$                        \\
\hline
\end{tabular}
\end{table}

\begin{table}[!htbp]
\begin{center}
\caption{
Compilation of $B^0$ baryonic branching fractions
(in units of $10^{-6}$). Upper limits are at 90\% CL.
values in {\red red} ({\blue blue}) are new {\red published} 
({\blue preliminary}) result since PDG2004  [as of August 25th, 2004].
}
\vskip 0.25cm
\end{center}
% \hspace*{1.0cm} \vspace*{-1.0cm}
\vspace*{-1.0cm}
\scriptsize
\begin{tabular}{|lcccccc|} 
\sgline
RPP\#   & Mode & PDG2004 Avg. & BABAR & Belle & CLEO & New Avg. \\
\sgline
%SCRIPTMARKER -- DO NOT ALTER THIS LINE
%-----------------------------------------------------
% p pbar
%-----------------------------------------------------
212                                               & %         RPP\#
\ppbar                                            & %         Mode
$< 1.2$                                           & %         PDG2004 Avg.
{\red $<0.27$}                                    & %   [10]  BABAR
{\blue $<0.50$}                                   & %   [1]  Belle
 $<1.4$                                           & %   [6]  CLEO
{ $<0.27$}                                        \\

% Our World Average

%-----------------------------------------------------
% p p K^0
%-----------------------------------------------------
214                                               & %         RPP\#
$p \overline{p} K^0$                              & %         Mode
$<7.2$                                            & %         PDG2004 Avg.
\nodata                                           & %   [0]  BABAR
{\red $\aerr{1.88}{1.06}{0.60}{0.23}$}            & %   [2]  Belle
\nodata                                           & %   [0]  CLEO
$\cerr{1.88}{1.08}{0.64}$                         \\

% Our World Average

%-----------------------------------------------------
% p p K^*0
%-----------------------------------------------------
$~-$                                              & %         RPP\#
$p \overline{p} K^{*0}$                           & %         Mode
New                                               & %         PDG2004 Avg.
\nodata                                           & %   [0]  BABAR
{\red $<7.6$}                                     & %   [2]  Belle
\nodata                                           & %   [0]  CLEO
{ $<7.6$}                                         \\
\hline

\end{tabular}
\end{table}

%\vspace*{-1.0cm}
\subsection{Charge asymmetries}

\begin{sidewaystable}[!htbp]
\parbox{8.2in}{\caption{
Compilation of charmless hadronic $CP$ asymmetries for charged $B$ decays.
Values in {\red red} ({\blue blue}) are %\linebreak
new {\red published} 
({\blue preliminary}) result since PDG2004  [as of August 25th, 2004].
}}

\vspace*{ 0.5cm}
\scriptsize
\begin{tabular}{|lccccccc|} 
\hline
RPP\#   & Mode & PDG2004 Avg. & BABAR & Belle & CLEO & CDF & New Avg. \\
\sglinespb
%SCRIPTMARKER -- DO NOT ALTER THIS LINE

%------------------------------------------------------------------
%                             K^0\pi^+
%------------------------------------------------------------------
117                                               & %         RPP\#
$K^0\pi^+$                                        & %         Mode
$0.03\pm 0.08$                                   & %         PDG2004 Avg.
{\blue$\err{-0.087}{0.046}{0.010}$}               & %   [1]  BABAR
{\blue$\err{0.05}{0.05}{0.01}$}                   & %   [25]  Belle
$\err{0.18}{0.24}{0.02}$                          & %   [41]  CLEO
\nodata                                           & %   [0]  CDF
$-0.020 \pm 0.034$                                \\

%------------------------------------------------------------------
%                             K^+\pi^0
%------------------------------------------------------------------
118                                               & %         RPP\#
$K^+\pi^0$                                        & %         Mode
$-0.10\pm 0.08$                                   & %         PDG2004 Avg.
{\blue $\err{0.06}{0.06}{0.01}$}                  & %   [12]  BABAR
{\blue $\err{0.04}{0.05}{0.02}$}                  & %   [35]  Belle
$\err{-0.29}{0.23}{0.02}$                         & %   [41]  CLEO
\nodata                                           & %   [0]  CDF
$0.04 \pm 0.04$                                   \\
\end{tabular}
\end{sidewaystable}

\begin{sidewaystable}[!htbp]
\begin{center}
\parbox{7.5in}{\caption{
Compilation of charmless hadronic $CP$ asymmetries for $B^\pm/B^0$ admixtures.
Values in {\red red} ({\blue blue}) are new {\red published} 
({\blue preliminary}) result since PDG2004  [as of August 25th, 2004].
}}

\vspace*{0.5cm}   %- { 0.4cm}
\scriptsize
%\hspace*{-0.4cm}
\begin{tabular}{|lccccccc|}
\hline
RPP\#    & Mode & PDG2004 Avg. & BABAR & Belle & CLEO & CDF & New Avg. \\
\sglinespb
%SCRIPTMARKER -- DO NOT ALTER THIS LINE
~56                                               & %         RPP\#
$K^*\gamma$                                       & %         Mode
$-0.01\pm 0.07$                                   & %         PDG2004 Avg.
{\blue $-0.013\pm 0.036\pm 0.010$}                & %   [3]  BABAR
{\red $-0.015\pm 0.044\pm 0.012$}                 & %   [26]  Belle
$0.08\pm 0.13\pm 0.03$                            & %   [42]  CLEO
\nodata                                           & %   [0]  CDF
$-0.010 \pm 0.028$                                \\

%         New Avg.

%

~67                                               & %         RPP\#
$s\gamma$                                         & %         Mode
$\err{-0.079}{0.108}{0.022}$                      & %         PDG2004 Avg.
\red$\err{0.025}{0.050}{0.015}$                   & %   [2]  BABAR
\red$\err{0.002}{0.050}{0.026}$                   & %   [31]  Belle
$\err{-0.079}{0.108}{0.022}$                      & %   [44]  CLEO
\nodata                                           & %   [0]  CDF
$0.005 \pm 0.036$                                 \\

%         New Avg.

103                                               & %         RPP\#
$s\ell\ell$                                       & %         Mode
New                                               & %         PDG2004 Avg.
\red$\err{-0.22}{0.26}{0.02}$                     & %   [14]  BABAR
\nodata                                           & %   [0]  Belle
\nodata                                           & %   [0]  CLEO
\nodata                                           & %   [0]  CDF
$-0.22 \pm 0.26$                                  \\

% New Avg.

%SCRIPTMARKER -- DO NOT ALTER THIS LINE
\hline
\end{tabular}
\end{center}

\vskip -1cm
\begin{center}
\parbox{8.5in}{\caption{
Compilation of charmless hadronic $CP$ asymmetries for neutral $B$ decays.
Values in {\red red} ({\blue blue}) are new {\red published} 
({\blue preliminary}) result since PDG2004  [as of August 25th, 2004].
}}

\vspace*{0.5cm}
\scriptsize
\begin{tabular}{|lccccccc|} 
%\begin{tabular}{lccccc} 
\sgline
RPP\#   & Mode & PDG2004 Avg. & BABAR & Belle & CLEO & CDF & New Avg. \\
\sglinespb
%SCRIPTMARKER -- DO NOT ALTER THIS LINE

%------------------------------------------------------------------
%                             K^+\pi^-
%------------------------------------------------------------------
123                                               & %         RPP\#
$K^+\pi^-$                                        & %         Mode
$-0.09\pm 0.04$                                   & %         PDG2004 Avg.
{\red $\err{-0.133}{0.030}{0.009}$}               & %   [6]  BABAR
{\blue $\err{-0.101}{0.025}{0.005}$}              & %   [35]  Belle
$\err{-0.04}{0.16}{0.02}$                         & %   [41]  CLEO
{\blue $\err{-0.04}{0.08}{0.006}$}                & %   [46]  CDF
$-0.109 \pm 0.019$                                \\

%------------------------------------------------------------------
%                             K^0\pi^0
%------------------------------------------------------------------
124                                             & %         RPP\#
$K^0\pi^0$ $\dag$                                 & %         Mode
\nodata                                           & %         PDG2004 Avg.
\nodata                                           & %   [0]  BABAR
\nodata                                           & %   [0]  Belle
\nodata                                           & %   [0]  CLEO
\nodata                                           & %   [0]  CDF
\nodata                                           \\
\end{tabular}
%\vskip 0.5cm
\end{center}

\hspace{0.8cm}
\parbox{8.5in}{\dag~Measurements of time-dependent $CP$ asymmetries are listed on 
the Unitarity Triangle home page. (http://www.slac.stanford.edu/xorg/hfag/triangle/index.html) }
%{\normalsize All entries are time-integrated except where 
%indicated by labelled coefficients $S_{xy}$ and $C_{xy}$.
%Note 
%$$S_{xy}\equiv
%\frac{2Im(\lambda)}{1+|\lambda|^2}$$
%and
%$$C_{xy}\equiv
%\frac{1-|\lambda|^2}{1+|\lambda|^2}
%\equiv -A_{xy}$$
%where $\lambda = {\cal A}(\bar B\to f_{CP})/{\cal A}(B\to f_{CP})$.}

\end{sidewaystable}

\newpage \normalsize
%\section{ Summary }
%\documentclass[12pt]{article}
%\begin{document}

\section{Summary }
\labs{summary}

 This article provides the updated world averages for 
$b$-hadron properties as of 2004 summer conferences (ICHEP04 and FPCP04).
A brief summary of the results described in Secs. 
\ref{sec:life_mix}-\ref{sec:rare} is given in 
Table~\ref{tab_summary}.

\begin{table}
\caption{ Brief summary of the world averages as of 2004 summer conferences.}
\label{tab_summary}
\begin{center}
\begin{tabular}{|l|c|}
\hline
 {\bf\boldmath \b-hadron lifetimes} &   \\
 ~~$\tau(\Bd)$  & \HFAGtauBd \\
 ~~$\tau(\Bu)$  & \HFAGtauBu \\
 ~~$\tau(\Bs)$  & \HFAGtauBs \\
 ~~$\tau(\Bc)$  & \HFAGtauBc \\
 ~~$\tau(\Lb)$  & \HFAGtauLb \\
\hline
 {\bf\boldmath \b-hadron fractions} &   \\
 ~~$f^{+-}/f^{00}$ in \Ups\ decays  & \HFAGfplusfzeroWorld \\ %\HFAGfplusfzero
 ~~$\fBd=\fBu$ at high energy & \HFAGfBd \\
 ~~\fBs\ at high energy  & \HFAGfBs \\
 ~~\fbb\ at high energy  & \HFAGfbb \\
\hline
 {\bf\boldmath \Bd\ and \Bs\ mixing parameters} &   \\
 ~~\dmd &  \HFAGdmd \\
 ~~$|q/p|_{\particle{d}}$ & \HFAGqp  \\
 ~~\dms  &  $>\HFAGdmslimCL$ \\
%-remove from table till updated results with CDF comes
%- ~~\DGGs  & $<\HFAGDGGsconslimCL$ \\
\hline
 {\bf\boldmath Semileptonic \B\ decay parameters} &   \\
 ~~${\cal B}(\BzbDstarlnu)$  & $( 5.33 \pm 0.20)\%$ \\
 ~~${\cal B}(\BzbDplnu)$      & $( 2.13 \pm 0.20)\%$ \\
 ~~${\cal B}(\Bb\to  X\ell\nub)$   & $(10.90 \pm 0.23)\%$ \\
 ~~$|V_{\particle{cb}}|\ (\BzbDstarlnu)$   
       & $ [41.4 \pm 1.0({\rm exp}) \pm 1.8({\rm theo})] \times 10^{-3}$ \\
 ~~$|V_{\particle{cb}}|\ (\BzbDplnu)$   
       & $ [40.4 \pm 3.6({\rm exp}) \pm 2.3({\rm theo}) ] \times 10^{-3}$ \\
 ~~$|V_{\particle{ub}}|$ (inclusive)    & $ (4.70 \pm 0.44 ) \times 10^{-3}$ \\
\hline
 {\bf\boldmath $\CP(t)$ and Unitarity Triangle angles } &   \\
 ~~$\stwob(\phi_1)$ (all charmonium) & $0.726 \pm 0.037$ \\
 ~~$\stwob(\phi_1)_{\rm eff}$ (all $b\to s$ penguin) & $0.41 \pm 0.07$ \\
\hline
 {\bf\boldmath Rare \B\ decays} &   \\
 ~~$A_{\CP}(\particle{\Bd\to K^+\pi^-})$ & $-0.109 \pm 0.019$ ($5.7\,\sigma$) \\
\hline
\end{tabular}
\end{center}
\end{table}

The HFAG provided the first averages at 2003 winter conferences
and have given updates at summer and winter conferences, as
well as annual PDG averages.

 The accuracies of $b$-hadron lifetimes and \Bd\ mixing parameters 
have been considerably improved with the asymmetric \B factory
results compared to the previous LEP working group 
averages~\cite{LEPHFS}. 
% However, compared to the previous HFAG
% averages at 2004 winter, no substantial changes happened for this
% update. 
In 2004 summer, the \babar and \belle collaborations reported the
simultaneous measurements of $B$ lifetime and $\Delta m_d$ with 
improved precision using increased data samples.
Because of correlation between lifetime and $\Delta m_d$ which 
complicates the average procedures, these new results are not included
in current averages and there are no substantial changes from
previous averages at 2004 winter.  It is planned to incorporate them
in the next update.
In the previous update, new results of \b-hadron fractions from
CDF are included and they remain the same in this update.  
The average of $f^{+-}/f^{00}$ was newly added to HFAG average items
from the previous update.  With new results from \babar, the accuracy
has been improved to $\sim 3\%$ and the average value is consistent
with 1.
In this update, the average of $|q/p|_d$ is newly added.

For $|V_{\particle{cb}}|$ and $|V_{\particle{ub}}|$ average values,
various new measurements have been available from CLEO, \babar, and Belle.
Accordingly, the statistical uncertainties of averages are significantly 
reduced from LEP working group averages~\cite{LEPHFS}.
Considerable progress has been also made in theoretical side, but the
reduction of uncertainty is somewhat slower in some cases (\eg\
$|V_{\particle{cb}}|$ in exclusive modes).
The determination of $|V_{\particle{cb}}|$ from the inclusive semileptonic
branching fraction in combination with hadronic mass and lepton energy 
moments is under discussion and foreseen in the future update.
% the errors of branching fraction measurements and extraction of 
% $|V_{\particle{cb}}|$ and $|V_{\particle{ub}}|$ are 
% dominated by theoretical uncertainties and no large reduction of errors
% is seen from LEP working group averages~\cite{LEPHFS} in spite of 
% the considerable progresses of the theoretical works.
The average of  $|V_{\particle{ub}}|$ using inclusive 
\particle{b\to u\ell\bar{\nu}} is provided from the previous update.
In this update, a substantial change has occurred with the usage of
the Belle photon spectrum measurement in $b \to s \gamma$ decays.
This is used for the determination of shape function parameters
which enter the $|V_{ub}|$ calculation in inclusive charmless semileptonic
\B\ decays.  The new shape function parameters result in 
substantially smaller error in $|V_{ub}|$ than previous one.

Measurements by \babar\ and \belle 
of the time-dependent $\CP$ violation parameter
$S_{b \to c\bar c s}$ in \B decays to charmonium and a neutral kaon
have established $\CP$ violation in $\B$ decays,
and allow a precise extraction of 
the Unitarity Triangle parameter $\stwob / \sin\! 2\phi_1$.
The $\B$ factories have also provided various measurements of
time-dependent $\CP$ asymmetries in hadronic $b \to s$ penguin decays,
establishing $\CP$ violation in these modes. 
Intriguingly, the measured parameters exhibit deviations 
from the Standard Model expectation.
The significance of this effect depends on the treatment of the
theoretical error.
Results from time-dependent analyses 
with the decays $\Bz \to \pi^+\pi^-, \rho^\pm\pi^\mp$ and $\rho^+\rho^-$ 
allow, via various methods, 
constraints on the Unitarity Triangle angle $\alpha/\phi_2$.
% Details can be found in Sec.~\ref{sec:cp_uta:uud}.
% At present there is no HFAG average for $\alpha/\phi_2$.
Constraints on the third Unitarity Triangle angle $\gamma/\phi_3$
have been obtained by \babar\ and \belle,
using $\Bm \to \DorDstar \Km$ decays with Dalitz plot analysis of
the subsequent $D \to \KS\pi^+\pi^-$ decay.

For the rare \B\ decays, branching fractions and charge asymmetries of many 
new decay modes have been measured recently, mostly by \babar\ and Belle.
Since there are several hundred measurements in the tables in \Sec{rare},
we highlight only the measurement of $A_{\CP}(\particle{\Bd\to K^+\pi^-})$,
which provides the observation of direct \CP violation ($5.7\,\sigma$) in 
in the $B$ meson system.

\section*{ Acknowledgements }

We would like to thank collaborators of \babar, Belle, CDF, CLEO, \dzero,
LEP, and SLD experiments who provided fruitful results on \b-hadron
properties and cooperated with the HFAG for averaging.
These results are thanks to the excellent operations of the 
accelerators and collaborations with experimental groups by the 
accelerator groups of PEP-II, KEKB, CESR, Tevatron, LEP, and SLC.
Some of the averages have been obtained based on the discussions 
between theorists for better understanding and improvement on the
theoretical uncertainties.  
% Among them, we would like to thank:
% M.~Beneke, I.I.~Bigi, Z.~Ligeti, M.~Neubert, and N.~Uraltsev.

%\end{document}

\end{thebibliography}

% \appendix
% \section{any ?}

\end{document}